%% file: ms.tex
\documentclass[useAMS,usenatbib]{mn2e}
\usepackage{graphicx, widetext}
\usepackage{amssymb, amsmath}
\usepackage{mathptm}
\usepackage{color}
\usepackage{lscape}
\usepackage[section]{placeins}
\usepackage{epstopdf}
\usepackage{epsf}
\usepackage{ulem}
\usepackage{hyperref}
\usepackage{array}
\usepackage{multirow}
\usepackage{mydefs}
\begin{document}

\title{Theoretical Systematics of Future Baryon Acoustic Oscillation Surveys}
\author[Ding et al.]
{\parbox{\textwidth}{Zhejie Ding$^{1}$\thanks{E-mail: \texttt{zd585612@ohio.edu}}, Hee-Jong Seo$^{1}$\thanks{E-mail: \texttt{seoh@ohio.edu}}, Zvonimir Vlah$^{2,3,4}$, Yu Feng$^{5}$, Marcel Schmittfull$^{6}$, Florian Beutler$^{7,8}$}\vspace{0.4cm}\\
\parbox{\textwidth}{$^{1}$ Department of Physics and Astronomy, Ohio University, Clippinger Labs, Athens, OH 45701, USA\\
$^{2}$ Stanford Institute for Theoretical Physics and Department of Physics, Stanford University,
Stanford, CA 94306, USA\\
$^{3}$ Kavli Institute for Particle Astrophysics and Cosmology, SLAC and Stanford University,
Menlo Park, CA 94025, USA\\
$^{4}$ Theory Department, CERN, CH-1211 Geneve 23, Switzerland\\
$^{5}$ Berkeley Center for Cosmological Physics, University of California at Berkeley, Berkeley CA 94720, USA\\
$^{6}$
Institute for Advanced Study, Einstein Drive, Princeton, NJ 08540, USA\\
$^{7}$ Institute of Cosmology \& Gravitation, Dennis Sciama Building, University of Portsmouth, Portsmouth, PO1 3FX, UK\\
$^{8}$ Lawrence Berkeley National Lab, 1 Cyclotron Rd, Berkeley CA 94720, USA
}}

\date{\today} 
\maketitle
\label{firstpage}

\begin{abstract}
Future Baryon Acoustic Oscillation surveys aim at observing galaxy clustering over a wide range of redshift and galaxy populations at great precision, reaching tenths of a percent, in order to detect any deviation of dark energy from the $\LCDM$ model.
We utilize a set of paired quasi-\Nb\, FastPM simulations that were designed to mitigate the sample variance effect on the BAO feature and evaluated the BAO systematics as precisely as $\sim 0.01\%$. We report anisotropic BAO scale shifts before and after density field reconstruction in the presence of redshift-space distortions over a wide range of redshift, galaxy/halo biases, and shot noise levels. We test different reconstruction schemes and different smoothing filter scales, and introduce physically-motivated BAO fitting models. For the first time, we derive a Galilean-invariant infrared resummed model for halos in real and redshift space.
We test these models from the perspective of robust BAO measurements and non-BAO information such as growth rate and nonlinear bias. We find that pre-reconstruction BAO scale has moderate fitting-model dependence at the level of $0.1\%-0.2\%$ for matter while the dependence is substantially reduced to less than $0.07\%$ for halos. We find that post-reconstruction BAO shifts are generally reduced to below $0.1\%$ in the presence of galaxy/halo bias and show much smaller fitting model dependence. Different reconstruction conventions can potentially make a much larger difference on the line-of-sight BAO scale, upto $0.3\%$. Meanwhile, the precision (error) of the BAO measurements is quite consistent regardless of the choice of the fitting model or reconstruction convention.
\end{abstract}

\begin{keywords}
cosmological parameters, distance scale, large-scale structure, theory
\end{keywords}

\section{Introduction}
\label{sec:intro}

Baryon acoustic oscillations (BAO) are a feature imprinted in the large scale structure of the universe by the primordial spherical sound waves that propagated in the hot plasma of photons and baryons as a result of an interplay between pressure and gravity. The largest distance that sound wave could propagate before the epoch of recombination is called the sound horizon scale $r_s$ at recombination, which corresponds to  about $110 \hMpc$ today. 

The idea of using Baryon Acoustic Oscillations for probing the expansion history of the Universe is based on the standard ruler test. By observing the apparent size of the BAO feature in the galaxy distributions at various epochs of the Universe, we can estimate the Hubble parameter and the angular diameter distances to different epochs of the Universe which depend on the expansion history of the Universe, thereby testing dark energy \citep[e.g.,][]{Eisenstein_Hu_Tegmark_98}. 
What makes the BAO a robust cosmological standard ruler is that the size of this feature, i.e., the sound horizon scale at recombination $r_s$ is precisely determined from an independent probe, i.e., from the cosmic microwave background data \citep[e.g.,][]{Planck_14,Planck_16}.

In real surveys, multiple nonlinear effects are present, which can affect the standard ruler technique \citep[see, e.g.,][]{Jain94,Meiksin99,Springel05,Angulo05,Seo_Eisenstein_05,White05,Jeong06,Crocce06b,Eisenstein_etal_07a,Huff07,smith07,Seo_etal_08, Angulo08,Crocce08,Matsubara2008,Taka08,Taruya09,Tassev12a}. Among those potential systematic errors, the main source is nonlinear structure growth, which is caused by the bulk flow on large scales due to the gravitational attraction between dark matter halos. The net effect is to smear and shift the location of the BAO peak relative to the BAO position in the linear mode. The smearing makes the feature less distinct, worsening the measurement precision. The shift of the feature introduces a bias on the BAO scale and therefore a bias on the final cosmological parameters in the process of comparing the observed BAO scale from galaxy surveys with that inferred from the CMB data. 

Another source of systematics is redshift-space distortions (RSD). Individual galaxies have peculiar velocities due to their falling into gravitational potential wells, and the line-of-sight velocity component distorts the physical distance of the galaxies along the line of sight derived from the observed redshift. 

With galaxy surveys we do not directly observe the mass/matter distribution but a biased version of it. The BAO feature in the galaxy distribution is found largely consistent with the BAO in the matter distribution with additional smearing and shifts. Extensive studies on such effects have been done in the literature utilizing cosmological simulations and perturbation theories \citep[e.g.,][]{Seo_Eisenstein_05,Sherwin12, vlah2015, Senatore15,
zv16, Blas16, Noda17}.

A breakthrough in BAO science was made when  \citet{Eisenstein_etal_07b} developed a density field reconstruction method that can largely undo the degradation due to the various effects listed above. The simple idea of this scheme is to estimate the gravitational infall and the peculiar velocity field using the observed galaxy density distribution as a tracer for the underlying gravitational potential field (i.e., from the Poisson equation and the continuity equation) and reverse the process. 
This operation returns a reconstructed density field with a BAO feature closer to its original location and strength. That is, the nonlinear effect of smearing and shift is greatly reduced by this process~\citep[e.g.,][]{Seo_etal_08,Seo_etal_10,Schmittfull_etal_15}.
The improvement by the density field reconstruction corresponds to quadrupling the survey size with little extra cost assuming a reasonable redshift/survey condition. This method has become a standard tool for the BAO analysis.\citep[e.g.,][]{Pad2012,Anderson_etal_12,Anderson_etal_14,Kazin2014,Alam2016}.

Since the first convincing detection of BAO from Sloan Digital Sky Survey (SDSS) II \citep{Eisenstein_etal_05} and 2-degree Field Galaxy Redshift Survey (2dFGRS) \citep{Cole_etal_05}, multiple large-scale galaxy surveys have been carried out to improve the detection precision, e.g., 6-degree Field Galaxy Survey (6dFGS) \citep{Beutler_etal_11}, WiggleZ dark energy survey \citep{Blake_etal_11}, SDSS-III Baryon Oscillation Spectroscopic Survey (BOSS) \citep[e.g.,][]{DawsonBOSS,Anderson_etal_14,Aubourg2015}. Recently, the completed SDSS-III Baryon Oscillation Spectroscopic Survey (Data Release 12), as the largest galaxy spectroscopic survey, has achieved a 1 per cent distance measurement in two independent redshift bins \citep[][and supporting papers]{Alam2016}.
As a successor of BOSS, the ongoing extended Baryon Oscillation Spectroscopic Survey \citep[eBOSS;][]{DawsonEBOSS} as a part of SDSS IV~\citep{Blanton2017} is conducting cosmological observations over redshift range $0.8 < z < 2.2$ using the traditional BAO tracer, luminous red galaxies (LRG), as well as using new tracers such as emission line galaxies (ELG), and quasars~\citep{Ata2017}.  

The upcoming future Stage IV ground-based or space-based surveys such as the Dark Energy Spectroscopic Instrument \citep[DESI;][]{Levi2013, DESI_16}, Euclid \citep{Laureijs_2011}, and Wide-Field Infrared Survey Telescope \citep[WFIRST;][]{Spergel_2013a, Spergel_2013b} are going to observe galaxy clustering to higher redshift at greater precision than the previous surveys in order to detect any deviation of dark energy from the $\LCDM$ model. For example, DESI will observe 10 times more targets than those from BOSS by covering a wide redshift range and utilizing a variety of galaxy/quasar populations that are different from those in previous surveys, aiming at distance measurement precision better than $0.3\%$ \citep{DESI_16}. With such small statistical errors, it is critical to control the systematic errors at the level of $\sim 0.1\%$ over the range of redshift and target populations. 

In this paper, we utilize a set of simulations that were designed to mitigate the sample variance effect on the BAO feature, following \citet{Prada2016,Schmittfull_etal_15}, and study the expected systematics on the BAO feature before and after density field reconstruction. The difference of this study from previous analyses is that we are revisiting such analysis with high precision, i.e., only a fraction of 0.1\%, which is necessary to be useful for future surveys, and we are measuring the systematics on the BAO scales along and across the line of sight separately in the presence of redshift-space distortions. 

Previous studies suggest that the shift of the BAO scale due to nonlinear effects is expected to decrease from 0.2-0.5\% to less than 0.1\% with the BAO reconstruction technique ~\citep[e.g,][]{Eisenstein_etal_07a,Padmanabhan_etal_09,Seo_etal_10,Mehta_etal_11}. While testing the post reconstruction residual systematics is of key importance, controlling the pre-reconstruction BAO systematics has continued being important since the current analyses of redshift-space distortions and the Alcock-Pazynski test heavily utilize the pre-reconstruction BAO information~\citep[e.g.,][]{Beutler17RSD}. Currently, models of such tests are built from sophisticated perturbation theories which supposedly account for the nonlinear effects on the BAO feature to some extent~\citep[e.g,][]{TNS2010,Beutler17RSD}. However, the accuracy of such models on the nonlinear BAO position is difficult to determine. Even without redshift-space distortion measurements and the Alcock-Pazynski test, the pre-reconstruction BAO analysis continues being a nominal choice for an early stage of a galaxy survey when its low signal-to-noise does not yet allow effective density reconstruction~\citep[e.g.,][]{Ata2017}.
For these reasons, we examine both pre- and post-reconstruction BAO systematics.

This study covers a wide range of redshift between $0\le z \le 2.5$ and galaxy/halo bias that is applicable to future dark energy survey designs such that one can calibrate future measurements against the BAO systematics. 
We also compare the BAO systematics for different reconstruction conventions that were introduced in previous studies~\citep{Pad2012,Cohn2016,Seo_etal_16} and test the smoothing filter scale dependence. We introduce physically motivated pre- and post-reconstruction fitting models based on the infrared (IR) 
resummation procedure on the BAO feature detailed in \citep{zv16} 
and based on the Lagrangian Effective Field Theory (LEFT) description of galaxy statistics \citep{VCW_16}. We  test the BAO systematics with these various fitting models. Our result, in turn, tests the validity of these EFT bias models in the context of the BAO feature alone, which would be more challenging than in the context of the global clustering feature. 
For comparison, extensive BAO systematics tests of the BOSS data release 12 catalog have been conducted in \citet{VM16}. \citet{Prada2016} first introduced the sample variance cancellation idea for the BAO systematics tests focused on $z<1$ and a spherically averaged BAO feature. As described above, the extent of our study differs in many ways from these earlier studies. We also note that we use the halo occupation distribution (HOD) approach to populate galaxies while \citet{Prada2016} (and also \citet{Angulo14}) derives galaxy populations based on the catalog of distinct halos and subhalos taking advantage of their full \Nb\ simulations with higher force and time resolution.

The paper is structured as follows. In \S~\ref{sec:methods}, we explain the \Nb\ simulations and the sample-variance cancellation arrangement, mock halo samples, the density field reconstruction parameters we adopt, and the fitting models we use.  In  \S~\ref{sec:results}, we present the pre- and post-reconstruction results for different redshift, halo/galaxy bias, and shot noise levels. We also present dependences of the BAO measurements as well as of growth rate $f$ and nonlinear bias effect on reconstruction conventions, fitting models, and smoothing scales. Finally, in \S~\ref{sec:con}, we conclude. In appendices we present the derivation of the BAO fitting models based on the EFT as well as supplementary figures.

\section{Method}\label{sec:methods}

\subsection{Sample-variance cancellation simulations}

We use a suite of $2\times 40$ cosmological N-body simulations produced using FastPM \citep{yu16} with a fiducial flat $\LCDM$ cosmology model based on \citet{Planck_16}. The cosmology parameters are $\Omega_m=0.3075$, $\Omega_{\Lambda}=0.6925$, 
$\Omega_b h^2=0.0223$, $h=0.6774$ and $\sigma_8=0.8159$. The simulation box has the size $1380 \hMpc$ per side, with total $2048^3$ particles inside, giving a particle mass of $2.611\times 10^{10}  h^{-1} {\rm M_\odot}$. 

We follow the method described by \citet{Prada2016} to reduce the number of simulations that are necessary to suppress the sampling variance. We use a paired set of simulations with initial conditions generated from identical white-noise fields (i.e., the same random phases), but sourced by a paired set of slightly different initial power spectra: one with the ordinary BAO feature (`wiggled', subscript `wig') and the other without the BAO feature (`dewiggled', subscript `now')\footnote{The initial condition without the BAO feature is generated by smoothing the BAO wiggles in the input linear power spectrum with spline fitting~\citep{zv2015, zv16}.}. By differencing power spectra of the matter or halo density field of each pair of simulations, we remove the sample variance imprinted from the broadband (nowiggle) component of the Fourier modes because these are the same in wiggle and nowiggle simulations. The remaining variance is due to the BAO wiggle component of the Fourier modes in the wiggle simulation, stochastic effects during structure growth, and any shot noise remaining in the simulation difference.

\subsection{Quasi N-body simulation}
The quasi \Nbody\ simulation scheme FastPM \citep{yu16} is used to model the evolution of dark matter particles. Quasi \Nbody\ simulation schemes such as COLA and FastPM employ a Particle-Mesh solver with finite number of time steps to model the evolution of dark matter non-perturbatively \cite{icecola}. We employ 40 time steps uniformly distributed in the scaling factor $a=1 / (z+1) $, starting from 2LPT initial conditions at redshift $z=9$ and ending at $z=0$. The force is calculated on a $4096^3$ mesh. 

We find good convergence at $z=0$ comparing simulations of 10 time steps and 40 time steps in the BAO damping profile measured from cross-correlation between the initial field and final field. We output six snapshot density fields at $z=2.5, \, 2.0, \, 1.5, \, 1.0, \, 0.6$, and $0$. The time stepping scheme ensures that at $z=2.5$ we have about 10 time steps. A minimum of 10 time steps is necessary to reproduce halos with accurate mass labels in a quasi-nbody simulation~\citep{cola}.

\begin{figure*}
\includegraphics[width=0.45\linewidth]{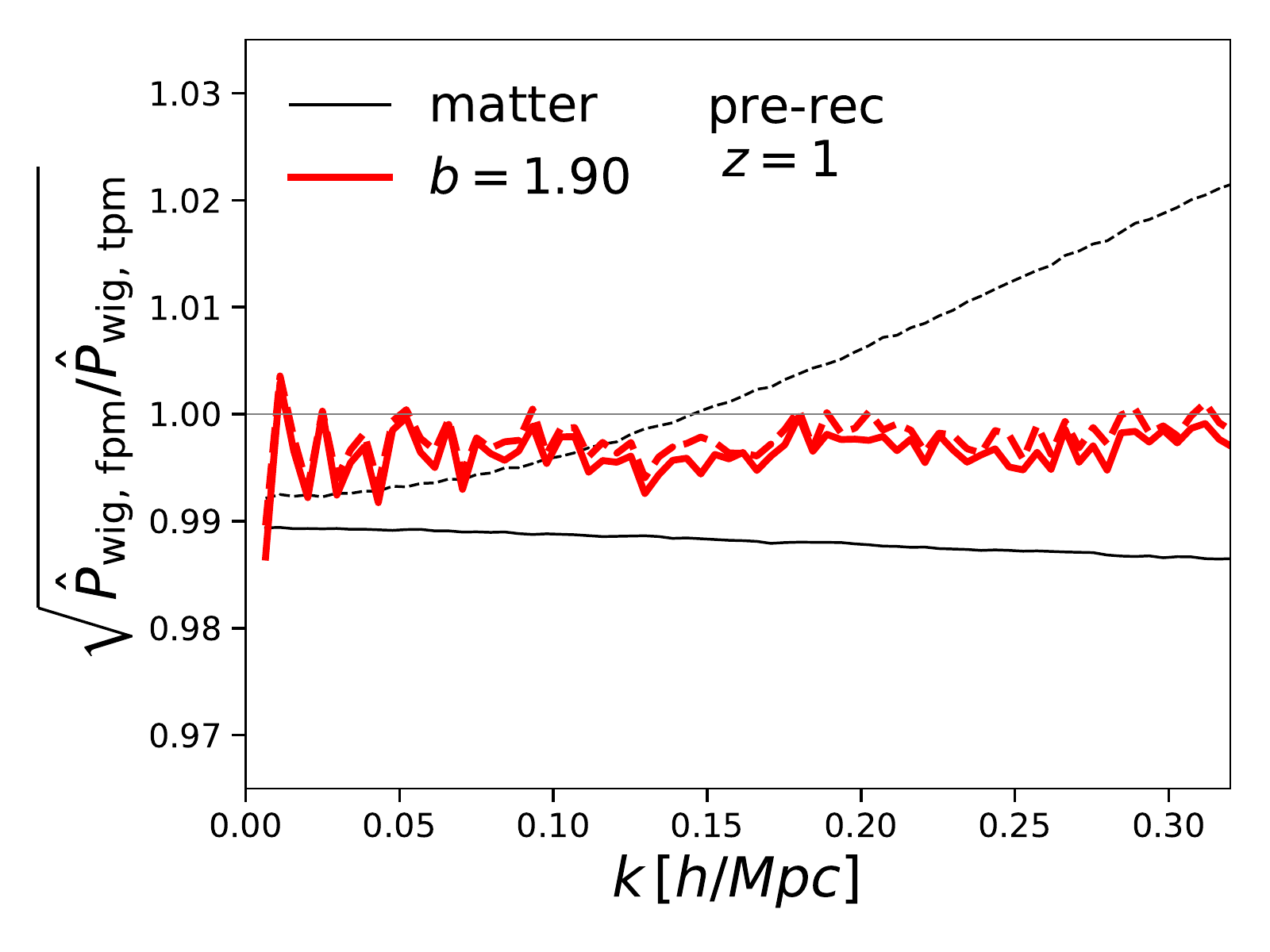}
\includegraphics[width=0.45\linewidth]{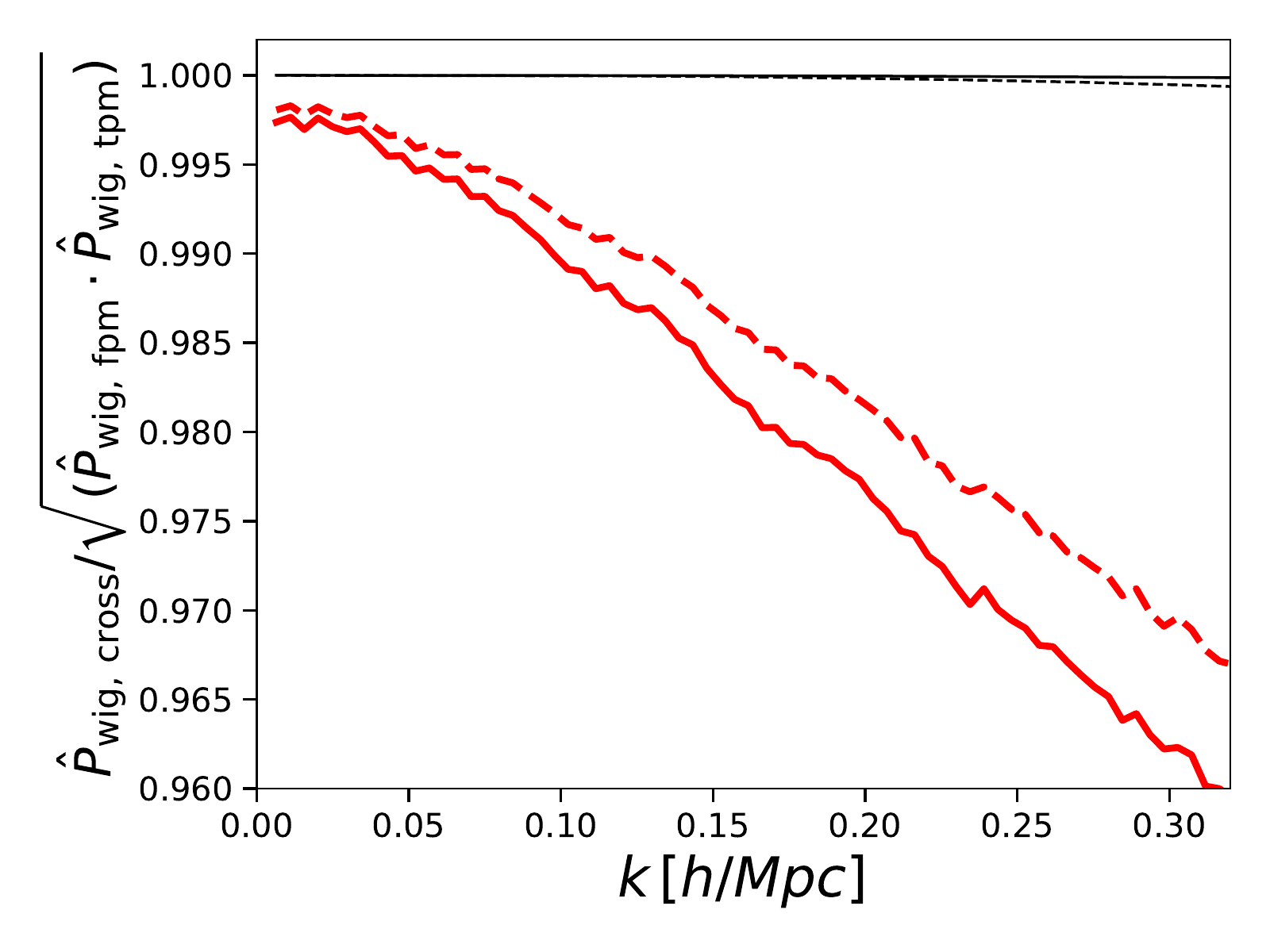}
\includegraphics[width=0.45\linewidth]{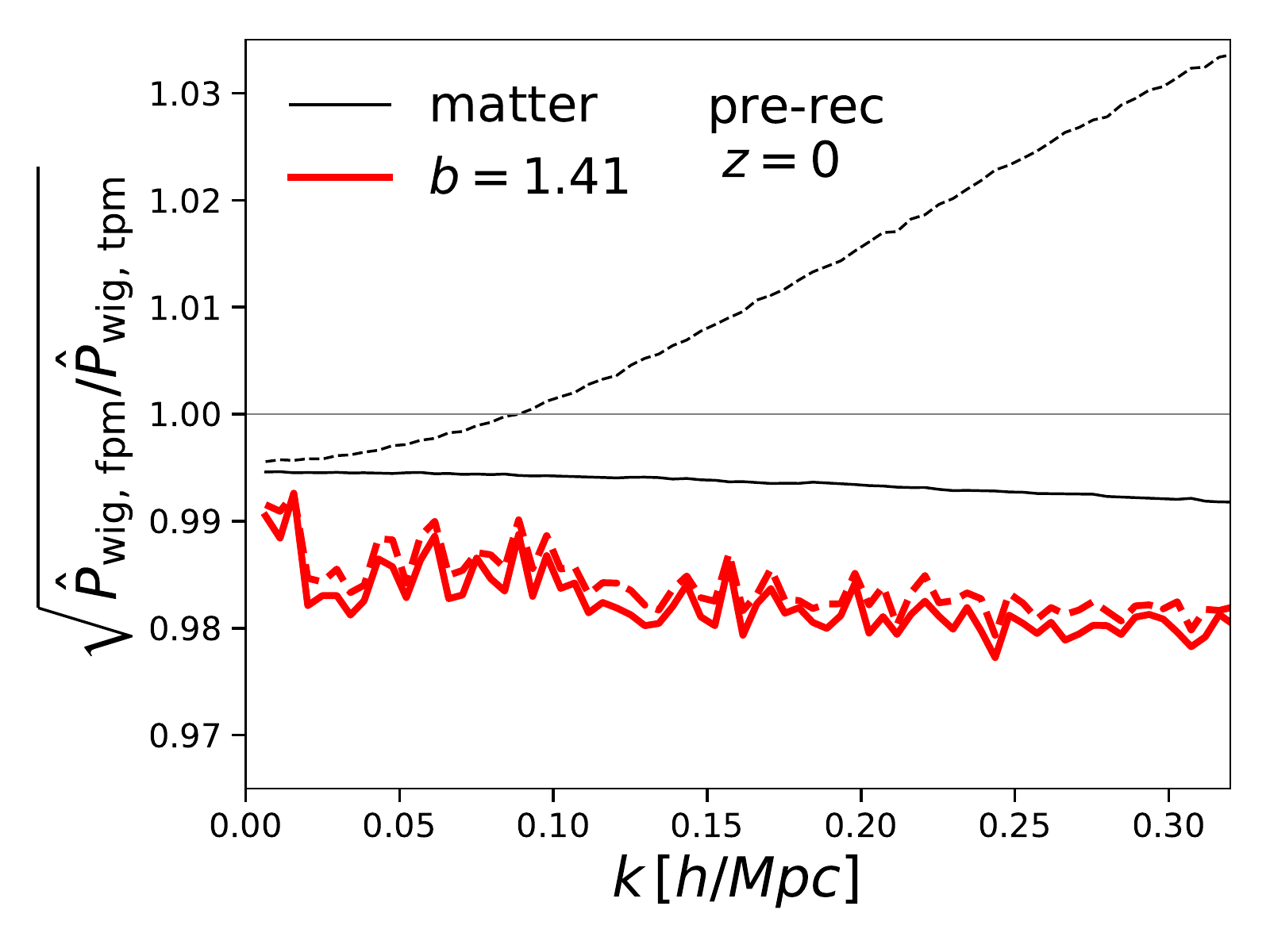}
\includegraphics[width=0.45\linewidth]{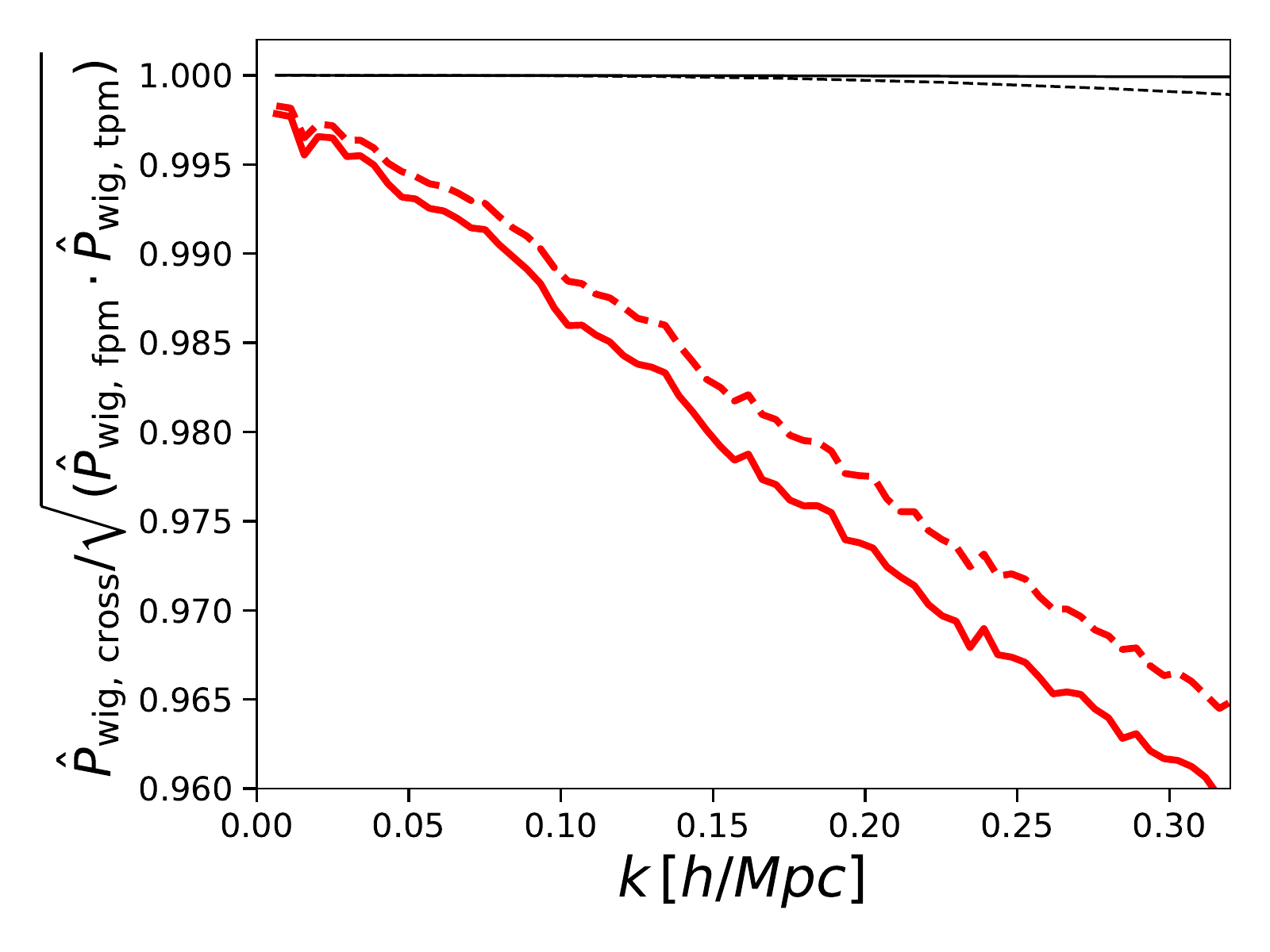}
\caption{Transfer function (left panels) and cross correlation coefficient (right panels) between wiggled FastPM and MP-Gadget simulations of the same initial random seed.  Upper panels show matter (thin black lines) and halos (thick red lines) at redshift $z=1$ and low panels at $z=0$. Solid lines are for real space and dashed lines for redshift space. Galaxy biases are noted in the left panels.}\label{fig:benchmarks}
\end{figure*}

\begin{figure*}
\includegraphics[width=0.45\linewidth]{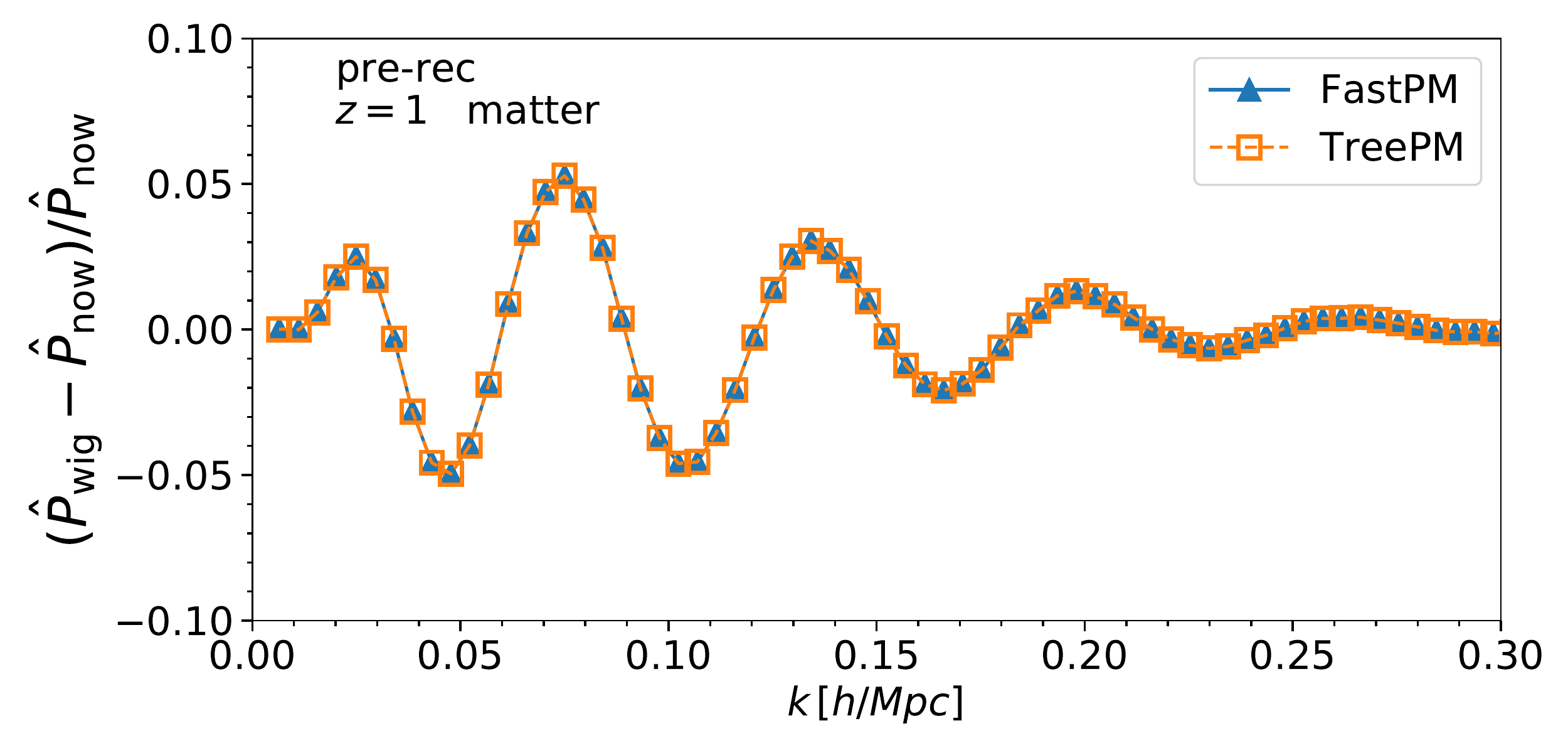}
\includegraphics[width=0.45\linewidth]{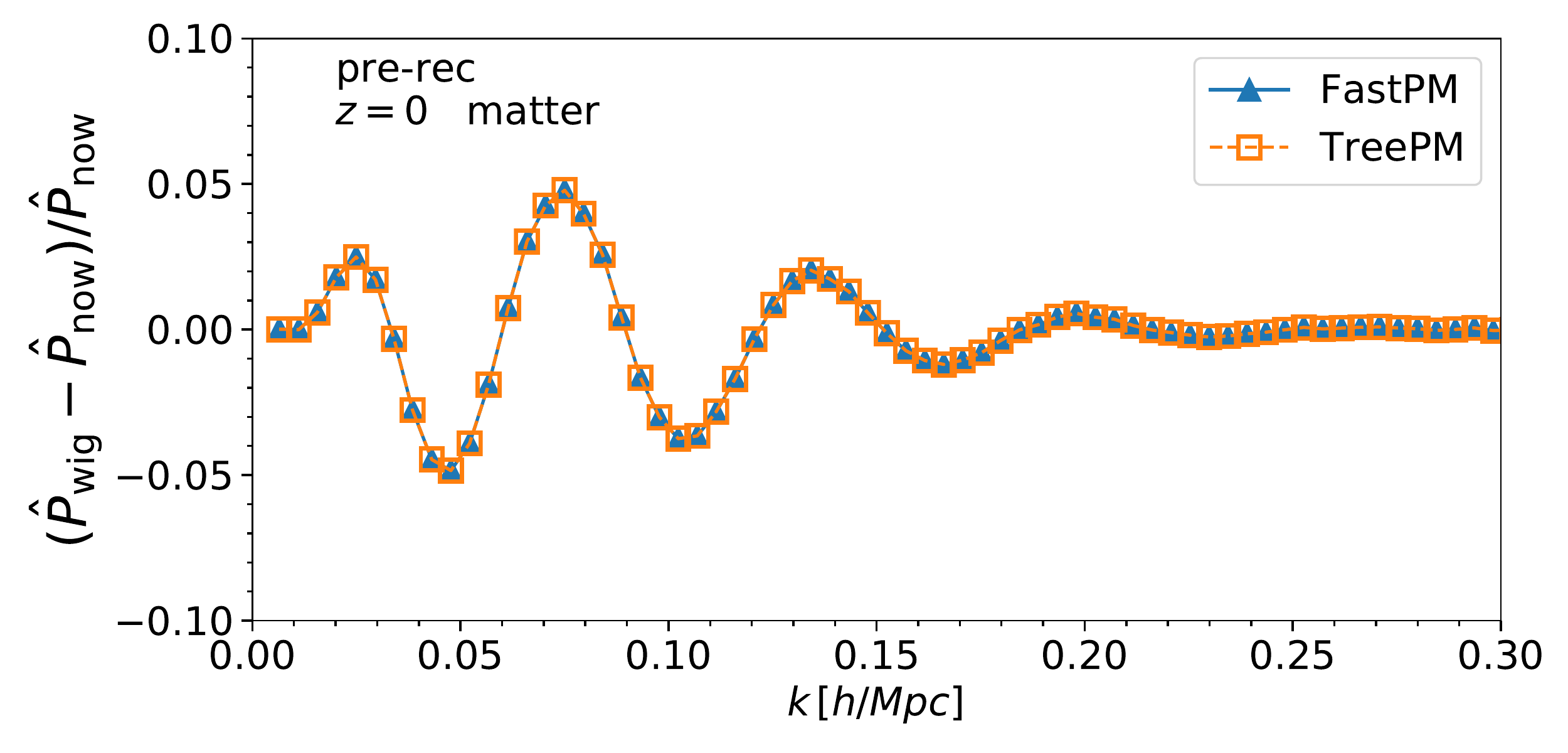}
\includegraphics[width=0.45\linewidth]{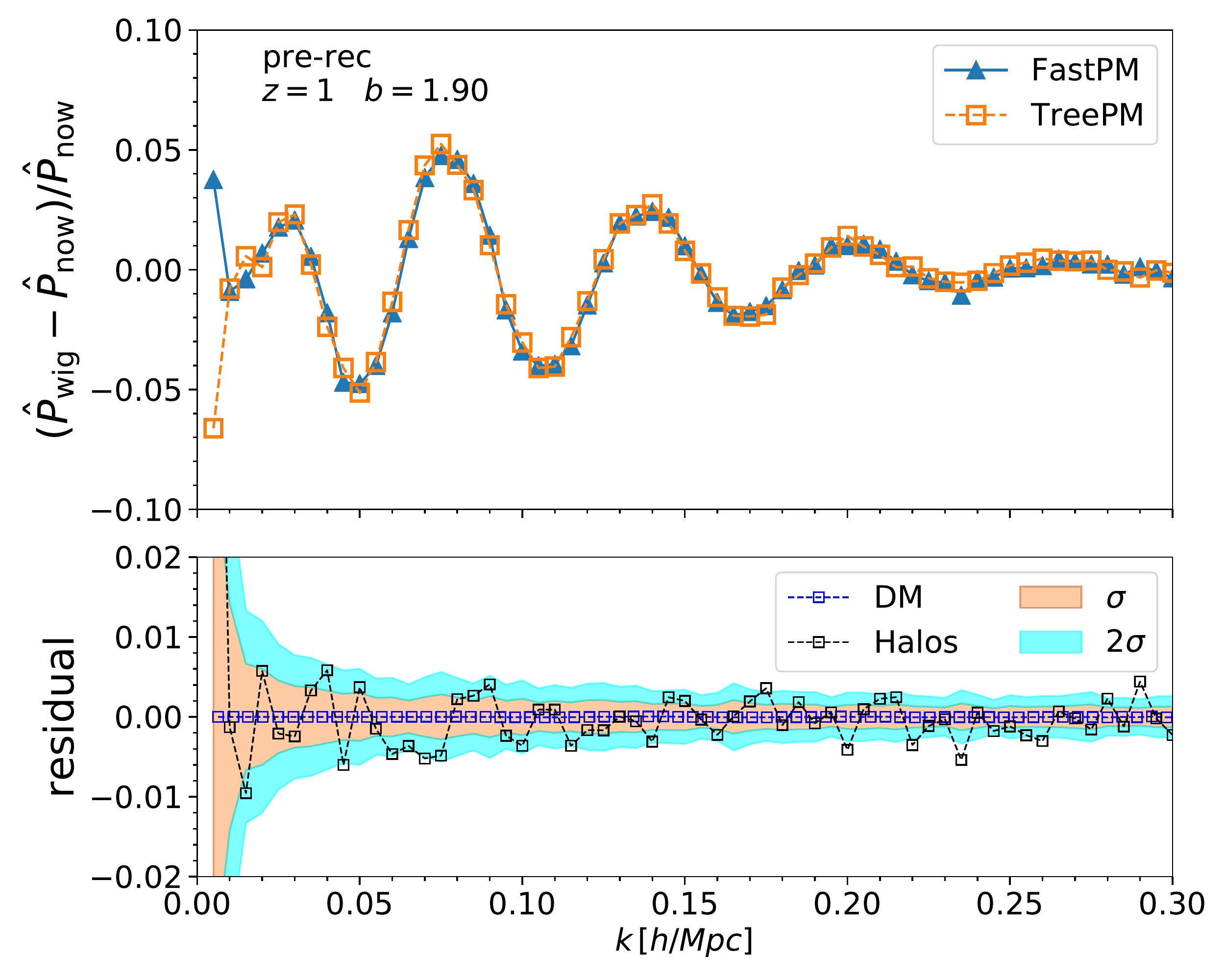}
\includegraphics[width=0.45\linewidth]{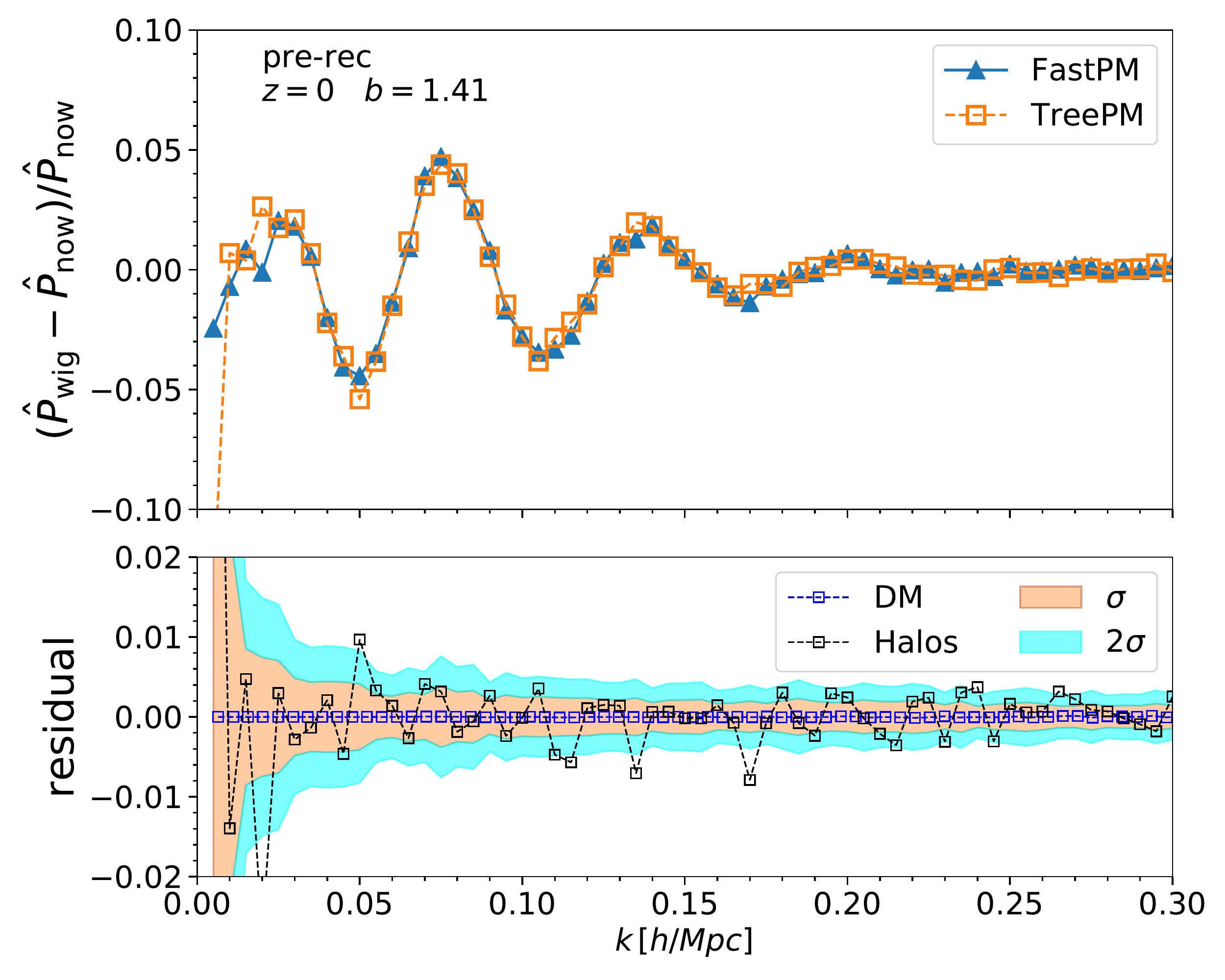}
\includegraphics[width=0.45\linewidth]{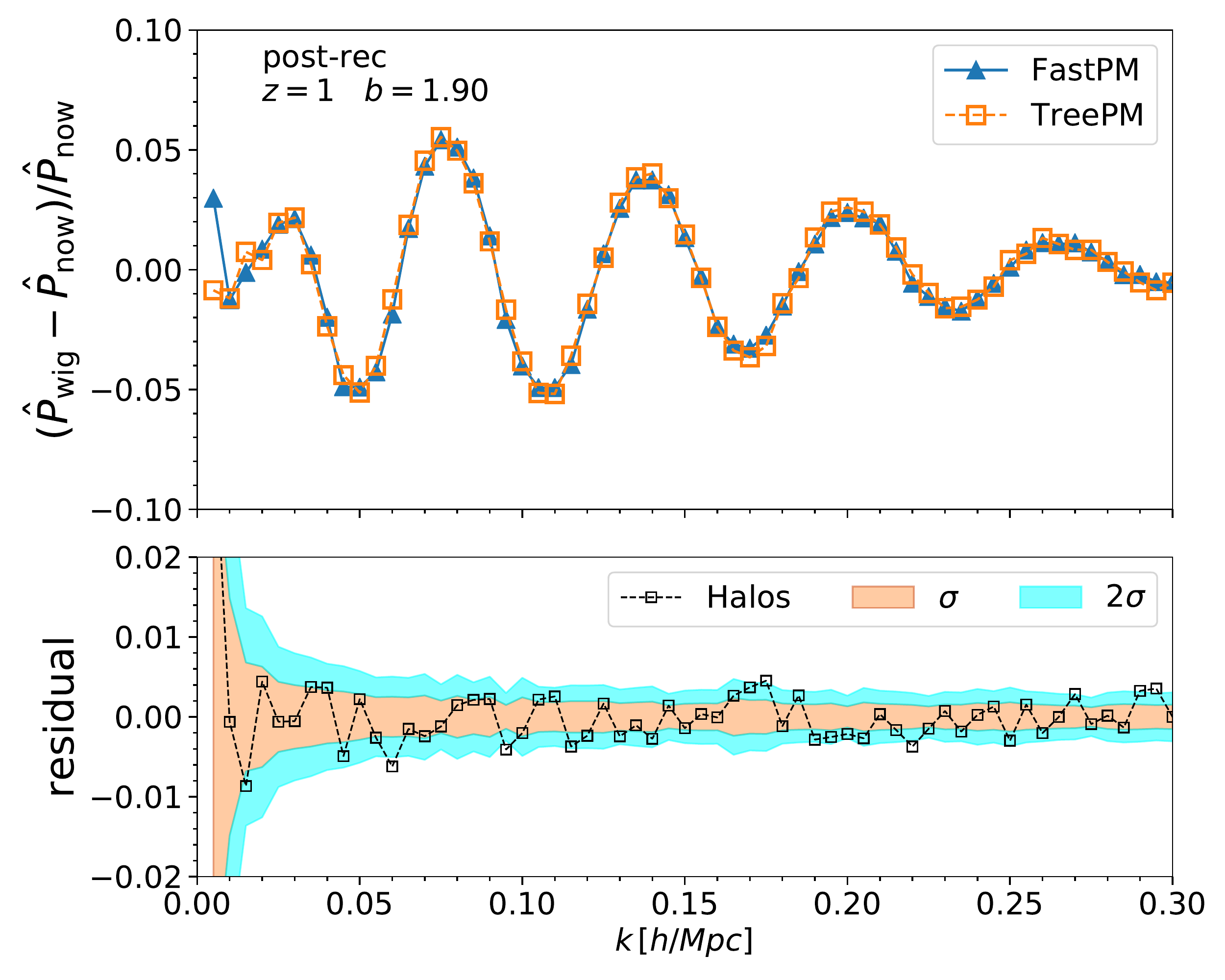}
\includegraphics[width=0.45\linewidth]{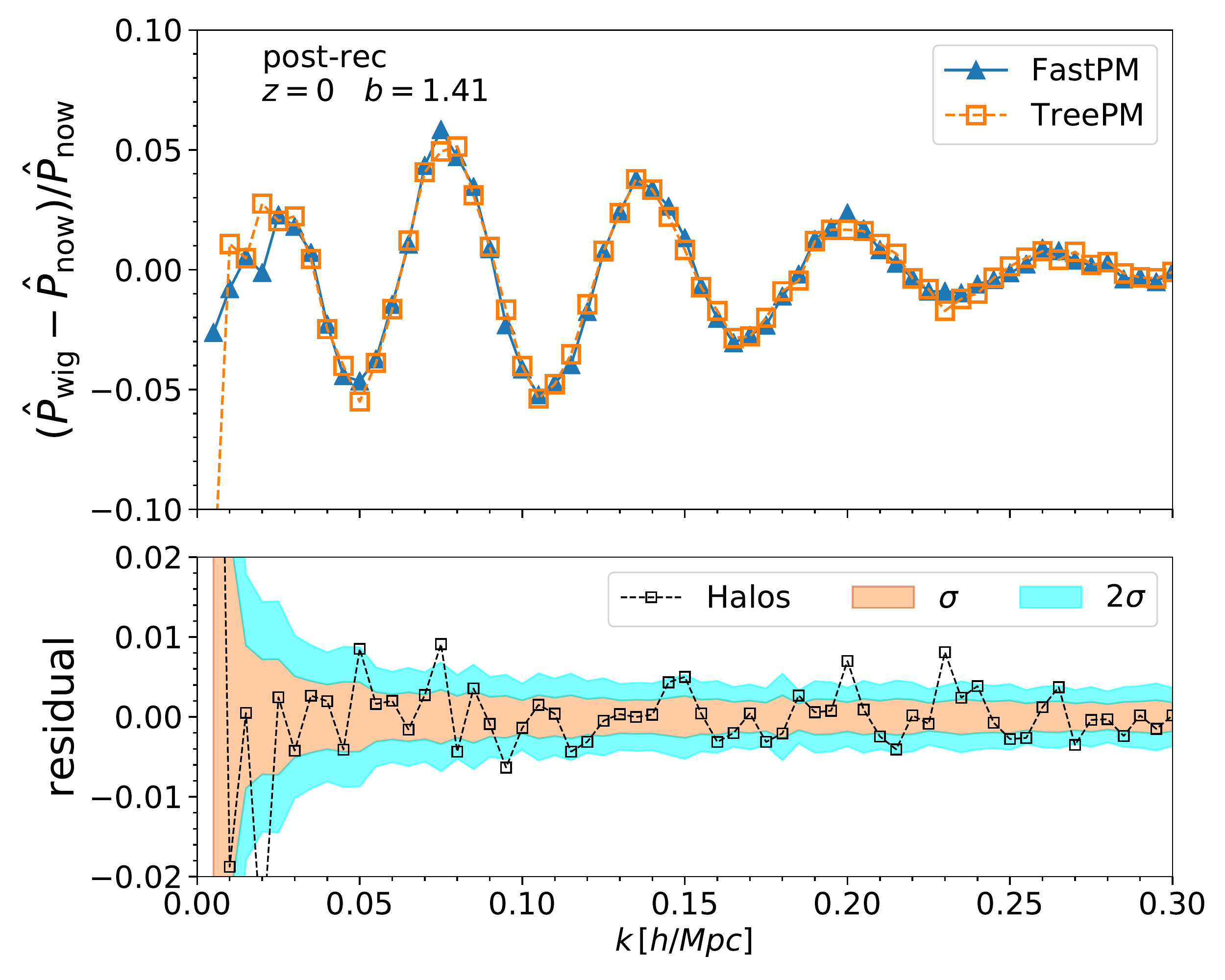}
\caption{The BAO signature from MP-Gadget (TreePM) and FastPM simulations in redshift space, as well as the residual of the signature between the two. The top row is for pre-reconstruction matter and the other rows are for pre- and post-reconstruction halos. At $z=1$ (left) and $z=0$ (right). In the third row, we compare the residual of BAO signature (blue squares for matter and the black squares for halos) between the two sets of sample-variance canceled simulations along with $1$, $2-\sigma$ dispersion (color region) of $\hat{P}_{\text{wig}}-\hat{P}_{\text{now}}$ from FastPM. We find that the residual from the matter field (blue squares) is negligible compared to halos (black squares). For halos, we observe residuals at the level of 1, $2-\sigma$, likely due to stochasticity from the halo finder algorithm. After reconstruction (bottom two rows), we observe a similar level of residuals.  }\label{fig:BAOtpm}
\end{figure*}

Previous work usually measured the BAO shifts from a single or a few full \Nbody\ simulations. We compare our FastPM scheme against a full \Nbody\ scheme (MP-Gadget) by computing the transfer function and cross-correlation coefficient of one pair of wiggle and dewiggled simulations using the same random seed. We define transfer function and cross-correlation coefficient as
\begin{align}
T(k) = \sqrt{P_{\rm fpm}(k)/P_{\rm tpm}(k)},
\end{align}
and
\begin{align}
r(k) = P_{\rm fpm, tpm}(k)/\sqrt{P_{\rm fpm}(k) P_{\rm tpm}(k)},
\end{align}
where $P_{\rm fpm}(k)$ and $P_{\rm tpm}(k)$ is the spherically averaged power spectrum of FastPM and MP-Gadget, abbreviated as fpm and tpm, respectively. $P_{\rm fpm, tpm}(k)$ is the cross-power spectrum between the two.
We measure both functions at redshift 1.0 and 0 for dark matter as well as halos of an intermediate bias, as shown in Fig.~\ref{fig:benchmarks}. The MP-Gadget simulation pair were ran with identical white-noise fields to the FastPM pair, but at slightly different output redshifts by 2\%. The small difference in the snapshot redshift manifests as a 0.5-1\% offset in the matter power spectrum as shown in the transfer function in the left panel. For friends-of-friends halos defined with the same number of particles, we observe slightly different bias at the level of 2\% at $z=0$. The matter transfer function in redshift space shows a 3.5\% scale dependency up to $k < 0.3\ihMpc$, while for halos the dependency drops to 1\%. In the the right panels, we find that the cross-correlation coefficient is very close to $1.0$ for matter and $> 96\%$ for halos for $k<0.3 \ihMpc$ for both redshifts.

Our sample-variance cancellation method differences the wiggled and dewigged power spectrum, and therefore strongly suppresses the difference in shape (e.g., in redshift space) and amplitude. The upper two rows of Figure~\ref{fig:BAOtpm} compare the sample-variance-canceled BAO features of matter (first row) and halos (second row) from paired simulations of FastPM and MP-Gadget at $z=1$ (left) and at $z=0$ (right). In the third row, we show the residual of BAO signature from the top two rows with blue squares for matter and black squares for halos, compared to the typical dispersion among sample-variance-canceled pairs of the given halos. The residual for matter is negligible compared to the relevant errors, implying a high level of agreement between FastPM and MP-Gadget simulations when focused on the BAO feature. On the other hand, for halos, once the sample variance is canceled, the residual of the BAO signature is comparable to the typical dispersion between independent sample-variance-canceled pairs, for both pre-reconstruction and post-reconstruction. We note that this residual appears uncorrelated with the BAO pattern and therefore it will unlikely cause a systematic BAO offset between the two pairs of simulations. We believe that this uncorrelated residual is due to nonlinear stochasticity that arises in halo identification, as we do not observe similar residual in the case of the matter field (blue squares in the third row) and that the bias in the estimator due to this stochasticity is small once the variance of stochasticity is reduced by averaging over multiple mocks. From the bottom two rows, we observe a similar level of stochasticity in the post reconstruction halo BAO feature, which implies that the difference between the two sets of simulations do not noticeably affect the reconstruction process.  Confirming this will however require rerunning a comparable set of MP-Gadget simulations which is beyond the scope of our current analysis. Nevertheless, we highlight that it can be a substantial improvement to follow up this work with a similar study on a large set of full \Nbody\ simulations.

\subsection{Calculating power spectra}
The publicly available module \textbf{nbodykit}\footnote{https://github.com/bccp/nbodykit} is used to produce the halo catalogs and subsamples of dark matter particles. For each snapshot, we produce friends-of-friends halo catalogs with linking length $b=0.2$. We cut the halo catalogs by friends-of-friends mass to produce biased tracer populations.
The minimum group multiplicities are $\sim 35$ particles for $z \le 1$, corresponding to  $9.1\times 10^{11}  h^{-1} {\rm M_\odot}$,  and 16 particles for $1 < z \le 2.5$, corresponding to $4.2\times 10^{11}  h^{-1} {\rm M_\odot}$. 
We generate halo catalogs by taking the position and the velocity of the center of mass of such individual halos. 
In addition to the halo catalog, we produce 1\% subsamples of the matter particles. These subsamples are used to compare with the mock halo/galaxy clustering. In order to reduce the variance introduced by subsampling, the same subsampling random seed is used in any pairs of wiggled and dewiggled distribution \citep{Schmittfull_etal_15}.

\begin{figure*}
\includegraphics[width=0.45\linewidth]{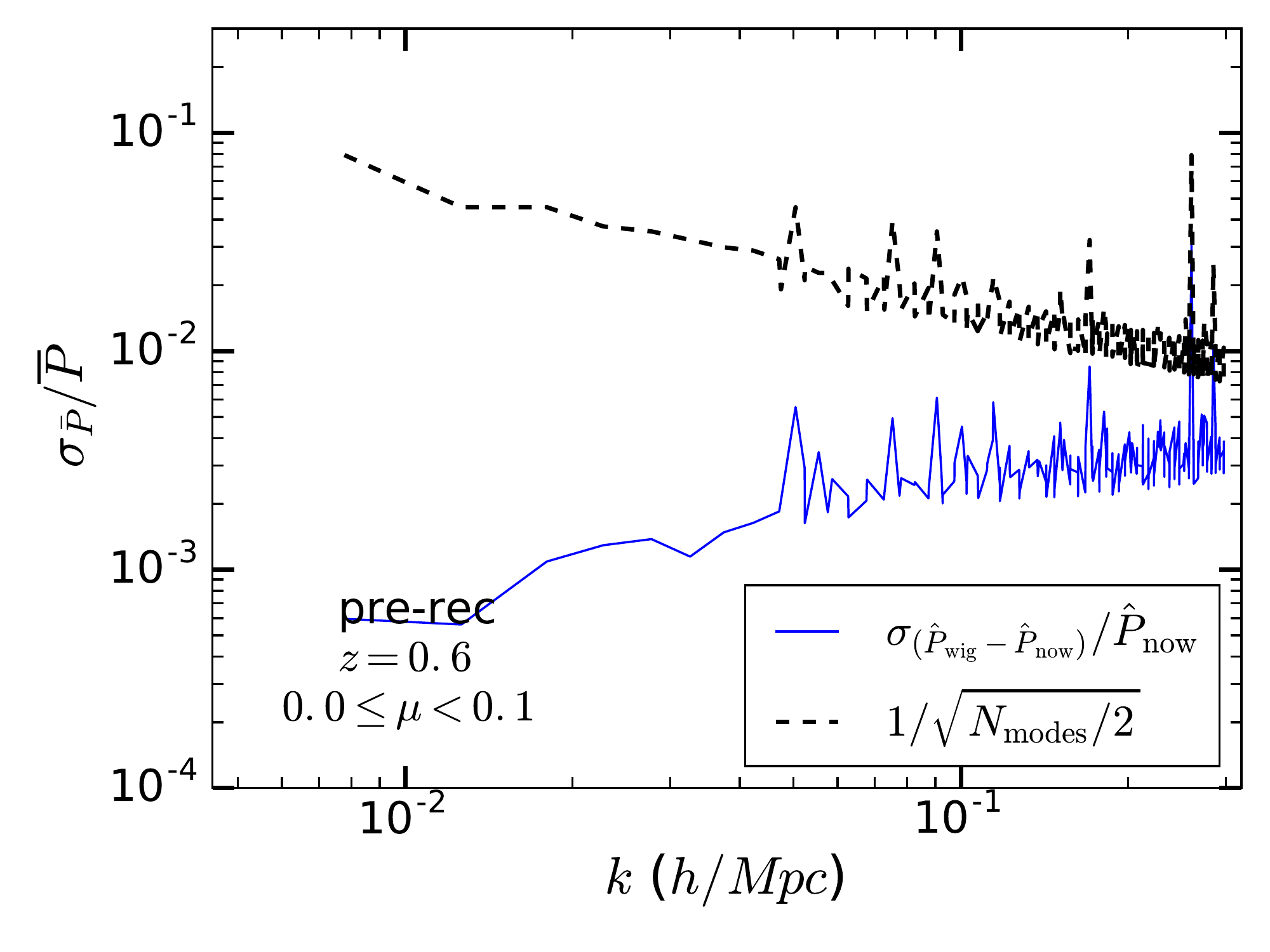}
\includegraphics[width=0.45\linewidth]{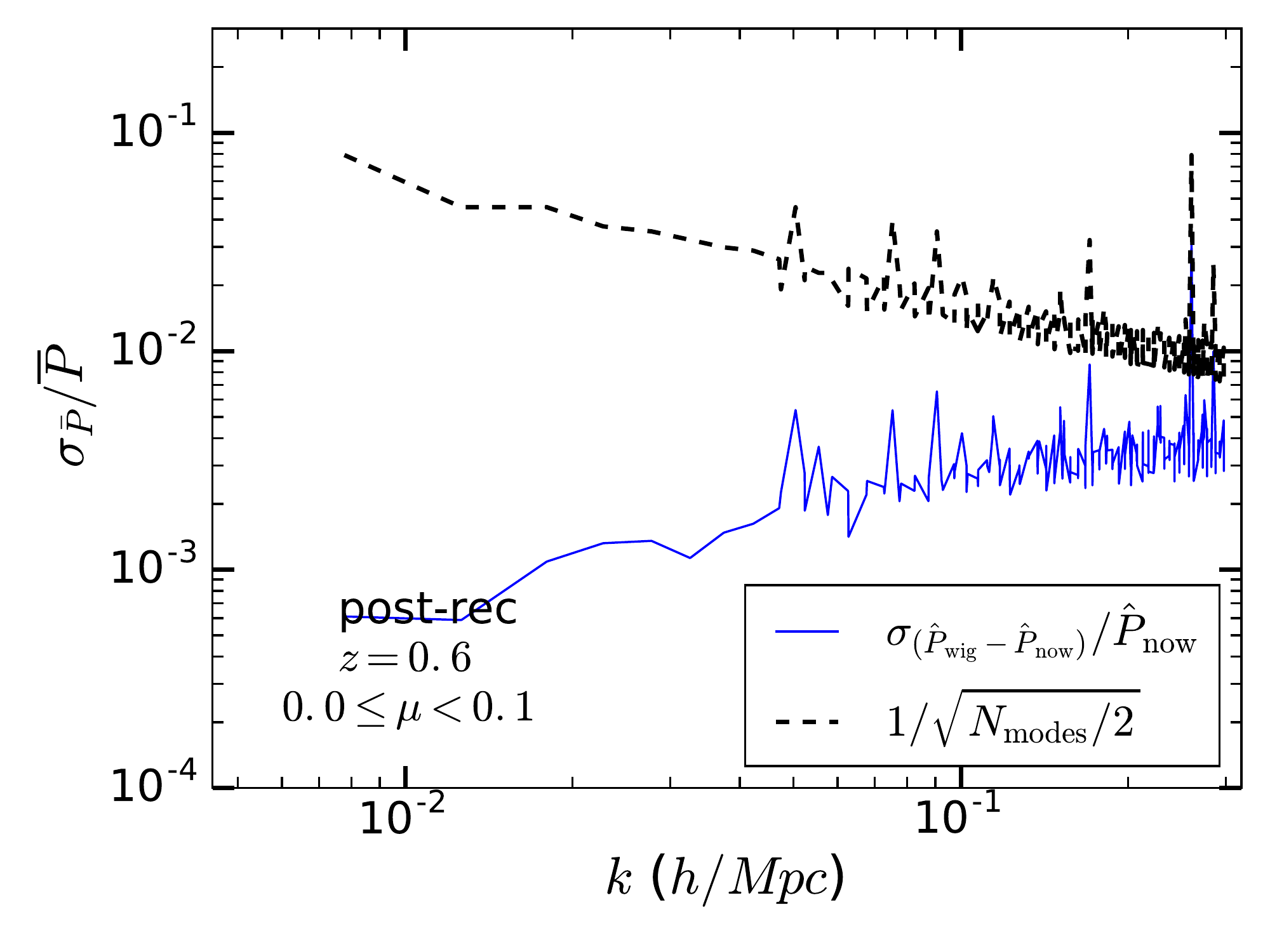}
\includegraphics[width=0.45\linewidth]{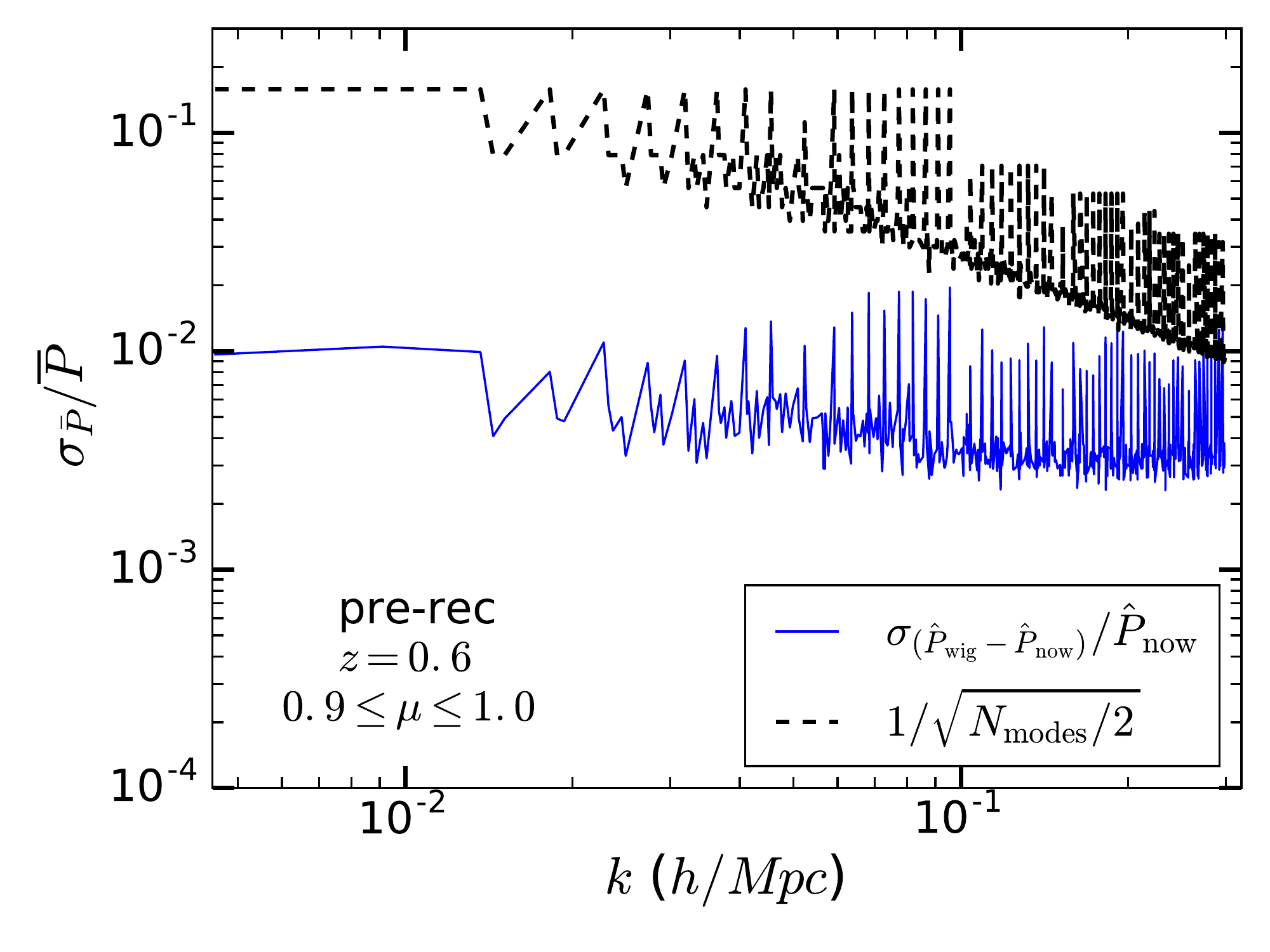}
\includegraphics[width=0.45\linewidth]{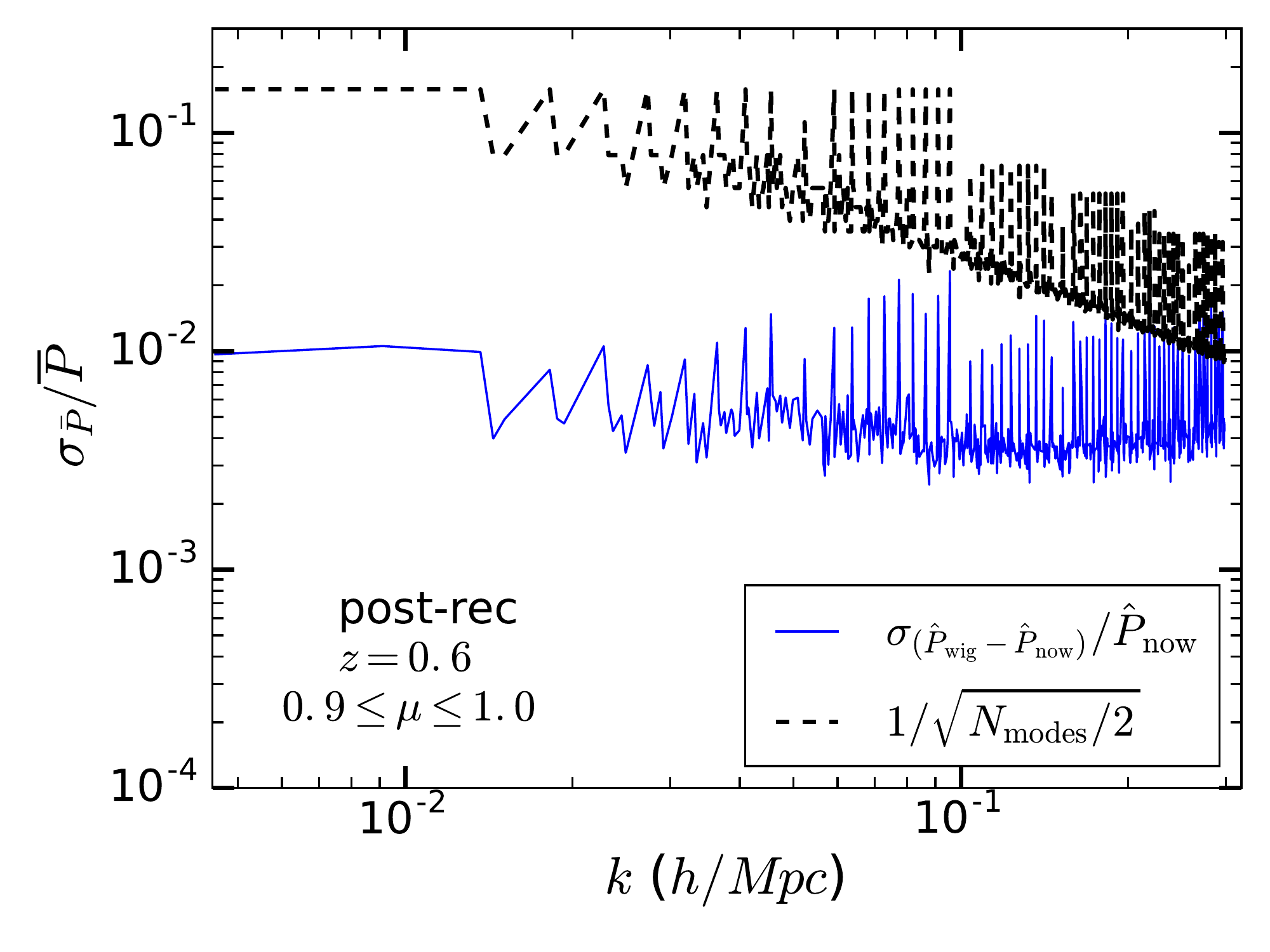}
\caption{The measured $\sigma_{\bar{P}_{\mathrm{wig}}-\bar{P}_{\mathrm{now}}}$ divided by $\bar{P}_{\mathrm{now}}$ of the sample variance cancellation simulations in comparison to the expectation (dashed line) from Gaussian error without sample variance cancellation (i.e., $1/\sqrt{N_{\rm modes}/2}$), for matter at $z=0.6$.
Both errors are divided by $\sqrt{40}$ to represent the errors of the mean. 
\textit{Left panels:} pre-reconstruction. \textit{Right panels:} post-reconstruction. \textit{Upper panels:} modes perpendicular to the line-of-sight. \textit{Lower panels:} modes along the line-of-sight. The figure implies that the gain in signal-to-noise due to the sample variance cancellation is a factor of 10 near $k\sim 0.05\ihMpc$ and a factor of 5 near $k\sim 0.1\ihMpc$, although it depends on redshift as well as galaxy bias.}\label{fig:covdia}
\end{figure*}

For each particle field output, we compute the density and the displacement field for density field reconstruction in a $512^3$ mesh  with the cloud-in-cell (CIC) scheme, and generate an anisotropic power spectrum that is subsequently deconvolved against the CIC window function effect. Each power spectrum is bin-averaged for 100 equal $\mu$ bins at each $k$ bin with $dk=0.005\ihMpc$ over the fitting range of $4.55\times 10^{-3}\ihMpc \le k<0.300\ihMpc$, where $\mu$ is the cosine of the angle between the line of sight and the wave number $\vec{k}$. The maximum $k$ in the fitting range is about four times smaller than the Nyquist frequency. The total number of power spectrum data points is $4342$.

\subsection{Mock halo/galaxy samples in comparison to future survey targets}

By applying different mass cuts, we generate mock halo samples over a broad range of halo/galaxy bias and redshift in order to represent targets of various upcoming galaxy redshift surveys. As an example, DESI will target three different large-scale density tracers. The first type among them is luminous red galaxies (LRGs) over $0.6< z < 1$ that typically reside in very massive halos (i.e., highly biased) with low star formation rates for a few billion years~\citep[e.g.,][]{Wake2006}. Although the exact bias and number density needs to be updated as the survey progresses, Table 4 of \citet{FontRibera2014} quotes $b_{\rm LRG} \sim 2$ and number density $n$ that gives $\nPt \sim 2-3$.
The second class of targets by DESI is Emission Line Galaxies (ELGs) for $0.6<z<1.6$ 
that are less biased than LRGs. From \citet{FontRibera2014}, $b_{\rm ELG}(Z)D(z)=0.84$, giving $b_{\rm ELG}\sim 1.5$ at $z\sim 1.2$.  
The third type of targets by DESI is QSOs between $2<z<3.5$: $b_{\rm QSO}(z)D(z)=1.2$, giving $b_{\rm QSO}\sim 3.3$ at $z\sim 2.5$.

\citet{FontRibera2014} crudely approximates the survey parameters for EUCLID\footnote{https://www.euclid-ec.org} and WFIRST\footnote{https://wfirst.gsfc.nasa.gov/index.html} as well. The targets of EUCLID can be modeled with $b(z)D(z)=0.76$ between $0.6 < z < 2$ and the targets of WFIRST for $1<z<2.8$ with $b(z)D(z)=0.76$. Our halo catalogs are designed to cover the relevant halo bias range for these surveys. Table~\ref{tab:fof_pre_rec} lists the bias and corresponding number density for our halo samples. 

\subsection{Estimating power spectrum covariance matrix}\label{subsec:calcov}

We obtain the covariance matrix of sample-variance canceled simulations by
\begin{align}
C_{ij} = \frac{1}{N_s-1} \sum_{n=1}^{N_s} \Bigl[\delta P_n(k_i, \mu_i) - \delta\bar{ P}(k_i, \mu_i) \Bigr] \Bigl[\delta P_n(k_j, \mu_j) \nonumber\\ - \delta\bar{P}(k_j, \mu_j) \Bigr],
\end{align}
where $N_s=40$ is the total number of  realization pairs, $i$ and $j$ denote the indices of $(k, \mu)$ bins, $\delta{P}_n={\Pwig}_n-{\Pnow}_n$ is the difference of wiggle and nowiggle power spectra in the $n$th simulation, and $\delta\bar{P}=\sum_n\delta{P}_n/N_s$ is its mean.

The $40$ pairs of realizations we use are certainly not enough to accurately invert the non-diagonal covariance matrix with $4342$ power spectrum data points~\citep{Hartlap07,Percival14}. We therefore heuristically assume the covariance matrix is diagonal and use the inverse dispersions of band powers when calculating best fits. In order to minimize the effect of the assumption of diagonal covariance matrix, we derive the errors on our parameters from the dispersions among individual $40$ best fits rather than from the likelihood surface of the best fit of the mean power spectrum \citep[also see][for the same approach]{Schmittfull_etal_15,Schmittfull_etal_17}.

As a caveat, for non-BAO parameters such as growth rate or scale-dependent bias, we occasionally use the best fit to the mean difference power spectrum rather than the average of the individual best fits when individual best fits are poorly constrained for such parameters due to large degeneracy among them. We will note whenever that is the case.

To check the efficiency of the sample-variance cancellation scheme, 
Figure \ref{fig:covdia} shows the measured error on the difference band power signal (blue solid lines) at $z=0.6$ for matter. We overplot the expected Gaussian error for $40$ sets of simulations without sample variance cancellation (dashed lines). The figure implies that the gain in signal-to-noise due to the sample-variance cancellation is at least a factor of $10$ near $k \sim 0.05\ihMpc$ and at least a factor of 5 near $k\sim 0.1\ihMpc$. That is, at our $40$ pairs of simulations correspond to $\sim 4000-1000$ sets of nominal simulations at this specific redshift. Such gain depends on redshift and galaxy/halo samples such that the gain decreases as redshift and bias increases. 

\subsection{Density field reconstruction}

The nonlinear structure growth damps and shifts the BAO peak dramatically at low redshift. 
\citet{Eisenstein_etal_07b} proposed a density field reconstruction formalism that reverses a good portion of nonlinear degradation and restores the BAO peak. This can potentially improve the measurement precision by a factor $\sim 2$ for densely populated targets at low redshift.
The density field reconstruction technique has now become a standard tool to improve the BAO scale measurement from both simulations and real surveys \citep[e.g.,][]{Seo_etal_08, Padmanabhan_etal_09, Seo_etal_10, Noh_etal_09, Mehta_etal_11, Anderson_etal_12, Anderson_etal_14, Beutler_etal_16}. 

Historically, there are a few reconstruction conventions that were tested. In general, the reconstruction process involves smoothing the observed nonlinear real-space or redshift-space galaxy/matter density field with a filter (typically, a Gaussian filter), deriving the displacement field $\tilde{\bf \chi}^r(\bk)$ assuming the Zeldovich approximation or a linear continuity equation, and moving the observed matter/galaxy particles (or meshes~\citep{SeoHirata15,Obuljen16}) by this displacement field. By moving the particles by $\tilde{\bf \chi}^r(\bk)$, we reverse the particles close to where they were before their nonlinear evolution.
Since that also removes the linear information that we are after, we move a reference set of particles (or meshes) by the same displacement field, and construct the final reconstructed target density field by differencing the density field from the displaced matter/galaxies $\delta_{\rm d}$ and the density field from the displaced reference particles $\delta_{\rm s}$: $\delta_{\text{r}}=\delta_d-\delta_s$ \citep{Eisenstein_etal_07b,Padmanabhan_etal_09,Tassev2012}.

When operating in redshift-space, there are broadly two options/conventions to deal with RSD. One option attempts to remove the large-scale RSD, i.e., Kaiser effect \citep{Kaiser_87}, and the other option attempts to leave in the Kaiser effect. We test both cases in this paper, while setting the former as our default reconstruction scheme since it has been more widely used in analyzing galaxy surveys~\citep[e.g.,][]{Alam2016}. To be more specific, our default reconstruction scheme is noted as `Rec-Iso', following the notation in \citet{Seo_etal_16}, which attempts to remove the Kaiser effect and makes the clustering more isotropic. For the second option, which leaves in the Kaiser effect, we adopt the specific convention used in~\citet{White2015,Cohn2016}.

\begin{itemize}
\item {\bf Isotropic BAO reconstruction -- ``Rec-Iso'' }~\citep[e.g.,][]{Pad2012,Seo_etal_16} \\
The real-space displacement field $\tilde{\bf \chi}^r(\bk)$ is estimated from the observed redshift-space density field $\tilde{\delta}^s_{\rm nl}(\bk)$ by
\begin{eqnarray}
&&\tilde{\bf \chi}^r(\bk) = - \frac{i\bk}{k^2} \frac{\tilde{\delta}^s_{\rm nl}(\bk)}{b(1+\beta\mu^2)}S(k).\nonumber\\
\label{eq:recisoq}
\end{eqnarray}
The displacement field in configuration space ${\bf s}$ is then derived by Fourier-transforming $\tilde{\bf \chi}^r(\bk)$;
\begin{eqnarray}
&&\tilde{\bf \chi}^r(\bk) \xrightarrow{\text{Fourier Transform}} {\bf \chi}^r({\bf s}).\\
\end{eqnarray}
The two density fields $\delta_d({\bf s})$ and $\delta_s({\bf s})$ are then derived, respectively by
\begin{eqnarray}
&&\delta_d : \mbox{displacing the galaxies by } {\bf \chi}^s={\bf \chi}^r + f ({\bf \chi}^r\cdot\hat{\bz})\hat{\bz}\nonumber\\
&&\delta_s: \mbox{displacing the random particles by } 
{\bf \chi}^r , \label{eq:reciso}
\label{eq:recisoqt}
\end{eqnarray}
where $\hat{\bz}$ is the unit vector pointing along the line of sight and $f$ is the growth rate. 

\item{\bf Anisotropic BAO reconstruction -- ``Rec-Cohn" }~\citep[][]{White2015,Cohn2016}\\
The second reconstruction option leaves in RSD by also displacing the reference particles by $1+f$ along the line of sight, following the convention used in \citet{Cohn2016}, rather than the anisotropic convention from \citet{Seo_etal_16} due to the modeling simplicity of the former.
The procedure is the same as `Rec-Iso' except for;
\begin{equation}
~~\delta_s: \mbox{displacing the random particles by } {\bf \chi}^s={\bf \chi}^r + f ({\bf \chi}^r\cdot\hat{\bz})\hat{\bz}. 
\label{eq:recaniqt}
\end{equation}
That is, this scheme is equivalent to setting $\lambda_d=\lambda_s=f$ in Appendix of \citet{Seo_etal_16}.

The smoothing filter $S(k)$ is assumed to be Gaussian: 
\begin{eqnarray}
&&S(k) =  \exp{\left[ -0.25k^2\Sigsm^2\right]}.\label{eq:sk}
\end{eqnarray}

For matter, we used $\Sigsm = 10\hMpc$ for all redshifts. For biased cases, we modified the smoothing scale $\Sigsm$ based on the signal to shot-noise of each case with an attempt to mimic Wiener filtering~\citep{SeoHirata15}:
\begin{align}
\exp{\left[ -0.25k^2\Sigsm^2\right]}=\exp{\left[ -0.25k^2 {\Sigma_{\rm sm,0}}^2\right]}\frac{1}{1+1/[nP(0.2)]},\label{eq:Sigsmscale}
\end{align}
where $n$ is the mean number density of the sample (Table~\ref{tab:fof_pre_rec}) and $P(0.2)$ is the galaxy/halo power spectrum at $k=0.2\ihMpc$ at given redshift. We set $\Sigma_{\rm sm,0} = 7\hMpc$ following the convention in ~\citet{SeoHirata15}. 

The smoothing scales for all cases are listed in Table~\ref{tab:fof_post_rec} and the number density of each halo sample is listed in Table~\ref{tab:fof_pre_rec}. In \S~\ref{subsec:Sigsm} we increase all smoothing scales by $\sqrt{2}$ and test the effect of the smoothing scale on post-reconstruction parameters. 
\end{itemize}

\subsection{Fitting models}
Since we are interested in the difference between the power spectrum with BAO wiggles and that without BAO wiggles, our model starts from the difference of the linear power spectrum,
\begin{align}
\delta P_L(k, z) = P_L(k, z)-P_{\mathrm{now},L}(k, z),\label{eq:plin}
\end{align}
where $P_L$ is the theoretical linear matter power spectrum with BAO and $P_{\mathrm{now},L}(k, z)$ is the corresponding dewiggled power spectrum~\citep{zv16}.
The difference power spectrum observed in the simulations,
\begin{equation}
\delta \hat{P}(k', \mu', z) = \hat{P}_{\text{wig}}(k', \mu', z)-\hat{P}_{\text{now}}(k', \mu', z),
\end{equation} 
is then fit using the basic template model of Eq.~\ref{eq:plin}, rescaled by corrections that we motivate and describe in the subsections below.

For each model, the observed coordinate $(k', \mu')$ of the observed power spectrum has the relationship with the reference coordinate  $(k, \mu)$  of the template power spectrum as
\begin{align}
k &= k' \times \frac{1}{\alpha_{\perp}}\sqrt{1+\mu'^2(\alpha^2_{\perp}/\alpha^2_{\|}-1)},\\
\mu &= \mu' \times \frac{1}{\alpha_{\|}/\alpha_{\perp}\times \sqrt{1+\mu'^2(\alpha^2_{\perp}/\alpha^2_{\|}-1)}}. 
\end{align}
Here $\alpha_{\|}$ and $\alpha_{\bot}$ account for the BAO scale shift along and perpendicular to the line-of-sight, respectively, relative to the isotropic BAO peak in the template power spectrum. They store information of Hubble parameter and angular diameter distance $D_A$, i.e.,
\begin{align}
\alpha_{\|} &= \frac{H^{\text{fid}}(z) r_s^{\text{fid}}(z_d)}{H(z)r_s(z_d)},\\
\alpha_{\perp} &= \frac{D_A(z) r_s^{\text{fid}}(z_d)}{D_A^{\text{fid}}(z) r_s(z_d)},
\end{align} 
where the superscript `fid' refers to the fiducial quantity derived from the template/fiducial cosmology model.\\

Based on this, we choose a set of nuisance parameters to construct the full fitting models. By focusing on the difference power spectrum, we are pardoned to omit nuisance parameters that account for any additive contribution to the broad-band shape of power spectrum. The effect of multiplicative contributions from nonlinear structure growth and nonlinear bias would still remain. In order to better account for such scale-dependent effects, in addition to testing the fitting model derived in~\citep{Seo_etal_16}, we also introduce physically motivated pre- and post-reconstruction models motivated by the Lagrangian Effective Field Theory (LEFT) ~\citet{vlah2015,VCW_16}. The full derivation based on~\citet{zv16, VCW_16} is presented in Appendix \ref{sec:EFTderivation}.

Parameters of our fitting models are summarized in Table~\ref{tab:model}. For each case, we use a slightly different set of free parameters and fixed parameters. We explicitly list the fitting model formulas below.

\begin{table*}
\centering
\caption{Parameters of the BAO fitting models.}
\begin{tabular}[c]{m{2.0cm} m{6.0cm} m{6.0cm}}
\hline
 & Pre-reconstruction & Post-reconstruction\\
\hline
\multirow{2}{*}{EFT0 model} & Free: $\alpha_{\bot}$, $\alpha_{\|}$, $f$, $b_1$, $b_{\partial}$.
&
Free: $\alpha_{\bot}$, $\alpha_{\|}$,  $f$, $b_1$, $b_{\partial}$.
\\
\multirow{2}{*}{} & 
Fixed: $\Sigxy$. For matter, $b_1=1$. &
Fixed: $\Sigsm$ (subsequently all the damping terms, e.g., $\Sigdd$, $\Sigss$, $\Sigsd$). For matter, $b_1=1$.
\\
\hline
\multirow{2}{*}{EFT1 model} & Free:  $\alpha_{\bot}$, $\alpha_{\|}$, $\Sigxy$, $f$, $b_1$, $b_{\partial}$.
& Free: $\alpha_{\bot}$, $\alpha_{\|}$, $\Sigsm$, $f$, $b_1$, $b_{\partial}$. \\ 
\multirow{2}{*}{ } & Fixed: for matter, $b_1=1$. 
& Fixed: for matter, $b_1=1$.\\ 
\hline
\multirow{2}{*}{SBRS model} & Free: $\alpha_{\bot}$, $\alpha_{\|}$, $\Sigma_{\mathrm{fog}}$, $f$, $b_1$. &
Free: $\alpha_{\bot}$, $\alpha_{\|}$, $\Sigma_{\mathrm{fog}}$, $f$, $b_1$. \\
\multirow{2}{*}{} & Fixed: $\Sigma_{xy}$, $\Sigma_{z}(= (1+f_{\rm fid})\Sigxy)$. For matter, $b_1=1$. &
Fixed: $\Sigxy$, $\Sigz(= (1+f_{\rm fid})\Sigxy)$. For matter, $b_1=1$.\\
\hline
\hline
\end{tabular}\label{tab:model}
\end{table*}

\subsubsection{Pre-reconstruction}
\begin{itemize}
\item{EFT0 model and EFT1 model\footnote{Labeling these two models EFT0 and EFT1 might be somewhat misleading in the sense that in this paper we mostly stick to the leading order correction and the leading correction coming from EFT shows up at the one loop level. The main part of the derivation and analysis in these models is the resummation of IR modes, here performed following the prescription in~\citet{zv16}. In that sense, perhaps more accurate labels would be IR0 and IR1, but for simplicity we will stick to the prior labels.}} 

As explained in detail in Appendix~\ref{sec:EFTderivation}, these models include the leading nonlinear bias contribution to the difference power spectrum, including the linear bias $b_1$ and the leading derivative bias $b_\partial$, which is proportional to $k^2$, while neglecting quadratic bias $b_2$, tidal bias $b_{s^2}$, and additional higher-order (loop) terms. By ignoring the higher-order terms, these models are designed to account only for the scale-dependent bias effect but not to model the nonlinear shifts of the BAO feature. 
Explicitly,
\begin{align}
\delta \hat{P}(k', \mu', z)
=\biggl((b_1+f\mu^2)^2+b_{\partial}\frac{k^2}{k^2_L} (b_1 + f\mu^2)\biggr)C_G^2\,\delta P_L(k, z),  \label{eq:EFT_pre}
\end{align}
where $f$ is the growth rate, $z$ is the redshift of the observation, and $C_G^2$ denotes the Gaussian damping factor of BAO wiggles~\citep{Eisenstein_etal_07a,Matsubara2008}:
\begin{align}
C_G^2 (k,\mu,z) = \exp\Big[-k^2(1-\mu^2)\Sigxy^2/2 - k^2 \mu^2 (1+f)^2\Sigxy^2/2\Big].\label{eq:cG}
\end{align}
The difference from the earlier results is the 
Galilean invariant (GI) derivation presented in Appendix~\ref{sec:EFTderivation}, where the exponential term is stable under coordinate transformation and can not be removed away by shifting the frame.
The leading order prediction for the value of 
the dispersion 
(see again Appendix~\ref{sec:EFTderivation}) is 
\begin{eqnarray}
&&\Sigxy^2(z) = 2 \int \frac{dp}{6\pi^2} (1-j_0(qp)) P_L(p, z), \label{eq:Sigxy}
\end{eqnarray}
with $q = 110\hMpc$ which is approximately the BAO scale~\footnote{Damping scales are often calculated with $\Sigxy^2(z) = \frac{1}{3}\int \frac{dp}{2\pi^2} P_L(p, z)$ only accounting for displacement correlation at a single position as in~\citet{Matsubara2008}. Such value would be $\sim 15\%$ larger relative to Eq.~\ref{eq:Sigxy}. 
}
and $j_0$ is the $\ell=0$ spherical Bessel function of the first kind.
With $\alpha_{\|}$, $\alpha_{\bot}$, $f$, $b_1$ and $b_{\partial}$ as free parameters (we fix $k_L = 1\,\ihMpc$), it yields the EFT0 model. This is our default fitting model.\\

With $\Sigxy$ as an additional free parameter, it yields the EFT1 model. This yields the results for the $C_G^2$ factor mathematically equivalent to the earlier models
(see e.g. ~\citep{Eisenstein_etal_07a,Matsubara2008}). \\

Alternatively, the role of the $b_\partial$ bias parameter above can  also be understood more generally than purely derivative bias and can be thought of as an estimate of the importance of all higher order terms (including biasing and dynamics corrections). In other words, when the performance of the model starts to significantly rely on the $b_\partial$ parameter, this can be understood as a sign of the importance of higher order corrections and that we are approaching the regime where these corrections should be added to the model above.\\

\item{SBRS model}

This fitting model adopts the nominal fitting formula for pre-reconstruction power spectrum that is widely used in the literature~\citep[e.g.,][]{Seo_etal_16}. 

\begin{align}
\delta \hat{P}(k', \mu', z)=\big(b_1+f \mu^2\big)^2 C_G^2 \, F(k,\mu,\Sigma_{\mathrm{fog}}) \, \delta P_L(k, z),\label{eq:SBRS_pre}
\end{align}
where the nonlinear 
Finger-of-God effect is modeled by 
\begin{align}
F(k, \mu, \Sigma_{\mathrm{fog}}) = \frac{1}{(1+k^2 \mu^2 \Sigma_{\mathrm{fog}}^2)^2}.
\label{eq:FOG}
\end{align}
For this model, $\alpha_{\|}$, $\alpha_{\bot}$, $\Sigma_{\mathrm{fog}}$, $f$ and $b_1$ are free parameters. 

For the damping term of all SBRS cases, unlike the EFT models, we fix $f=\ffid$ in the Gaussian damping factor and the effect of $f$ varies only the amplitude:
\begin{align}
C_G^2 (k,\mu,z) = \exp\Big[-k^2(1-\mu^2)\Sigxy^2/2 - k^2 \mu^2 (1+\ffid)^2\Sigxy^2/2\Big],\label{eq:cGsbrs}
\end{align}
To compare with the EFT0 model, we fix the damping scale $\Sigxy$ using Eq. \ref{eq:Sigxy}.
\end{itemize}

\begin{figure*}
\centering
\includegraphics[width=0.45\linewidth]{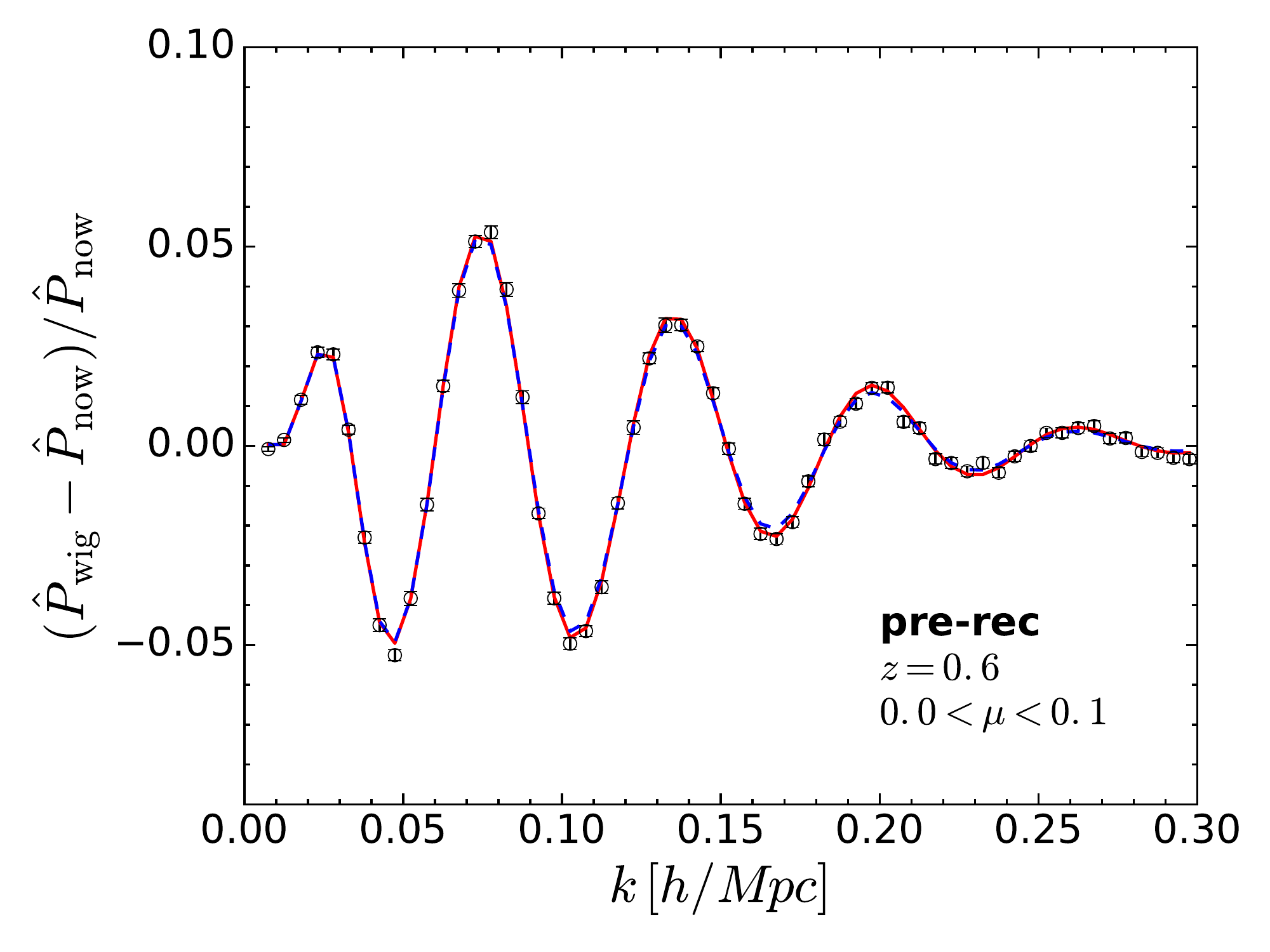}
\includegraphics[width=0.45\linewidth]{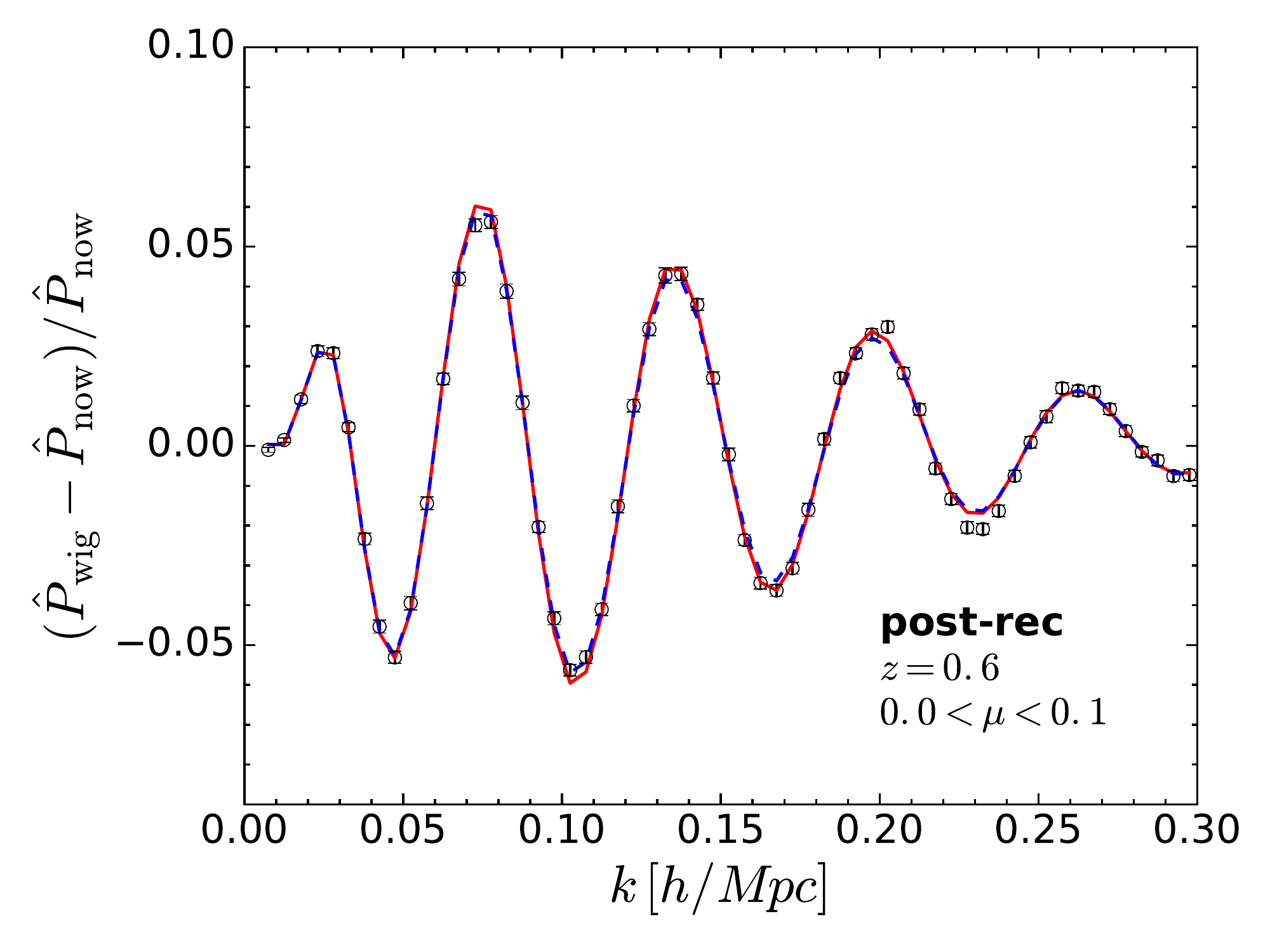}
\includegraphics[width=0.45\linewidth]{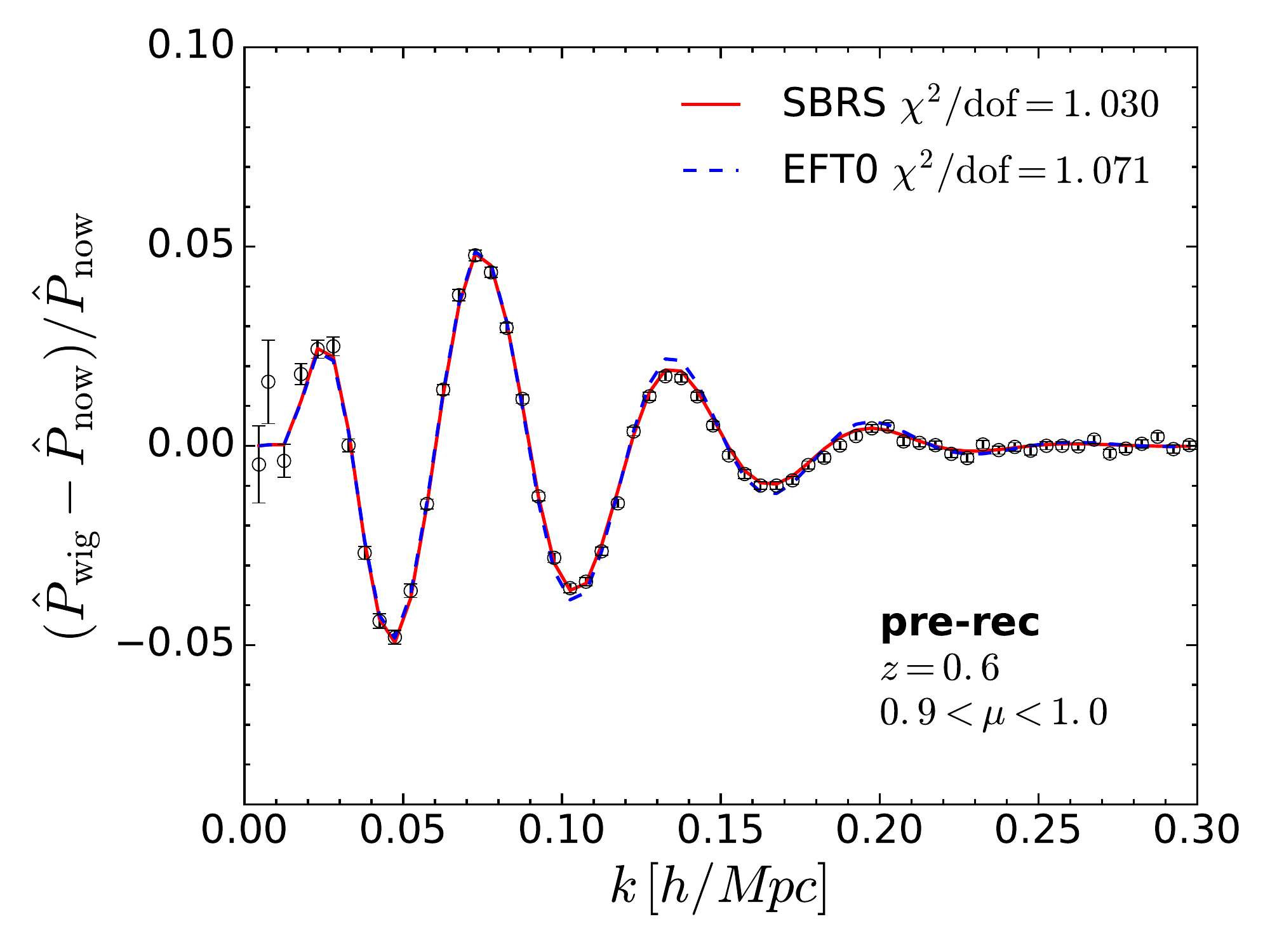}
\includegraphics[width=0.45\linewidth]{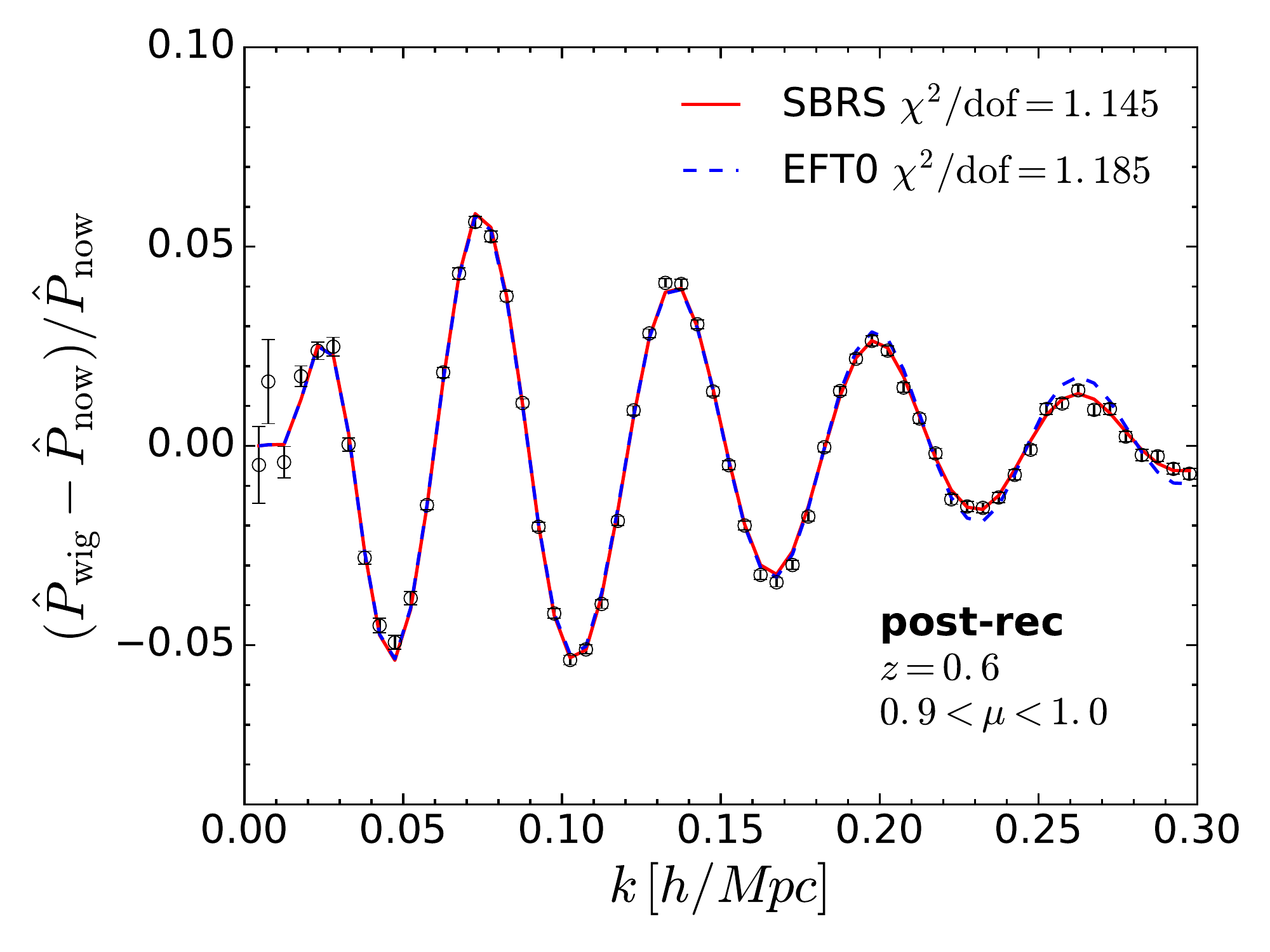}
\caption{Average difference matter power spectra compared with best fit models from the SBRS model and the EFT0 model. We calculate the mean matter power spectrum $\hat{P}_{\text{wig}}-\hat{P}_{\text{now}}$ and $\hat{P}_{\text{now}}$ over $40$ realizations at redshift $0.6$. The over-plotted error bars are very small due to the sample variance cancellation. \textit{Left panels:} pre-reconstruction. \textit{Right panels:} post-reconstruction. \textit{Top panels:} modes of power spectrum perpendicular to the line-of-sight. \textit{Bottom panel:} modes along the line-of-sight. Red solid line shows the fitting from the SBRS model, and blue dashed line shows the result from the EFT0 model. The reduced $\chi^2$ from fitting all the modes of the averaged power spectrum is shown for each model. Here the data points are from re-binned ($10\, \mu$ bins) power spectra to reduce fluctuations for clarity while the best fit models and the reduced $\chi^2$ are from fitting the original $100\,\mu$ bins.
}\label{fig:Pdiff}
\end{figure*}

\subsubsection{Post-reconstruction for `Rec-Iso'}
The post-reconstruction EFT fitting models are derived in Appendix~\ref{subsec:eftpost}. To summarize, 
\begin{itemize}
\item{EFT0 model}
\begin{flalign}
\delta \hat{P}_{\mathrm{rec}}(k', \mu', z) =  \delta P_{\rm dd} - 2\delta P_{\rm sd}+ \delta P_{\rm ss}, 
\end{flalign}
where models for individual terms, i.e., $\delta \Pdd$ for the difference power spectrum from the displaced galaxy particles,  $\delta \Pss$ for the difference power spectrum from the displaced reference particles, $\delta \Psd$ for the difference cross power spectrum, are written as:
\begin{flalign}
&\delta \Pdd(k', \mu', z) = e^{-k^2(1+f(2+f)\mu^2)\Sigdd^2} \left[ \Bigl(b_1 - S(k) \right. &&\nonumber\\  
+ f\mu^2 
&\left. \bigl(1-S(k)\bigr) \Bigr)^2  + b_{\partial} \Bigl(b_1 - S(k) + f\mu^2 (1-S(k))\Bigr)\frac{k^2}{k^2_L}\right] \delta P_L(k, z) 
, && \nonumber\\ 
&\delta \Psd(k', \mu', z) = -e^{-k^2(1+f\mu^2)\Sigsd^2}\left[ b_1 - S(k) + f \mu^2 (1-S(k))\right.  && \nonumber \\
&\left. \qquad \qquad \qquad + \frac{1}{2} b_{\partial} \frac{k^2}{k^2_L}\right] S(k)  \delta P_L(k, z) 
, &&\nonumber \\
&\delta \Pss(k', \mu', z) = e^{-k^2 \Sigss^2}S(k)^2  \delta P_L(k, z) 
. && \label{eq:EFT_post}
\end{flalign}
We ignore higher order perturbation terms.
The damping parameters are 
\begin{flalign}
\Sigma_{dd}^2(q, z) = &\frac{1}{3}\int \frac{dp}{2\pi^2} (1-j_0(qp))(1-S(p))^2 P_L(p, z). && \label{eq:Sigma_dd}\\
\Sigma_{sd}^2(q, z) = &\frac{1}{3}\int \frac{dp}{2\pi^2} \biggl( \frac{1}{2} \Bigl(S(p)^2 + (1-S(p))^2 \Bigr) -\nonumber\\
&j_0(qp)(1-S(p)) S(p) \biggr) P_L(p, z). && \\
\Sigma_{ss}^2(q, z) = &\frac{1}{3}\int \frac{dp}{2\pi^2} (1-j_0(qp))S(p)^2 P_L(p, z).\label{eq:Sigma_ss}
\end{flalign}
We derive these damping scales by setting $q = 110\hMpc$ and fix them in the EFT0 model. For the EFT0 model, we therefore have $\alpha_{\|}$, $\alpha_{\bot}$, $f$, $b_1$ and $b_{\partial}$ as free parameters as in the pre-reconstruction EFT0 model. 

We also prepare a model with a moderate freedom to perturb the damping scales. Rather than varying $\Sigdd$, $\Sigsd$, and $\Sigss$ individually, we allow $\Sigsm$ in $S(k)$ to vary; this will perturb nonlinear damping factors in a coherent way through Eqs, \ref{eq:Sigma_dd} -- \ref{eq:Sigma_ss}, but as a caveat it will also modulate the amplitude in an angle-dependent way through $S(k)$ in Eqs. \ref{eq:EFT_post}. With $\alpha_{\bot}$, $\alpha_{\|}$, $f$, $b_1$ and $b_{\partial}$, and $\Sigsm$ as free parameters, we call this the EFT1 model even though this model has a different parameter freedom from the pre-reconstruction EFT1 model.  \\
\end{itemize}

\begin{itemize}
\item{SBRS model}\\

We adopt the approximate fitting formula derived in \citet{Seo_etal_16} for `Rec-Iso' convention. 
The model is the same as the pre-reconstruction case (in Eq.,\ref{eq:SBRS_pre}) but with a modified Kaiser factor and damping scale:
\begin{align}
\big(b_1+f\mu^2\big)^2 \rightarrow \big[b_1+f \mu^2(1-S(k))\big]^2.
\end{align}\label{eq:recSigsm}
We fix $\Sigxy^2$ to be $2\times \Sigdd^2$ derived from Eq.~\ref{eq:Sigma_dd} to construct $C_G$ of Eq.~\ref{eq:cGsbrs}. Free parameters are $\alpha_{\|}$, $\alpha_{\bot}$, $\Sigma_{\mathrm{fog}}$, $f$ and $b_1$. 
\end{itemize}

\subsubsection{Post-reconstruction for `Rec-Cohn'}

\begin{itemize}
\item{EFT0 and EFT1 models}

The EFT0 and EFT1 models for `Rec-Cohn' are very similar to the `Rec-Iso' cases, except for modifications in $\delta \Psd$ and $\delta \Pss$:
\begin{flalign}
&\delta \Psd(k', \mu', z) = -e^{-k^2(1+f(2+f)\mu^2)\Sigsd^2}\left[ b_1 - S(k) + f \mu^2 (1-S(k))\right.  && \nonumber \\
&\left. \qquad \qquad \qquad+ \frac{1}{2} b_{\partial} \frac{k^2}{k^2_L}\right] \bigl(1+f\mu^2\bigr)  S(k)  \delta P_L(k, z) 
, &&\nonumber \\
&\delta \Pss(k', \mu', z) = e^{-k^2(1+f(2+f)\mu^2) \Sigss^2}(1+f\mu^2)^2 S(k)^2  \delta P_L(k, z) 
. && \label{eq:EFT_postcohn}
\end{flalign}
Damping parameters are derived from Eqs~\ref{eq:Sigma_dd}--\ref{eq:Sigma_ss}. 
Again, free parameters for the EFT0 model are $\alpha_{\bot}$, $\alpha_{\|}$, $f$, $b_1$ and $b_{\partial}$; for the EFT1 model, free parameters are $\alpha_{\bot}$, $\alpha_{\|}$, $f$, $b_1$ and $b_{\partial}$, $\Sigsm$.
\\
\item{SBRS model}\\
We derived the corresponding fitting formula from \citet{Seo_etal_16} Appendix A1 with $\lambda_s=f$ in their Eq A13. Then, the fitting model becomes identical to Eq.~\ref{eq:SBRS_pre} without $1-S(k)$, i.e., without the explicit anisotropic amplitude dependence on the smoothing scale. 

The smoothing scale affects the damping scales through Eq.~\ref{eq:Sigma_dd}. Again, we fix $\Sigxy^2$ to be $2\times \Sigdd^2$ derived from Eq.~\ref{eq:Sigma_dd} and construct $C_G$ of Eq~\ref{eq:cGsbrs}.
Free parameters are $\alpha_{\|}$, $\alpha_{\bot}$, $\Sigma_{\mathrm{fog}}$, $f$ and $b_1$. 
\end{itemize}

\vspace{0.5cm}
In all models discussed above, we fix $b_1=1$ for matter. 

\begin{figure*}
\centering
\includegraphics[width=0.45\linewidth]
{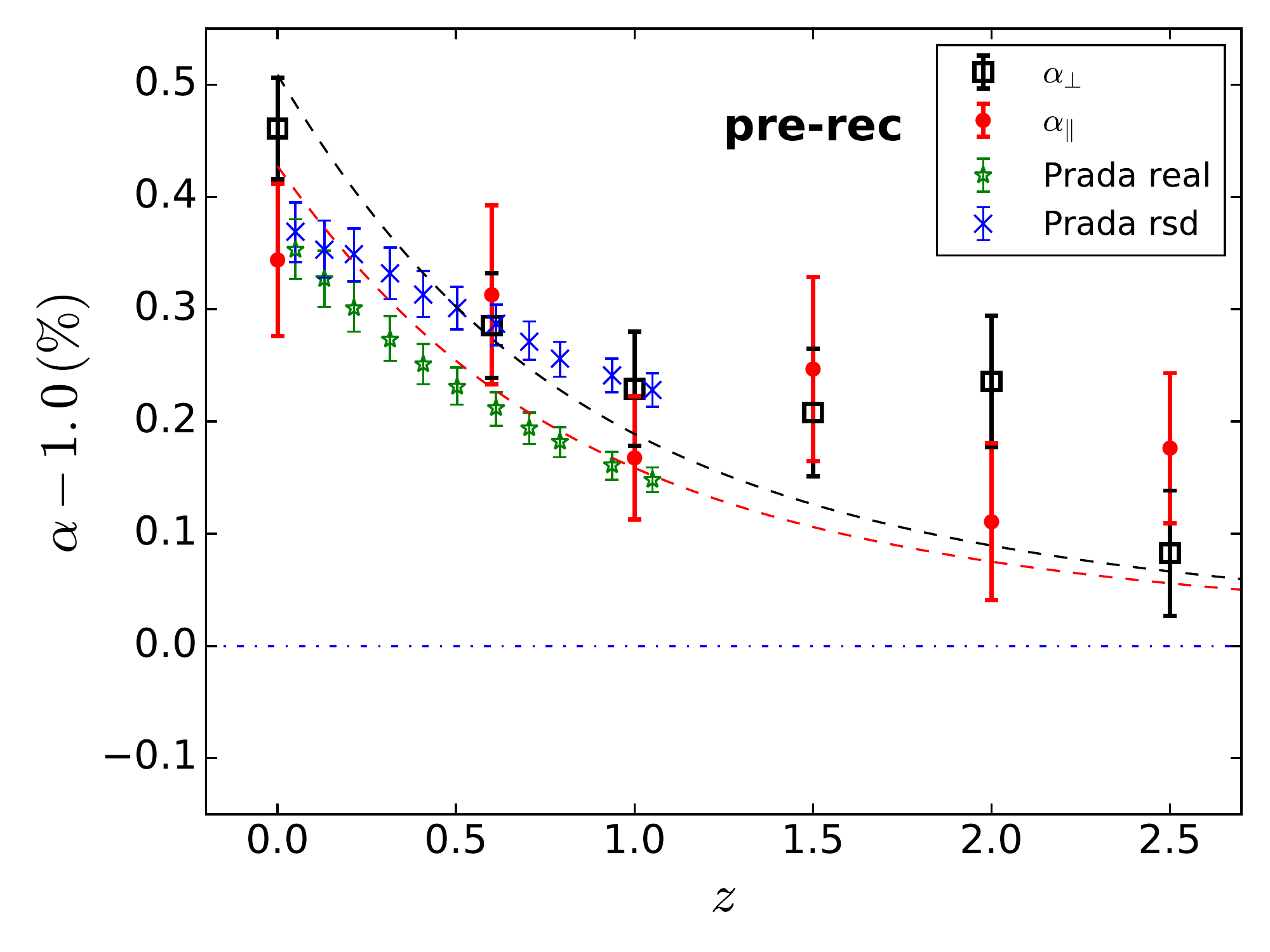}
\includegraphics[width=0.45\linewidth]{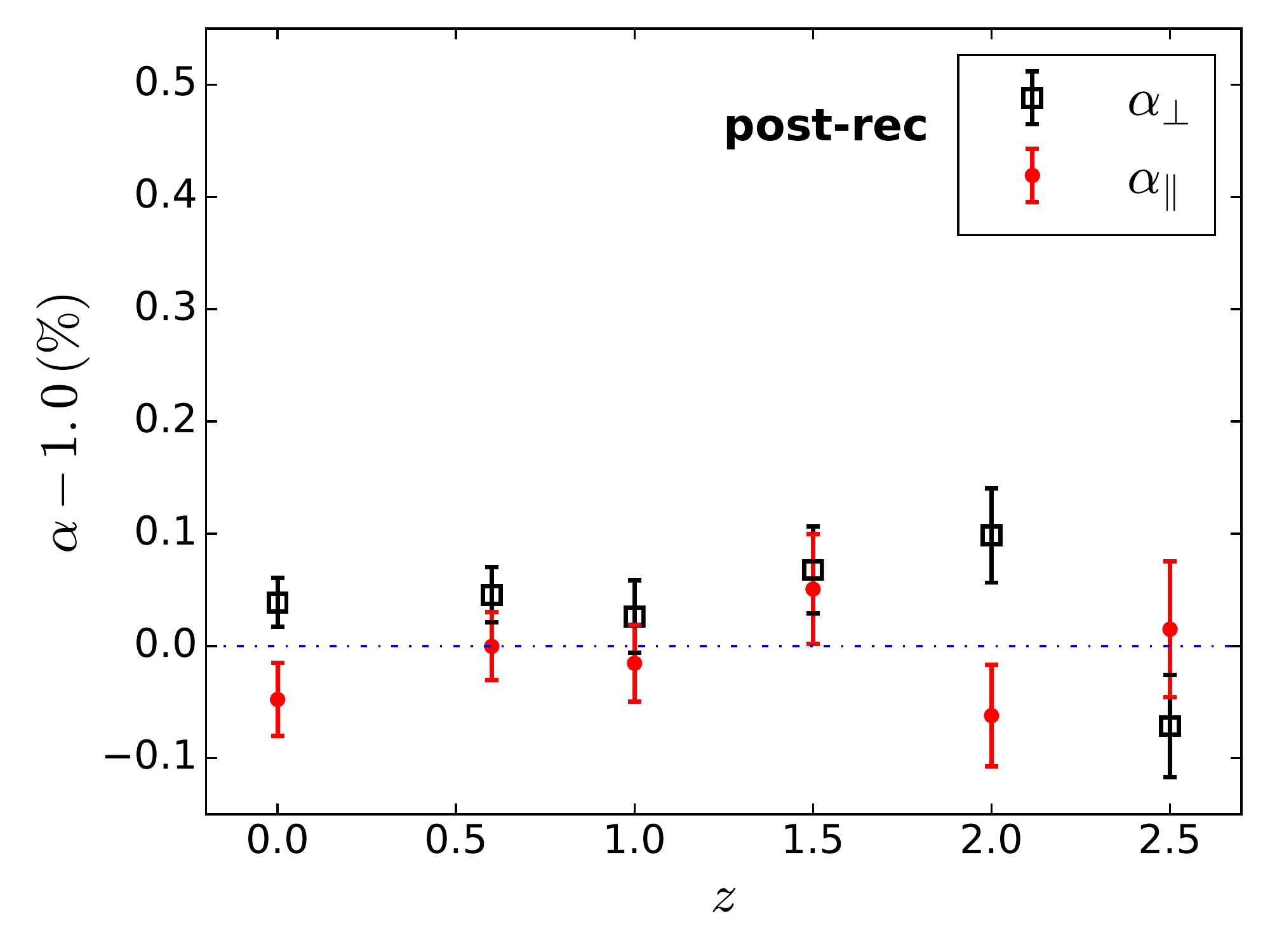}
\caption{BAO scale shifts in the matter distribution. We use our default model, the EFT0 model. The black and red data points show the means and the standard deviations of the means of $\alpha_{\bot}$ (black empty squares) and $\alpha_{\|}$ (red solid circles) over $40$ individual best fits. \textit{Left panel:} pre-reconstruction. \textit{Right panel:} post-reconstruction. We use the same scale on the left and right panels in order to ease the comparison between pre- and post-reconstruction. The values can be found in Table \ref{tab:alpha_statistics}.
Before reconstruction, nonlinear BAO shift has a decreasing trend as redshift increases, as expected. For this EFT0 fitting model, we find that the BAO shifts on $\alpe$ and $\alpa$ are very similar. We fit the trend for pre-reconstruction $\alpe$ as $0.51\%[D(z)/D(0)]^2$ (black dashed line), and $0.43\%[D(z)/D(0)]^2$ (red dashed line) for pre-reconstruction $\alpa$. The fitting lines tend to lie below the points due to the restricted form of $[D(z)/D(0)]^2$. We also overplot the isotropic BAO scale measurements reported in \citet{Prada2016} in real space (green stars) as well as the ones in redshift space (blue crosses). We slightly shifted their points horizontally (by 0.05) to avoid overlaps with our results. Our precision for the matter distribution is worse than \citet{Prada2016} probably because we use 1\% subsample of the matter particles. Their redshift-space points can be compared to our estimates of $\alpe$ and $\alpa$ while taking into account differences in the fitting formula and the shift estimator as well as in the fiducial cosmology. The right panel shows that reconstruction dramatically decreases the shift to within $\pm 0.1\%$.}\label{fig:alpha_ZV}
\end{figure*}

\subsection{MCMC fitting}

To find the best fitting parameters, we use \textbf{emcee} \footnote{http://dan.iel.fm/emcee/current/} \citep{Foreman-Mackey_etal_13} which implements Markov Chain Monte Carlo (MCMC) affine-invariant ensemble sampler \citep{Goodman_Weare_10}. We set each parameter to have a uniform prior distribution. For pre-reconstruction power spectrum, in EFT models the prior distributions of parameters are $\al\in(0.9,\, 1.1)$, $\Sigxy\in (0,\, 100)$, $f\in (0, \, 4)$, $b_1\in(0,\, 6)$, and $b_{\partial}\in (-1000, \, 1000)$. In the SBRS model, we add $\Sfog\in(0, \,20)$ while other priors are the same as EFT0 model. For post-reconstruction, $\Sigsmsm \in(0, \, 600)$ in EFT1 model.    

\section{Results}\label{sec:results}

In this section, we present various tests of the BAO scale systematics as well as other fitting parameters such as growth rate $f$, the nonlinear damping scales, and scale-dependent bias as a function of galaxy/halo bias and redshift utilizing the mitigated sample variance. As a comparison to the results we will present, 
\citet{Alam2016} assumes BAO systematics error of $0.3\%-0.5\%$ on the BAO scales for both pre- and post-reconstruction density fields of the BOSS DR12 galaxy data.

\def\aperp{\alpha_{\perp}}
\def\apar{\alpha_{\parallel}}

\subsection{BAO scales in matter distribution}\label{subsec:BAOmatter}

We first exclude the galaxy/halo bias effect by focusing on clustering of sub-sampled matter particles in redshift-space. The transverse BAO scale $\aperp$ should closely represent what we would observe in the absence of redshift-space distortions. 

Figure \ref{fig:Pdiff} shows the averaged difference matter power spectrum from our simulations $\dP=\Pwig-\Pnow$ at selected redshift output z=0.6 with the corresponding best fit models. 
The corresponding reduced $\chi^2$ of the averaged power spectrum from using our two fitting models is between $1-1.2$, which is much better than the model performance evaluated in \citet{Seo_etal_16} despite much smaller error we are dealing with in this paper.
It is likely that the models work better due to the removal of additive nonlinear effects that enter both wiggle and nowiggle simulations. The average of the reduced $\chi^2$ of individual pairs is even closer to unity.

\subsubsection{Pre-reconstruction}

Figure \ref{fig:alpha_ZV} shows the percentage offset between the measured BAO scale and the input BAO scale as a function of redshift for the matter field. We used the EFT0 model (Eq.~\ref{eq:EFT_pre}, Table \ref{tab:model}) to fit individual two dimensional difference power spectra of 40 pairs and averaged the best fits. The error bars represent the error of the mean and are derived from the dispersions among the 40 individual best fits, rescaled by $1/\sqrt{40}$. Left panel shows the pre-reconstruction data and the right panel shows the post-reconstruction data. Black and red points show the BAO scale in transverse $\aperp$ and along the line of sight $\apar$, respectively.  In the fitting, we fix the parameter $\Sigxy$ based on Eq.~\ref{eq:Sigxy} and set linear bias $b_1=1.0$, but keep the nonlinear scale-dependent bias effect $b_{\partial}$ as a free parameter in order to parametrically account for nonlinear structure growth.

From the figure, we find the expected increase of the nonlinear BAO shift with decreasing redshift before reconstruction: we find $0.1-0.2\%$ of shift at $z=2.5$ and $\sim 0.3\%$ shift at $z=0.6$. Previous literature typically showed greater BAO shifts for spherically averaged power spectra in redshift space relative to the ones in real space~\citep{Seo_etal_10,Prada2016}, implying greater shift on $\apar$ relative to $\aperp$ in redshift space. In the results with the EFT0 model, we do not observe an obvious excess shift on $\apar$ relative to that on $\aperp$.

\begin{table*}
\centering
\caption{The pre and post-reconstruction BAO scale shifts for matter in redshift space using the EFT0 model. We used the `Rec-Iso' scheme for the density field reconstruction.  By default, we calculate the mean and dispersions of $\alpha_{\bot}$ (the transverse BAO scale) and $\alpha_{\|}$ (the line-of-sight BAO scale) of the 40 best fits to the individual difference power spectrum $(P_{\text{wig}}-P_{\text{now}})$. We derive the error of the mean by dividing the dispersion by $\sqrt{40}$. As a comparison, we also show the best fit and the corresponding 68\% likelihood error using the averaged difference power spectrum inside parentheses. The reduced $\chi^2$ shown in the last column is using the best fit of the averaged power spectrum. }\label{tab:alpha_statistics}
\begin{tabular}[c]{m{1.4cm} m{1.8cm} m{1.8cm} m{1.8cm} m{1.8cm} m{1.5cm}}
\hline
\hline
Redshift & $\alpha_{\bot}-1\,(\%)$ & $\sigma_{\alpha_{\bot}}(\%)$ & $\alpha_{\|}-1\,(\%)$ & $\sigma_{\alpha_{\|}}(\%)$ & $\chi^2/\text{d.o.f.}$\\
\hline
\multicolumn{6}{c}{Pre-reconstruction}\\
\hline
0.0 & 0.461 (0.467) & 0.045 (0.042) & 0.344 (0.336) & 0.068 (0.067) & 1.126\\
0.6 & 0.285 (0.284) & 0.047 (0.043) & 0.313 (0.318) & 0.080 (0.067) & 1.071 \\
1.0 & 0.229 (0.229) & 0.051 (0.044) & 0.168 (0.167) & 0.055 (0.067) & 1.056\\
1.5 & 0.208 (0.209) & 0.057 (0.049) & 0.247 (0.241) & 0.082 (0.069) & 1.133\\
2.0 & 0.236 (0.237) & 0.059 (0.051) & 0.111 (0.104) & 0.070 (0.070) & 1.030\\
2.5 & 0.083 (0.079) & 0.056 (0.057) & 0.176 (0.175) & 0.067 (0.075) & 1.079 \\
\hline
\multicolumn{6}{c}{Post-reconstruction}\\
\hline
0.0 & 0.039 (0.035) & 0.022 (0.023) & -0.048 (-0.041) & 0.033 (0.035) & 1.305\\
0.6 & 0.046 (0.047) & 0.025 (0.027) & 0.000 (-0.003) & 0.030 (0.038) & 1.185\\
1.0 & 0.026 (0.022) & 0.032 (0.031) & -0.015 (-0.010) & 0.034 (0.041) & 1.066\\
1.5 & 0.068 (0.065) & 0.039 (0.036) & 0.051 (0.047) & 0.049 (0.045) & 1.102\\
2.0 & 0.098 (0.098) & 0.042 (0.042) & -0.062 (-0.066) & 0.045 (0.051) & 1.070 \\
2.5 & -0.071 (-0.074) & 0.045 (0.047) & 0.015 (0.013) & 0.060 (0.056) & 1.086\\
\hline
\end{tabular}
\end{table*}

Based on perturbation theory, we expect the nonlinear BAO shift to be proportional to $[D(z)/D(0)]^2$ where $D(z)$ is the growth factor~\citep[e.g.,][]{Eisenstein_etal_07a,Padmanabhan_etal_09,Sherwin12}. We overplot the fitting models for $\aperp$ and $\apar$ in Figure \ref{fig:alpha_ZV}, which are $\aperp-1=(0.51\pm 0.05)\% [D(z)/D(0)]^2$ and $\apar=0.43\pm 0.07\% [D(z)/D(0)]^2$.
This is somewhat greater than the predictions/measurements from \citet{Sherwin12,PadBAOshift,Seo_etal_10} which derived $\alpha-1 \sim 0.3\% [D(z)/D(0)]^2$ for real-space matter. If we return to our real space $\alpha$ that we do not show explicitly in the paper, we derive $\aperp - 1 = (0.42 \pm  0.05)\% [D(z)/D(0)]^2$ and  $\apar -1 = (0.37 \pm 0.08)\% [D(z)/D(0)]^2$, which are slightly lower than the redshift-space quantities and closer to the values from the previous literature. That is, although we naively expect that our redshift-space $\aperp$ would represents the real-space $\aperp$ and $\apar$, the EFT0 model produces rather isotropic shifts in redshift-space that are slightly greater than the real-space counterparts.

The greater shift we observe relative to the previous literature could be partly due to the difference in fiducial cosmology models;  the listed references assumed WMAP-like cosmology~\citep{KomatsuWMAP5} with 10-20\% lower $\Omega_m$ than our fiducial model which is based on ~\citep{Planck_16}. \citet{Prada2016} used mocks based on the Planck13 cosmology and measured real-space $\alpha-1 = 0.37\% [D(z)/D(0)]^2$, which is closer to our estimates. We overplot their results for the isotropic $\alpha$ in real as well as in redshift space in the left panel of Figure \ref{fig:alpha_ZV}.

\begin{figure*}
\includegraphics[width=0.45\linewidth]{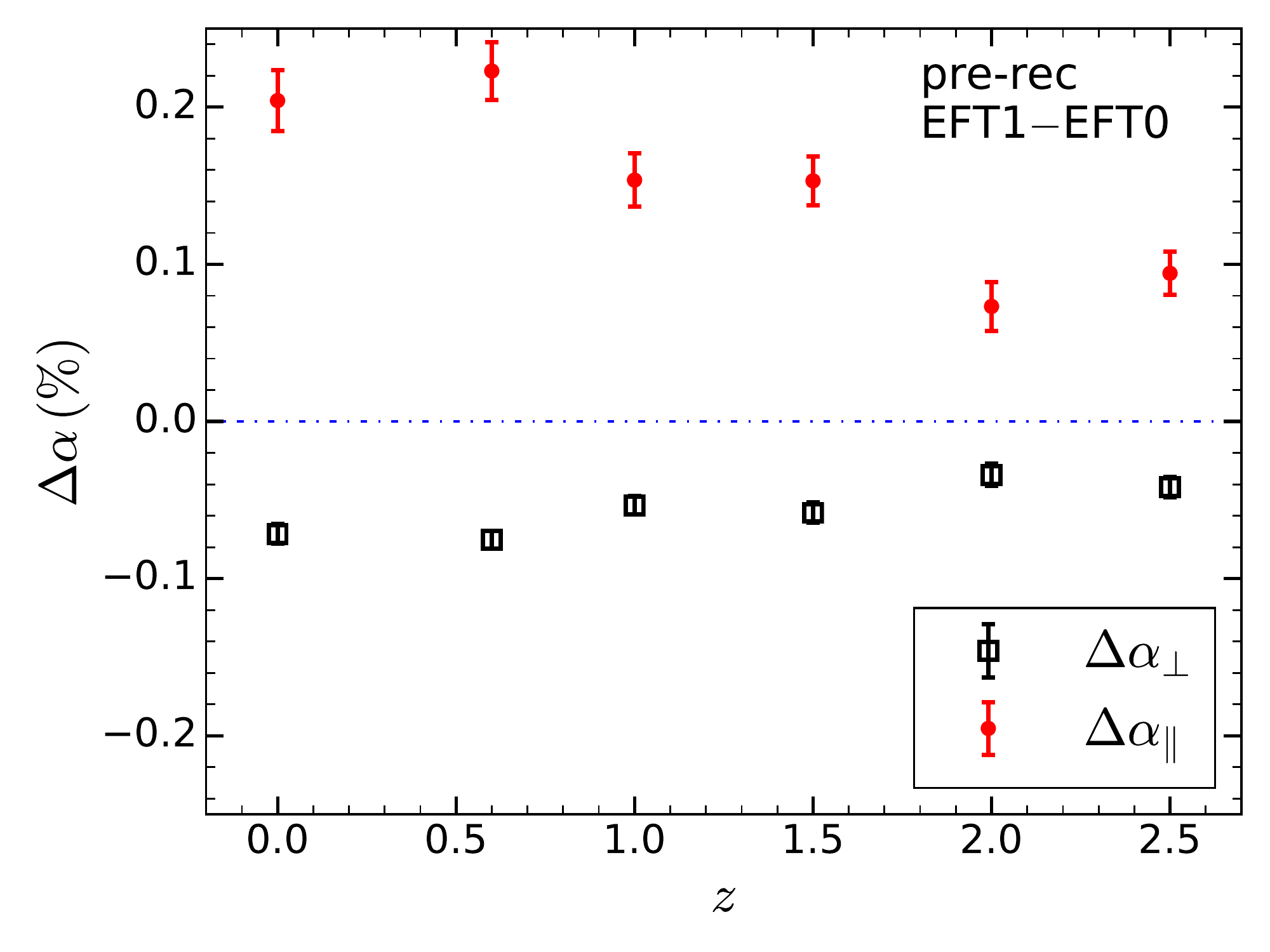}
\includegraphics[width=0.45\linewidth]{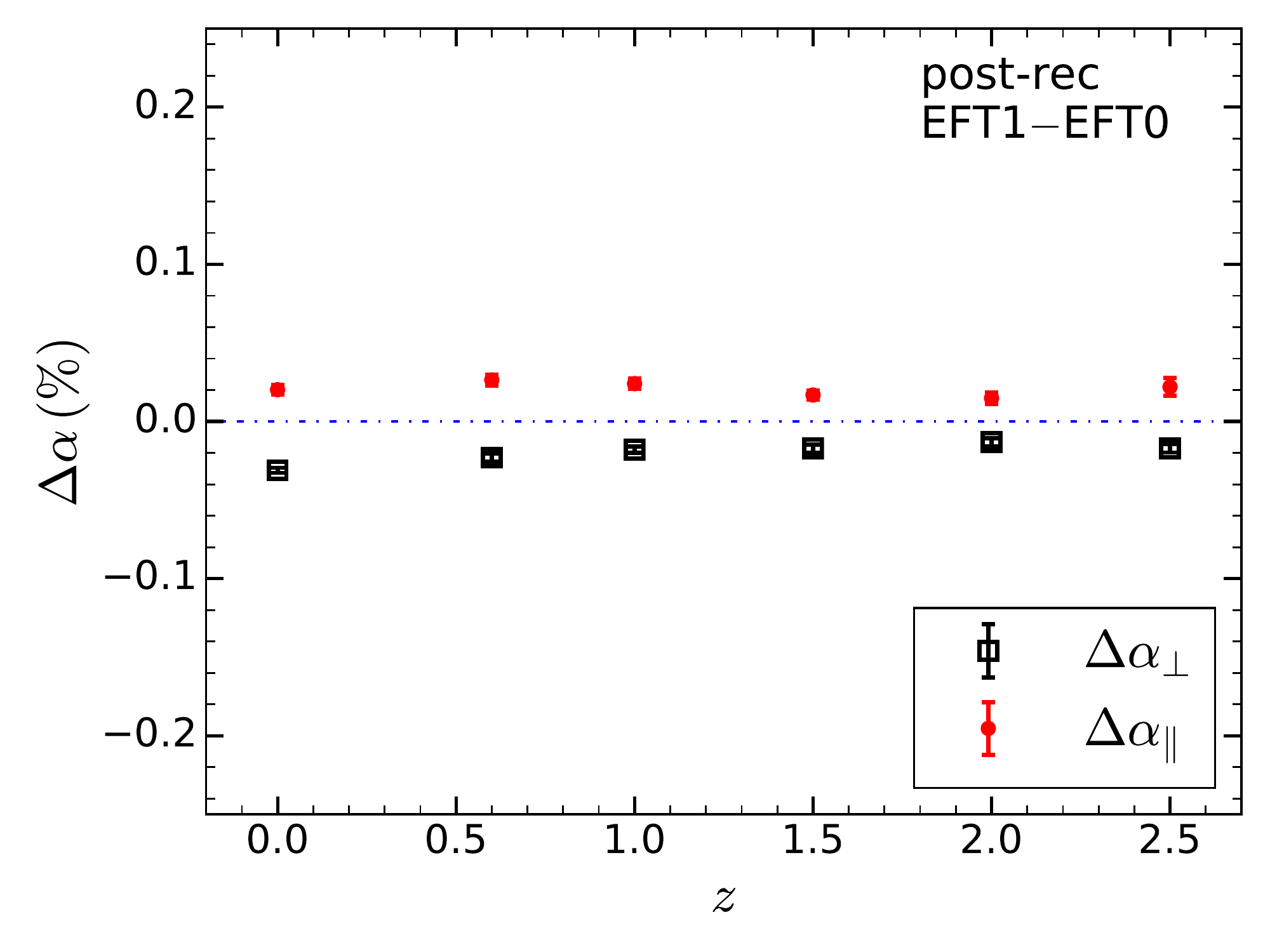}
\includegraphics[width=0.45\linewidth]{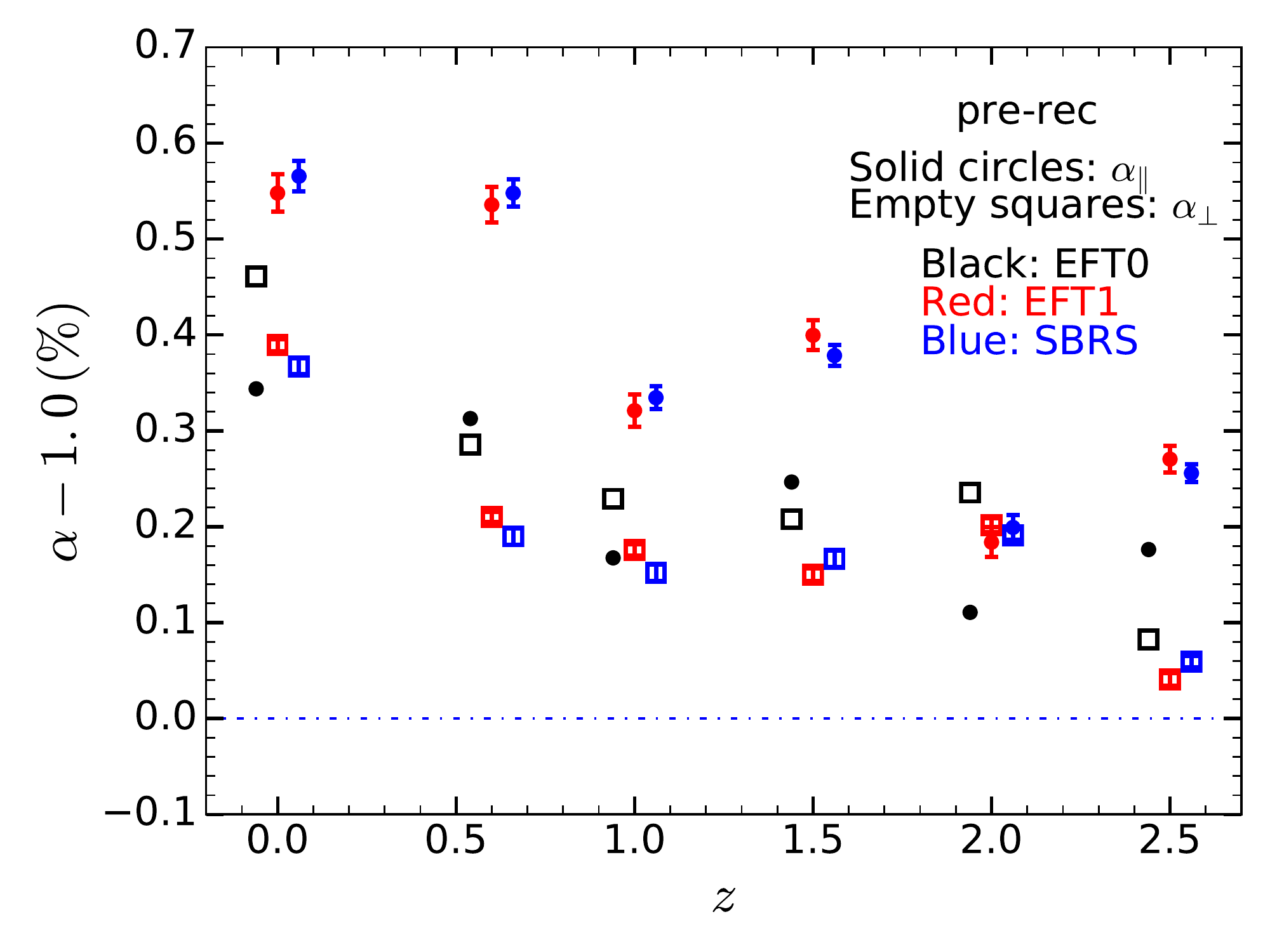}
\includegraphics[width=0.45\linewidth]{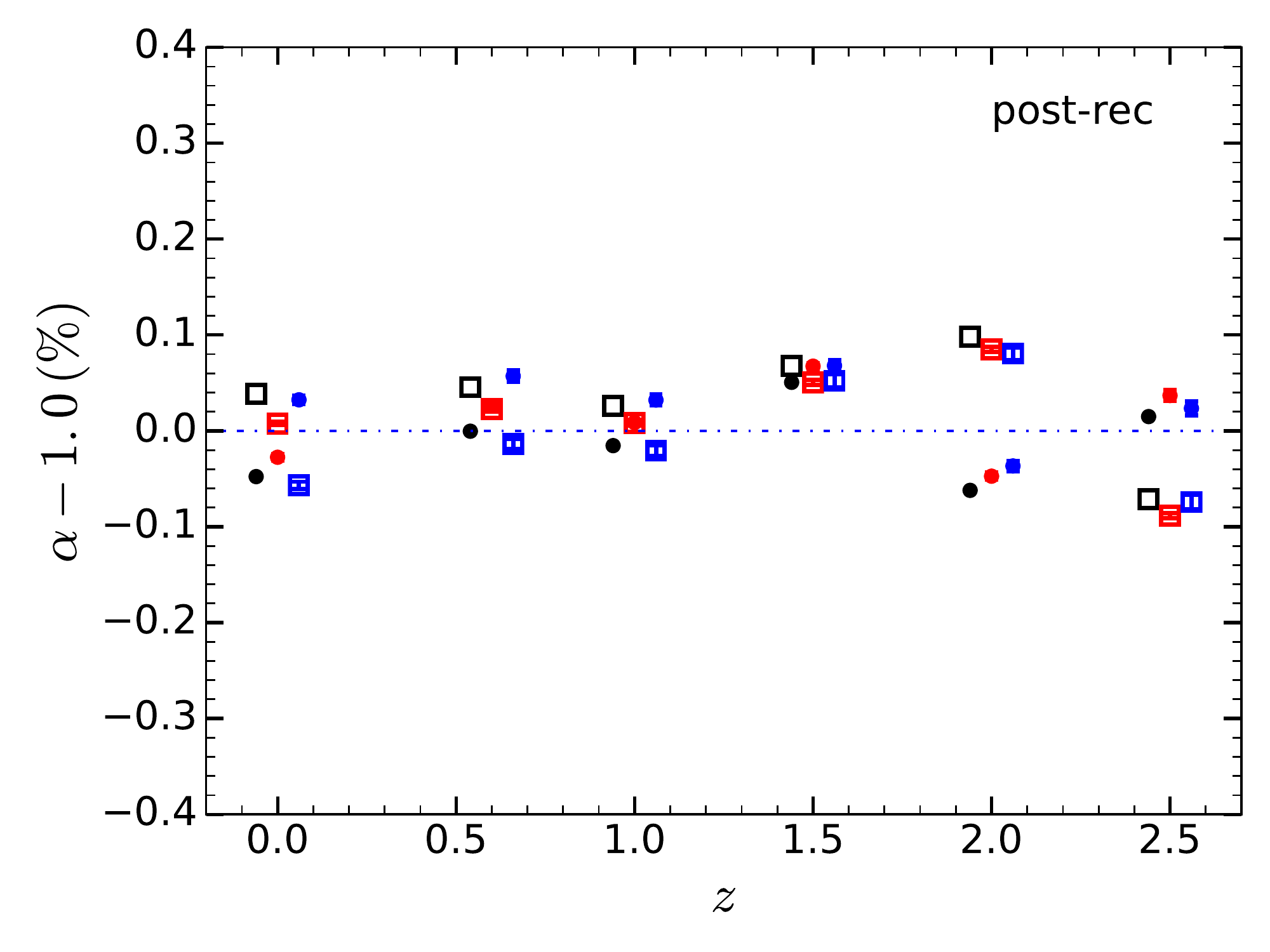}
\includegraphics[width=0.45\linewidth]{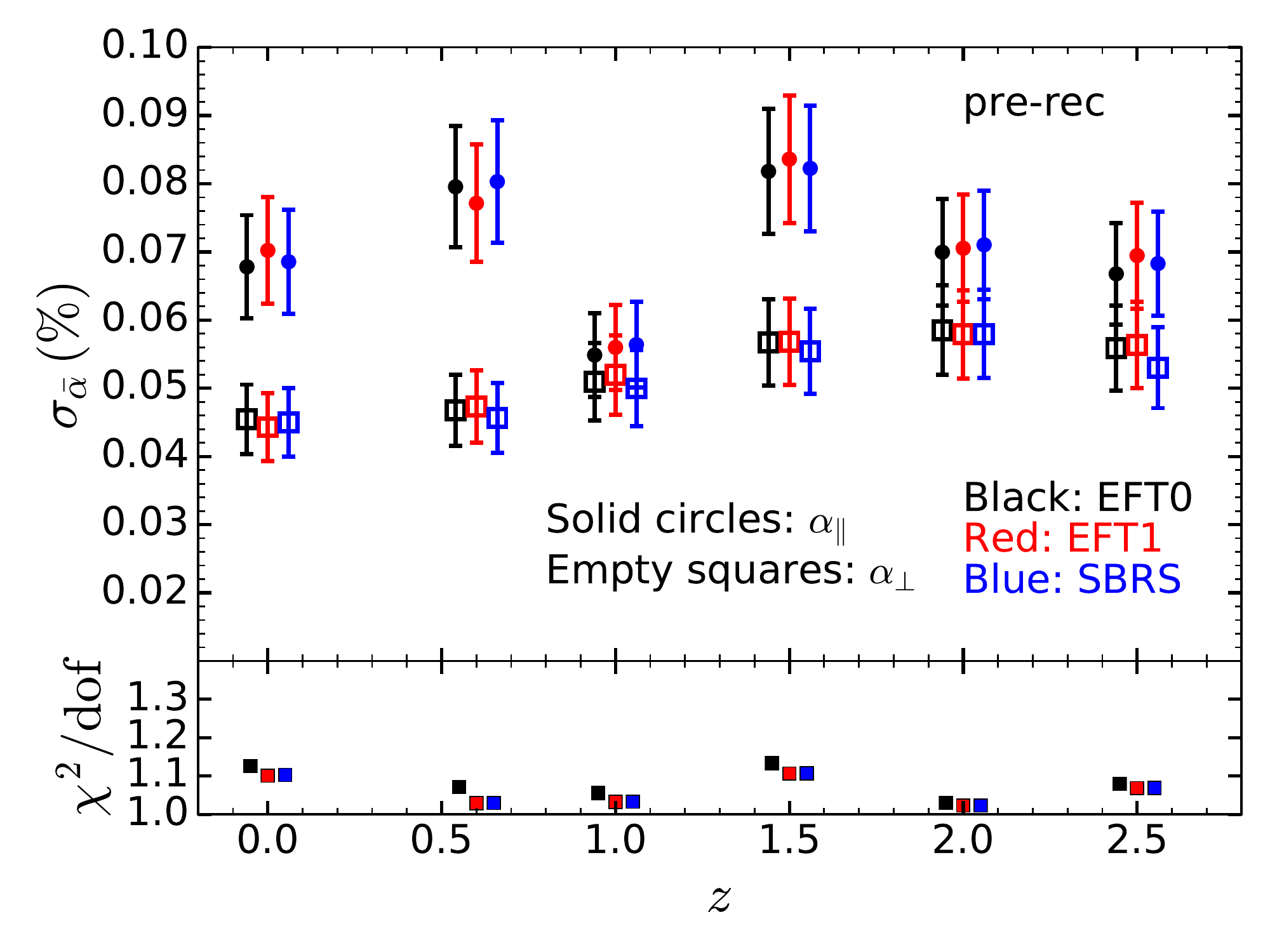}
\includegraphics[width=0.45\linewidth]{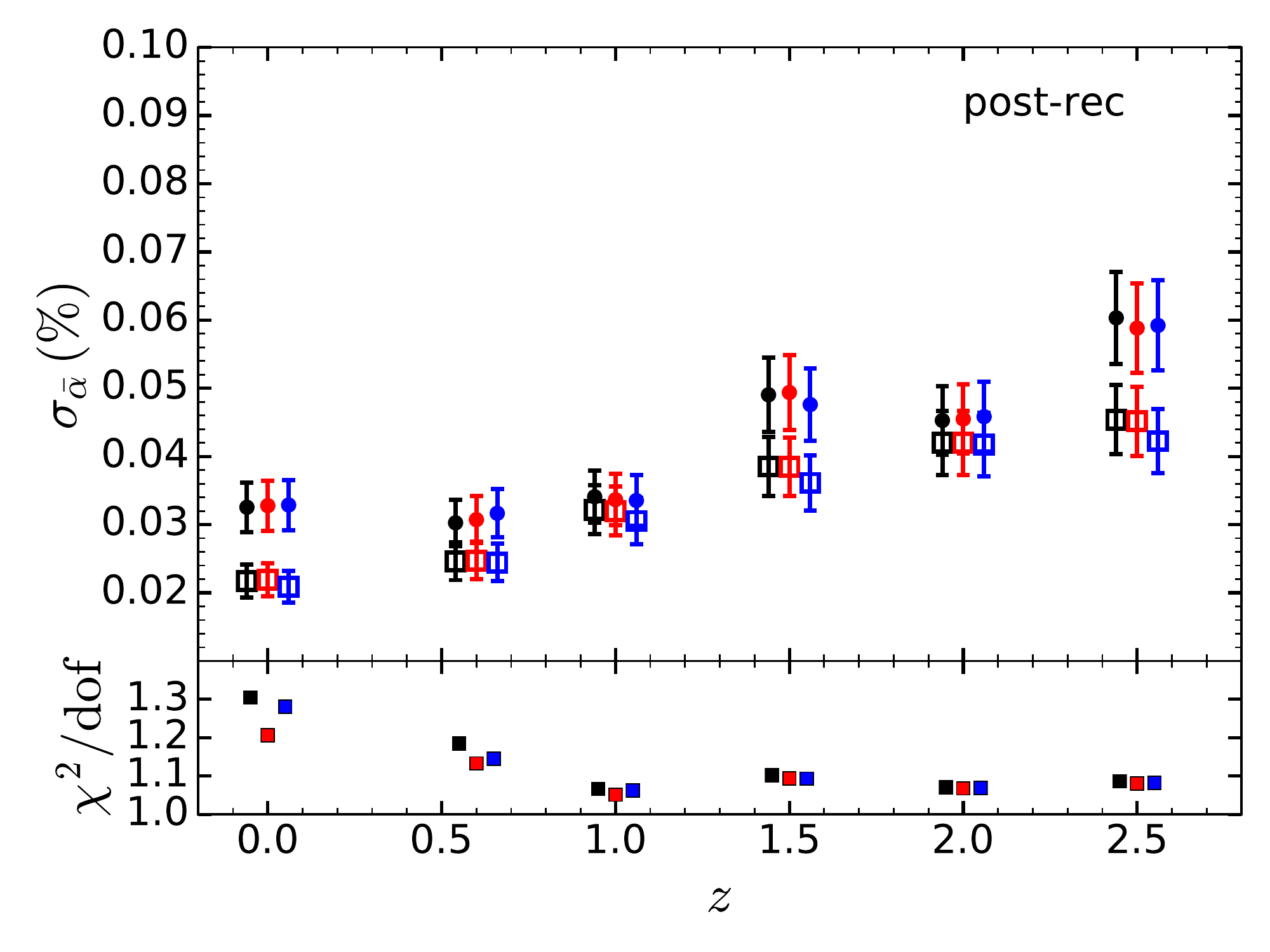}
\caption{The effects of different fitting models on BAO scale systematics. We focus on matter power spectrum. \textit{Left panels:} pre-reconstruction. \textit{Right panel:} post-reconstruction. The top and the middle panels show the effect on the mean $\alpha$, and the bottom panels show the effect on the errors on $\alpha$.
Empty square points denote for $\alpe$ and solid circular points for $\alpa$. Top panels show relative $\alpha$ {\it differences} between the EFT1 and the EFT0 model (i.e., $\Delta \alpha =\alpha_{\rm EFT1}-\alpha_{\rm EFT0}$) with errors derived from the dispersions of the $\alpha$ differences divided by $\sqrt{40}$; the error bars on these differences are much smaller than the nominal errors on $\alpha$'s in the bottom panels because the fitting models are applied to the same sets of simulations, mitigating most of the sample variance effect by taking differences.
The middle panels compare the BAO shifts from the EFT0 model (black points) with those from the EFT1 (red points) and the SBRS models (blue points). The error bars on the EFT1 and the SBRS points represent the dispersion on the mean $\Delta \alpha$. No error bars are plotted for the EFT0 model as we focus on the statistical significance of the differences. Note that the EFT1 and the SBRS models systematically produce greater $\apar$ than $\aperp$ before reconstruction.
The bottom panels compare nominal errors on the mean $\alpha$ (i.e., not errors on $\Delta \alpha$ ) using the EFT0 model (black points, from Table~\ref{tab:alpha_statistics}) with errors using other fitting models. The error bars on $\sigma_{\alpha}$ are derived using Gaussian prediction of error on error which is $\sigma_\alpha/\sqrt{2\times 40}\sim 0.11\sigma_\alpha$. We find that the dependence of $\sigma_\alpha$ on fitting models is small, well within their expected statistical dispersions. The bottom panels also include the reduced $\chi^2$ of the mean power spectra for different fitting models. }
\label{fig:alpha_ZVS}
\end{figure*}

\begin{figure*}
\includegraphics[width=0.45\linewidth]{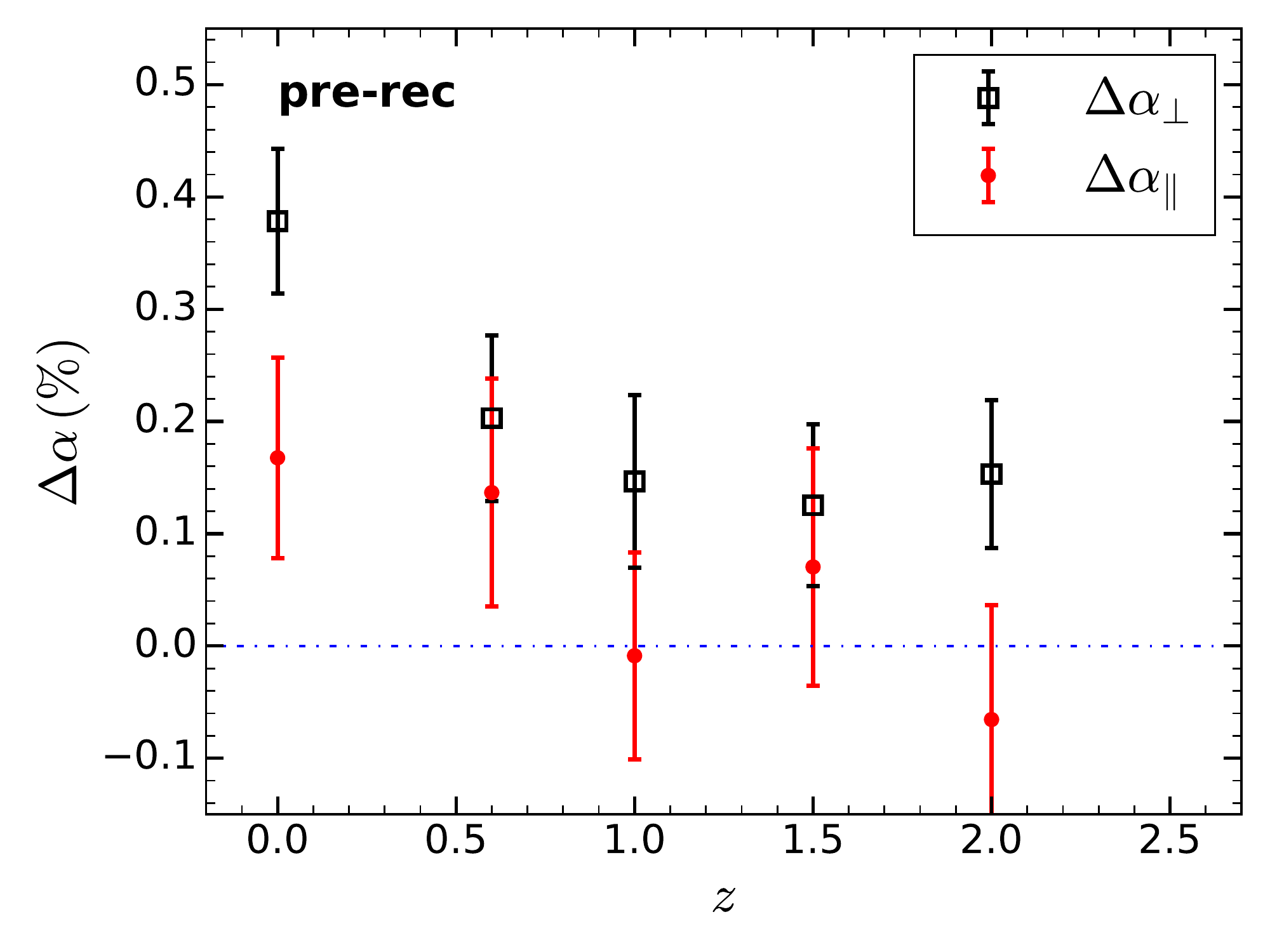}
\includegraphics[width=0.45\linewidth]{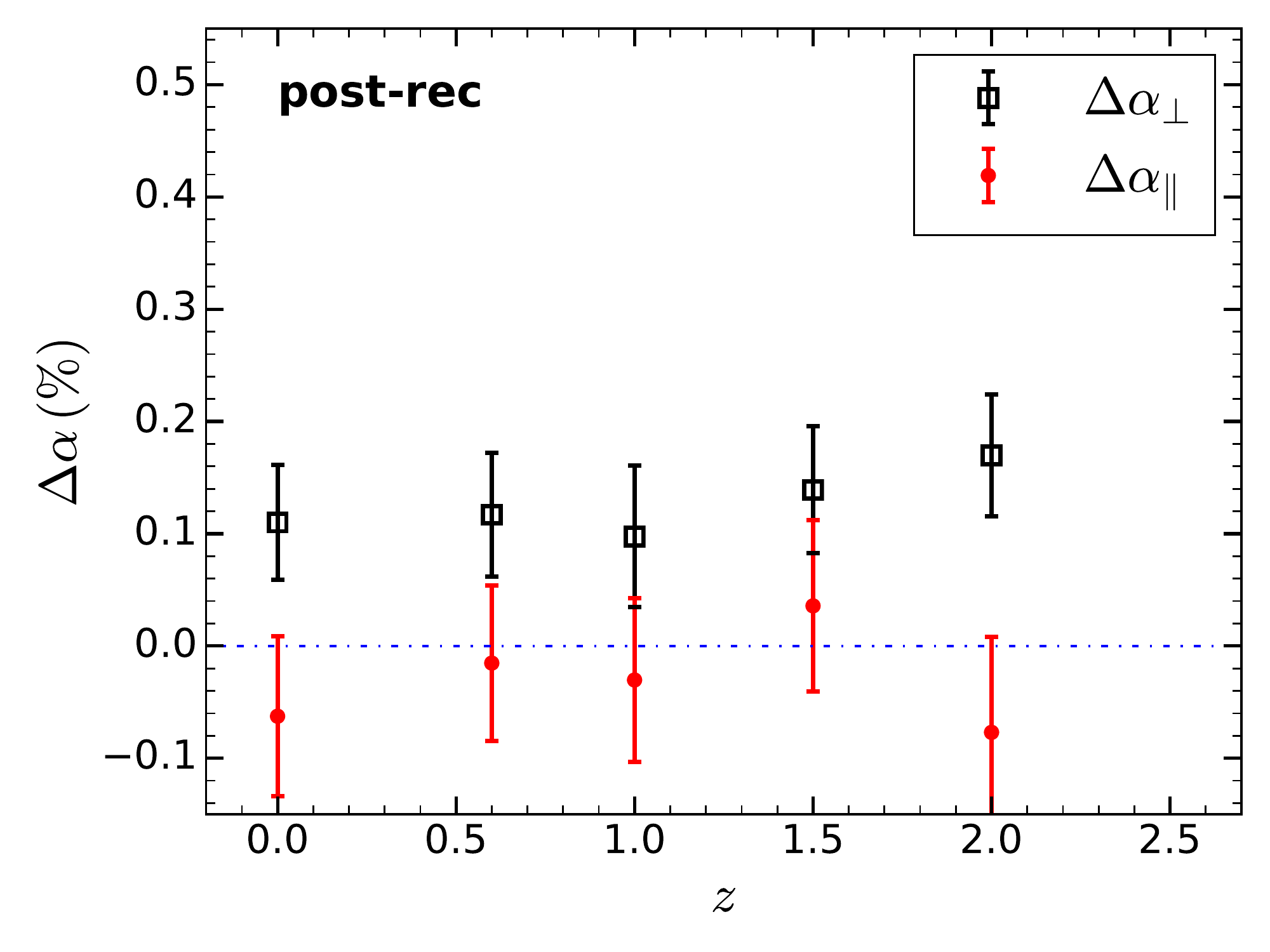}
\includegraphics[width=0.45\linewidth]{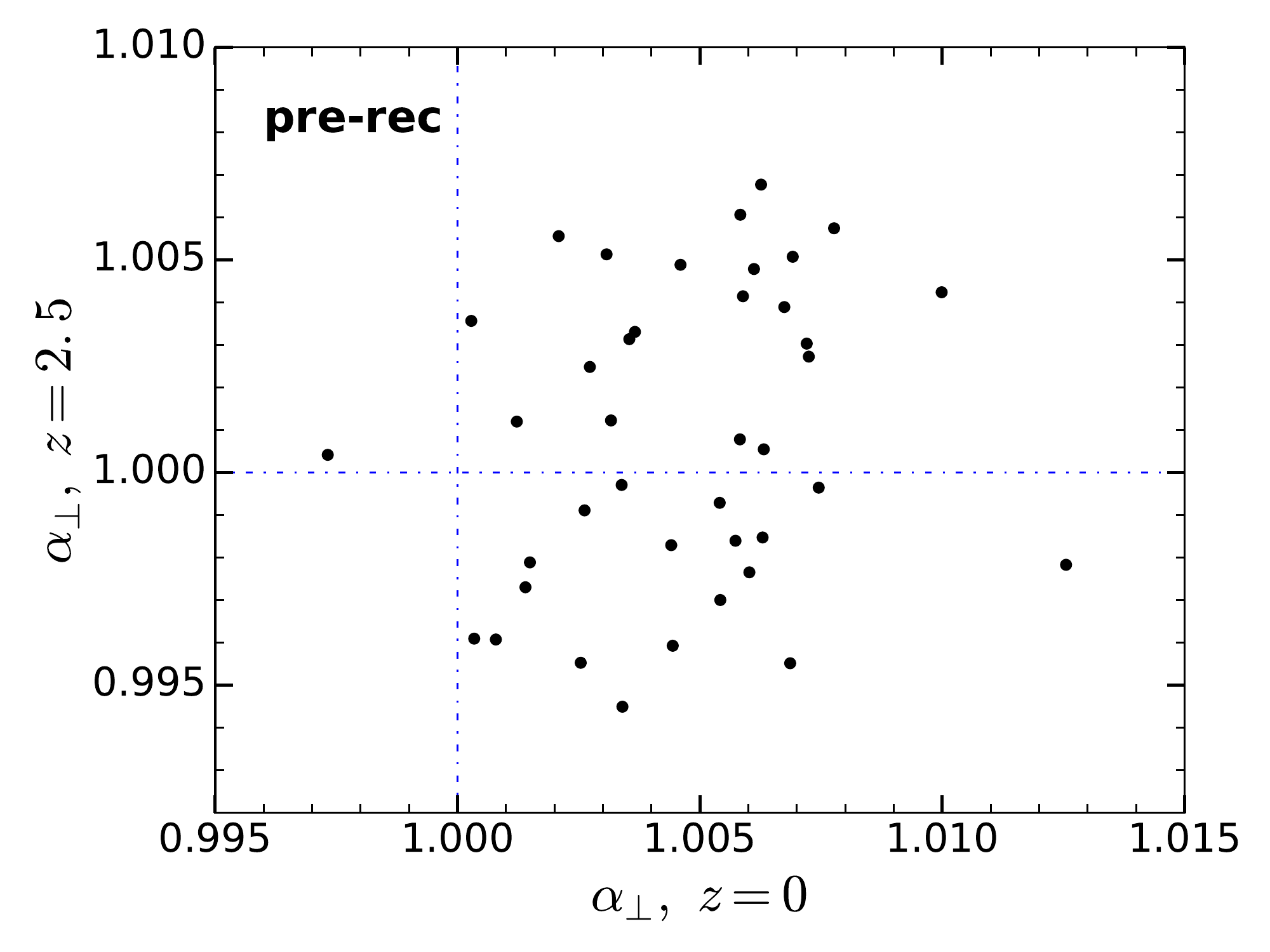}
\includegraphics[width=0.45\linewidth]{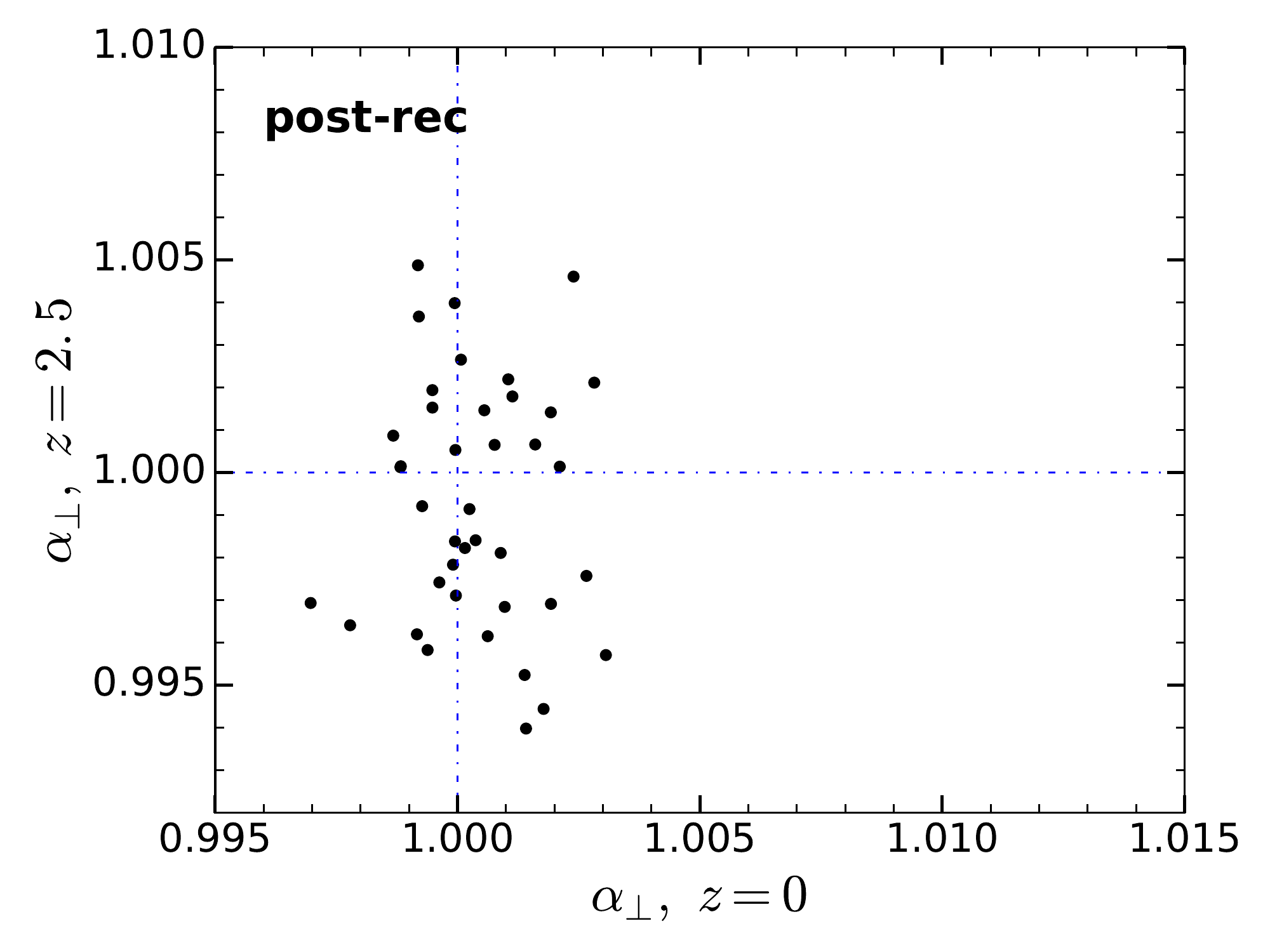}
\caption{Top panels: the mean and dispersions of difference of $\alpha_{\bot}$ and $\alpha_{\|}$ in Figure \ref{fig:alpha_ZV} between redshift $2.5$ and a lower redshift to further remove any remaining correlated variance between high redshift and low redshift. The errors increase in comparison to Figure \ref{fig:alpha_ZV}, implying that there are little correlations remaining between redshifts.   Lower panels: scattering of $\alpha_{\bot}$ at $z=0$ vs at $z=2.5$. These again show that once cosmic variance is reduced by the sample variance cancellation, there is almost no correlation between the nonlinear BAO shifts between high and low redshifts before and even after reconstruction.
}\label{fig:scatteralpha}
\end{figure*}

The discrepancy appears to be sourced by the BAO fitting models to a greater extent. 
Figure \ref{fig:alpha_ZVS} shows dependence of the results on fitting models. We take the differences in the best fit $\alpha$'s when varying fitting models and/or fitting parameters and calculate the means and dispersions among 40 differenced $\alpha$'s. The resulting errors (the dispersions divided by $\sqrt{40}$) are significantly reduced in this difference statistics since we are using the identical simulations for different fitting models. 
Top panels compare the EFT0 case with the EFT1 model (i.e. the same as EFT0, but with a free damping parameter $\Sigxy$, see Table \ref{tab:model}). The result implies that if we let $\Sigxy$ vary, we would find slightly smaller $\aperp$ (by $< 0.1\%$) and larger $\apar$ (by $\sim 0.1-0.2\%$), giving larger $\apar$ than $\aperp$ progressively at lower redshift as shown in the middle panel. That is, if we used the EFT1 model, we would find shifts on $\aperp$ that are consistent with our real-space $\alpha$'s and systematically larger shifts for $\apar$ than $\aperp$ as shown in the middle panels, which would agree with the previous literature that showed the spherically averaged $\alpha-1$ being greater in redshift space than in real space.

The middle panels overplot the EFT1 and the SBRS results with the EFT0 result from Figure~\ref{fig:alpha_ZV}. In order to show the significance of the difference between the models, we remove the error bars on the reference EFT0 model and denote the errors on the mean differences (as shown in the top panels) on the EFT1 and the SBRS points. We find that the SBRS model (with a fixed $\Sigxy$ but a free $\Sigma_{\rm fog}$ parameter) also returns anisotropic BAO shifts that are very similar to the EFT1 cases. With the EFT1 model, we derive fits of $\aperp -1 = (0.41\pm 0.04)\% [D(z)/D(0)]^2$ and $\apar-1=(0.72\pm 0.13)\% [D(z)/D(0)]^2$. With the SBRS model, we derive $\aperp -1 = (0.39\pm 0.04)\% [D(z)/D(0)]^2$ and $\apar-1=(0.73\pm 0.12)\% [D(z)/D(0)]^2$.
Top panels of Figure~\ref{fig:SBRSmodel} in Appendix explicitly show the difference between the EFT0 model and the SBRS model that is equivalent to the top panels of Figure~\ref{fig:alpha_ZVS}. If we fix $\Sigma_{\rm fog}=0$ for the SBRS model, we find that this tendency of anisotropic shift decreases (bottom panels of Figure \ref{fig:SBRSmodel} in Appendix). We suspect that the anisotropic shift could be related to the freedom in the damping scale such as $\Sigxy$ or a term that can mimic a perturbation in the damping scale, e.g., $\Sigma_{\rm fog} $.  For example, \citet{Seo_etal_10,Prada2016} found a greater shift in $\alpha$ in redshift-space than in real space indeed adopted a free parameter for the Finger-of-God effect or BAO damping scales.

\subsubsection{Post-reconstruction}

We next discuss post-reconstruction results in the right panels of Figure \ref{fig:alpha_ZV} -- \ref{fig:alpha_ZVS}. Figure \ref{fig:alpha_ZV} and Table \ref{tab:alpha_statistics} confirm that with density-field reconstruction, the nonlinear shift on the BAO scale substantially reduces to $< 0.1\%$. 
Overall, the fitting model dependence is also substantially reduced after reconstruction; the remaining difference is $\sim 0.02\%$ between the EFT0 and EFT1 models.  Between the SBRS model and the EFT0 model (Figure~\ref{fig:SBRSmodel}), the maximum difference reaches as much as $0.1\%$ at $z=0$, but decreases quickly at higher redshift. Note that the post-reconstruction EFT1 model is slightly different from the pre-reconstruction EFT1 model in that the former allows $\Sigma_{\rm sm}$ (instead of $\Sigxy$) to vary such that we can slightly perturb nonlinear damping factors as well as the redshift-space distortion effect away from the default EFT0 model. 

\subsubsection{Precisions and the reduced $\chi^2$}

In terms of errors, Table \ref{tab:alpha_statistics} and the bottom panels of Figure \ref{fig:alpha_ZVS} present errors on the mean of the measured $\alpha$'s using the EFT0 model. The table lists errors for the EFT0 model defined from the dispersions of the best fits as well as defined by $68\%$ of the posterior from MCMC using the mean power spectra. The two error estimates are similar within 20\%. Since we assume the covariance matrix of the difference power spectrum to be diagonal before and after reconstruction, we believe that the dispersions of the best fits are more conservative estimators than the $68\%$ of the posterior range from MCMC that depends more on the assumed covariance matrix. From Figure \ref{fig:alpha_ZVS}, one finds that the error is not necessarily monotonic with redshift. Given the cancellation of the sample variance, it is not straightforward to predict the errors as a function of redshift. The post-reconstruction errors decrease by $\sim 50\%$ at $z=0$ and $\sim 20\%$ at $z=2.5$ for $\aperp$; by $\sim 50\%$ at $z=0$ and $\sim 10\%$ at $z=2.5$ for $\apar$.  At high redshift the gain from reconstruction is small because the BAO signature is close to linear even without reconstruction.  As a caveat, these numbers are derived for matter with negligible shot noise given the default smoothing scale $\Sigsm$. 

The bottom panels of Figure \ref{fig:alpha_ZVS} also compare the EFT0 errors with errors using different fitting models. 
We find that the dependence of $\sigma_\alpha$ on fitting models is small and well within the expected Gaussian random dispersion~\footnote{The Gaussian prediction of dispersion on error is calculated from $\sigma_\alpha/\sqrt{2\times 40}\sim 0.11\sigma_\alpha$.}, despite the difference in the number of free parameters. We confirm that the $\alpha$ precision is robust against a freedom in the nonlinear damping scale, as shown in previous studies~\citep[e.g.,][]{Ata2017}. 

The bottom panels of Figure \ref{fig:alpha_ZVS} also include the reduced $\chi^2$ for different fitting models. The values   do not appear strongly dependent on the fitting models we test.
With one more parameter, the EFT1 model and the SBRS models tend to produce slightly smaller reduced $\chi^2$ than the EFT0 model. 
We find no clear advantage of the EFT1 model over the SBRS model, although 
the EFT1 model tends to work better than the others. 

\begin{figure*} 
\centering
\includegraphics[width=0.45\linewidth]{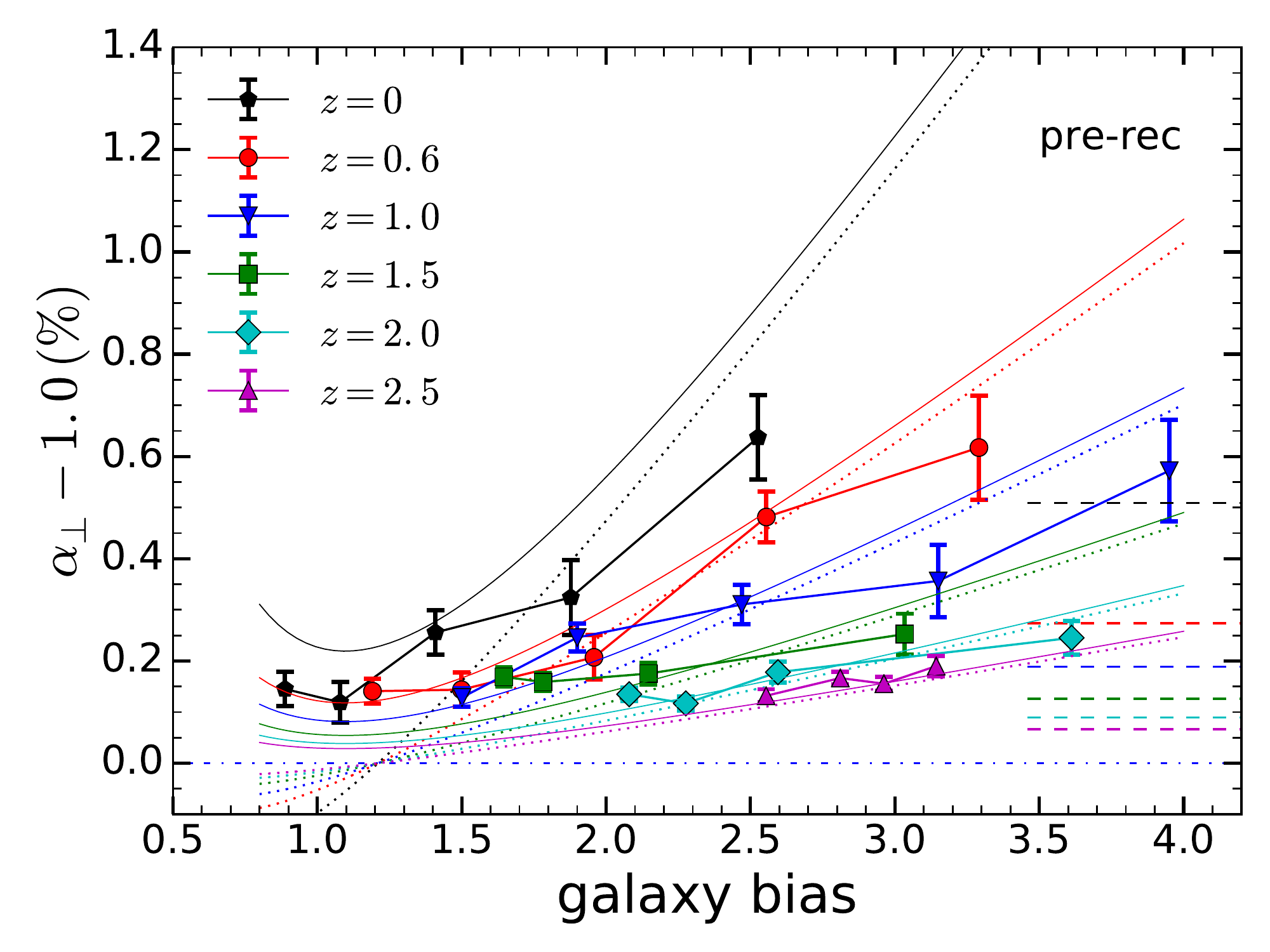}
\includegraphics[width=0.45\linewidth]{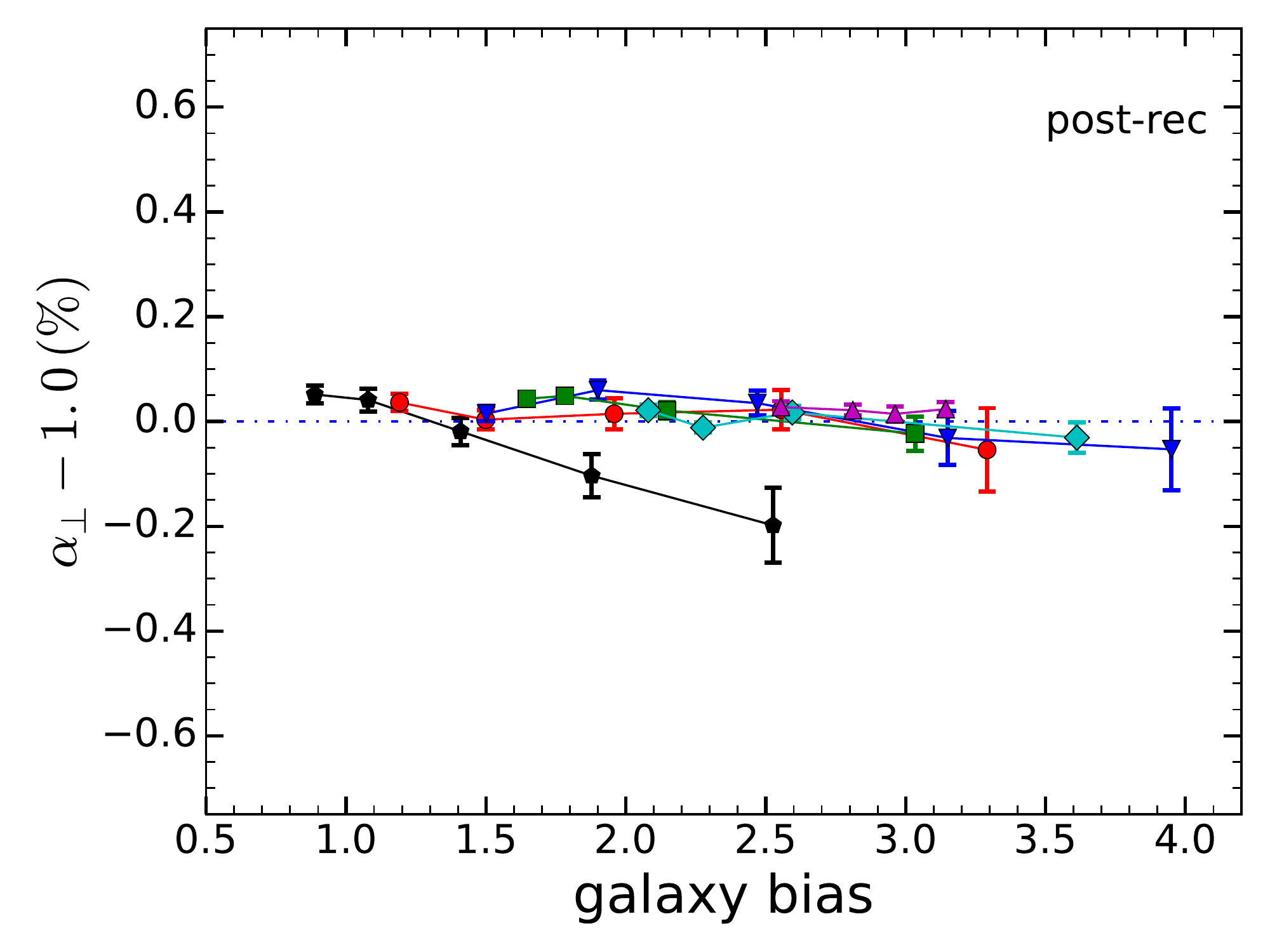}
\includegraphics[width=0.45\linewidth]{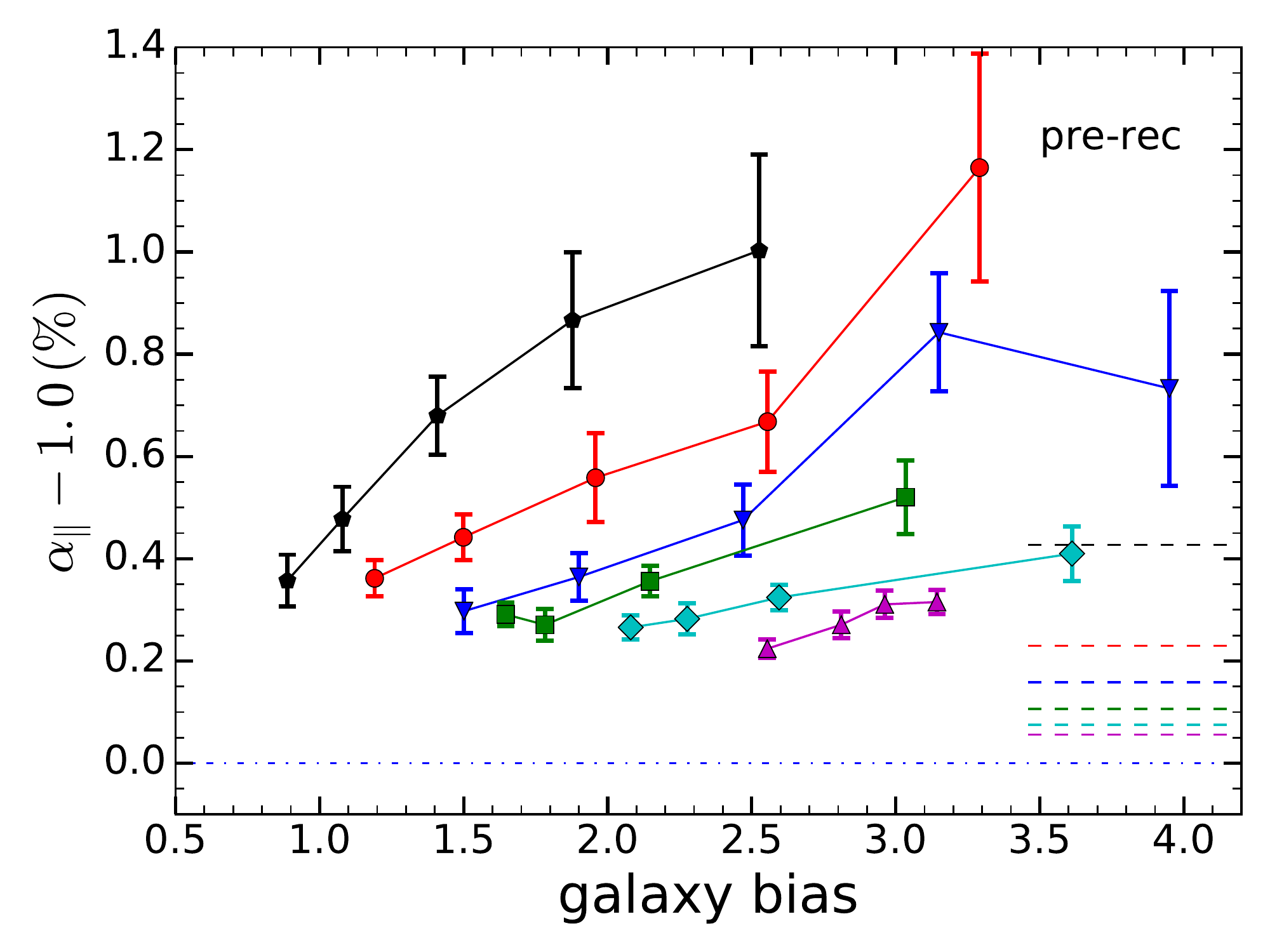}
\includegraphics[width=0.45\linewidth]{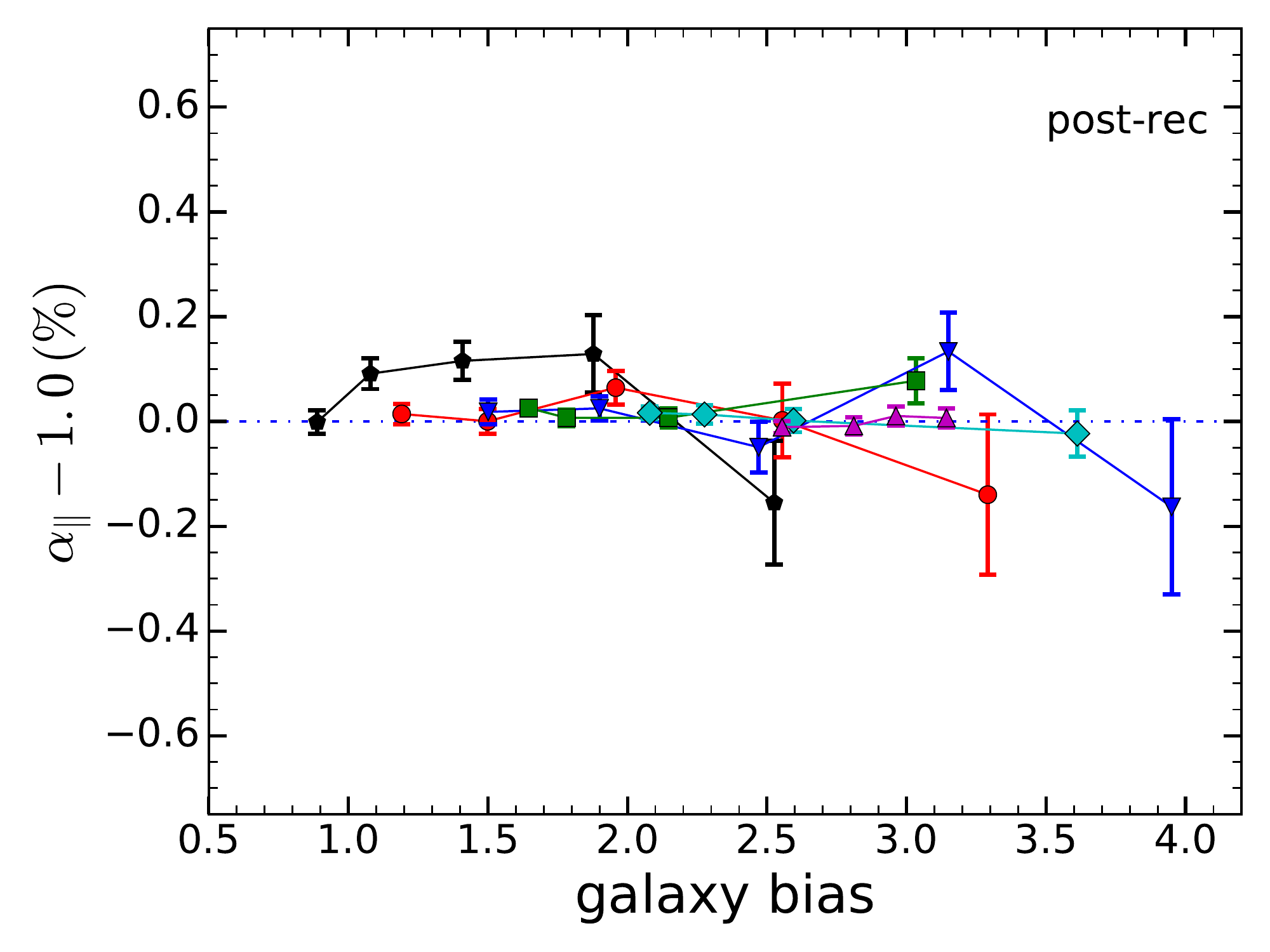}
\caption{BAO scale shifts in the presence of galaxy/halo biases: $\alpe$ (upper panels) and $\alpa$ (lower panels) fitted using the EFT0 model in redshift space. \textit{Left panels:} pre-reconstruction. \textit{Right panels:} post-reconstruction. We again use the same scale on the left and right panels in order to ease the comparison between pre- and post reconstruction. The values can be found in Table \ref{tab:fof_pre_rec} and \ref{tab:fof_post_rec}. A different colour denotes a different redshift, and at each redshift we show the BAO shift measurements as a function of galaxy bias. Horizontal dashed lines on the right hand side in the left panels correspond to the values for matter power spectrum, which are derived using the fitting in Fig.~\ref{fig:alpha_ZV}. The thin solid curves and thin dotted curves in different colours correspond to the predictions based on $b_2(b_1)$ from \citet{Localbias2016}
and \citet{PadBAOshift}, respectively, at the corresponding redshift.
We estimate galaxy bias ($b_1$) from the ratio of galaxy--matter cross-power spectrum over matter auto-power spectrum at large scales $k\leq0.02\ihMpc$. 
 }\label{fig:alpha_bias}
\end{figure*}

\subsubsection{Correlation of nonlinear shifts between different redshifts}

The upper panels of Figure \ref{fig:scatteralpha} show the differences on the best fits $\alpha_{\bot}$ and $\alpha_{\|}$ (from Figure \ref{fig:alpha_ZV})
 between redshift $2.5$ and a lower redshift to further remove any remaining correlated variance between high redshift and low redshift. The errors that were calculated from dispersions of the differences tend to increase in comparison to Figure \ref{fig:alpha_ZV}, implying that the effect of the initial cosmic variance has been mostly canceled between different redshifts and there are little remaining correlations. The lower panels show the scatter of best fit $\alpha$'s at $z=0$ versus at $z=2.5$. Again, the plot implies that once cosmic variance is reduced by differencing wiggle and de-wiggled power spectra, there is little correlation between the nonlinear BAO shift at low redshift and at high redshift before as well as after reconstruction. That is, the nonlinear shift term (from $P_{22}$) does not appear to be correlated between different redshifts. We find a slightly stronger correlation between closer redshift bins, e.g., between $z=2$ and $z=2.5$.

\subsubsection{Summary}

In summary, we find results that are largely consistent with the previous literature while we find that the fitting models with free $\Sigxy$ or $\Sfog$ tend to generate greater shift on $\apar$ than on $\aperp$ before reconstruction; the best fit $\apar$ tend to be more sensitive to the fitting models before reconstruction, as much as 0.2\% at $z \sim 0$ while the best fit values of $\aperp$ appear more robust, i.e., within $0.1\%$ before reconstruction. Such model dependence is substantially reduced after reconstruction, generally within $\sim 0.02\%$, but it could still account as much 0.1\% at $z=0$. 
We also find that the precision (error bar) of the BAO scale does not strongly depend on the choice of models we test, especially on the freedom in the nonlinear damping scale.

\begin{figure*}
\includegraphics[width=0.4\linewidth]{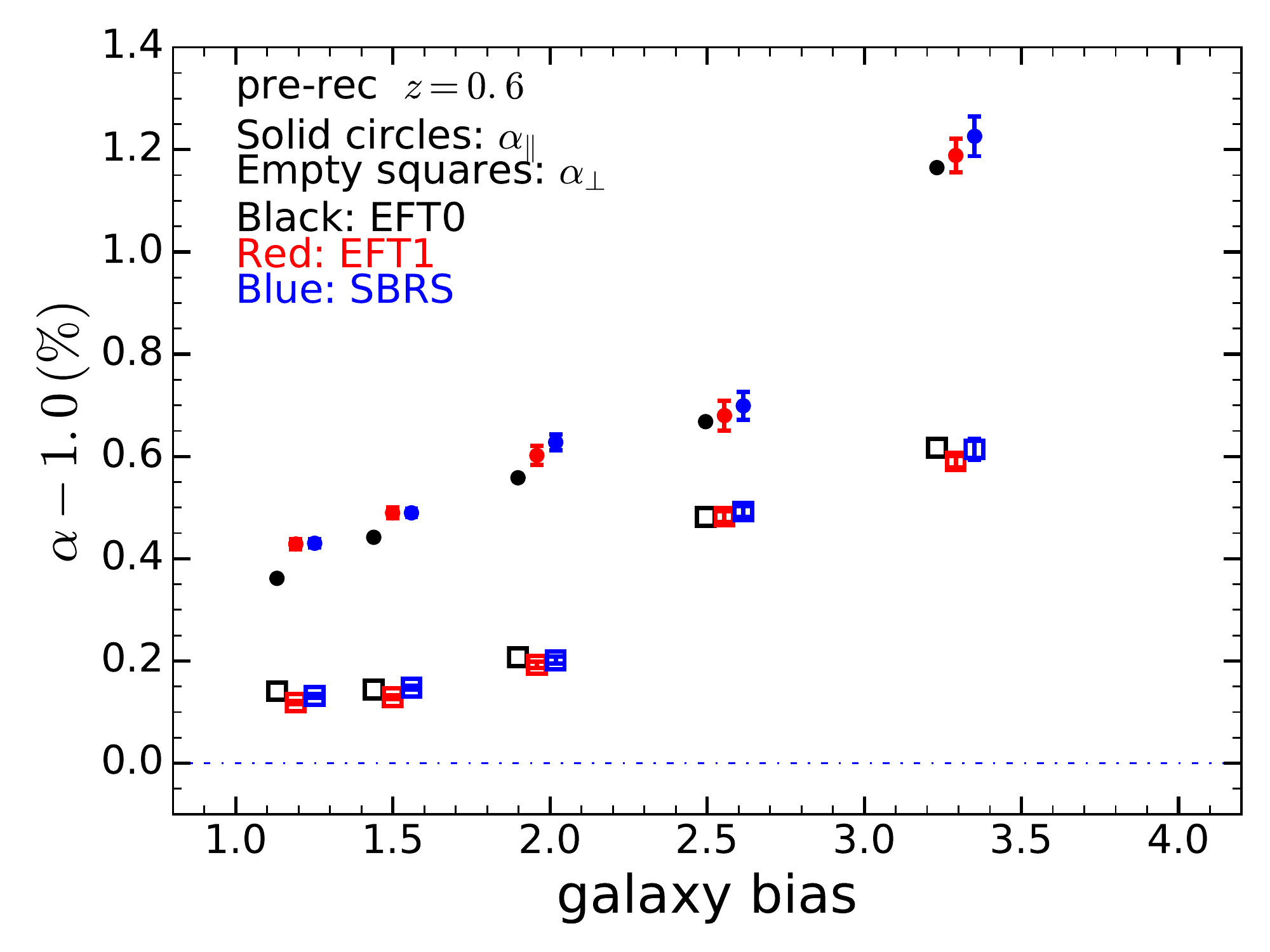}
\includegraphics[width=0.4\linewidth]{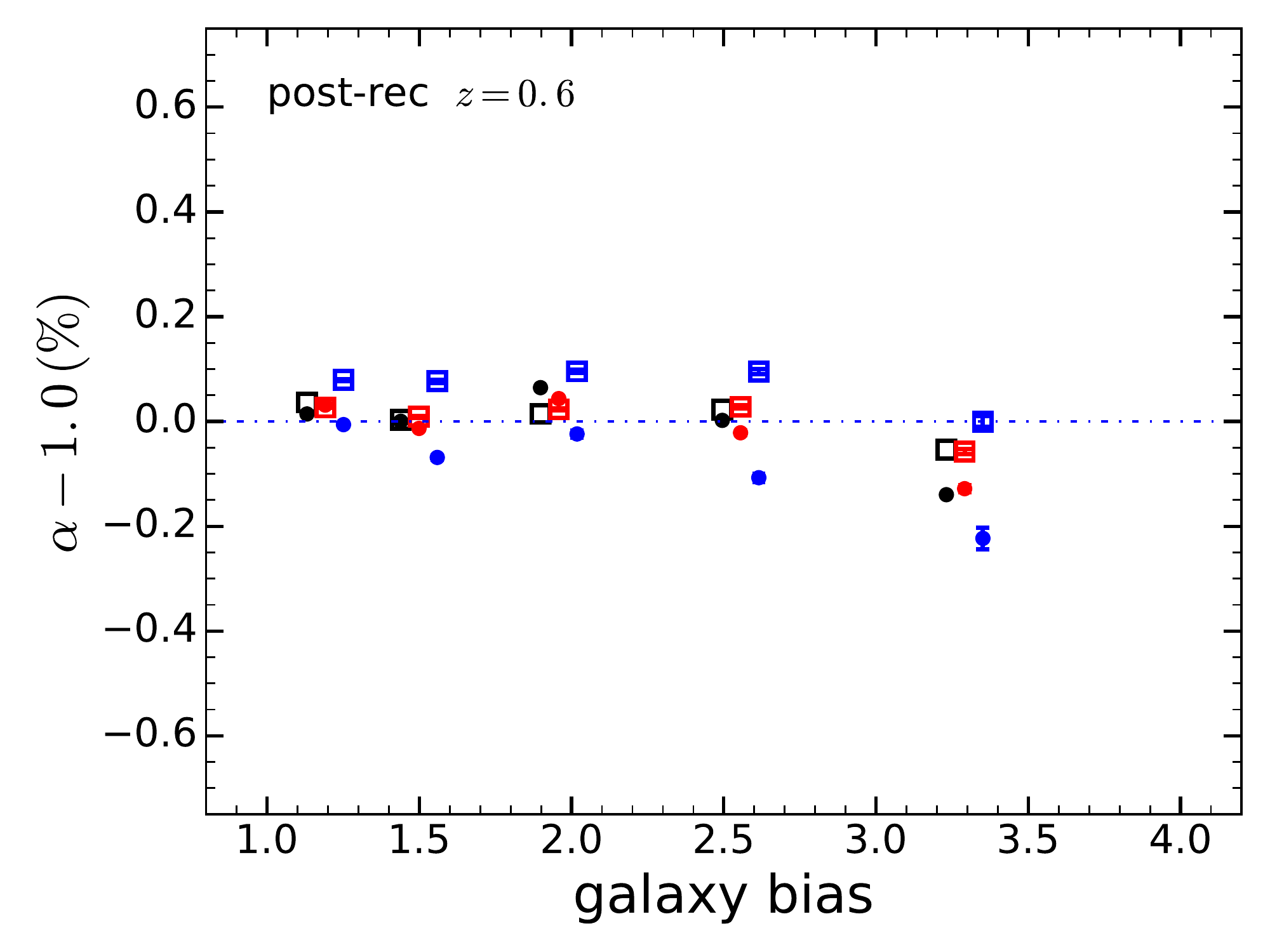}
\includegraphics[width=0.4\linewidth]{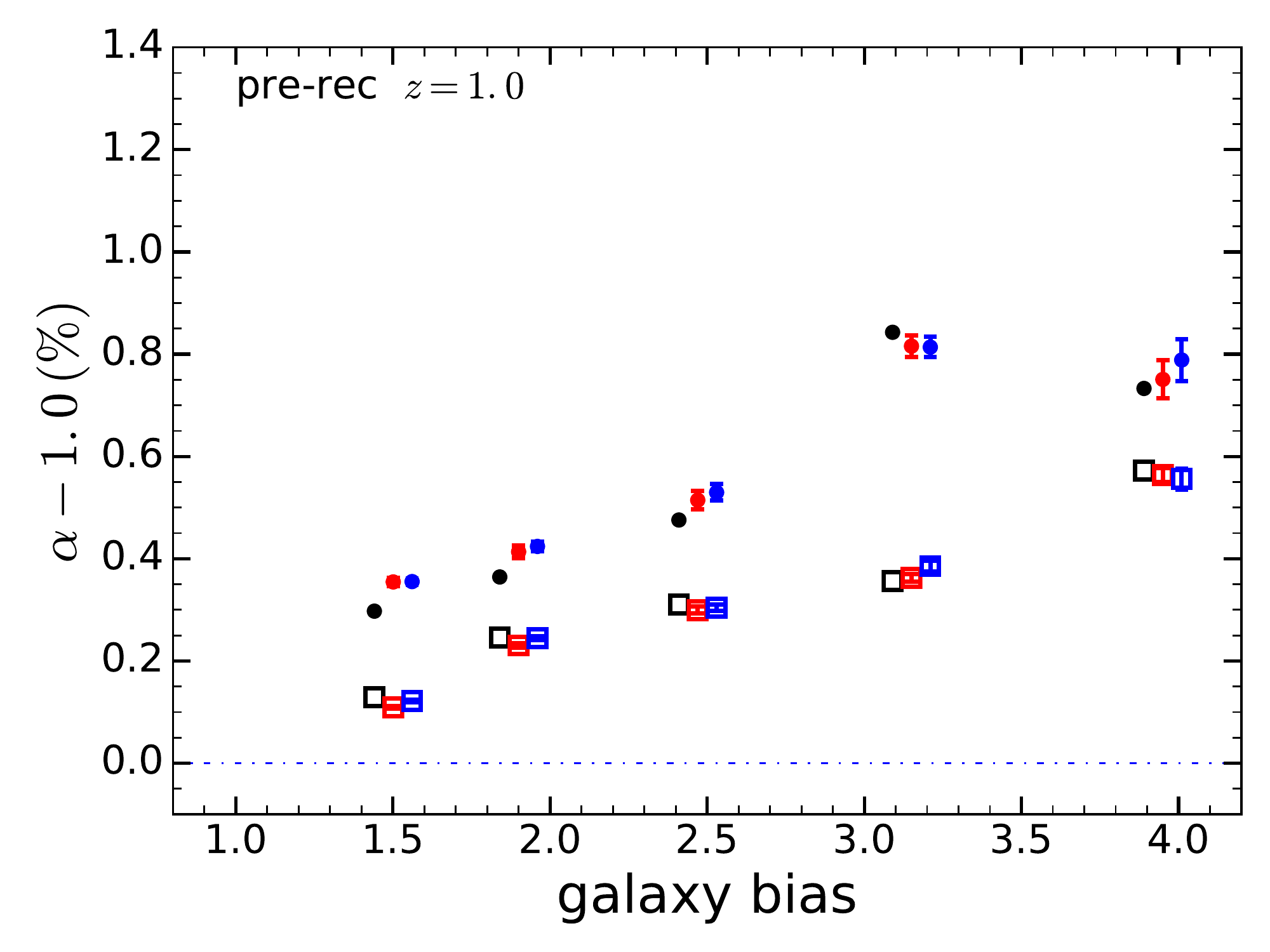}
\includegraphics[width=0.4\linewidth]{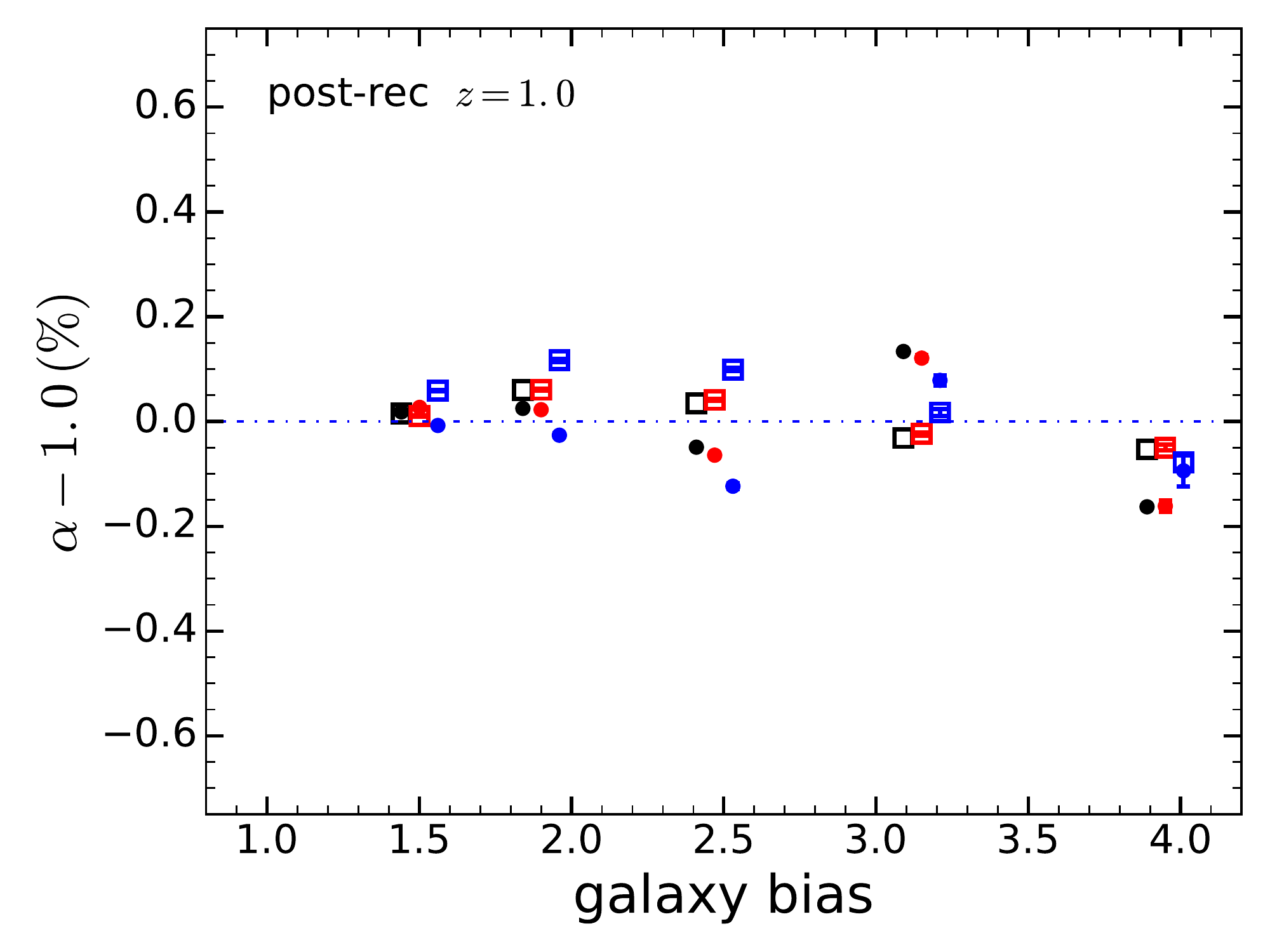}
\includegraphics[width=0.4\linewidth]{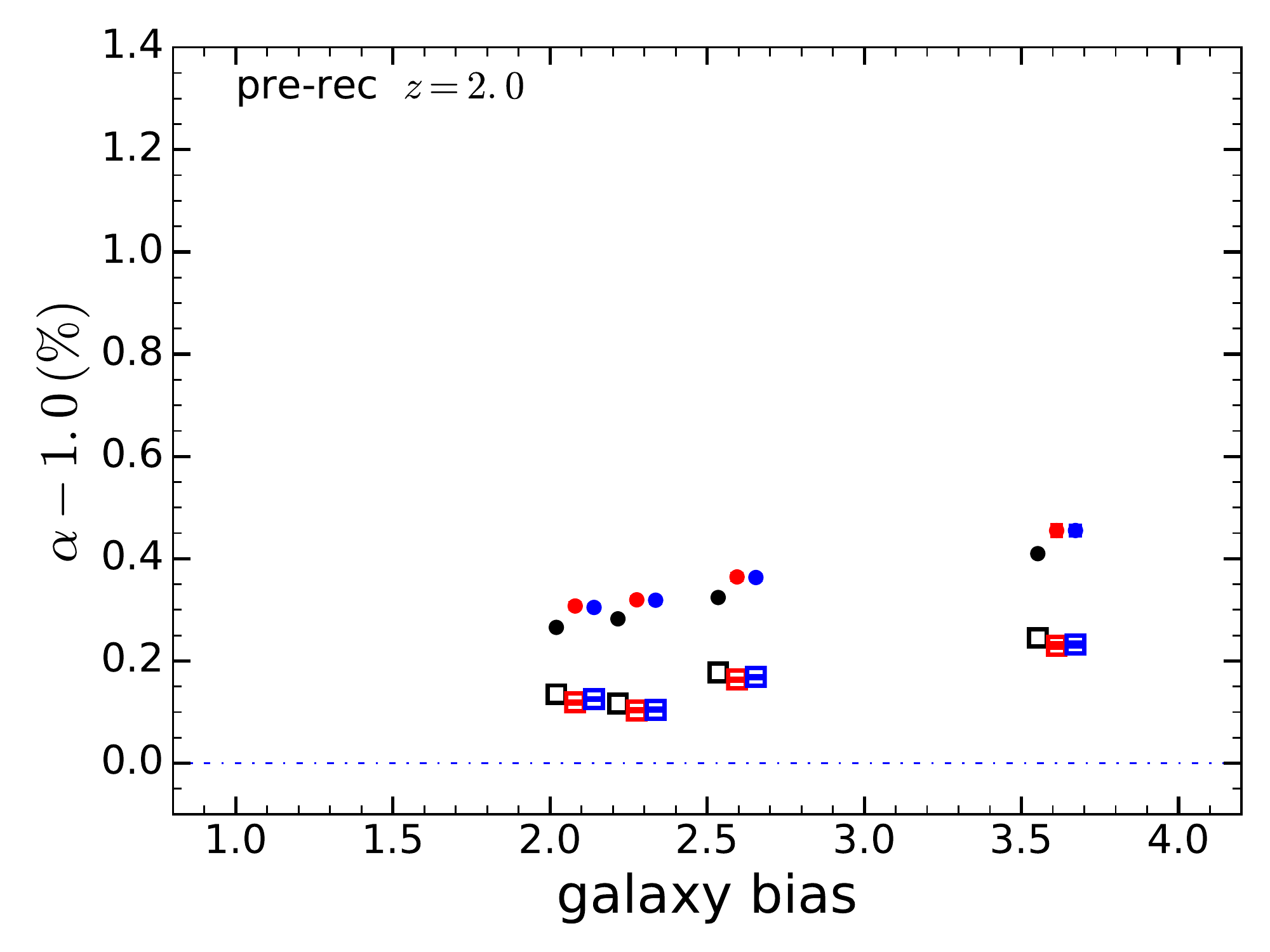}
\includegraphics[width=0.4\linewidth]{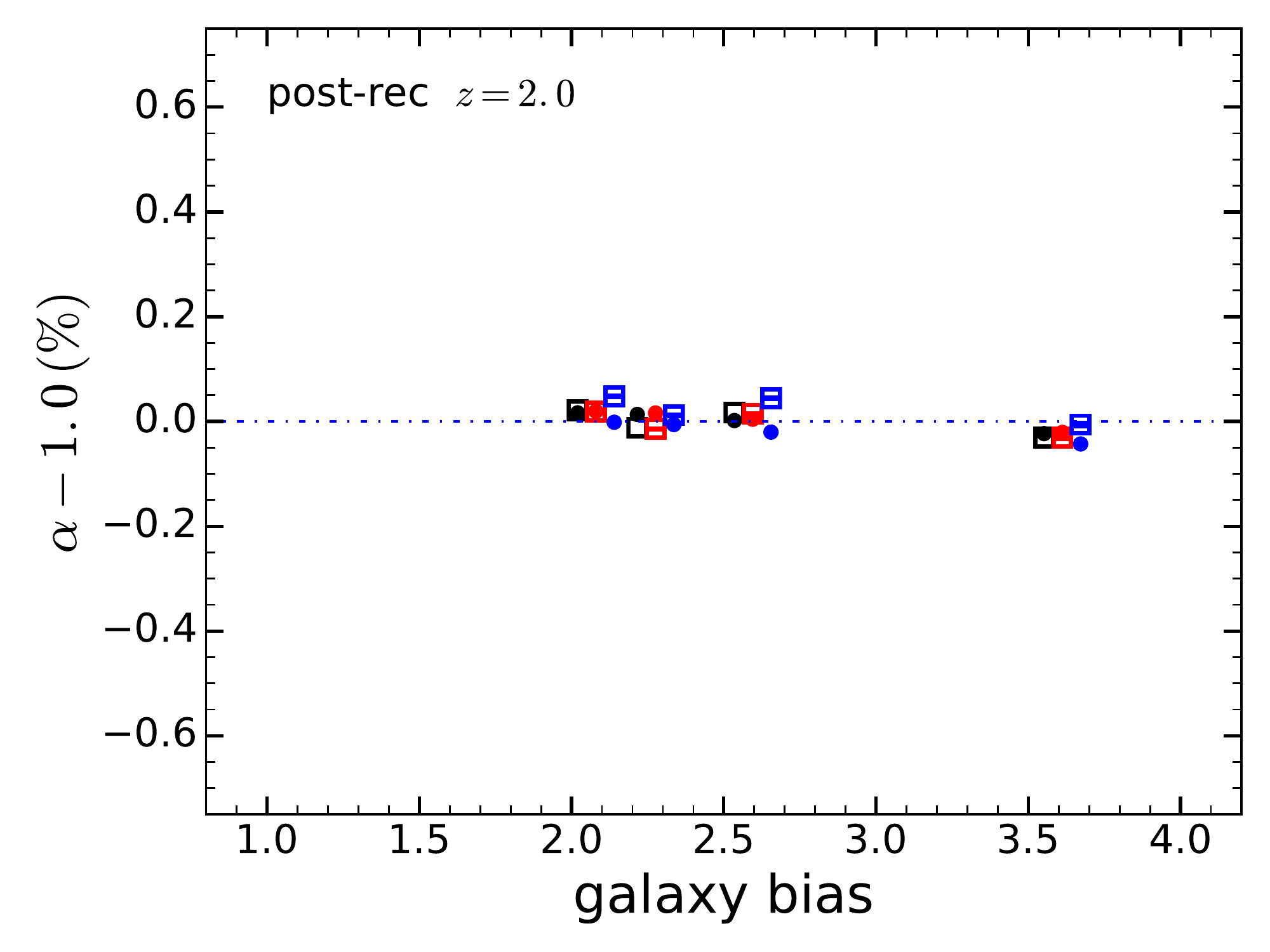}
\includegraphics[width=0.4\linewidth]{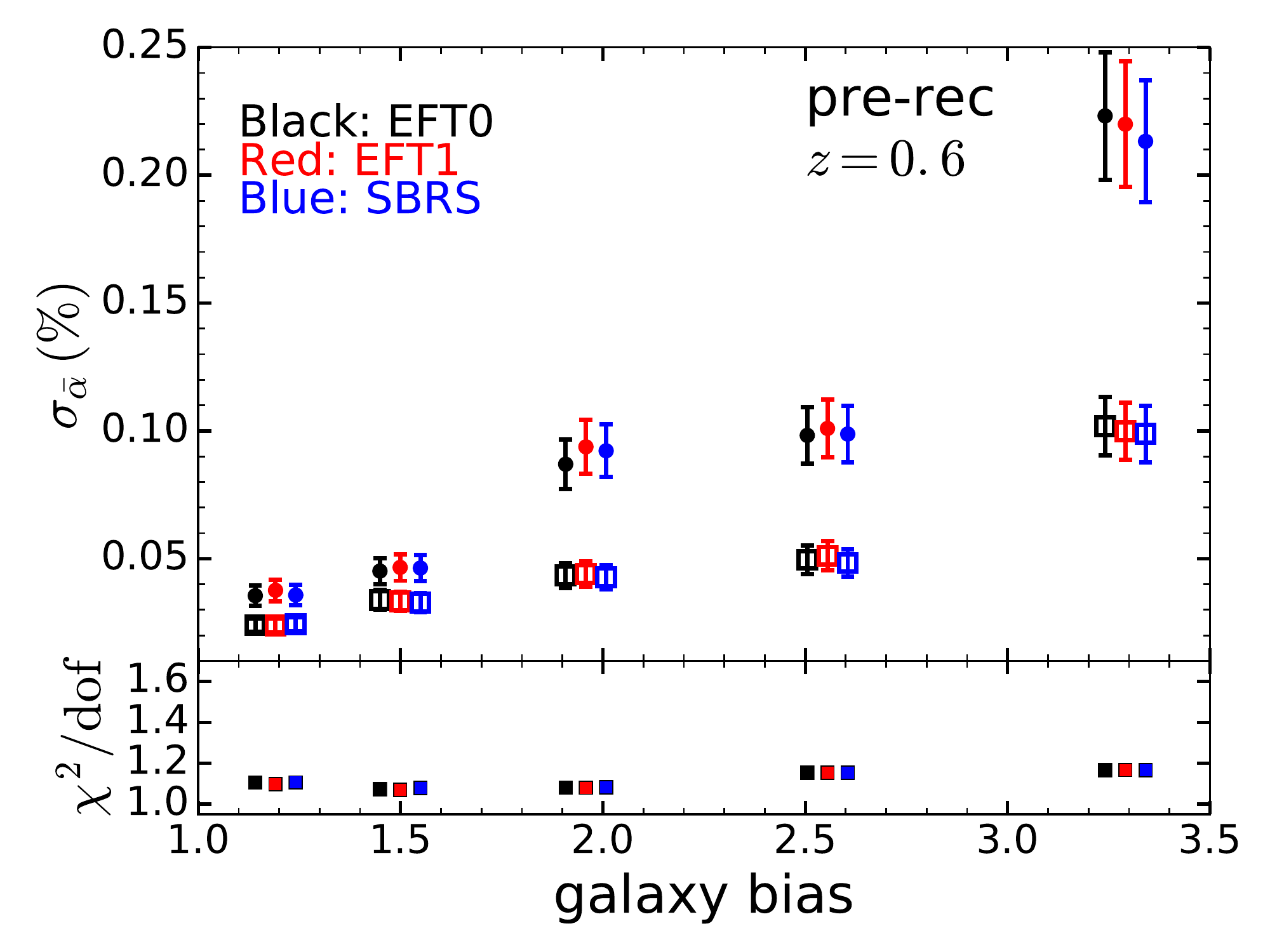}
\includegraphics[width=0.4\linewidth]{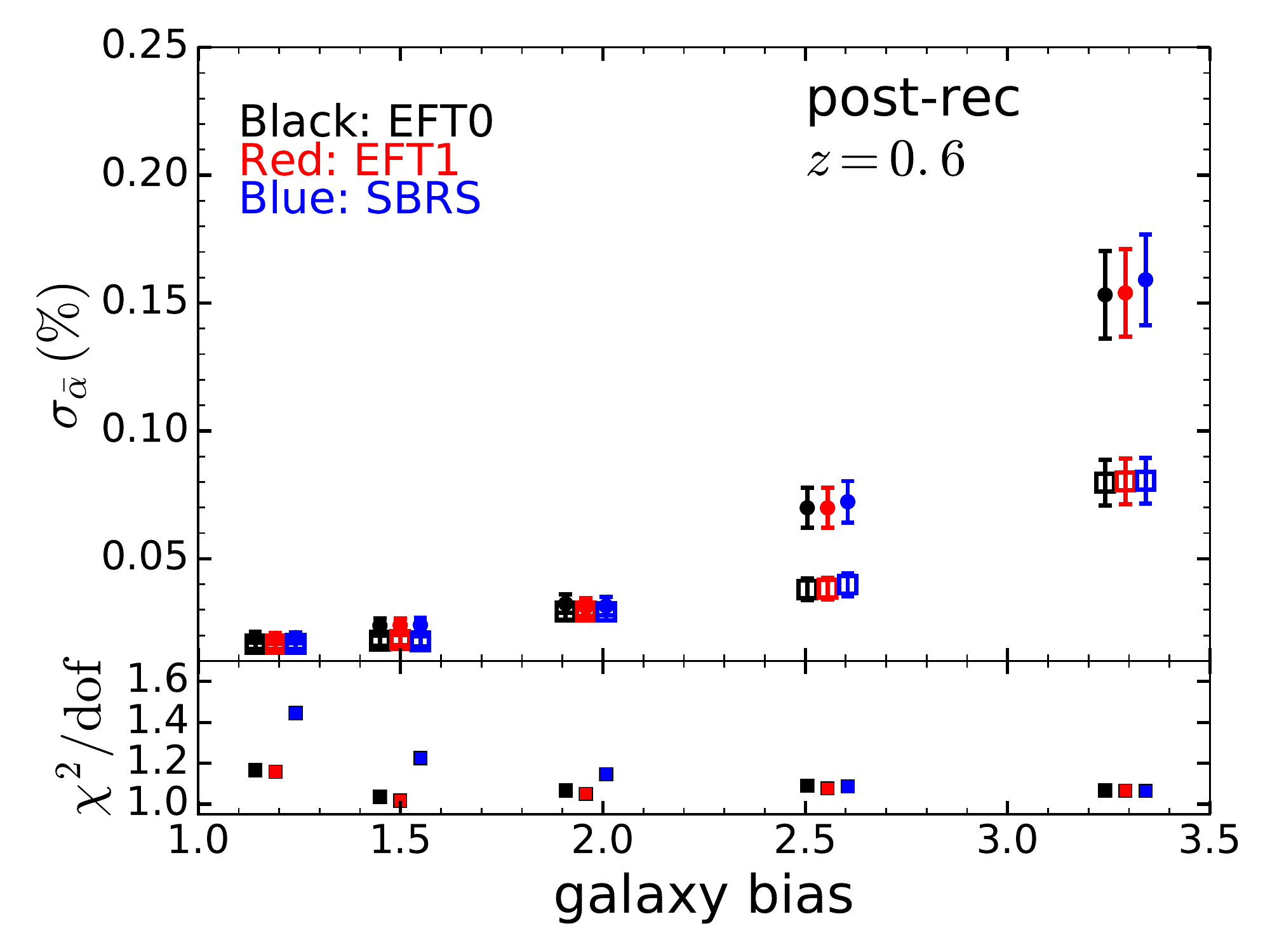}
\caption{The effects of different fitting models for samples with galaxy/halo bias. \textit{Left panels:} pre-reconstruction. \textit{Right panel:} post-reconstruction.
The top three rows show the effect on the mean $\alpha$ ($z=0.6$ for the top row, $z=1.0$ for the second row and $z=2.0$ for the third row) and the bottom panels show the effect on the errors on $\alpha$. 
Empty square points denote for $\alpe$ and solid circular points for $\alpa$. The top three rows compare the BAO shifts from the EFT0 model (black points) with those from the EFT1 (red points) and the SBRS models (blue points). Again, the error bars on the EFT1 and the SBRS points represent the dispersion on the mean $\Delta \alpha$ with no error bars plotted on the EFT0 model. Errors on the $\Delta \alpha$ are negligible and hard to be seen in the figures. 
The bottom panels compare precisions on the mean $\alpha$ using the EFT0 model (black points, from Table~\ref{tab:fof_pre_rec} and ~\ref{tab:fof_post_rec}) with errors using other fitting models at $z=0.6$. The bottom panels also show the reduced $\chi^2$ from the best fit of the mean differenced power spectrum for the three fitting models. }\label{fig:delta_alpha_std} 
\end{figure*}

\begin{figure*}
\includegraphics[width=0.45\linewidth]{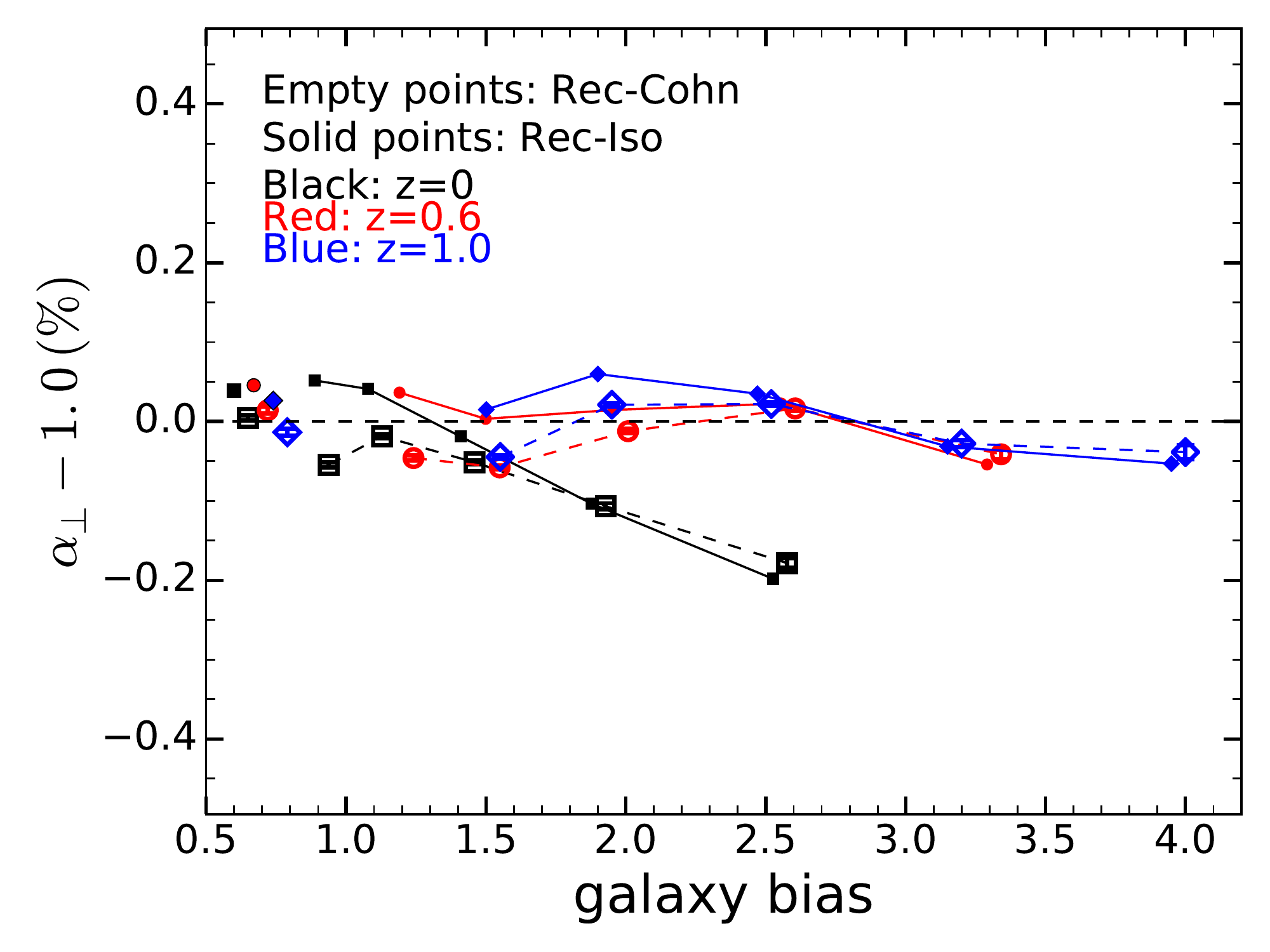}
\includegraphics[width=0.45\linewidth]{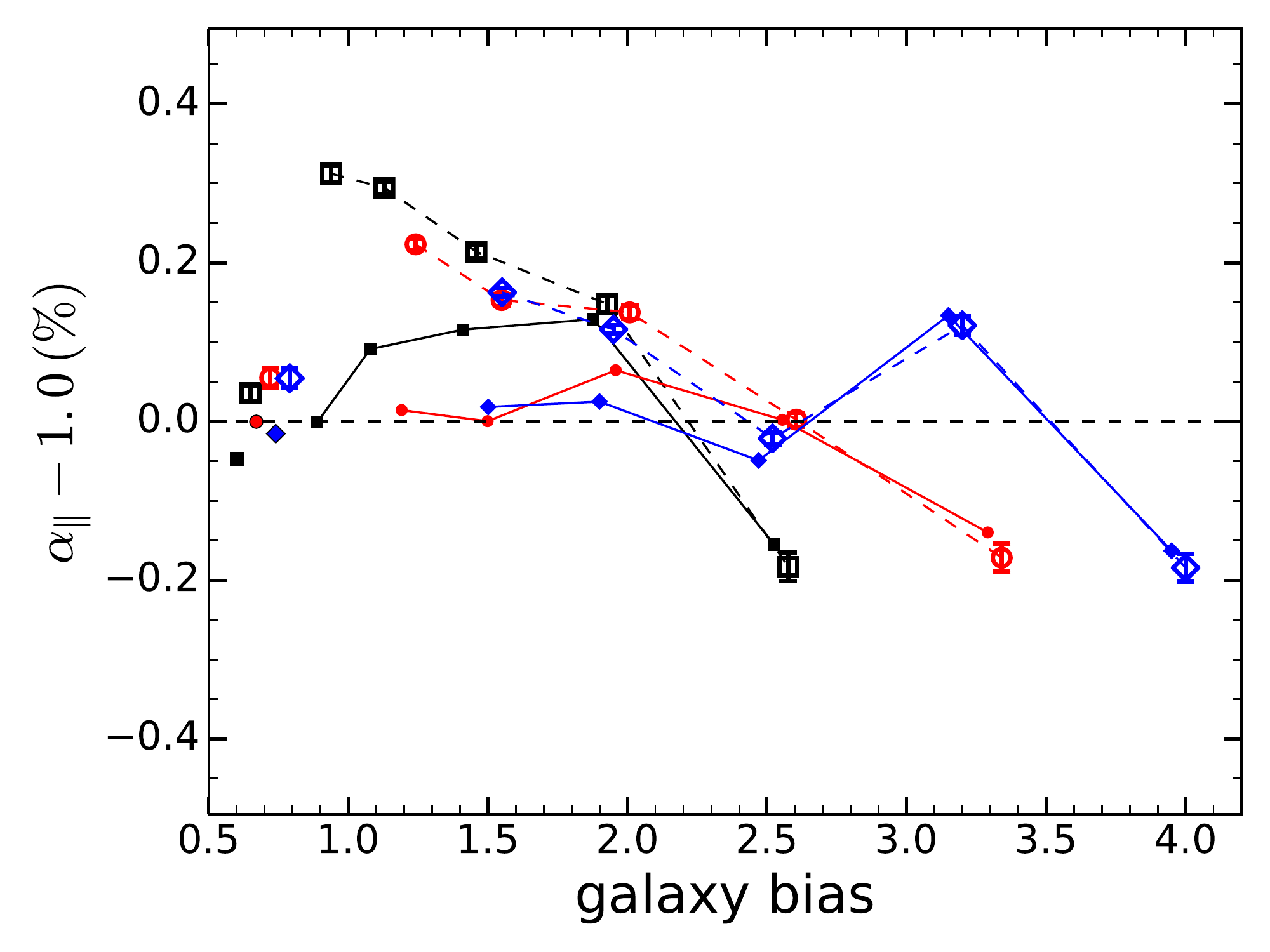}
\includegraphics[width=0.45\linewidth]{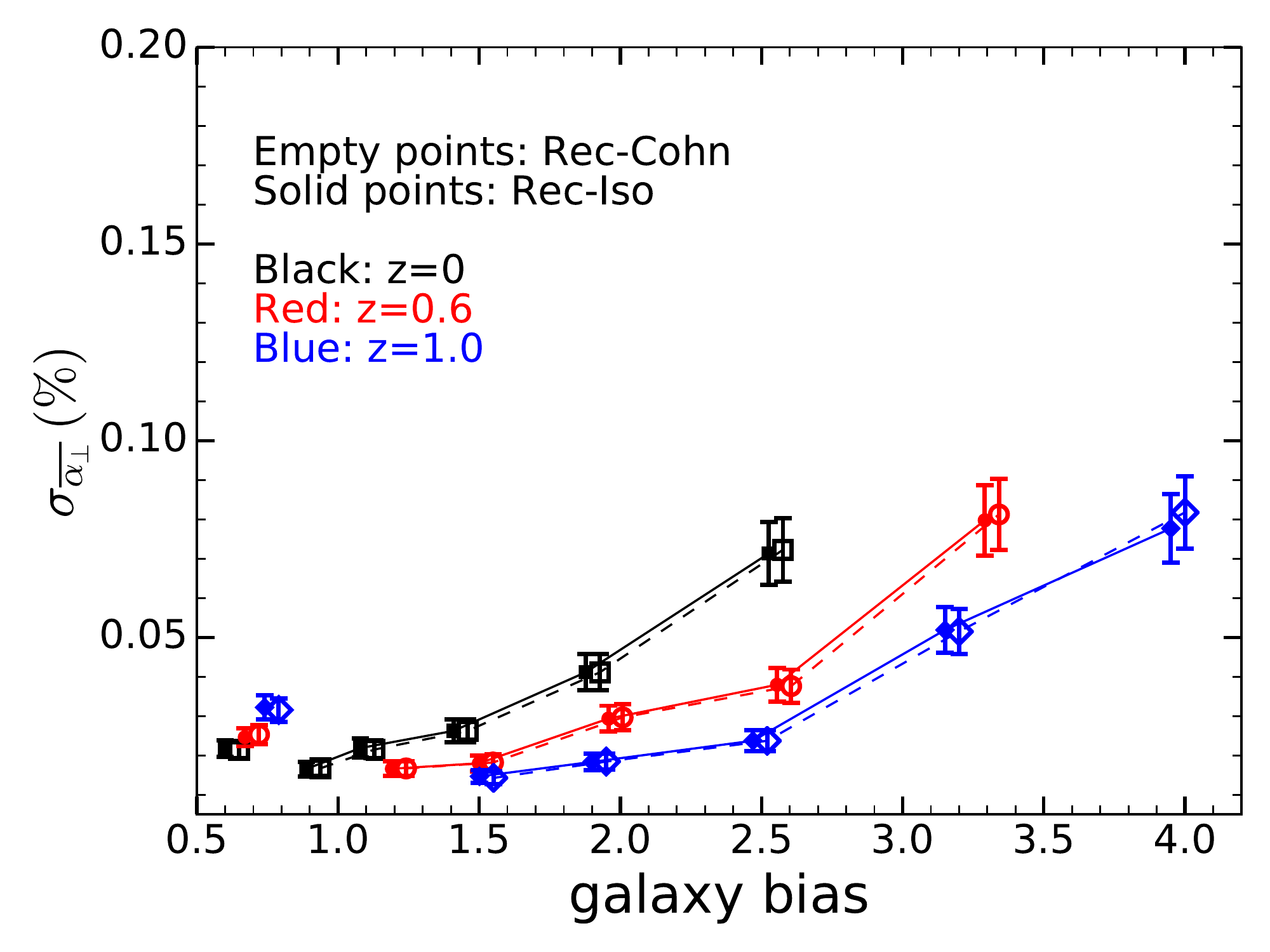}
\includegraphics[width=0.45\linewidth]{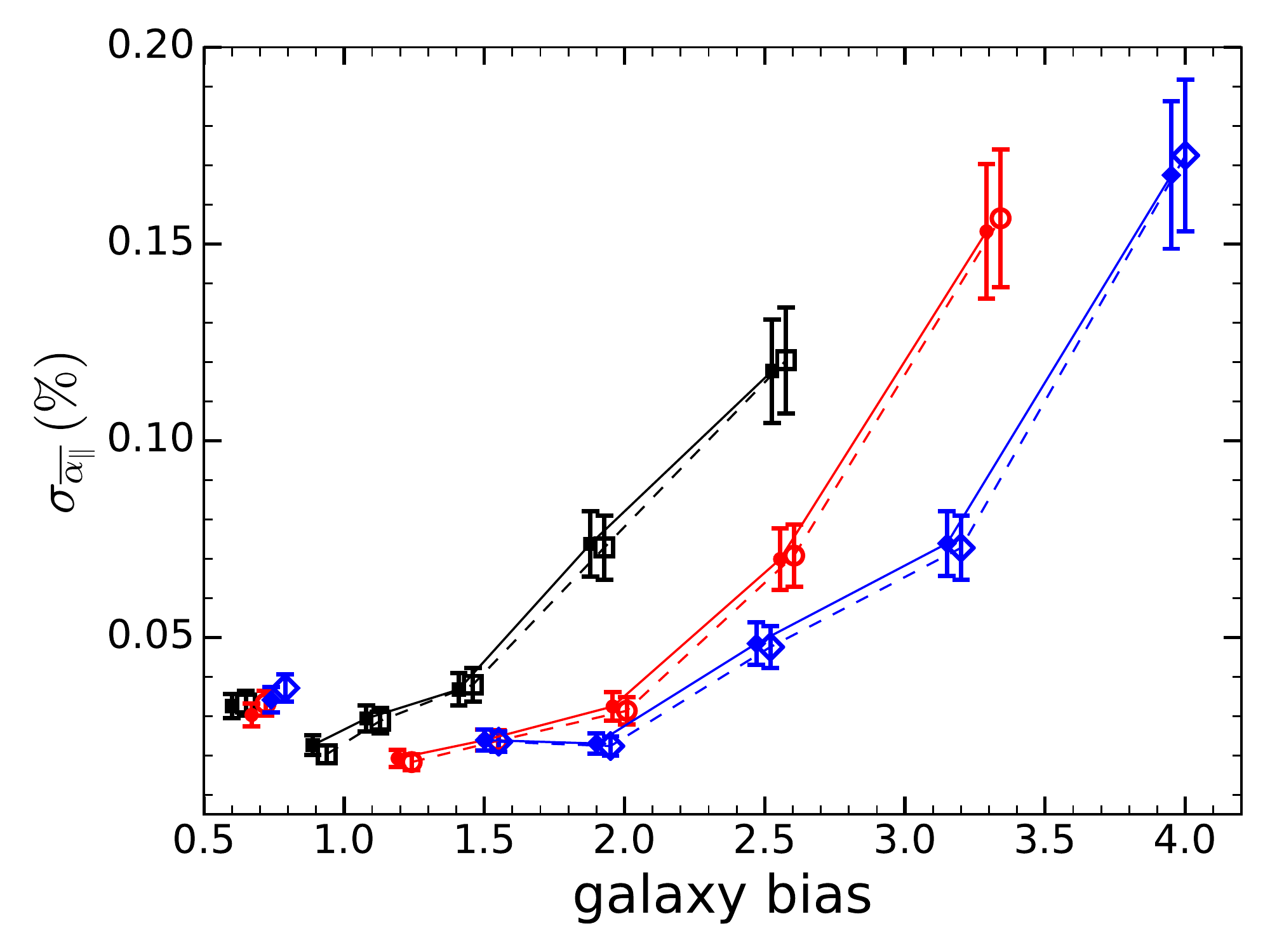}
\caption{The BAO scale shifts for different reconstruction conventions. We compare the default reconstruction scheme `Rec-Iso' (solid points with solid lines) with the `Rec-Cohn' scheme (empty points with dashed lines) using the EFT0 model. The error bars on the `Rec-Cohn' cases in the top panels represent the error on the mean difference between the two conventions, derived from the dispersions of individual difference $\alpha$'s. \textit{Left panel:} $\alpha_{\bot}$. \textit{Right panel:} $\alpha_{\|}$. Different colours represent different redshifts: black for $z=0$, red for $z=0.6$ and blue for $z=1.0$. Singular points denote for matter and they are shifted from galaxy bias of 1.0 not to overlap with data points for galaxy. Empty points (`Rec-Cohn') are also slightly shifted from the original bias values to avoid overlaps with the `Rec-Iso' cases. The bottom panels compare precisions on $\alpha$ using the two reconstruction schemes. They are very consistent. }\label{fig:Cohn_fit_alpha}
\end{figure*}

\begin{table*}
\centering
\caption{The pre-reconstruction BAO scale shifts in the presence of galaxy/halo bias in redshift space using the EFT0 model. In each case, we show the mean galaxy biases derived from the galaxy-matter cross power spectrum and particle number densities $\bar{n}$ over $40$ realizations. The unit of $\bar{n}$ is $10^{-3}~\itrihMpc$. Again, the values of $\alpha$ and $\sigma_{\alpha}$ outside the parentheses are the mean and the error of the mean derived from 40 best fits while the values inside the parentheses are from the best fit of the averaged difference power spectrum.  }\label{tab:fof_pre_rec}
\begin{tabular}[c]{m{1.0cm} m{1.5cm} m{1.0cm} m{1.6cm}  m{1.6cm} m{1.6cm}  m{1.6cm} m{1.0cm}}
\hline
\hline
Redshift & galaxy bias & $\bar{n}$ & $\alpha_{\bot}-1\,(\%)$ & $\sigma_{\alpha_{\bot}}(\%)$ & $\alpha_{\|}-1 \,(\%)$ & $\sigma_{\alpha_{\|}}(\%)$ & $\chi^2/\text{d.o.f.}$ \\
\hline
\multirow{5}{*}{0} & 0.89 & 4.0 & 0.145 (0.147) & 0.033 (0.031) & 0.357 (0.354) & 0.051 (0.051) & 1.134\\
& 1.08 & 1.2 & 0.119 (0.118) & 0.039 (0.038) & 0.478 (0.471) & 0.063 (0.066) & 1.104\\
& 1.41 & 0.37 & 0.256 (0.259) & 0.044 (0.049) & 0.680 (0.673) & 0.076 (0.087) & 1.061\\
& 1.88 & 0.11 & 0.324 (0.342) & 0.074 (0.070) & 0.87 (0.82) & 0.13 (0.12) & 1.135\\
& 2.53 & 0.035 & 0.64 (0.67) & 0.083 (0.11) & 1.00 (0.92) & 0.19 (0.20) & 1.119\\
\hline
\multirow{5}{*}{0.6} & 1.19 & 4.0 & 0.141 (0.142) & 0.024 (0.023) & 0.362 (0.357) & 0.036 (0.039) & 1.109\\
& 1.50 & 1.2 & 0.144 (0.147) & 0.034 (0.031) & 0.442 (0.435) & 0.045 (0.054) & 1.074\\
& 1.96 & 0.38 & 0.207 (0.209) & 0.043 (0.042) & 0.558 (0.546) & 0.087 (0.075) & 1.082\\
& 2.56 & 0.12 & 0.482 (0.487) & 0.050 (0.060) & 0.668 (0.63) & 0.098 (0.11) & 1.153\\
& 3.29 & 0.036 & 0.62 (0.62) & 0.10 (0.10) & 1.16 (1.06) & 0.22 (0.19) & 1.167\\
\hline
\multirow{5}{*}{1.0} & 1.50 & 3.8 & 0.129 (0.130) & 0.018 (0.020) & 0.297 (0.290) & 0.043 (0.034) & 1.107\\
& 1.90 & 1.2 & 0.246 (0.249) & 0.028 (0.027) & 0.364 (0.358) & 0.047 (0.045) & 1.130\\
& 2.47 & 0.36 & 0.311 (0.313) & 0.039 (0.039) & 0.476 (0.464) & 0.070 (0.069) & 1.125\\
& 3.15 & 0.11 & 0.356 (0.363) & 0.071 (0.063) & 0.84 (0.80) & 0.12 (0.11) & 1.098\\
& 3.95 & 0.034 & 0.572 (0.58) & 0.099 (0.11) & 0.73 (0.64) & 0.19 (0.19) & 1.122\\
\hline
\multirow{4}{*}{1.5} & 1.65 & 7.0 & 0.169 (0.168) & 0.019 (0.015) & 0.291 (0.291) & 0.023 (0.025) & 1.168\\
& 1.78 & 4.5 & 0.159 (0.158) & 0.017 (0.016) & 0.271 (0.273) & 0.031 (0.027) & 1.175\\
& 2.15 & 1.9 & 0.175 (0.176) & 0.021 (0.021) & 0.356 (0.352) & 0.030 (0.035) & 1.091\\
& 3.03 & 0.34 & 0.253 (0.258) & 0.040 (0.038) & 0.520 (0.507) & 0.072 (0.065) & 1.161\\
\hline
\multirow{4}{*}{2.0}& 2.08 & 5.7 & 0.135 (0.135) & 0.013 (0.013) & 0.266 (0.264) & 0.024 (0.022) & 1.164\\
& 2.28 & 3.6 & 0.117 (0.117) & 0.013 (0.015) & 0.282 (0.278) & 0.030 (0.025) & 1.153\\
& 2.60 & 1.9 & 0.178 (0.179) & 0.020 (0.018) & 0.324 (0.323) & 0.025 (0.031) & 1.131\\
& 3.61 & 0.34 & 0.245 (0.248) & 0.032 (0.036) & 0.410 (0.403) & 0.054 (0.062) & 1.114\\
\hline
\multirow{4}{*}{2.5}& 2.55 & 4.4 & 0.132 (0.132) & 0.013 (0.013) & 0.224 (0.222) & 0.018 (0.021) & 1.180\\
& 2.81 & 2.6 & 0.167 (0.166) & 0.012 (0.015) & 0.271 (0.269) & 0.026 (0.025) & 1.111\\
& 2.96 & 2.0 & 0.155 (0.157) & 0.014 (0.017) & 0.311 (0.309) & 0.027 (0.028) & 1.101\\
& 3.14 & 1.5 & 0.190 (0.190) & 0.020 (0.018) & 0.315 (0.311) & 0.023 (0.031) & 1.103\\
\hline
\hline
\end{tabular}
\end{table*}

\begin{table*}
\centering
\caption{The post-reconstruction BAO scale shifts in the presence of galaxy/halo bias in redshift space using the EFT0 model. We used the `Rec-Iso' scheme. We also show the fiducial smoothing scales $\Sigsm$ for the density field reconstruction.  }\label{tab:fof_post_rec}
\begin{tabular}[c]{m{1.0cm} m{1.5cm} m{1.0cm} m{1.8cm} m{1.6cm} m{1.8cm} m{1.6cm} m{1.0cm}}
\hline
\hline
Redshift & galaxy bias & $\Sigma_{\text{sm}}$ & $\alpha_{\bot}-1\,(\%)$ & $\sigma_{\alpha_{\bot}}$ & $\alpha_{\|}-1 \,(\%)$ & $\sigma_{\alpha_{\|}}$ & $\chi^2/\text{d.o.f.}$ \\
\hline
\multirow{5}{*}{0} & 0.89 & 8.60 & 0.052 (0.053) & 0.017 (0.016) & -0.001 (-0.003) & 0.023 (0.023) & 1.262\\
& 1.08 & 9.85 & 0.041 (0.043) & 0.022 (0.020) & 0.091 (0.086) & 0.029 (0.031) & 1.235\\
& 1.41 & 11.45 & -0.019 (-0.019) & 0.026 (0.027) & 0.116 (0.114) & 0.037 (0.043) & 1.185\\
& 1.88 & 12.98 & -0.104 (-0.100) & 0.041 (0.041) & 0.129 (0.113) & 0.074 (0.068) & 1.122\\
& 2.53 & 14.83 & -0.198 (-0.181) & 0.071 (0.075) & -0.16 (-0.19) & 0.12 (0.13) & 1.073\\
\hline
\multirow{5}{*}{0.6} & 1.19 & 8.12 & 0.036 (0.036) & 0.017 (0.014) & 0.014 (0.015) & 0.019 (0.020) & 1.167\\
& 1.50 & 8.72 & 0.003 (0.005) & 0.018 (0.019) & 0.000 (-0.002) & 0.024 (0.028) & 1.035\\
& 1.96 & 9.85 & 0.015 (0.014) & 0.029 (0.026) & 0.065 (0.063) & 0.032 (0.042) & 1.066\\
& 2.56 & 11.27 & 0.023 (0.022) & 0.038 (0.043) & 0.002 (-0.003) & 0.070 (0.070) & 1.089\\
& 3.29 & 13.53 & -0.054 (-0.049) & 0.080 (0.080) & -0.14 (-0.15) & 0.15 (0.13) & 1.067\\
\hline
\multirow{5}{*}{1.0} & 1.50 & 7.87 & 0.015 (0.015) & 0.015 (0.013) & 0.018 (0.019) & 0.024 (0.018) & 1.124\\
& 1.90 & 8.72 & 0.060 (0.063) & 0.018 (0.018) & 0.025 (0.018) & 0.023 (0.026) & 1.060\\
& 2.47 & 9.90 & 0.035 (0.040) & 0.024 (0.027) & -0.049 (-0.060) & 0.048 (0.043) & 1.036\\
& 3.15 & 11.40 & -0.032 (-0.030) & 0.052 (0.046) & 0.134 (0.120) & 0.074 (0.075) & 1.059\\
& 3.95 & 13.11 & -0.053 (-0.065) & 0.078 (0.089) & -0.16 (-0.16) & 0.17 (0.15) & 1.078\\
\hline
\multirow{5}{*}{1.5} & 1.65 & 7.75 & 0.043 (0.042) & 0.010 (0.011) & 0.026 (0.027) & 0.014 (0.015) & 1.113\\
& 1.78 & 7.87 & 0.049 (0.049) & 0.013 (0.012) & 0.007 (0.006) & 0.016 (0.016) & 1.076\\
& 2.15 & 8.25 & 0.021 (0.022) & 0.016 (0.015) & 0.007 (0.004) & 0.018 (0.021) & 1.103\\
& 3.03 & 9.59 & -0.023 (-0.015) & 0.033 (0.029) & 0.078 (0.057) & 0.043 (0.042) & 1.057\\
\hline
\multirow{5}{*}{2.0} & 2.08 & 7.75 & 0.021 (0.021) & 0.011 (0.010) & 0.016 (0.017) & 0.013 (0.014) & 1.071\\
& 2.28 & 7.87 & -0.012 (-0.012) & 0.010 (0.011) & 0.013 (0.013) & 0.018 (0.017) & 1.110\\
& 2.60 & 8.25 & 0.017 (0.018) & 0.012 (0.014) & 0.002 (-0.001) & 0.022 (0.021) & 1.066\\
& 3.61 & 9.70 & -0.031 (-0.025) & 0.029 (0.029) & -0.023 (-0.031) & 0.044 (0.043) & 1.018\\
\hline
\multirow{5}{*}{2.5} & 2.55 & 7.87 & 0.027 (0.027) & 0.011 (0.010) & -0.011 (-0.010) & 0.011 (0.015) & 1.086\\
& 2.81 & 8.12 & 0.021 (0.022) & 0.010 (0.012) & -0.009 (-0.009) & 0.017 (0.018) & 1.110\\
& 2.96 & 8.25 & 0.014 (0.015) & 0.014 (0.014) & 0.011 (0.011) & 0.018 (0.020) & 1.066\\
& 3.14 & 8.49 & 0.023 (0.023) & 0.013 (0.015) & 0.007 (0.006) & 0.018 (0.022) & 1.086\\
\hline
\hline
\end{tabular}
\end{table*}

\subsection{BAO scales in the presence of galaxy/halo bias}

We repeat our analyses using halo catalogs with various mass cuts to simulate the effect of galaxy/halo bias in redshift space. Figure \ref{fig:alpha_bias} (and  Table \ref{tab:fof_pre_rec}) shows best fit $\alpha$'s as a function of galaxy/halo bias at different redshifts before (left panels) and after density field reconstruction (right panels). Relative to the biased cases (data points), we also overplot the corresponding shift model for the matter cases we derived earlier as horizontal dashed lines (i.e. $0.51\%[D(z)/D(0)]^2$ and $0.43\%[D(z)/D(0)]^2$ for top and bottom left panels on the right hand side, respectively). 

\subsubsection{Pre-reconstruction}

Focusing on the pre-reconstruction case, we find that the lower bias cases at all redshifts converge toward a BAO shift of $0.1\%$ on $\aperp$ regardless of the redshift of interest. This is much smaller and different from $\aperp$ of the matter case in Figure \ref{fig:alpha_ZV} which gradually increases from $0.1\%$ to $\sim 0.5\%$ from $z=2.5$ to $z=0$.  
As bias increases, we observe deviation from this low bias convergence and the deviation is faster at low redshift while is very slow at high redshift. For example, the nonlinear shift on $\aperp$ for $b_1 \sim 1.5$ at $z=1-1.5$ (e.g., DESI ELG-like) would be $\sim 0.12-0.16\%$ and the corresponding shift would be  $< 0.2\%$ even for high bias $b_1 \sim 3.1$ at $z\sim 2.5$ (e.g., DESI QSO-like). On the other hand, for targets at low redshift, e.g., $z\sim 0.6$, the shift will increase more rapidly with increasing bias: $b_1\sim 2$  at $z=0.6-1$ (e.g., DESI LRG-like), we expect shift of $\sim 0.2-0.25\%$. 
This trend is consistent with theoretical predictions based on perturbation theories from \citet{PadBAOshift,Sherwin12}. The BAO shift of biased sample is expected to $\propto D(z)^2(1+1.5 \frac{b_2}{b_1})$ depending on the linear bias $b_1$ and the second order bias $b_2$. The second order bias $b_2$ is expected to be tightly correlated with $b_1$~\citep[e.g.,][]{Localbias2016,Modi2017} and reaches minimum, negative trough for $b_1 \sim 1$. This explains both the smaller BAO shift of the low biased cases relative to the matter cases and the trough shape of the BAO shift curve at the low bias limit at given redshift. The $D(z)^2$ dependence explains the greater slope of the BAO shift curve with decreasing redshift. Figure 12 of \citet{PadBAOshift} shows a trend that is indeed very similar to our results in the top left panel of Figure \ref{fig:alpha_bias}.  

In order to compare our results with the PT-based prediction in detail, we adopt the $b_2$ ($b_1$) relation from ~\citet[][Eq. 5.2]{Localbias2016} and equivalently from ~\citet{Modi2017}: $b_2(b_1) = 0.412 - 2.143 b_1 + 0.929 b_1^2+0.008 b_1^3$. This relation is fairly universal for different cosmologies and redshifts. As a caveat, this relation was derived for the halos within individual mass bins while our galaxy/halo mocks correspond to halo centers with minimum mass thresholds; nevertheless, since bias of our mocks would be dominated by the halos near the mass threshold, we believe that this relation should be a good approximation to our case. Dotted lines from Figure \ref{fig:alpha_bias} correspond to $\aperp-1 = 0.51\%(1+1.5\frac{b_2}{b_1}) [D(z)/D(0)]^2$ where $0.51\%$ is fixed from the fit to the matter case. Although the prediction at low bias limit systematically underestimates the BAO shifts, the bias dependence and the redshift dependence qualitatively agree with the prediction and even quantitatively at higher redshift. 
The thin solid curves are predictions using the $b_1$ ($b_2$) fit from \citet{PadBAOshift} instead, i.e., $b_2(b_1)=1.09 b_1^2-2.77 b_1+1.31$, and this prediction overestimates the BAO shifts at low bias limit at $z=0$ such that our results tend to lie between the two predictions.

\begin{figure*}
\centering
\includegraphics[width=0.45\linewidth]{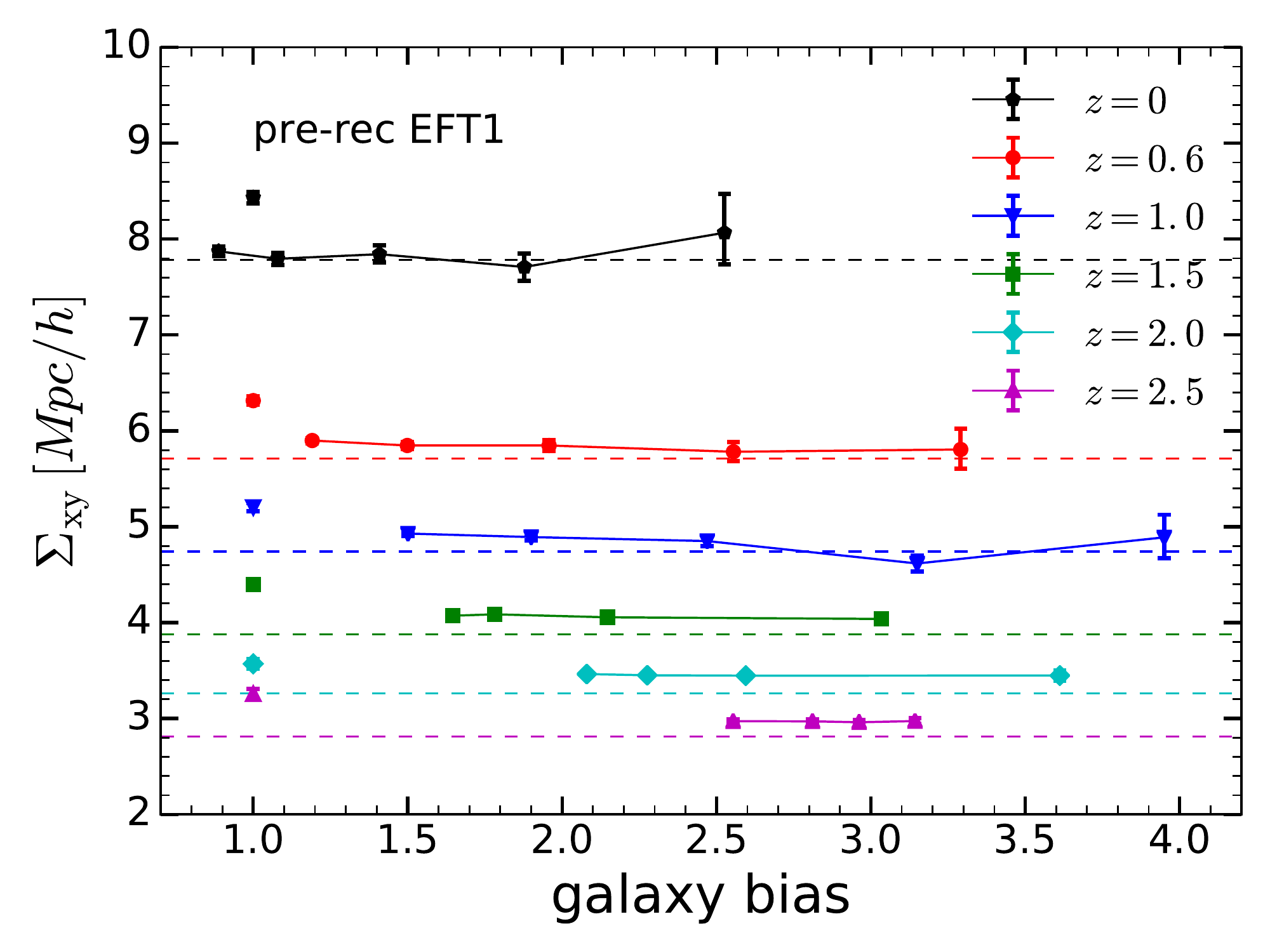}
\includegraphics[width=0.45\linewidth]{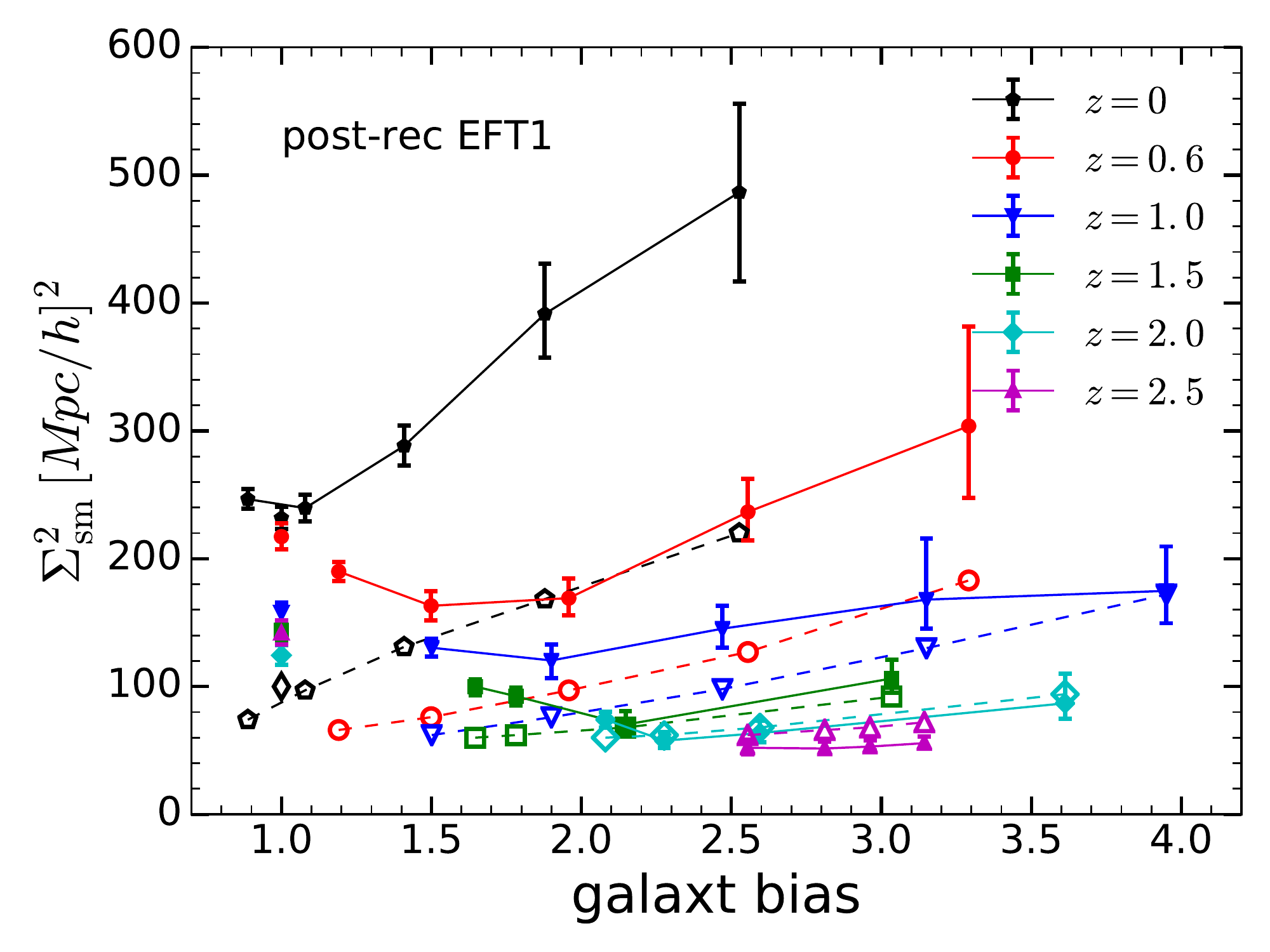}
\caption{
Nonlinear damping scales: the expected values (i.e, used for the pre-reconstruction EFT0 model, dashed lines) v.s., the measured values (data points) using the EFT1 model. The EFT1 model allows $\Sigxy$ to vary before reconstruction and allows $\Sigsm$ to vary after reconstruction such that it can allow perturbing nonlinear damping scales around the predicted value (dashed lines). We present the best fit of the averaged difference power spectra. Solid data points show the best fit $\Sigxy$ (left) or $\Sigsm$ (right) for both galaxy/halo (connected points) and matter (singular points) in the pre-reconstruction case (left) and the post-reconstruction case (right). Different colours represent different redshifts. Empty points (right) denote the input $\Sigsm$ in EFT0 model. For matter, we used $\Sigsm=10\,\hMpc$ (black open diamond point) at all redshifts. Error bars are the 68\% likelihood error. }\label{fig:Z1_fit_Sigma}
\end{figure*}

If halo populations included satellites, the clustering would be sourced from more massive halos and therefore we expect greater shift for the same underlying halo populations. Accordingly, a relation of $b_2$ ($b_1$) for a different choice of halo occupation distribution (HOD) can be derived by weighing $b_1$ and $b_2$ with the halo mass function and the halo occupation distribution. Figure 13 of \citet{PadBAOshift} shows that the difference due to different HOD models is a few sub-percent offset while following a similar bias dependence. \citet{Mehta_etal_11} empirically derived BAO shifts for various HOD models that range from 0  up to 10\% of the satellite fraction; their Figure 8 implies that the BAO shift can be described mainly by the dependence on $b_1$ with only moderate dependence on the HOD variations. To summarize, we believe that the HOD dependence would be moderate for a moderate satellite fraction and the resultant offset in the BAO scale shift from what we measured in this paper can be predicted analytically. 

The bottom left panel of Figure \ref{fig:alpha_bias} shows the BAO shift along the line of sight in redshift space before reconstruction. Unlike the $\aperp$ case, the low bias convergence appears to increase with decreasing redshift, i.e., from 0.22\% to 0.36\% from $z=2.5$ to $z=0$. Again, the deviation from the convergence as a function of bias rises more rapidly at lower redshift.  For example, the expected nonlinear shift on $\apar$ is $\sim 0.3\%$ for tracers with $b\sim 1.5$ at $z=1-1.5$ (e.g., DESI ELG-like) as well as tracers  with $b\sim 3$ at $z\sim 2.5$ (e.g., DESI QSO-like). For  $b\sim 2$  at $z=0.6$ and $z=1$, we expect $\sim 0.56\%$ and $\sim 0.4\%$, respectively. Comparing the top and the bottom panels, the values of $\apar$ are systematically greater than $\aperp$ even for the EFT0 model, which is contrary to the matter case in Figure \ref{fig:alpha_ZV}.

Figure \ref{fig:delta_alpha_std} shows the effect of different fitting models on the BAO scales as a function of galaxy/halo bias at three redshift bins, $z=0.6$ (top panels), 1 (second row panels), and 2 (third row panels).
For the top three rows, again, we show errors for $\Delta \alpha$'s between the EFT1 and the EFT0 models and between the SBRS and the EFT0 models and draw the derived errors on the EFT1 and the SBRS points without any error bars on the EFT0 points; such errors are negligible and hard to be identified in the figures. Figure~\ref{fig:delta_alpha_stdS} shows $\Delta \alpha$ between the EFT1 and the EFT0 models explicitly. 

Unlike the matter case (Figure \ref{fig:alpha_ZVS}), we observe much smaller dependence on the fitting models, i.e., less than $\pm \sim 0.03\%$ for $\aperp$ and $\pm \sim 0.07\%$ for $\apar$ before reconstruction.
 While for matter the anisotropic BAO scale shift was dependent on fitting models, the biased cases consistently seem to show anisotropic BAO scale shift for all fitting models. In detail, although insignificant, the EFT1 model and the SBRS model again tend to increase the shift on $\apar$ relative to $\aperp$. 

\subsubsection{Post-reconstruction}

The right panels of Figure \ref{fig:alpha_bias} show the post-reconstruction best fit $\alpha$'s. As expected, the nonlinear BAO shifts decrease well within $\pm 0.1\%$ in general except for a couple of high bias cases ($b>1.5$) at $z=0$. With low galaxy/halo bias samples that have higher signal-to-noise due to lower shot noise, we detect $< \pm 0.06\%$ of remaining shift on $\aperp$ after reconstruction. For $\apar$, we observe a level of $\pm 0.1\%$ remaining shift after reconstruction. 
The right panels of Figure \ref{fig:delta_alpha_std} (and \ref{fig:delta_alpha_stdS}) then show that the fitting model dependence is negligible after reconstruction: less than $\pm 0.01\%$ for $\aperp$ and 0.03\% for $\apar$.

\subsubsection{Precisions and the reduced $\chi^2$}

The bottom panels of Figure \ref{fig:delta_alpha_std} show precisions on $\alpha$ using the three fitting models at $z=0.6$ as an example. Again, the derived precisions both before and after reconstruction
are found to be quite insensitive to the fitting models as in the matter case. The post-reconstruction errors decrease by $\sim 30\%$ at $b=1.19$ and $\sim 20\%$ at $b=3.3$ for $\aperp$ and by $\sim 50\%$ at $b=1.19$ and $\sim 30\%$ at $z=3.3$ for $\apar$. The improvement decreases with increasing bias as a higher bias sample suffers higher shot noise which makes the reconstruction less efficient. Also, our results tend to show more improvement for $\apar$ in these biased cases. 
The bottom panels also compare the reduced $\chi^2$ of the three models. All three models perform similarly before reconstruction in terms of the goodness of the fit. After reconstruction, the SBRS model seems to perform worse than the other model for low bias cases, while the EFT0 and EFT1 perform similarly.

\begin{figure*}
\centering
\includegraphics[width=0.45\linewidth]{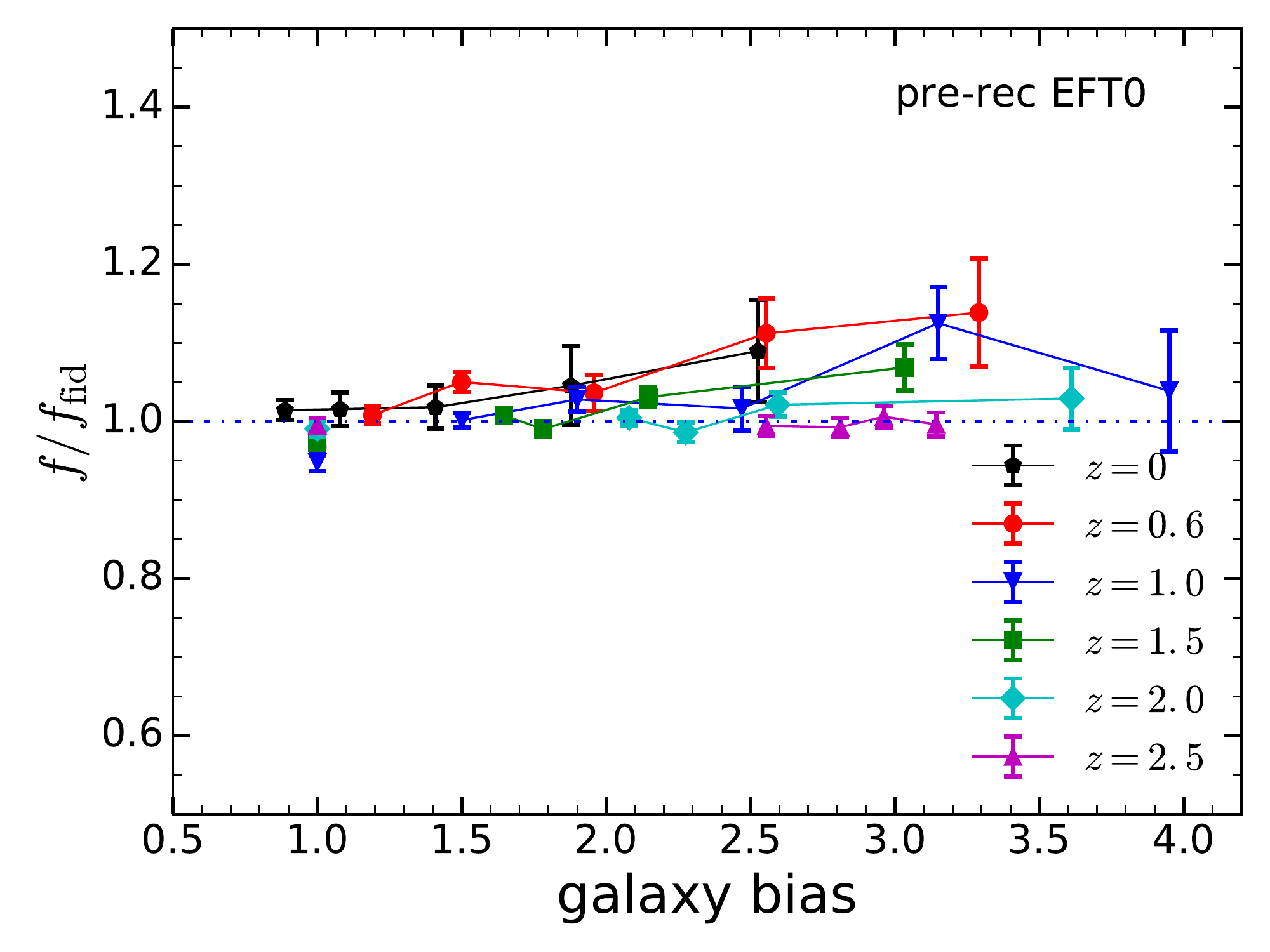}
\includegraphics[width=0.45\linewidth]{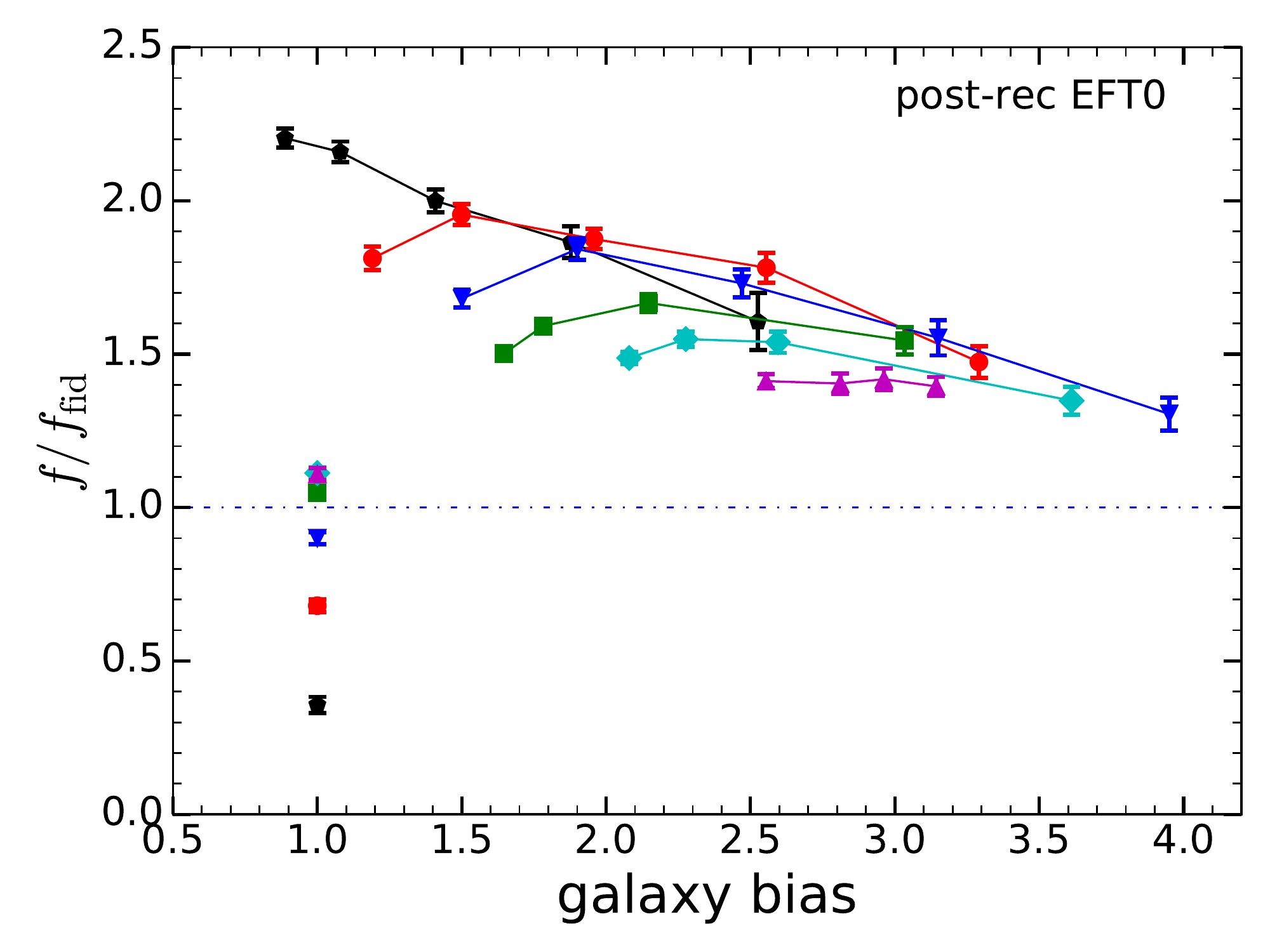}
\includegraphics[width=0.45\linewidth]{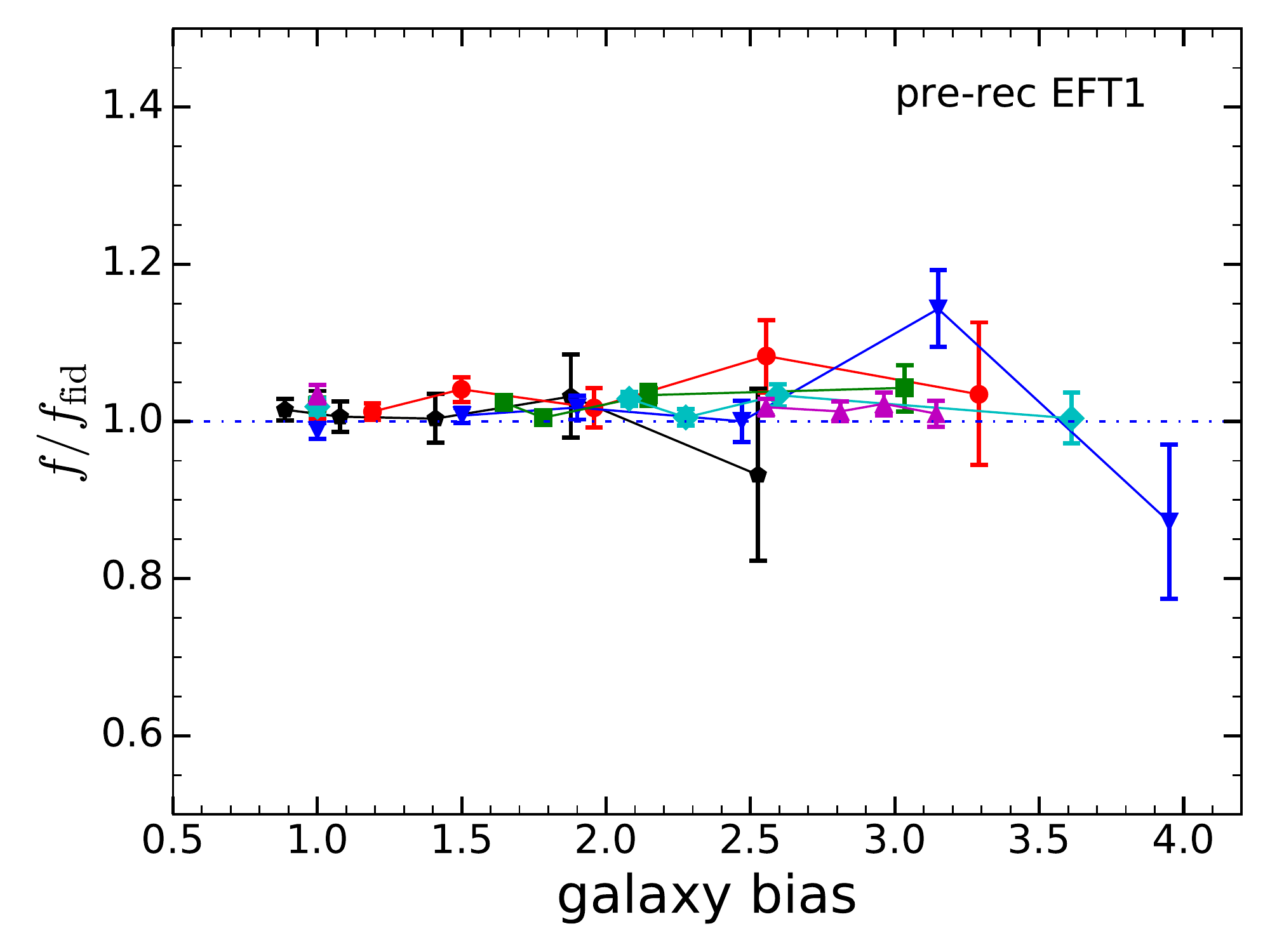}
\includegraphics[width=0.45\linewidth]{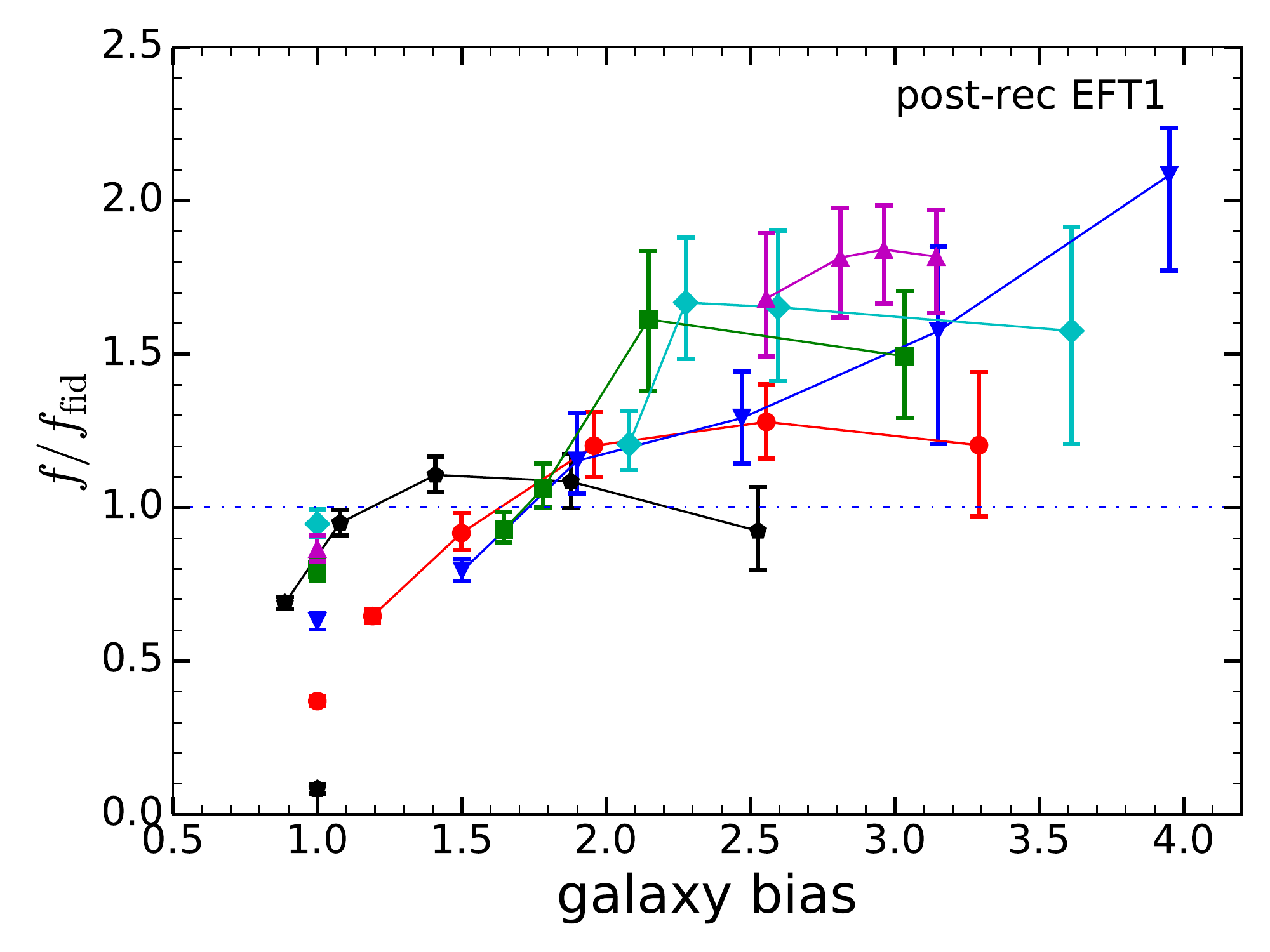}
\caption{Constraints on growth rate $f$ from the EFT0 (upper panels) and EFT1 model (lower panels). \textit{Left panels:} pre-reconstruction. \textit{Right panels:} post-reconstruction. The best fits $f$ (data points) are compared to the linear $f$ ($\Omega_m(z)^{0.56}$, dash-dotted lines). Single points denote for matter, and points with connected solid lines are for galaxies/haloes. Different colours denote different redshifts.}\label{fig:ZV_fit_f}
\end{figure*}

\begin{figure*}
\centering
\includegraphics[width=0.3\linewidth]{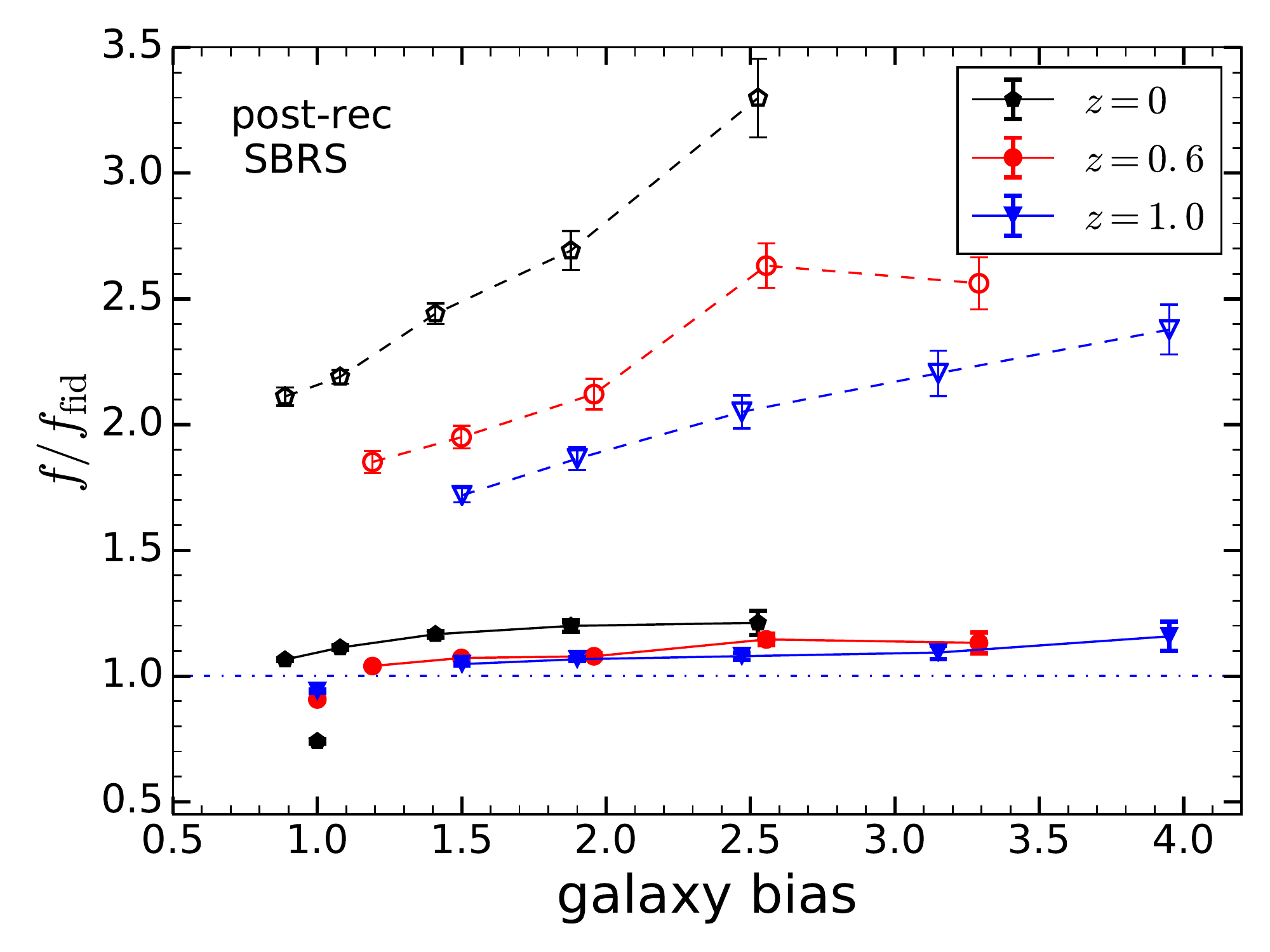}
\includegraphics[width=0.3\linewidth]{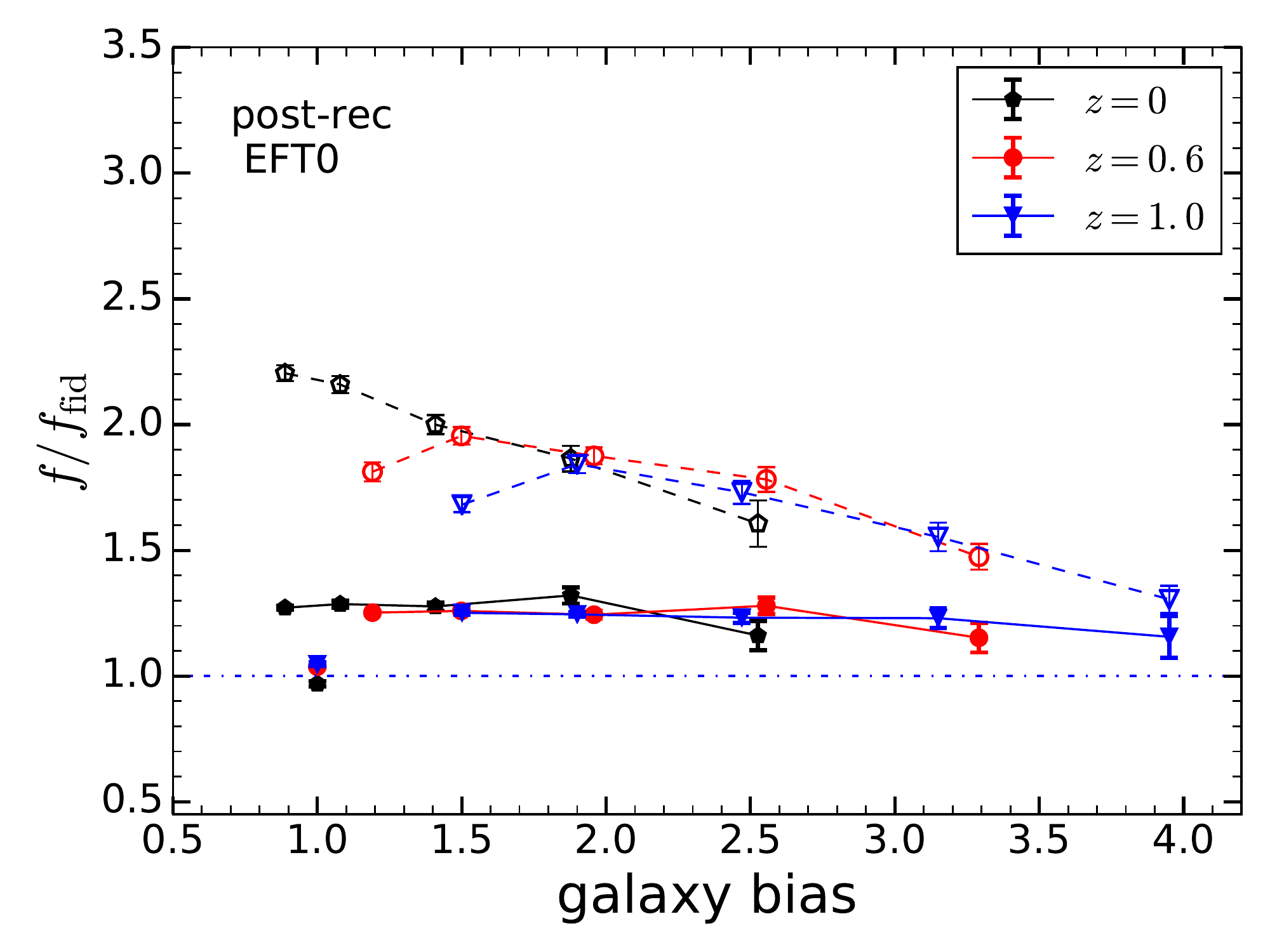}
\includegraphics[width=0.3\linewidth]{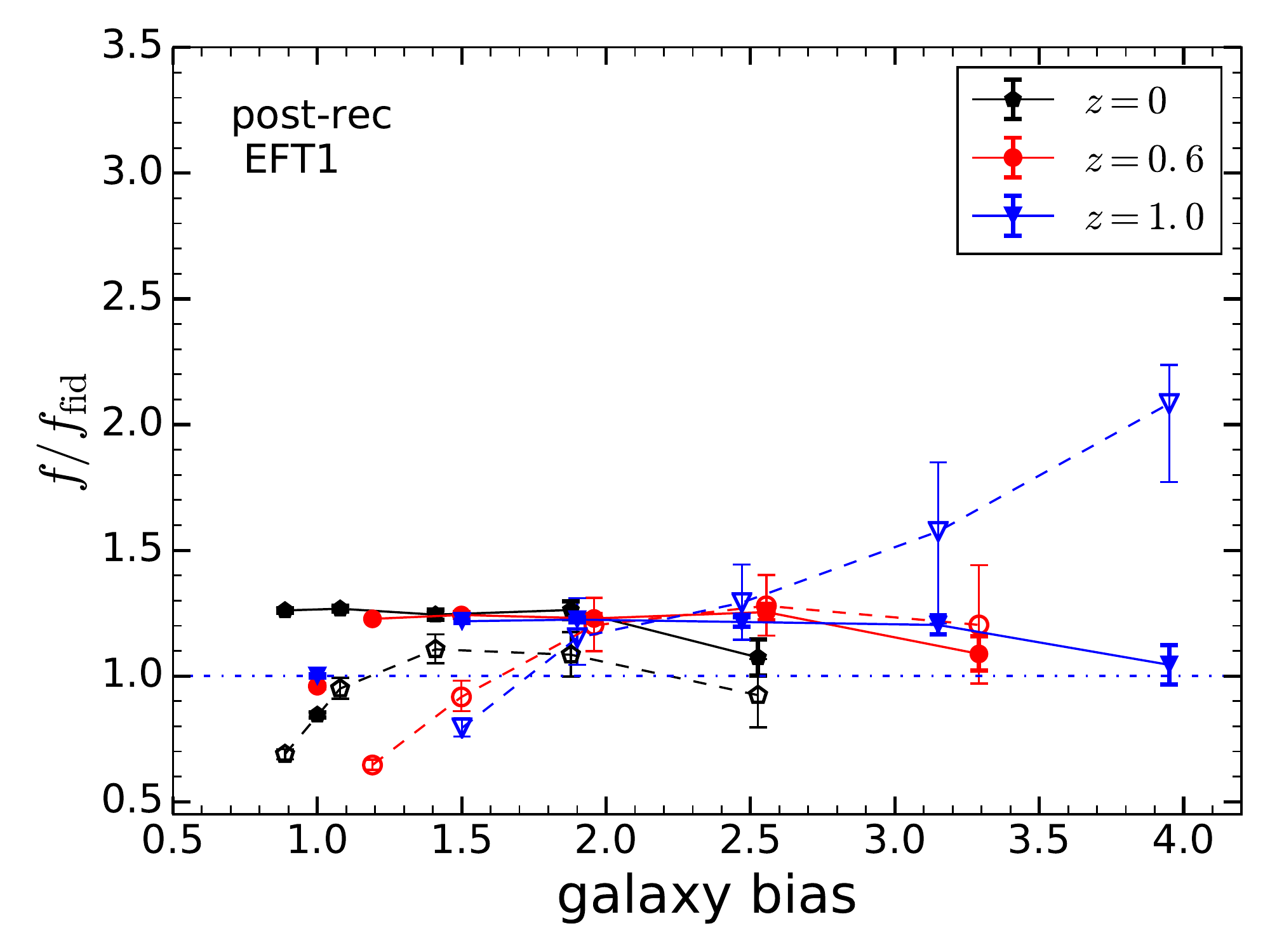}
\caption{Growth rate $f$ using different reconstruction conventions: `Rec-Cohn'(sold lines with data points) in comparison to `Rec-Iso' (dashed lines with data points). \textit{Left:} using the SBRS model. \textit{Middle:} using the EFT0 model,  \textit{Right:} using the EFT1 model. Dash-dotted horizontal lines denote the linear $f=\Omega^{0.56}_m(z)$. Singular points without line connection denote matter cases.
}\label{fig:Cohn_fit_f}
\end{figure*}

\subsubsection{Summary}

In summary, we find that before reconstruction the nonlinear shift of the transverse BAO scale $\aperp$ tends to converge toward $\sim 0.1\%$ for biased tracers in the low bias limit regardless of redshift. As bias increases, the expected shift increases, but very slowly at high redshift. Such tendency qualitatively agrees with predictions from perturbation theories. As a result, we expect a small shift on $\aperp$ at high redshift even for highly biased samples. The values of $\apar$ that account for the RSD effect show stronger redshift dependence while the bias dependence appears comparable to that of $\aperp$. After reconstruction, we find the shift reduces to less than $0.1\%$ except for high bias at low redshift. We also find that the shift measurements are more robust than the matter cases over the range of the fitting models we test, before and after reconstruction. Like the matter case, the precision of the BAO scale is robust against the choice of fitting model. 
In terms of the goodness of the fits, all three models perform similarly before reconstruction, while the EFT0 and the EFT1 perform better for low-bias cases after reconstruction. 

\subsection{BAO systematics dependence on reconstruction conventions}

We compare the measured BAO systematics for different reconstruction schemes: our default `Rec-Iso' (i.e., an isotropic reconstruction scheme) that we have tested in previous sections v.s., `Rec-Cohn' (i.e., an anisotropic reconstruction scheme) using the EFT0 model. The upper panels of Figure \ref{fig:Cohn_fit_alpha} compares the derived BAO scales for the two cases (solid points and lines for `Rec-Iso' and open points with dashed lines for `Rec-Cohn'). Since two operations differ only in the line-of-sight displacement, we expect that post-reconstruction $\aperp$ would be similar between the two cases. We indeed find more consistent results for $\aperp$ compared to the results for $\apar$. The error bars in the top panels denote dispersions on the mean $\Delta \alpha$ between the two reconstruction schemes and are plotted only for the `Rec-Cohn' cases.  In detail, the matter cases show relatively consistent results, i.e. $\sim 0.04-0.08\%$ offset in $\aperp$ and $\apar$, respectively; for the biased cases, the dependence curve as a function of bias is very similar between the two conventions, due to our operating on the same underlying simulations, while the difference increases at lower redshift and at low bias limit. We find that the offset is at the order of 0.1\% for $\aperp$ and 0.3\% for $\apar$ at $z=0$ for $b\sim 1$ in the worst case. This offset for $\apar$ (top right) is greater than the typical fitting formula dependence from the right panels of Figure~\ref{fig:delta_alpha_std} (or see Figure~\ref{fig:delta_alpha_stdS}).
When the offset is largest, the `Rec-Iso' scheme returns smaller post-reconstruction BAO shift on $\apar$ than the Rec-Cohn scheme, which can be one advantage of the `Rec-Iso' scheme compared to the `Rec-Cohn' scheme. We find the same trend when using the SBRS fitting model instead of the current EFT0 model.

The lower panels show that the precisions are very similar for the two reconstruction conventions. This is consistent with \citet{Burden2014,Seo_etal_16} that found that anisotropic and isotropic reconstructions return very similar signal to noise both in the transverse as well as along the line of sight.~\footnote{As a caveat, the anisotropic reconstruction they tested are not identical to the convention used in this paper, and their test is for a galaxy population with shot noise level of the BOSS CMASS sample. \citet{Seo_etal_16} implied that one may find difference at very low shot noise level, but we do not see such indication. } 

\subsection{Nonlinear damping scales}

We next investigate behaviors of parameters other than the BAO scale in order to test if our fitting models catch general features of the difference power spectrum correctly.

The first parameter is the nonlinear damping scale that describes the damping of the BAO on small scales. As will be explained in \S~\ref{subsec:rsd}, we switch to showing the best fit $\Sigxy$ and $\Sigsm$ of the mean difference power spectra instead of the average $\Sigxy$ and $\Sigsm$ of the best fits. 
The left panel of Figure \ref{fig:Z1_fit_Sigma} shows the average of the best fit damping scales in the transverse direction $\Sigxy$ for the EFT1 model relative to Eq.~\ref{eq:Sigxy} that was used for the pre-reconstruction EFT0 model; the right panel shows the best fit $\Sigsm$ relative to the input values of $\Sigsm$ after reconstruction. The singular points at $b_1=1$ denote the matter cases and the connected points denote the biased cases. Focusing on biased cases, we find that the pre-reconstruction best fit damping scales tend to be slightly greater than the predicted values 
which could be due to any additional implicit damping effects such as the sizes of the grids and the finger of God effect, etc. 
For matter, the best fit with the EFT1 model tends to return noticeably greater damping than what was predicted (by $9\%$ at $z=0$ and by $15$\% at $z=2.5$), which could be related to why we observed a greater sensitivity between models with and without free damping parameters for the matter case in Figure \ref{fig:alpha_ZV} compared to the biased cases. That is, the pre-reconstruction matter BAO feature is more smeared than the expectation while in the biased sample the one-loop EFT predicted value appears reasonable except for highly biased cases at each redshift. Although not shown in this paper, our inspection of propagators (i.e., the cross correlation between the initial field and the final nonlinear field) also indicates greater damping for the matter cases compared to the biased cases, more so along the line of sight due to the stronger finger of God effect; this likely drives the best fit $\Sigxy$ to larger values given the limited freedom in $\Sigz = (1+f)\Sigxy$ since $f$ is also constrained by the amplitude of the power spectrum.   

The right panel of Figure \ref{fig:Z1_fit_Sigma} shows the post-reconstruction cases. We have three damping scale parameters in this case, i.e., $\Sigdd$, $\Sigss$, and $\Sigsd$. Instead of setting all of these parameters as free parameters, we set the smoothing scale $\Sigsm$ as a free parameter and let it vary around the input smoothing scale that we actually used for reconstruction. Since $\Sigdd$, $\Sigss$, and $\Sigsd$ depend on $\Sigsm$ (Eq.~\ref{eq:Sigma_dd}-\ref{eq:Sigma_ss}), this allows us to coherently perturb these three parameters; on the other hand, $\Sigsm$ also adversely affects post-reconstruction RSD amplitude (Eq.~\ref{eq:EFT_post}). The best fit $\Sigsm$ deviates the most from the input at low redshift, implying either that the observed nonlinear damping is greater than a combination of the predicted values of $\Sigdd$, $\Sigss$, and $\Sigsd$ or that the observed amplitude anisotropy deviates from the $1-S(k)$ model at low redshift. When we inspect individual power spectrum components of the post-reconstruction power spectrum, i.e., $\Pdd$ and $\Pss$ and $\Psd$, we find that the best fit $\Sigsm$, while it giving a better fit to the overall $\Prec$, gives worse matches to individual components than the true input $\Sigsm$, probably due to its side affect on the amplitude anisotropy $1-S(k)$. That is, we believe that the deviation in $\Sigsm$ should be simply taken as a phenomenological indicator of our post-reconstruction model fairing worse at low-redshift rather than as an indicator for a physically motivated alternative model for $\Sigsm$.

\begin{figure*} 
\includegraphics[width=0.45\linewidth]{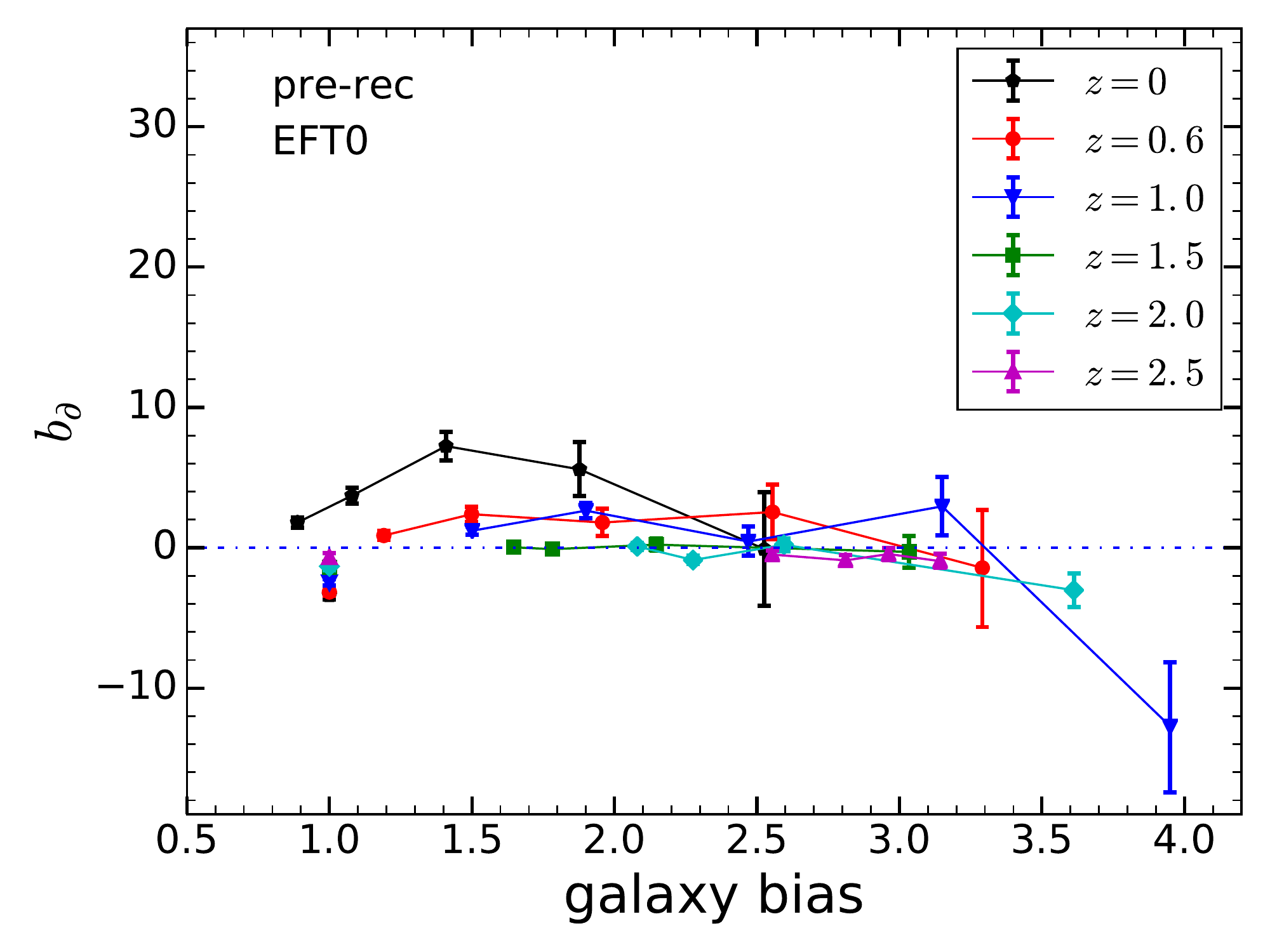}
\includegraphics[width=0.45\linewidth]{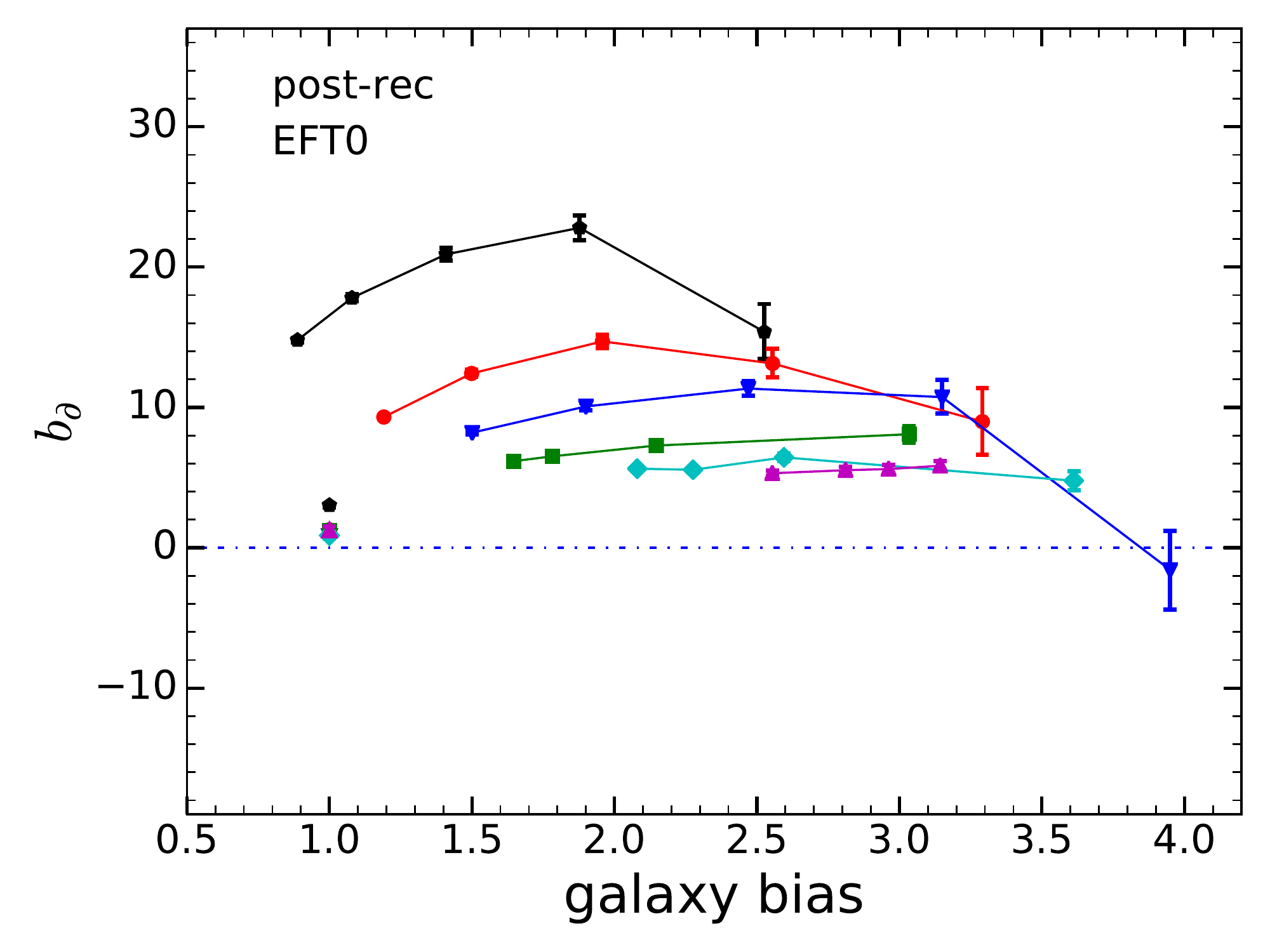}
\includegraphics[width=0.45\linewidth]{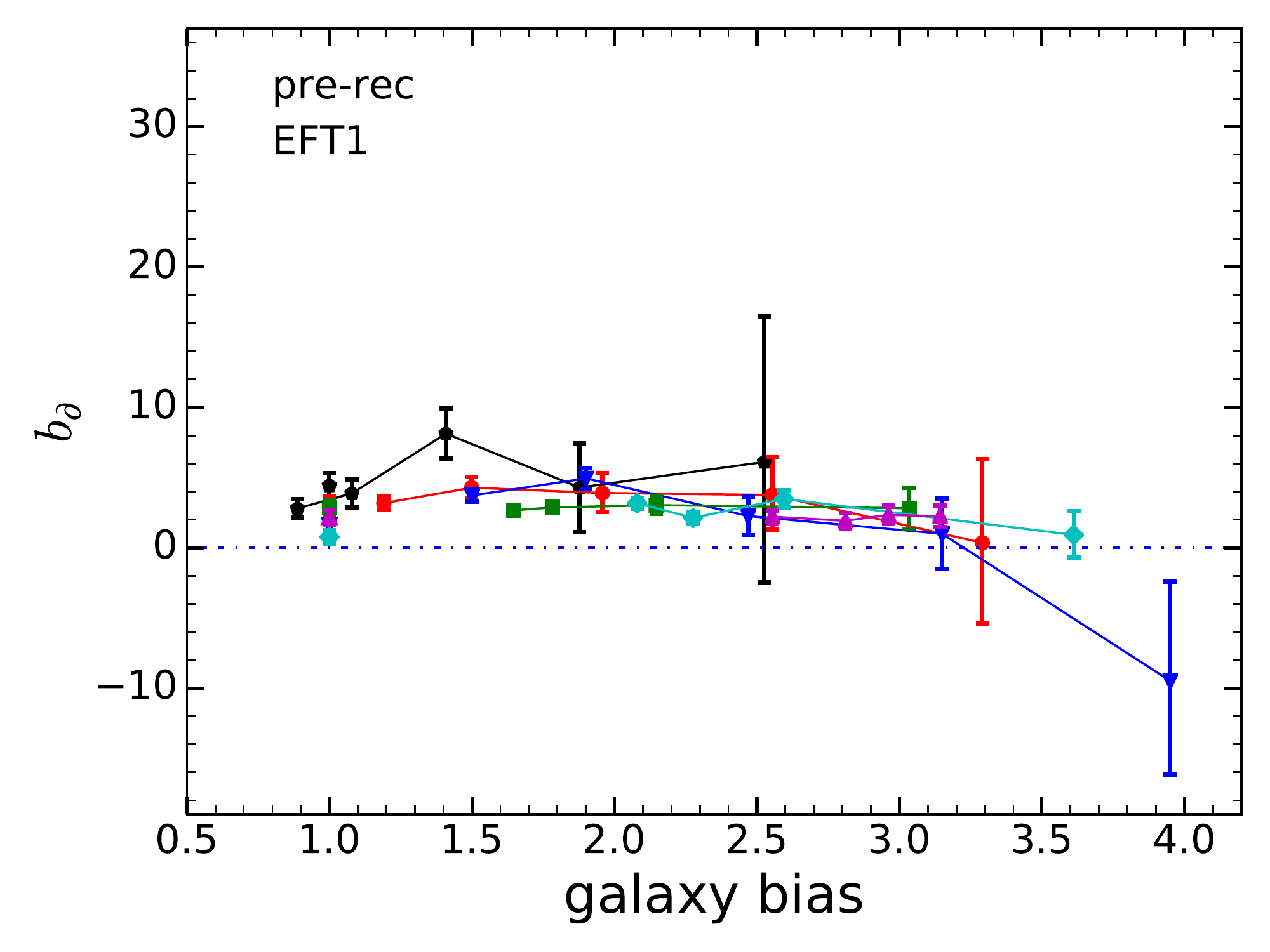}
\includegraphics[width=0.45\linewidth]{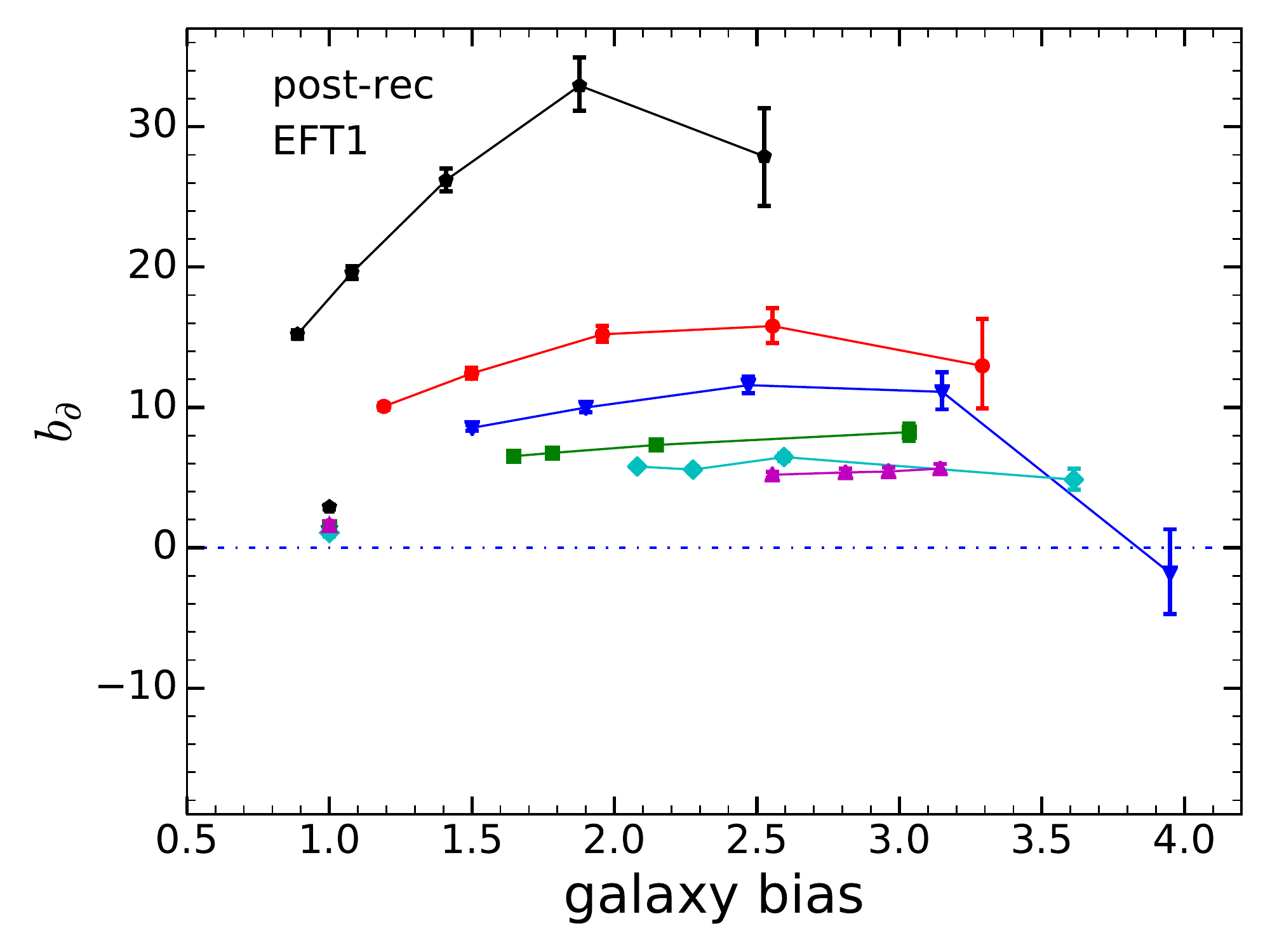}
\caption{Constraints on $b_{\partial}$. 
From the EFT0 model (upper panels) and the EFT1 model (lower panels). \textit{Left panels:} pre-reconstruction. \textit{Right panels:} post-reconstruction. Isolated singular points denote for the matter case and the connected points for galaxies/halos. Different colours represent different redshifts.}\label{fig:ZV_bpartial}
\end{figure*}

\subsection{Redshift-space distortion parameter--growth rate $f$}
\label{subsec:rsd}

Our operation of differencing a paired set of wiggled and dewiggled power spectra allowed us to minimize the number of nuisance parameters for the shape of the power spectrum. Although it is not the main focus of this paper, we examine our EFT models from the perspective of grasping the redshift-space distortion effect (i.e., growth rate $f$) and nonlinear scale-dependent bias effect $b_\partial$ from the amplitude of the BAO feature. That is, we are testing if these parameters grasp physics rather than being phenomenological. As a caveat, in the EFT models, the free parameter $f$ in our case determines the overall amplitude modulation of the BAO feature as well as the modulation of the damping of the BAO as a function of the line-of-sight angle; on the other hand, in the SBRS model, $f$ only modulates the amplitude while the damping modulation is fixed with $\ffid$. 
Also, because the information on $f$ in our case is derived purely from the damping and amplitude of the BAO feature alone rather than the modulation of the overall power spectrum and because our fitting range extends to $k=0.3\ihMpc$ which is much larger $k$ than typical redshift-space distortion analyses of $k<0.15-0.2\hMpc$, we do not expect high-accuracy determination of $f$. We indeed observe outliers and non-symmetric errors with discrepant estimates between the posterior mean and the maximum likelihood points when we derived the best fits of post-reconstruction individual power spectra using the EFT1 model. For these reasons, for the EFT1 model, we switch from showing the average $f$ of 40 best fit parameters to showing the best fit $f$ of the mean difference power spectrum for these non-BAO parameters both before and after reconstruction. 

\begin{figure*}
\centering
\includegraphics[width=0.425\linewidth]{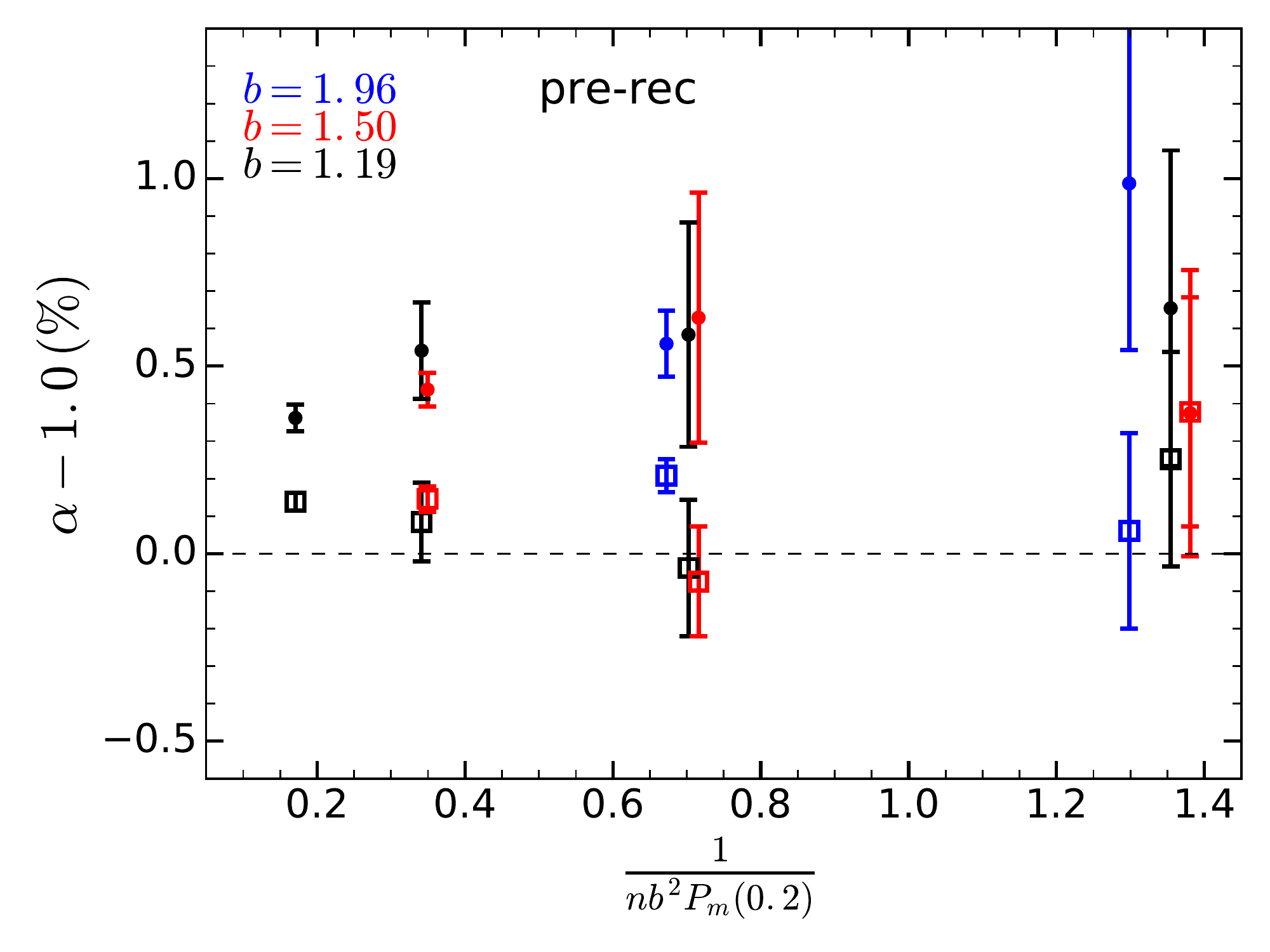}
\includegraphics[width=0.425\linewidth]{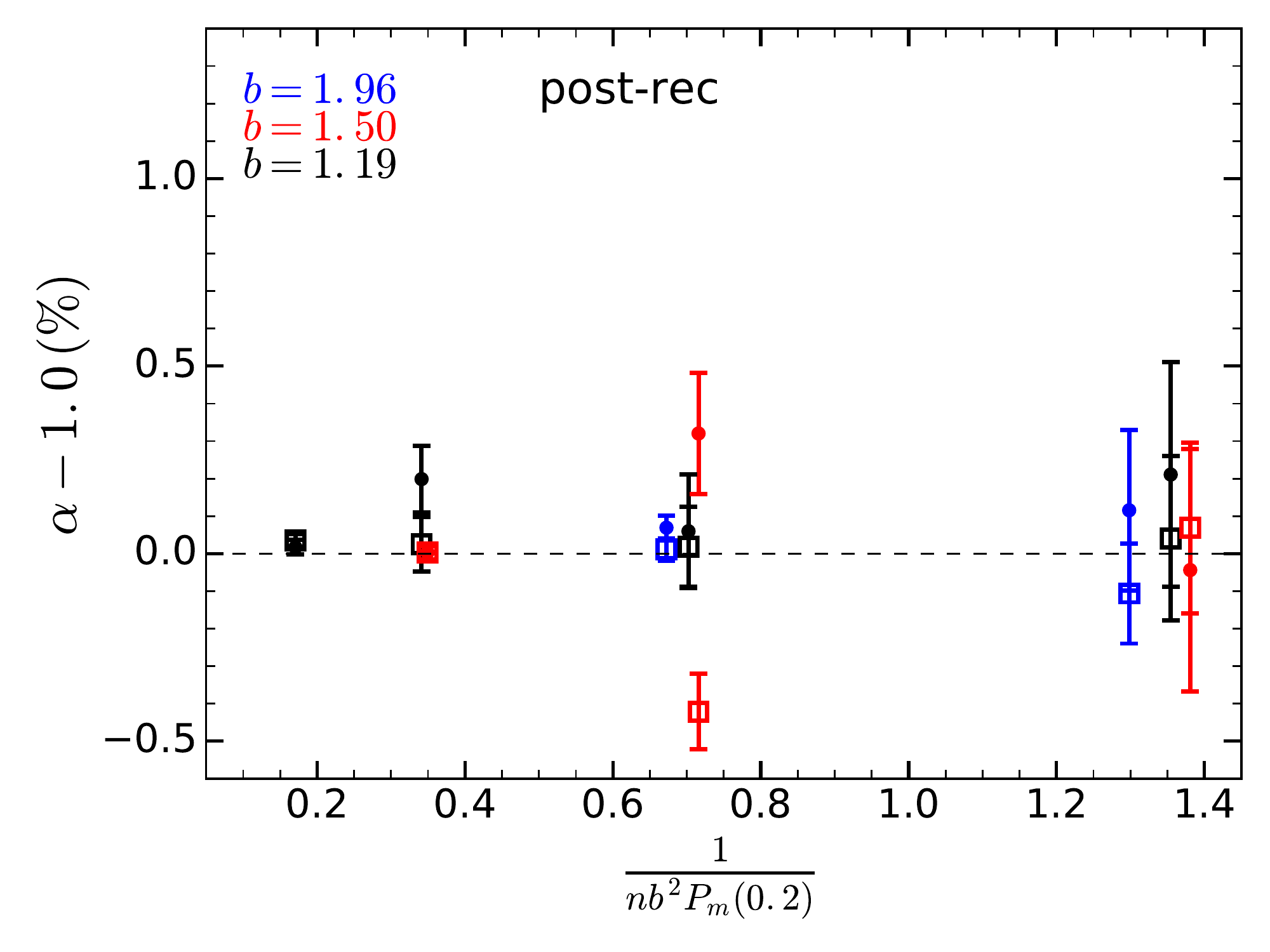}\\
\includegraphics[width=0.425\linewidth]{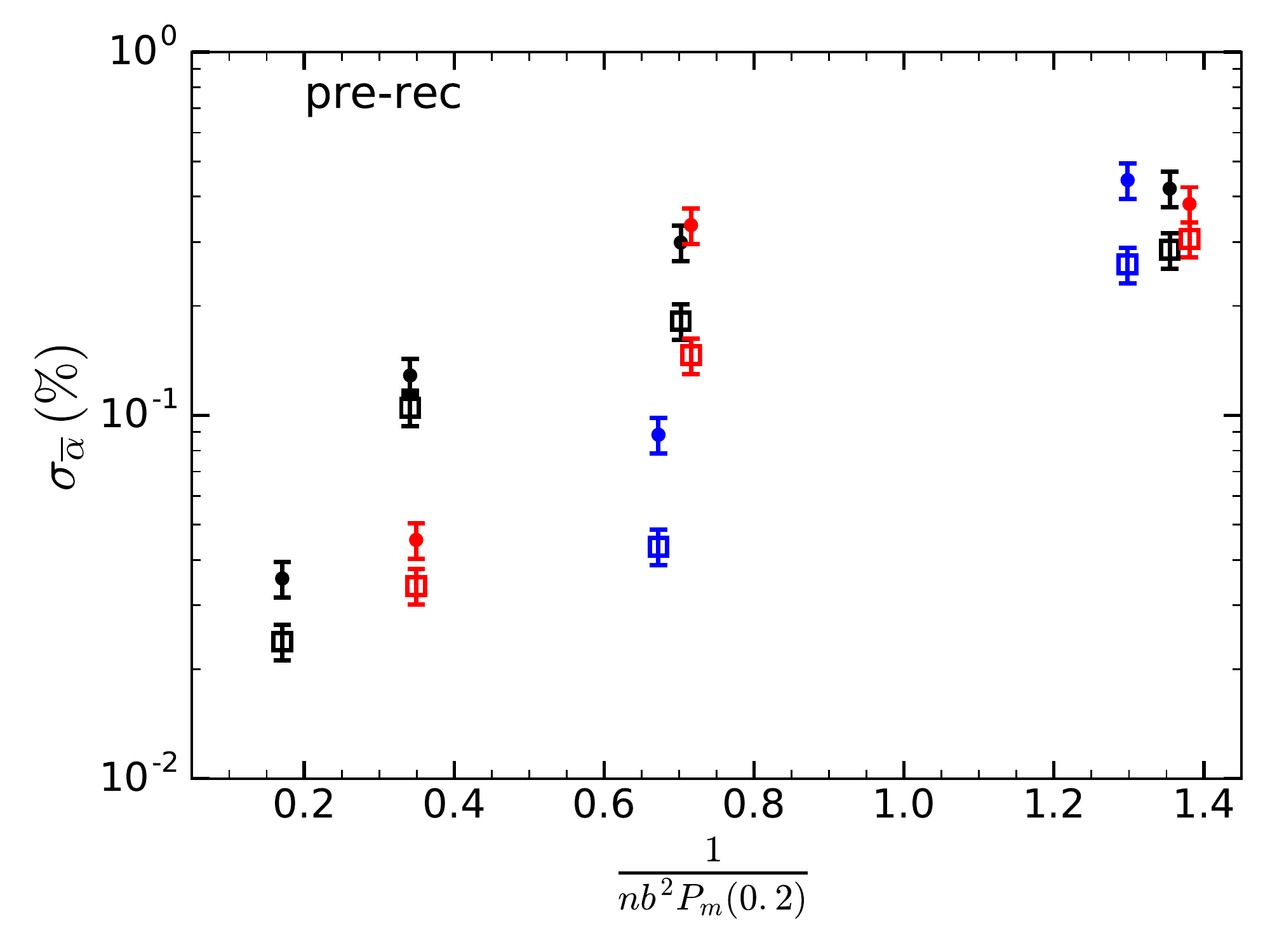}
\includegraphics[width=0.425\linewidth]{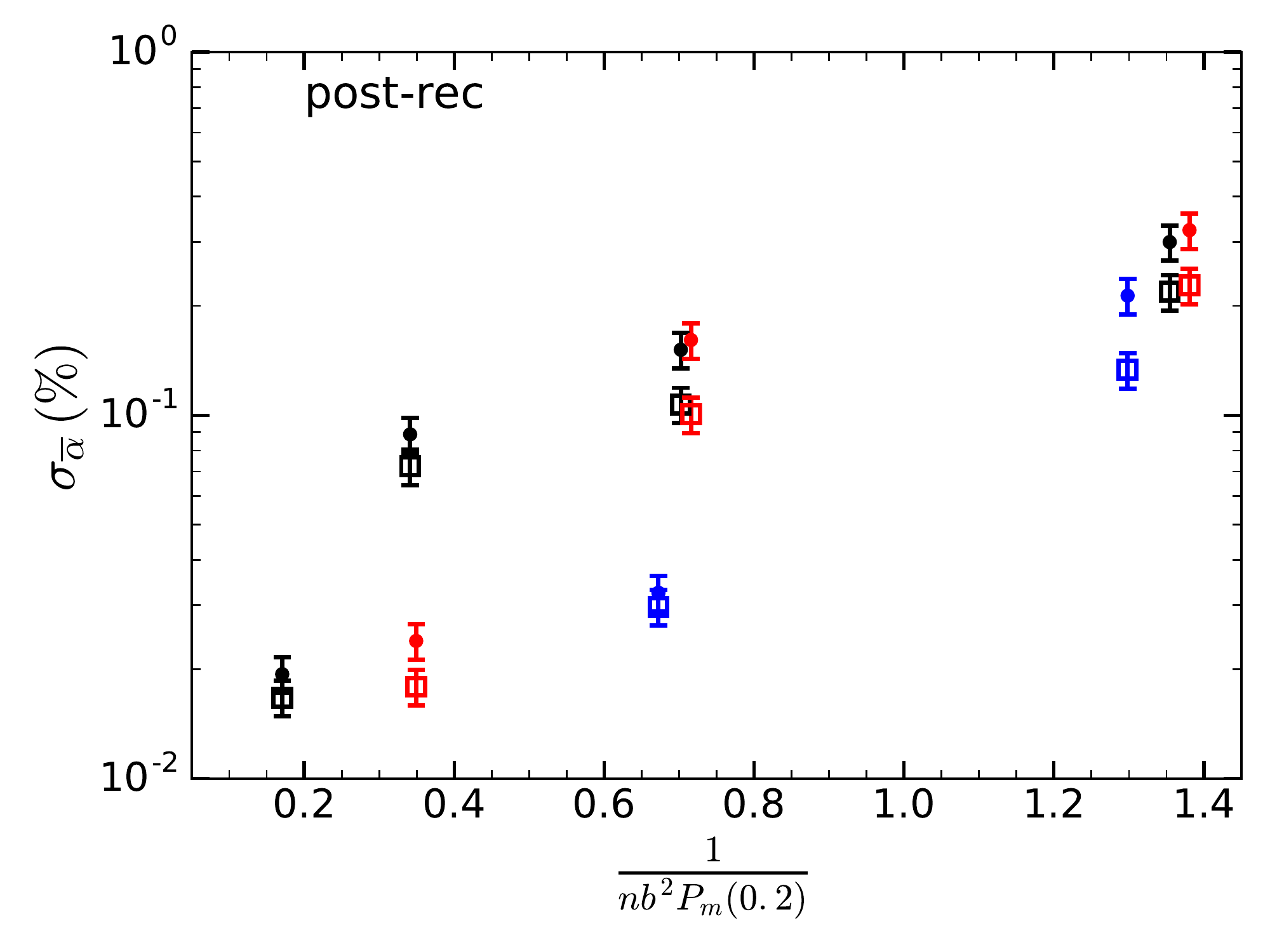}
\caption{The BAO scale measurements as a function of shot noise effect given fixed galaxy bias: $\alpha_{\bot}$ (empty squares) and $\alpha_{\|}$ (solid circles). We focus on $z=0.6$. The values of $1/[nb_1^2P_m(0.2\ihMpc)]$ represent the noise-to-signal level at $k=0.2\ihMpc$. We used the EFT0 model. A given colour denotes the same underlying galaxy population with different subsampling, i.e., shot noise. \textit{Left panels:} pre-reconstruction. \textit{Right panels:} post-reconstruction. \textit{Upper panels:} average of the best fit $\alpha$'s. \textit{Lower panels:} 1-$\sigma$ error on the mean $\alpha$. $1/[nb^2P_m(0.2\ihMpc)]$ is derived using the number density $n$ and the galaxy bias $b_1$  (Table~\ref{tab:fof_pre_rec}) with the real-space matter power spectrum at $k=0.2\ihMpc$, $P_m(0.2)$, at $z=0.6$. }\label{fig:shotnoise}
\end{figure*}

Figure \ref{fig:ZV_fit_f} shows the best fit growth rate $f$ for the pre-reconstruction cases (left) and the post-reconstruction cases (right panels). The top panels show the results using the default EFT0 model and the bottom panels show the EFT1 model.  Before reconstruction (left panels),  one finds that the derived $f$ values are fairly consistent with the expected linear $f$ values (i.e., $\Omega_m(z)^{0.56}$, dot-dashed lines at unity) especially at lower galaxy bias and at higher redshifts. With the EFT0 model (top left), we find $f$ derived for matter (isolated single points at $b = 1$) tend to be underestimated for $0.6 \le z \le 1.5$ compared to the EFT1 model. The EFT1 model that allows a free $\Sigxy$ gives a better match to the linear $f$ at low redshift. For the biased cases, the two models return similar results. 

The density field reconstruction modifies RSD in the post-reconstruction power spectrum in a manner that depends on the reconstruction convention. Our default reconstruction scheme (`Rec-Iso') attempts to remove large-scale velocity field based on one's guess of $f'$ on true $f$ (in our case $f'=f$). This will remove RSD on large scales. The remaining RSD information in this case therefore comes from quadrupole due to$~\sim f\mu^2(1-S(k))$ in Eq. \ref{eq:EFT_post}. This small quadrupole can in principle return a reasonable fit to $f$ if we use the entire power spectrum information~\citep{Seo_etal_16,Beutler_etal_16} in the presence of a large set of free parameters. However, in the case of the difference power spectrum we utilize in this paper, $f$ is constrained only based on the quadrupole in the {\it difference}, i.e., the BAO feature alone with a very limited freedom (i.e., $b_\partial$ and $b_1$).  As a result our post-reconstruction $f$ constraint appears sensitive to nonlinear terms in our fitting model and are largely off from the expected values as shown in Figure \ref{fig:ZV_fit_f}. In the bottom right panel, for matter, $f$ at low redshift tends to be substantially underestimated in the EFT1 model. For the biased case, by allowing perturbations on the damping scales and the anisotropic amplitude (through $\Sigsm$), the EFT1 model results are closer to the expected values for $z \le 0.6$. This implies that $f$ and $\Sigsm$ is correlated in the EFT1 model, as shown in the bottom right panels of Figure~\ref{fig:DM_likelihood} and \ref{fig:galaxyzone_likelihood}.

We believe that the observed model-sensitive deviation of $f$ is partly a result of the choice of the reconstruction scheme rather than the limit in the EFT0 model. We test this claim by choosing `Rec-Cohn' that leaves in the full RSD effect during reconstruction. Figure \ref{fig:Cohn_fit_f} shows the best fits $f$ using the `Rec-Cohn' scheme are much more consistent over different fitting models and returns $f$ closer to the linear values except for the EFT1 model, compared to the `Rec-Iso' case.

\begin{figure*}
\centering
\includegraphics[width=0.45\linewidth]{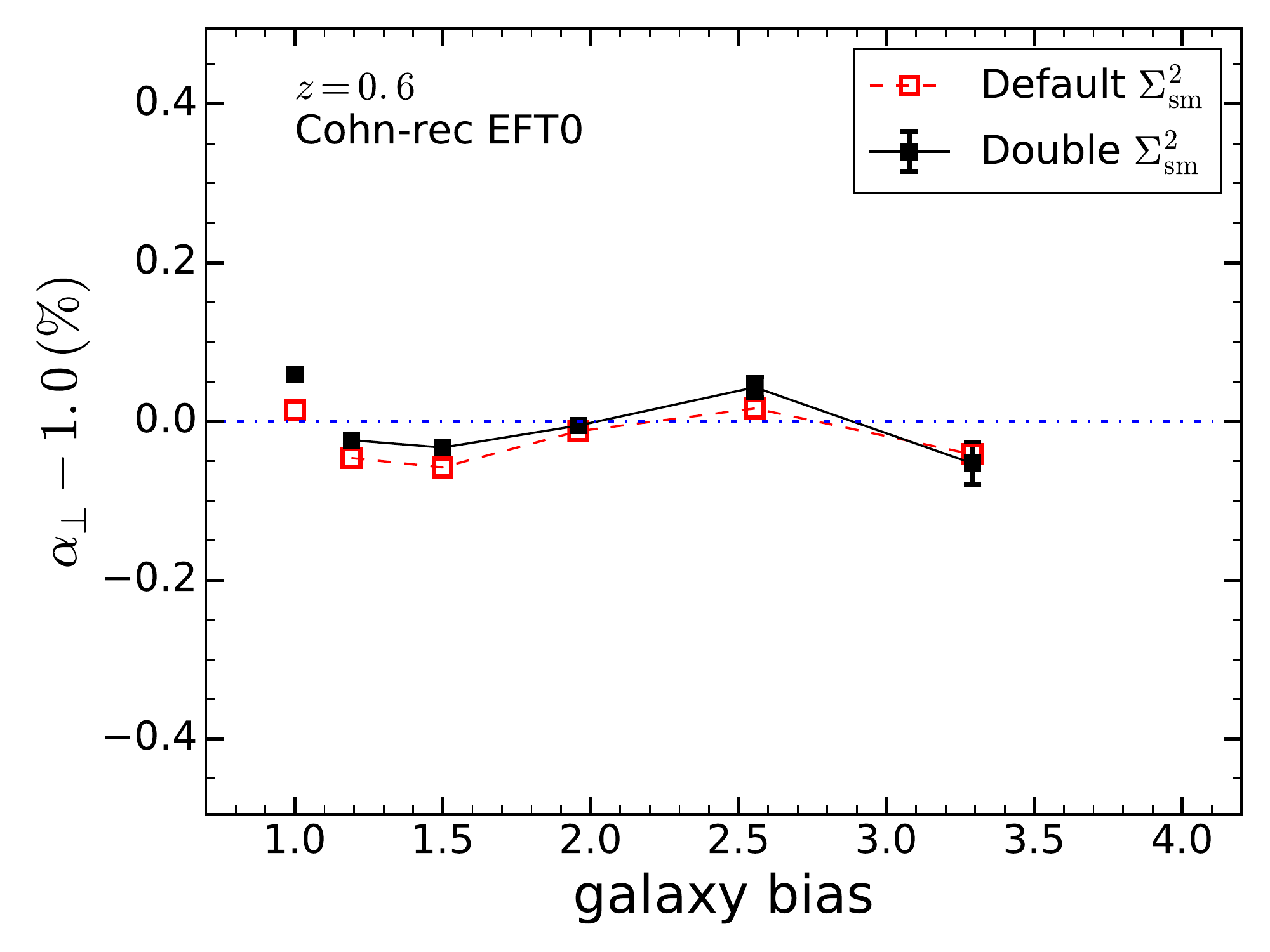}
\includegraphics[width=0.45\linewidth]{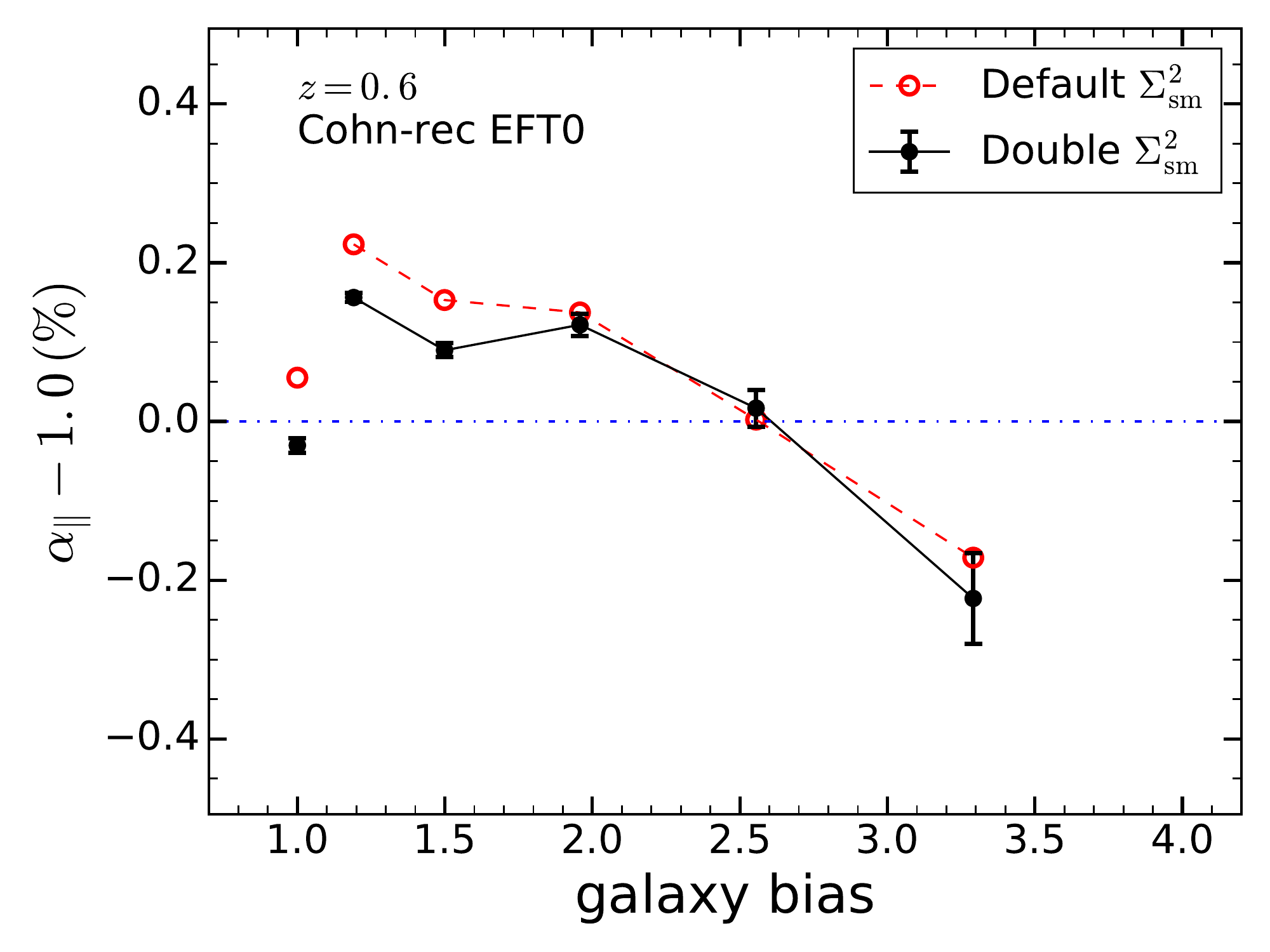}
\includegraphics[width=0.45\linewidth]{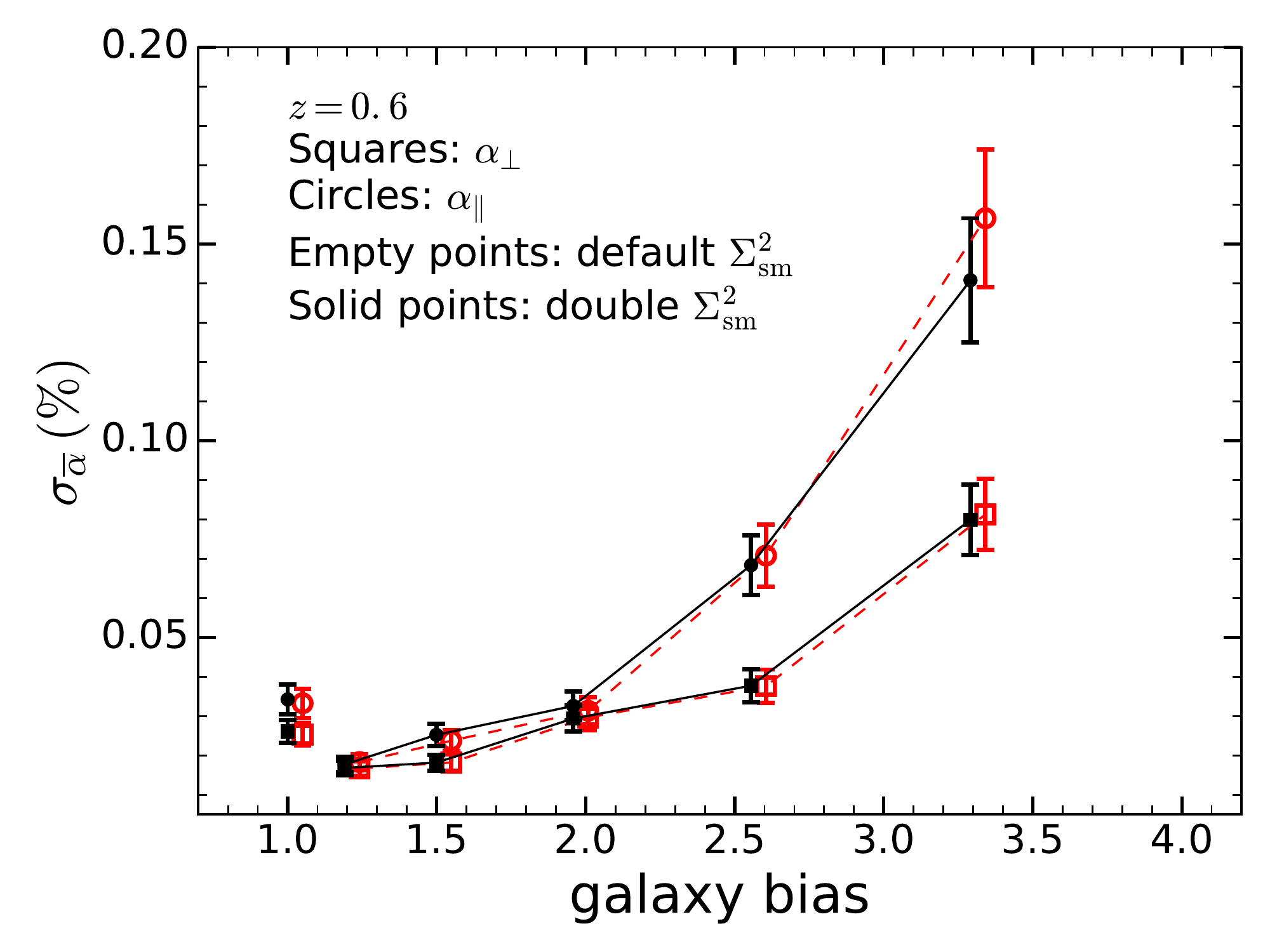}
\includegraphics[width=0.45\linewidth]{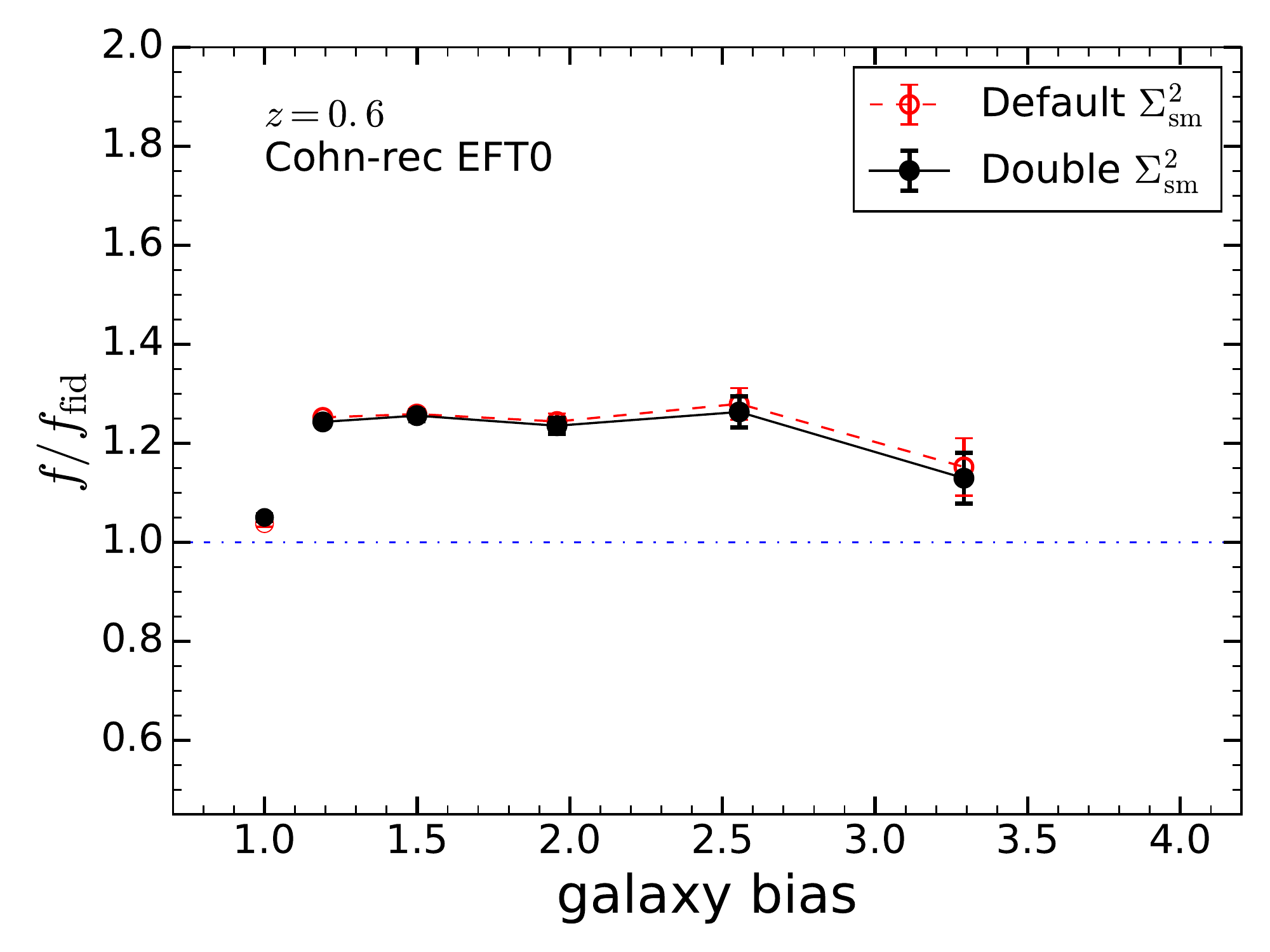}
\caption{The effect of the smoothing scale $\Sigsm$ on the BAO shift and growth rate $f$ in the `Rec-Cohn' convention. We focus on $z=0.6$ and use the EFT0 model. 
 The error bars in the top panels again represent the error on the mean difference between the two smoothing scales, derived from the dispersions of individual difference $\alpha$'s.
The upper panels show that increasing smoothing scale $\Sigsm$ by a factor of $\sqrt{2}$ decreases the residual post-reconstruction bias in $\alpa$ at low galaxy biases:  by~$0.07$\% for galaxy bias $1.2$. The bottom left panel shows increasing $\Sigsm$ by $\sqrt{2}$ does not obviously change the precisions of $\alpha$'s. The bottom right panel shows there is no improvement in bringing the growth rate $f$ closer to the linear prediction.
}\label{fig:RecCohnSigsm}
\end{figure*}

\begin{figure*}
\centering
\includegraphics[width=0.45\linewidth]{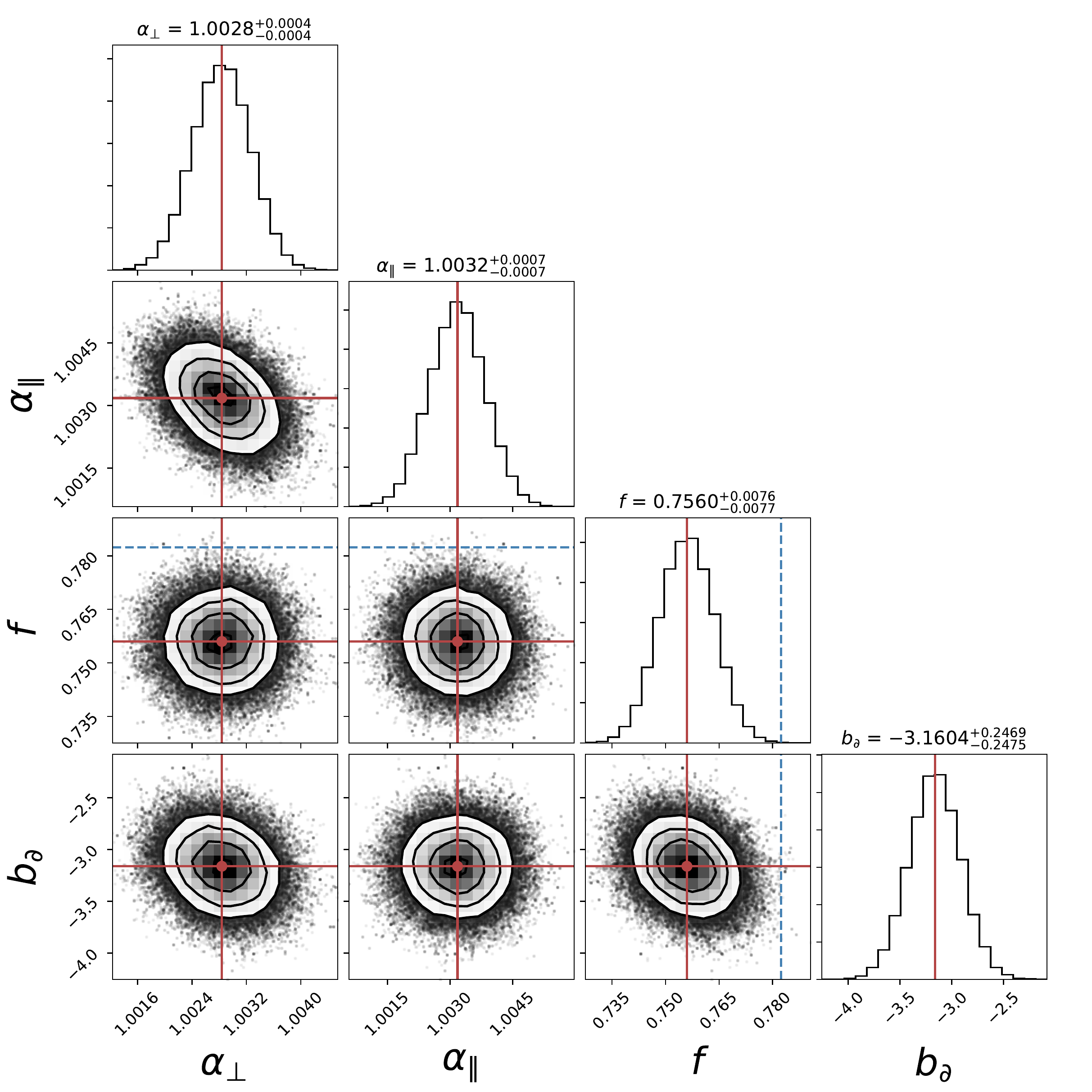}
\includegraphics[width=0.45\linewidth]{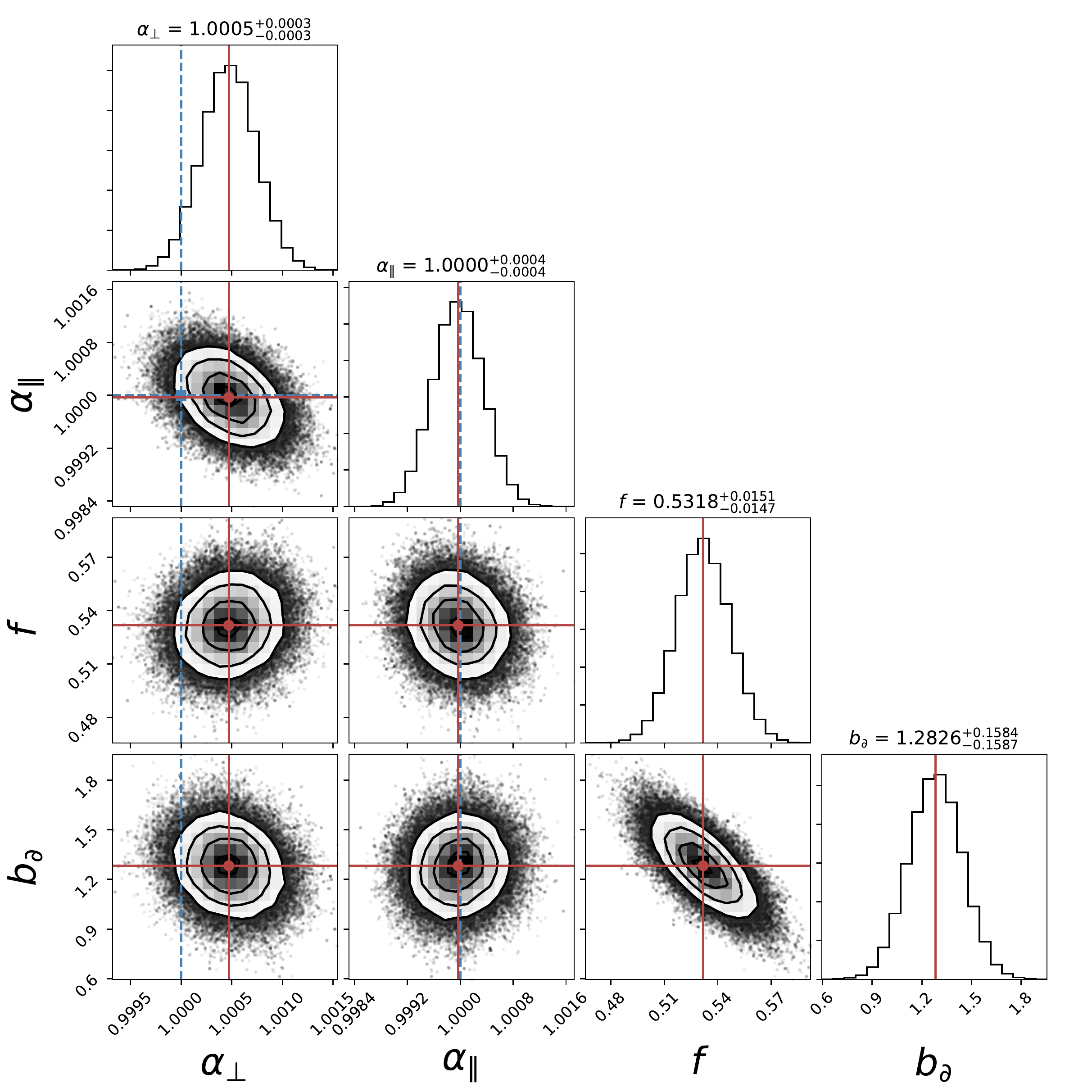}
\includegraphics[width=0.45\linewidth]
{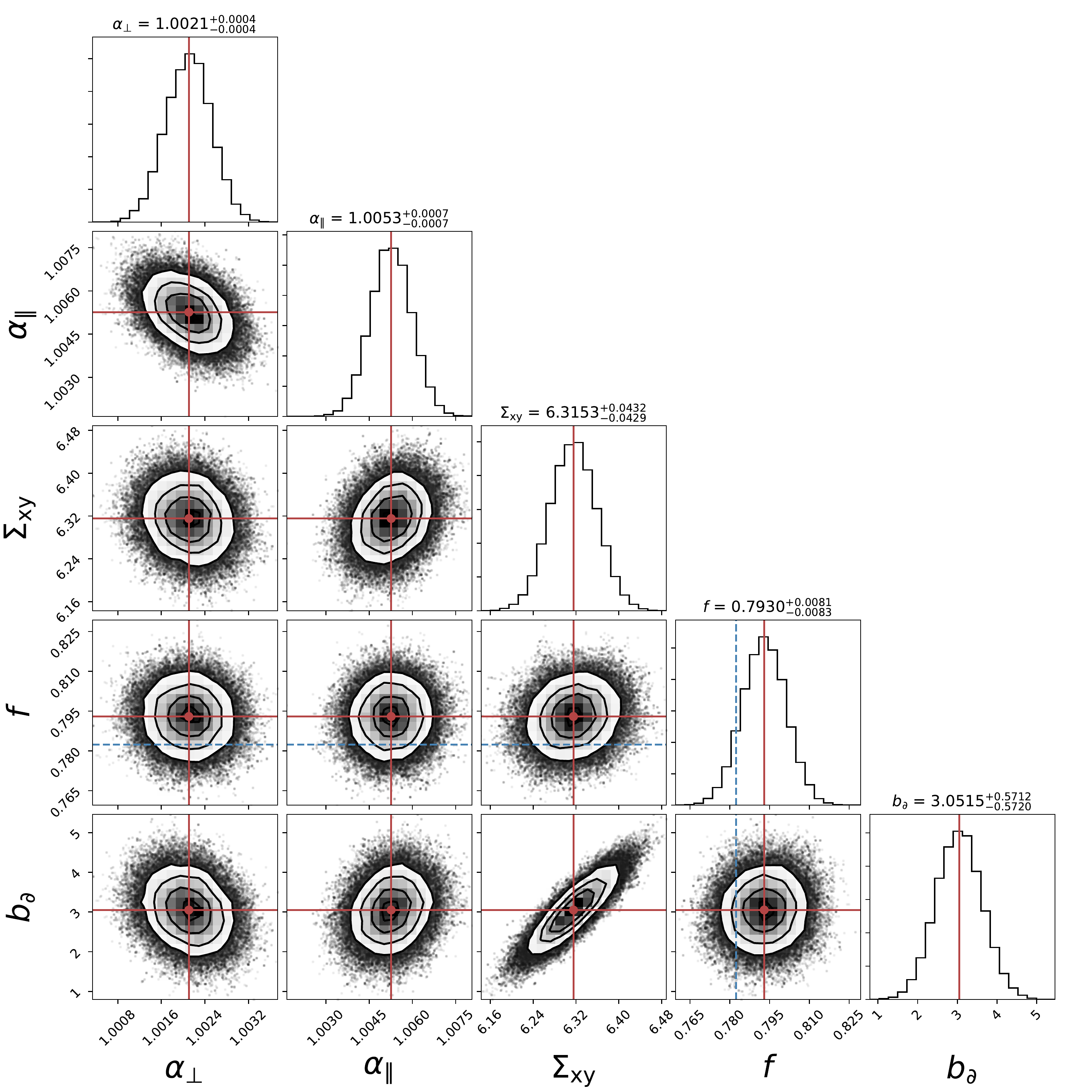}
\includegraphics[width=0.45\linewidth]{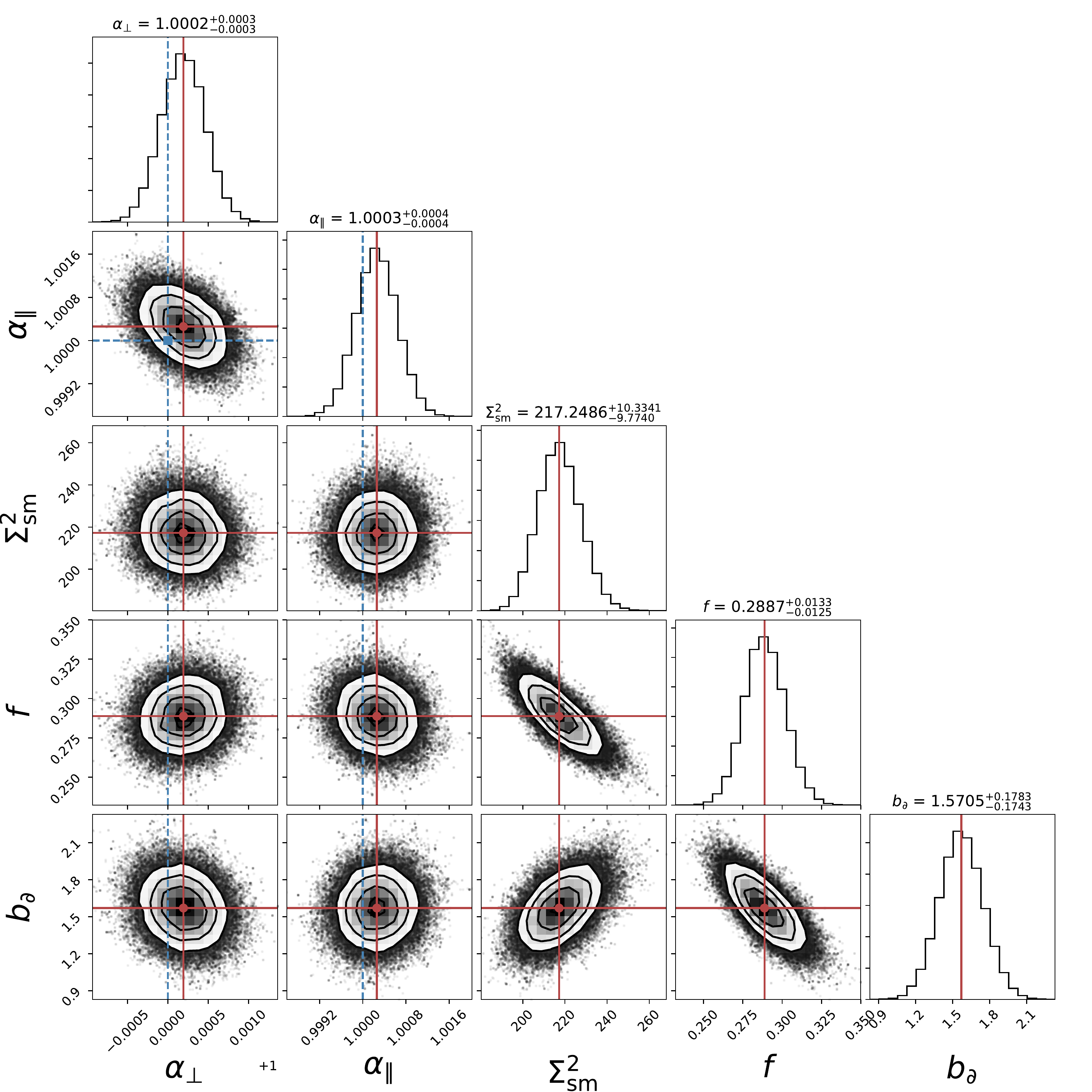}
\caption{The joint distributions of fitting parameters for the averaged difference matter power spectrum at $z=0.6$. \textit{Upper panels:} EFT0 model. \textit{Bottom panels:} EFT1 model. \textit{Left panels:} pre-reconstruction. \textit{Right panels:} post-reconstruction. In each panel, black dots represent steps from MCMC-chains implemented by \textbf{emcee}. Contours overlapping on dots show 1$-\sigma$, 2$-\sigma$ and 3$-\sigma$ confidence regions. The marginalized distribution for each parameter is shown at the top in each column; each red line denotes the best estimation of the parameter and each blue line denotes the corresponding input/linear-model value when available. Joint distribution shows the correlation of paired parameters. $\alpha_{\bot}$ and $\alpha_{\|}$ have anti-correlation of $\sim -0.4$ in both pre- or post-reconstruction as expected. 
}\label{fig:DM_likelihood}
\end{figure*}

\begin{figure*}
\centering
\includegraphics[width=0.45\linewidth]{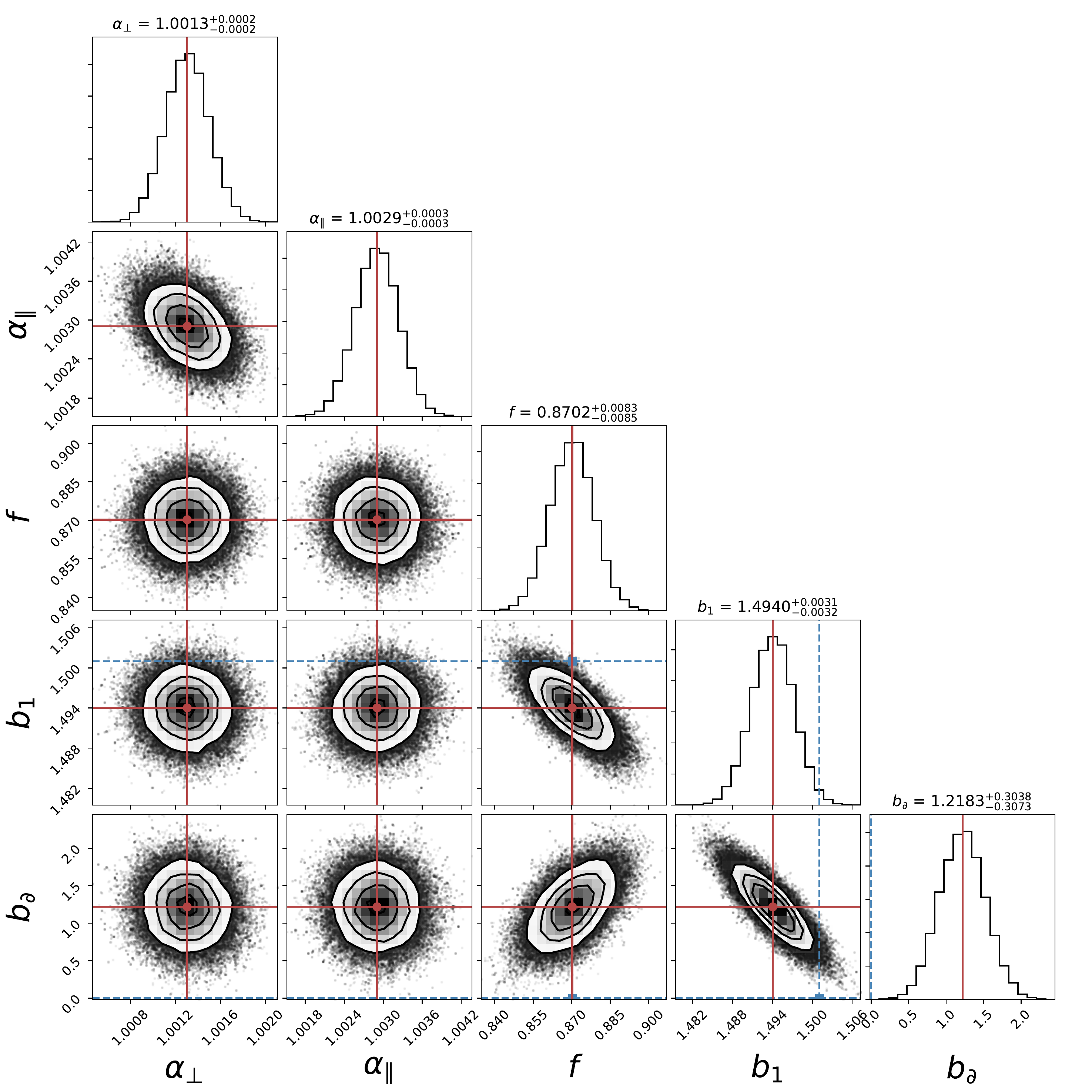}
\includegraphics[width=0.45\linewidth]{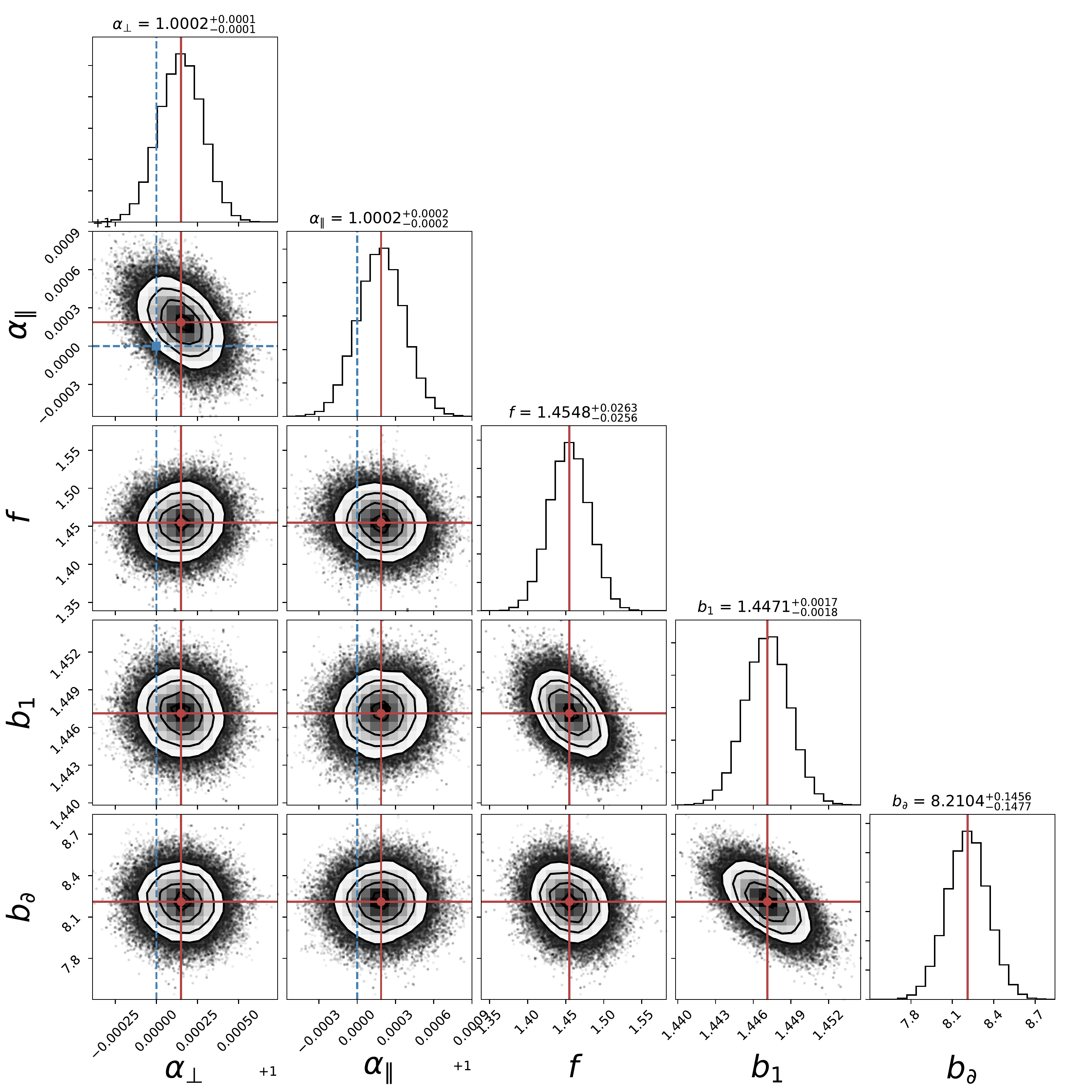}
\includegraphics[width=0.45\linewidth]{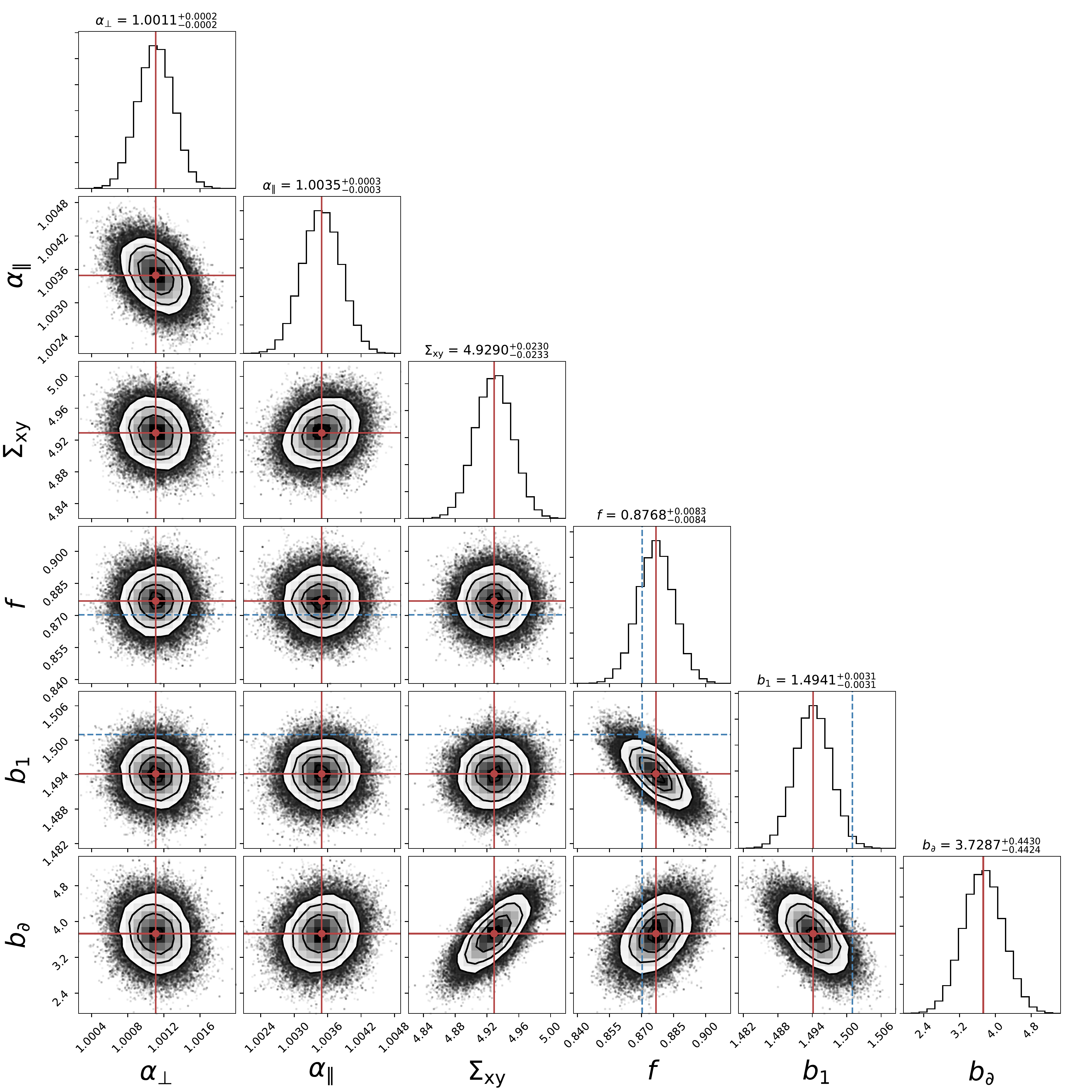}
\includegraphics[width=0.45\linewidth]{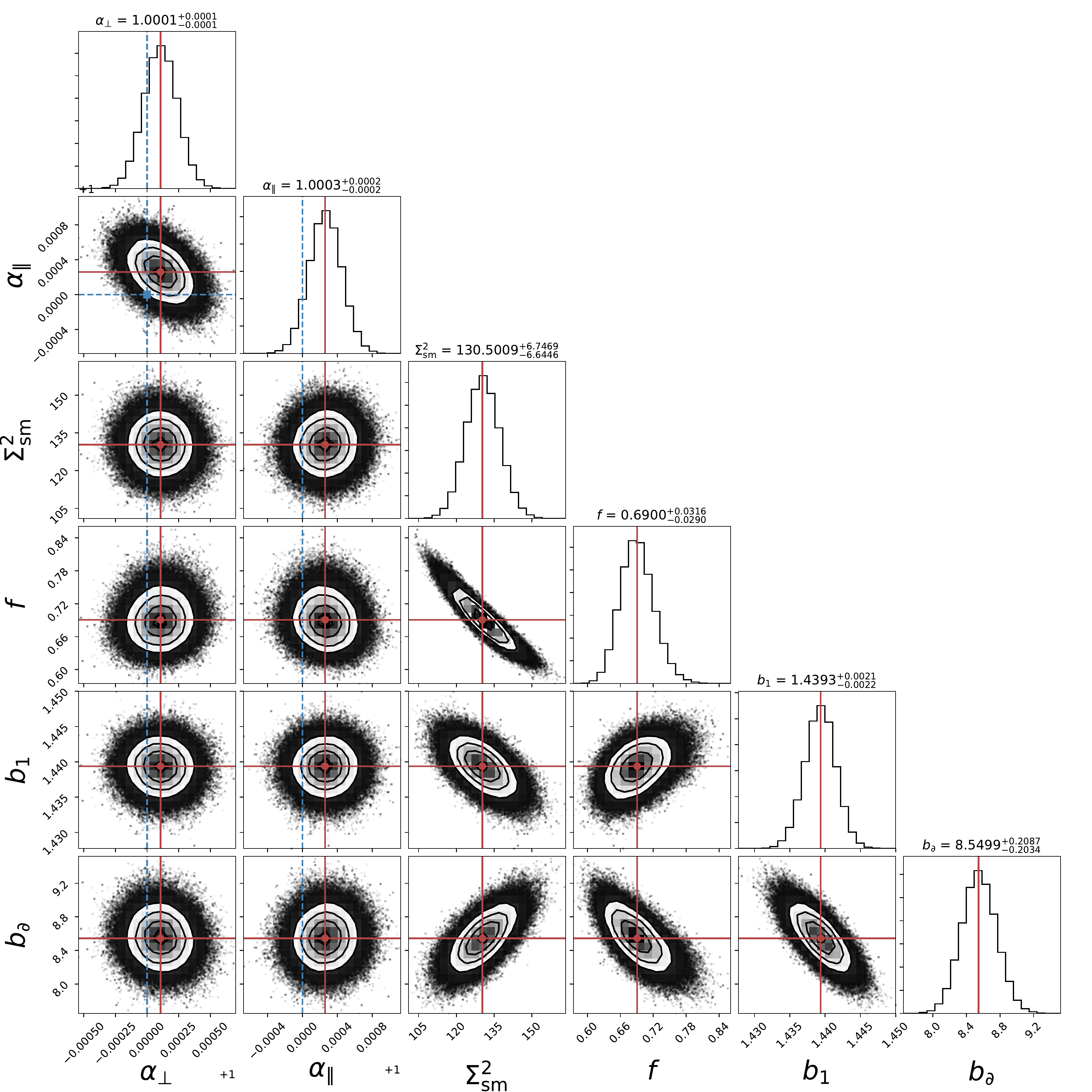}
\caption{Same as Fig.~\ref{fig:DM_likelihood} but for the averaged difference galaxy power spectrum at $z=1$ with fiducial galaxy/halo bias $1.5$. For the biased cases, $b_1$ is an additional free parameter. \textit{Upper panels:} EFT0 model. \textit{Bottom panels:} EFT1 model. \textit{Left panels:} pre-reconstruction. \textit{Right panels:} post-reconstruction.}\label{fig:galaxyzone_likelihood}
\end{figure*}

\subsection{Nonlinear scale-dependent bias $b_\partial$}

The first derivative bias $b_\partial$ in Eq.~\ref{eq:EFT_pre} and \ref{eq:EFT_post} accounts for nonlinear scale-dependent effect at one-loop level \citep{VCW_16, Desjacques_etal_16}, where we assumed the Lagrangian scale $k_{\rm L}=1\ihMpc$. Since this term accounts for higher order effect, we expect this term not to greatly exceed unity. In reality, this term can absorb higher order correction to the damping scale, any potential simulation effect as well as the dependence of $k_{\rm L}$ on halo mass.
Figure \ref{fig:ZV_bpartial} shows the best fit $b_\partial$ using the averaged power spectrum for the EFT0 and EFT1 cases. 
The figure shows that the pre-reconstruction cases (left) return smaller $b_\partial$ than the post-reconstruction cases (right). Interpreting $b_\partial$ as an indicator for various unaccounted higher order effects, this implies that our models give better physical description before reconstruction. 
Before reconstruction, for the matter case, the EFT1 model (lower left panel) gives slightly larger $b_\partial$ compared to the EFT0 model probably due to the freedom to modulate the damping scale $\Sigxy$.  Figure~\ref{fig:DM_likelihood} indeed shows a strong correlation between $b_\partial$ and $\Sigxy$ for the matter case.
For the biased cases, both models show similar results.

In summary, the large values of post-reconstruction $b_\partial$ (as well as the larger offset for $f$ measurements from \S~\ref{subsec:rsd}) implies that our models for the non-BAO information work better with the pre-reconstruction power spectrum while modeling the non-BAO information in the post-reconstruction power spectrum is more challenging.

\subsection{Shot noise dependence}

We have assumed no sub-sampling in our halo catalogs so far. The precision of the BAO measurements depends on the shot noise level \citep[e.g,][]{SE2007}. In our sample-variance cancellation scheme, we expect that a fair portion of the shot noise is canceled, mitigating the shot-noise dependent precision before reconstruction. On the other hand, post-reconstruction result is expected to be more sensitive to the shot noise effect since the process of reconstruction depends on the raw galaxy density field, although such sensitivity might be mitigated to some extent when differencing a pair of reconstructed power spectra. 
We test the BAO systematics as function of shot noise level in this subsection. The smoothing scale is accordingly increased from what is presented in Table~\ref{tab:fof_post_rec} using Eq.~\ref{eq:Sigsmscale}.
We focus at $z=0.6$, subsample halos at given mass cut~\footnote{We use the same random seed for wiggled and dewiggled halo catalogs that are ranked by mass.} to vary the level of shot noise $1/\nbar$ relative to the amplitude of the clustering $b_1^2$, and observe how pre- and the post-reconstruction BAO precision changes as a function of the effective noise to signal at $k=0.2\hMpc$, i.e., $1/[nb_1^2P_m(0.2)]$ in Figure \ref{fig:shotnoise}. The points with the same colour correspond to the same halo mass cut (i.e., the same galaxy bias) with different sub-sampling. Looking at the black points, which are galaxies/halos with $b_1=1.19$, we find consistent measurements of the BAO scale within $\sim 1-\sigma$ before and after reconstruction. For $b=1.96$, we also find that the pre and post-reconstruction shifts are consistent within $\sim 1-\sigma$ for the two different shot noise levels. As a reference, the DESI LRG is expected to have $1/[nb^2P_m(0.2)] \sim 0.4$.

The lower panels show a fast increase of errors with an increasing noise-to-signal ratio, implying that the sample variance cancellation is less effective as subsampling becomes more severe.  

\subsection{Smoothing scale $\Sigsm$ dependence}\label{subsec:Sigsm}

We investigate the effect of the reconstruction smoothing scale $\Sigsm$ on various post-reconstruction parameters. Too aggressive smoothing (i.e., too large $\Sigsm$) is expected to degrade the reconstruction of the BAO feature while a greater smoothing should also improve the performance of the fitting models by reducing the effect of nonlinearities and shot noise on the density field~\citep[e.g.,][]{Seo_etal_16}. We focus on $z=0.6$ and the `Rec-Cohn' model that was found to return a growth rate $f$ that is more consistent over different fitting models and closer to the linear values. We increase all smoothing scales by a factor of $\sqrt{2}$ and repeat reconstruction using the `Rec-Cohn' scheme.
Figure~\ref{fig:Cohn_fit_alpha} showed a tendency of greater post-reconstruction $\apar$ shift by the `Rec-Cohn' scheme compared to the `Rec-Iso' scheme. Figure \ref{fig:RecCohnSigsm} shows that a greater smoothing scale returns post-reconstruction $\apar$s that are less biased than the default $\Sigsm$ cases by $\sim 0.07\%$ in the low bias limit.  The effect of $\Sigsm$ is smaller for $\aperp$. In terms of post-reconstruction precision (in the bottom left panel), the precision does not change noticeably by increasing $\Sigsm$ by a factor of $\sqrt{2}$, which may be because our choice of smoothing scale corresponds to near the maximum reconstruction performance; \citet{White2010} shows that the shot noise and the default smoothing scale we used corresponds to near the maximum performance around which the reconstruction efficiency does not change rapidly. We expect that a much greater smoothing scale beyond the ones we tried would eventually degrade the reconstruction performance.
The bottom right panel of Figure~\ref{fig:Cohn_fit_alpha} shows that increasing the smoothing scale does not improve the best fit $f$ estimates in terms of bringing them closer to the Kaiser effect. Indeed, the reduced $\chi^2$ and therefore the goodness of the fit improved with a greater $\Sigsm$. In summary, moderately increasing the smoothing scale near our default choice improved the goodness of the fit as well as the post-reconstruction BAO scale shift while it does not help returning more accurate linear growth rate $f$.

\subsection{Parameter covariances}

In this subsection, we check the correlation between different parameters. 
Figure~\ref{fig:DM_likelihood} shows the multi-panel likelihood surface plots of fitting parameters from the MCMC using the averaged difference matter power spectrum at $z=0.6$. The top panels show the EFT0 model and the bottom panels show the EFT1 model. Again, the left panels are for pre-reconstruction and the right panels are for post reconstruction. We notice that for matter, for which we fix $b_1$, the pre-reconstruction shows no obvious parameter correlation in the EFT0 model case while the EFT1 model with a free $\Sigxy$ parameter shows strong positive correlation between $b_\partial$ and $\Sigxy$. 
That is, in the EFT1 model, a greater damping can be compensated by positive $b_\partial$, which is consistent with the EFT1 model returning larger $\Sigxy$ values than the theory (Figure~\ref{fig:Z1_fit_Sigma}) while returning larger $b_\partial$ than the EFT0 model (Figure~\ref{fig:ZV_bpartial}).

After reconstruction (right panels), the EFT models show moderate negative correlation between $b_\partial$ and $f$ for matter, which again is consistent with underestimation of $f$ and large $b_\partial$ we observed in Figure~\ref{fig:ZV_fit_f} and ~\ref{fig:ZV_bpartial}. If this mild degeneracy is determined by the damping feature, the two would have a positive correlation. The negative correlation seems to imply that this degeneracy is rooted on the post-reconstruction anisotropic amplitude term. 
For the EFT1 model (bottom right), the strong $f$ and $\Sigsm$ correlation is expected due to $\Sigsm$ changing the post-reconstruction anisotropic amplitude.

Figure~\ref{fig:galaxyzone_likelihood} shows the biased cases: we show $b=1.5$ at $z=1$ as example. For the biased cases, we allow $b_1$ to vary, and as a result, we notice additional, negative correlations between $b_1$ and $b_\partial$ and $f$, implying more freedom to deal with a deviation from our fitting model by freeing $b_1$.
Before reconstruction, we find moderate positive correlation between $b_\partial$ and $f$, which agrees with slight overestimation of $f$ and $b_\partial$ in Figure~\ref{fig:ZV_fit_f} and \ref{fig:ZV_bpartial} for the biased cases. 

After reconstruction (right panels), compared to the matter case, the correlation between $b_\partial$ and $f$ switches its sign back to positive with negative correlation between $b_1$ and $b_\partial$.  Given the positive $b_\partial$ from Figure~\ref{fig:ZV_bpartial}, we indeed find that the best fit $b_1$ derived from the fits tend to be lower than the estimates derived from cross-power spectrum between the galaxy-matter cross correlation as shown in the right panel of Figure~\ref{fig:Zmodel_b1}.

For the BAO scale parameters, we consistently observe the negative correlation of $\sim -0.4$ between $\aperp$ and $\apar$, which is very close to what is expected when $\alpe$ and $\alpa$ constraints are only from the BAO feature~\citep{SE2007}; the BAO scale parameters show no strong correlation with non BAO parameters. For the matter case, we found in \S~\ref{subsec:BAOmatter} that fitting models with free damping scales tend to produce slightly greater $\apar$. Looking at the correlation between $\Sigxy$ and $\apar$ in the lower left panel of Figure~\ref{fig:DM_likelihood}, we can indeed trace a slight positive correlation between $\apar$ and $\Sigxy$ for the EFT1 model, although it is rather inconspicuous.   

\section{Conclusion}
\label{sec:con}

We generated a set of paired FastPM simulations to study the BAO feature more accurately than conventional simulations that are limited by the number of modes in the simulated volume.
We use these simulations to test how well theoretical models for the BAO feature in the matter or galaxy/halo power spectrum capture the true BAO scale of the simulations.
The specific design of the simulations allows systematic shifts of the BAO scale as low as $0.01\%$ to be detectable (in the most precise case). This analysis is one of the most precise checks for BAO systematics carried out so far.
We derive BAO systematics before and after density field reconstruction, along and across the line of sight separately, and in the presence of redshift-space distortions. We tested two different density field reconstruction schemes that were introduced in previous literature and also tested the effect of the smoothing filter scale.
Our tests covered a wide range of redshift, galaxy/halo biases, and shot noise levels that are applicable to future galaxy/quasar survey designs. 
We also introduced new BAO fitting models with a physically motivated scale-dependent bias, and investigated the BAO systematics dependence on the choice of the fitting formula. 
For the first time, we derived in a fully Galilean invariant way the BAO dampening kernel for the halos in real and redshift space for both pre- and post-reconstruction.
We summarize our findings below:

First, for the matter density field, the nonlinear BAO scale shift is increasing as a function of decreasing redshift, as expected and shown in previous literature. However, we find that such shift measurements before density field reconstruction depend on the fitting formula especially at lower redshift; as much as 0.1\% for $\aperp$ and 0.2\% for $\apar$ within the models we tested. We find that the fitting models with a free damping scale tend to generate greater shift on $\apar$ than on $\aperp$, producing an anisotropic BAO shift. After reconstruction, the model dependence of $\apar$ is substantially reduced; we find model-dependence of less than  
0.1\% on the BAO scale after reconstruction.

In the presence of galaxy/halo bias, we find that before reconstruction the nonlinear shift on the transverse BAO scale $\aperp$ tends to converge toward $\sim 0.1\%$ for biased tracers in the low bias limit regardless of redshift, which is less than the shift measured from the matter cases. As bias increases, the expected shift increases, but very slowly at high redshift. Such tendency qualitatively agrees with predictions from perturbation theories. As a result, we expect a small shift on $\aperp$ at high redshift even for highly biased samples. The values of $\apar$ that account for the RSD effect show stronger redshift dependence while the galaxy/halo bias dependence appears comparable to that of $\aperp$. After reconstruction, we find the shift reduces to less than $0.1\%$ except for high bias and at low redshift. 
We also find that the shift measurements are more robust than the matter cases over the range of the fitting models we tested, both before and after reconstruction. 

In terms of the BAO measurement precision, all the fitting models we tested give quite consistent precision. In terms of reduced $\chi^2$, all models give reasonable fits while the EFT1 model tends to perform slightly better.

We compared the BAO systematics for different reconstruction conventions, one for isotropic reconstruction and the other for anisotropic reconstruction, depending on the treatment of the Kaiser effect. The offset in the BAO scale measurements is largest at $z=0$ and greater for $\apar$ than $\aperp$ and greater for the biased cases. The offset in $\apar$ at $z=0$ is $\sim 0.3\%$ which is much more than the level of fitting model-dependence for the biased cases while the offset in $\aperp$ is much smaller. We find that the `Rec-Iso' scheme tends to return smaller residual post-reconstruction BAO shift on $\apar$ than the Rec-Cohn scheme.  

By subsampling our halo catalogs, we tested the shot noise dependence of our results. As we subsample, the sample-variance cancellation becomes less effective, quickly degrading precision. We find no obvious dependence of the BAO scale shifts on the shot noise level within our precision. 

While it was not the main focus of this paper, we also examined our models from the perspective of grasping non-BAO information such as the growth rate $f$ and scale-dependent bias $b_\partial$. We find that while the pre-reconstruction cases return a more reasonable $f$ estimate as well as smaller $b_\partial$, the post-reconstruction models return systematically underestimated $f$ for matter and tend to overestimate $f$ for the biased cases in a strongly fitting-model dependent way. Using an anisotropic reconstruction convention such as the Rec-Cohn also returns an overestimated $f$, but closer to the prediction values. Overall, the EFT1 model tends to return post-reconstruction $f$ that is closer to the prediction at least at low redshift. We conclude that the post-reconstruction model needs to be further improved to accurately capture non-BAO information. This also indicates that additional effects beyond leading order could play an important role (e.g.~corrections due to modeling the displacement in Eulerian rather than Lagrangian coordinates; \citet{Schmittfull_etal_15}).
As a caveat, we tried to catch the non-BAO information only using the BAO amplitude modulations parallel v.s. transverse to the line of sight. Testing our models for the full post-reconstruction power spectrum \citep[e.g., similar to][]{Hikage07} would be interesting for future investigation.  

We investigated the effect of the reconstruction smoothing scale $\Sigsm$ on various post-reconstruction parameters using the Rec-Cohn convention at $z=0.6$.  Moderately increasing the smoothing scale near our default choice improves the goodness of the fit as well as the post-reconstruction BAO scale of the Rec-Cohn case while it does not help returning more accurate linear growth rate $f$.

As a caveat, the BAO systematics we studied in this paper only account for the theoretical budgets. Real galaxy surveys would also potentially suffer from observational systematics in the clustering measurements due to fiber assignment~\citep[e.g.,][]{Burden2017} and additional non-astrophysical observational systematics that need to be calibrated for~\citep[e.g.,][]{Wang2017}. Even among the theoretical budgets, we ignored higher order effects such as a potential BAO scale bias that can be induced by the supersonic streaming velocity of baryons at high redshift ~\citep[e.g.,][]{Dalal2010,Tseliakhovich10,Blazek15,Slepian16, Schmidt2016,Schmidt2017,BeutlerRbias}. 

There are additional aspects that are left for future studies. 
\citet{IsobaricREC17} and \citet{Schmittfull_etal_17} showed that alternative reconstruction methods can substantially improve the BAO reconstruction efficiency, with similar improvements also expected from maximum likelihood reconstruction \citep{2017arXiv170606645S}. 
Testing the post-reconstruction BAO systematics for galaxies/halos in redshift space using such improved reconstruction schemes would be an interesting future project. We also note that a similar study with a comparable  set of a full \Nb\ simulations would make a substantial improvement to our study that is done with quasi-\Nbody\ simulations by allowing calibrations of the BAO results between two sets of simulations.

\section*{Acknowledgements}
We thank Chirag Modi and Shun Saito for useful comments. 
The simulations and part of the analysis are performed on NERSC facilities, including Cori Phase-I, Edison, and the Jupyter Hub service through the CosmoSim and DESI project allocations.
This research has made use of NASA's Astrophysics Data System Bibliographic Services. We use corner.py \citep{Foreman-Mackey_16} to plot probability contours for fitting parameters. Z.D.~and H.-J.S.~are supported by the U.S.~Department of Energy, Office of Science, Office of High Energy Physics under Award Number DE-SC0014329. 
Z.V.~is supported in part by the U.S.~Department of Energy contract to
SLAC no.~DE-AC02-76SF00515.
M.S.~gratefully acknowledges support from the Bezos Fellowship.
F.B.~acknowledges support from the STFC and the Ernest Rutherford Fellowship scheme.

\clearpage
\appendix

\begin{widetext}
\section{Fitting models with scale-dependent bias effect based on the Effective Field Theory}\label{sec:EFTderivation}
\subsection{Pre-reconstruction difference power spectrum}\label{subsec:eftpre}
Mapping from real-space position $\vec x$ to redshift space $\vec s$ in
the plane-parallel approximation is given by:
\eeq{
\vec s=\vec x + {\hat z} \frac{v_\parallel(\vec x)}{\mathcal H} 
= \vec x + \vec R . \vec u,
}
where $\vec v(\vec x)$ is the peculiar velocity field. In the second equality we have 
defined the normalized velocity field $\vec u = \vec v / \mathcal H$ and 
introduced the redshift operator $R_{ij} = \hat z_i \hat z_j $ in the plane parallel approximation.
The density field in redshift space $\df_s(\vec s)$ can be obtained 
from the real-space density field $\df(\vec x)$ by requiring that 
the redshift-space mapping conserves mass:
\eeq{
(1+\df_{X,s}(\vec s, \tau )) d^3 s 
= (1+\df_X(\vec x)) d^3 x 
= (1+\df_X(\vec q, \tau_{\rm in}))d^3 q
}
where $\vec q$ is the Lagrangian coordinates 
$\vec x =\vec q + \psi (\vec q)$.
The same expression is valid for tracers $\df_X$ since the redshift-space mapping
should also preserve the number of tracers.
We can use the fact that $ d^3 s=J(\vec x) d^3 x$, where 
$J(\vec x) =|1-\nabla_z v_\parallel(\vec x)/{\mathcal H}| = |1 - R^{ij} \partial_j u_j(\vec x)|$ 
is the exact Jacobian of the mapping in the plane-parallel approximation. 
From the continuity equations we have
\eq{
1 + \df_{X}(\vec s) 
& =  \int d^3x ~ ( 1 + \df_X(\vec x) ) \df^D \big( \vec s - \vec x -  \vec R . \vec u \big) \non\\
&= \int d^3 q ~  \lb 1+\df_X(\vec q, \tau_{\rm in}) \rb 
\df^D \big( \vec s - \vec q - \psi (\vec q) - R^{ij} \psi'_j (\vec q) \big),
}
where $\psi'_i(\vec q)= \dot{\psi}_i(\vec q)/\mathcal H = \vec v (\vec x) /\mathcal H = \vec u (\vec x)$.
Fourier transforming the expression above we obtain: 
\eq{
(2\pi)^3\df^D(\vec k) + \df_s(\vec k) 
& = \int d^3 q ~ e^{ i \vec k \cdot \vec q} ~ \lb 1+\df_X(\vec q, \tau_{\rm in}) \rb e^{i \vec k \cdot  \psi(\vec q) 
+ i R^{ij} k_i \psi'_j (\vec q)}\non\\
& = \int d^3 q ~ e^{ i \vec k \cdot \vec q} ~ \lb 1+\df_X(\vec q, \tau_{\rm in}) \rb e^{i \vec k_i \psi^s_i(\vec q)}
}
where we introduced $\psi^s_i(\vec q) = \hat s^{ij} \psi_j(\vec q) $ and the redshift space operator $\hat s_{ij}(\tau) = \df^K_{ij} + R_{ij}\times\partial_\tau 
= \df^K_{ij} + z_i z_j \times\partial_\tau $. 
We can write the biasing model in Lagrangian space \citep[see e.g.][]{VCW_16, Desjacques_etal_16}
in the form
\eq{
\df_X(\vec q) &= \lb b_1 + b_{\partial} \partial_q^2 / k_L^2 \rb \df_L (\vec q) + \rm{h.o.} + {\rm ``stochastic"}  \non\\
&= -i \hat b_q \partial_\lambda e^{ i \lambda \df_L (\vec q)}\Big |_{\lambda=0}  + \rm{h.o.} + {\rm ``stochastic"} ,
}
where we assume a simplified form neglecting most of the higher order bias effects, and we keep just the 
linear bias $b_1$ and the first derivative bias term $b_\partial$.
Following \citet{VCW_16}, we obtain the power spectrum
\eq{
(2\pi)^3\df^D (k)+P(\vec k, \tau)&=\int d^3 q ~e^{-i\vec{q} \cdot \vec{k}} ~ 
 \Big( 1 -i \hat b_q (\partial_{\lambda_1} + \partial_{\lambda_2}) - \hat b^2_q \partial_{\lambda_1}\partial_{\lambda_2} \Big)
\exp {\lb - \frac{1}{2} A^s (\vec k, \vec{q},\tau)\rb} \non\\
&\hspace{8cm} + \rm{h.o.} + {\rm ``stochastic"},
}
where we introduced $A^s(\vec k, \vec{q}) = \llan \Big( \lambda_1 \df_L (\vec q_1) + \lambda_2 \df_L (\vec q_2) + \vec k \cdot \Delta^s(\vec q) \Big)^2 \rran_c$ and
$A^s_{ij}(\vec{q}) = \llan \Delta^s_i \Delta^s_j \rran_c$, where $ \Delta^s_i(\vec q) = \psi^s_i(\vec q_2) - \psi^s_i(\vec q_1) $.
Using the techniques presented in \cite{zv16} the expression for the power spectrum can be written in the form 
\eq{
P(\vec k, \tau)&=\int d^3 q ~e^{-i\vec{q} \cdot \vec{k}} ~   \Big( 1 - \rm{``bias"} \Big) \bigg[
e^{-\frac{1}{2} A^s(\vec k, \vec{q},\tau_1,\tau_2)} - e^{-\frac{1}{2} A^s(\vec k, \vec{q},\tau)_{q \to \infty}}
\bigg] + \rm{h.o.} + {\rm ``stochastic"}\non\\
&=\int d^3 q ~e^{-i\vec{q} \cdot \vec{k}} ~ e^{-\frac{1}{2} k_ik_jA^s_{ij}(\vec{q},\tau)}  \Big( 1 - \rm{``bias"} \Big)
\mathcal A^s(\vec k, \vec{q},\tau)  + \rm{h.o.} + {\rm ``stochastic"},
}
where we define 
\eq{
\mathcal A^s(\vec k, \vec{q}) = \frac{1}{2} A^s(\vec k, \vec{q})  - \frac{1}{2} A^s(\vec k, \vec{q}) \big |_{q \to \infty} ~~ \rm{and~thus}~~
\df \mathcal A^s(\vec k, \vec{q}) = \frac{1}{2} A^s_{w}(\vec k, \vec{q}) - \frac{1}{2} A^s_{nw}(\vec k, \vec{q}).
}
For the difference of the wiggle and no-wiggle power spectrum we then have
\eq{
\df P(\vec k, \tau)&=\int d^3 q ~ e^{-i\vec{q} \cdot \vec{k}} ~ e^{-\frac{1}{2} k_ik_jA^s_{\rm{nw}, ij}(\vec{q},\tau)}  \Big( 1 - \rm{``bias"} \Big)
\df \mathcal A^s(\vec k, \vec{q},\tau) + \rm{h.o.},
\label{eq:dps}
}
where we assumed that the stochastic contributions will, to a large extent, cancel in the wiggle and no-wiggle power spectrum difference.

Next, we focus on the exponent in the power spectrum expressions. Following the arguments presented in \citet{zv16}
for dark matter in real space we have
\eq{
-\frac{1}{2} k_ik_jA^s_{\rm{nw}, ij}(\vec{q},\tau) &= - k_ik_j \hat s_{in} \hat s_{jm}
\bigg[ \frac{1}{3}\df^K_{nm} \Big( \xi_0(0) - \xi_0(q) \Big) - \lb \hat{q}_n \hat{q}_m - \frac{1}{3} \df^K_{nm} \rb  \xi_2(q) \bigg] \non\\
&\simeq - k^2 \lb 1 + f ( 2 + f) \mu^2  \rb \Sigma^2(q) + \ldots
}
where we dropped the sub-leading contributions in the second expression. 
Above we introduced the first two spherical Bessel transforms of the two-point displacement power spectra 
\eq{
\xi_{0} (0) =  \int \frac{d k}{2\pi^2} ~ P_L(k), ~~ \xi_{0} (q) =  \int \frac{d k}{2\pi^2} ~ P_L(k)  j_0(qk), ~~ \xi_{2} (q) =  \int \frac{d k}{2\pi^2} ~ P_L(k)  j_2(qk).
}
Then $\Sigma^2(q) = \frac{1}{3} \lb \xi_{0} (0) - \xi_{0} (q) \rb = \frac{1}{3} \int \frac{d k}{2\pi^2} ~ \lb 1 - j_0(qk) \rb P_L(k)$. 
If we insist on resumming only the true infrared (IR) modes, the same procedure can be followed to end up with 
$\Sigma_{\rm{IR}}^2(q) = \frac{1}{3} \int \frac{d k}{2\pi^2} ~ \lb 1 - j_0(qk) \rb W(k, k_{\rm{IR}})P_L(k)$,
where we can take the filter to be a Gaussian, $W(k, k_{\rm{IR}}) = \exp \lb -k^2/k^2_{\rm{IR}} \rb$, with the IR scale taken to be the BAO peak scale, i.e.~$k_{\rm{IR}} \simeq 2 \pi/ 110\rm{Mpc}/h$.

Let us now look at the term 
\eq{
 \Big( 1 - \rm{``bias"} \Big) \df \mathcal A^s(\vec k, \vec{q},\tau) \bigg|_{\lambda_1=\lambda_2=0}
&=  \Big( 1 -i \hat b_q (\partial_{\lambda_1} + \partial_{\lambda_2}) - \hat b^2_q \partial_{\lambda_1}\partial_{\lambda_2} \Big) \\
&\hspace{3cm} \times \lb - \frac{1}{2} k_ik_j \df A^s_{ij} - \lambda_1\lambda_2 \df \xi_L - (\lambda_1 + \lambda_2 )k_i \df U^{s,10}_i \rb
 \bigg|_{\lambda_1=\lambda_2=0}  \non\\
& = - \frac{1}{2} k_ik_j \df A^s_{ij} + b_{1}^2 \df \xi_L + 2 i b_{1} k_i \df U^{s,10}_i
+ 2 i b_{\partial }  k_i \frac{\partial^2}{k_L^2} \df U^{s,10}_i
+ 2 b_{1}  b_{\partial}  \frac{\partial^2}{k_L^2} \df \xi_L . \non
}
Collecting all the above and following the argumentation presented in \citet{zv16} we can extract the exponential 
from the power spectrum expression by evaluating it at the $q_{\rm{max}}$ of the dominant contribution
\eq{
\df P(k, \mu , \tau)&\simeq e^{- k^2 \lb 1 + f ( 2 + f) \mu^2  \rb \Sigma^2(q_{\rm{max}})} \int d^3 q ~ e^{-i\vec{q} \cdot \vec{k}} ~ 
\Big( 1 - \rm{``bias"} \Big) \df \mathcal A^s(\vec k, \vec{q},\tau_1,\tau_2) \bigg|_{\lambda_1=\lambda_2=0} + \rm{h.o.}, \non\\
&\simeq e^{- k^2 \lb 1 + f ( 2 + f) \mu^2  \rb \Sigma^2(q_{\rm{max}})} 
\bigg( b_1^2 + 2 f b_1 \mu^2 + f^2 \mu^4  
+ b_{\partial} \lb b_1 + f \mu^2 \rb \frac{k^2}{k_L^2} \bigg)\df P_L(k,\tau)
+ \rm{h.o.}
}
where $q_{\rm{max}}$ is defined by the peak of the BAO~\citep[see][]{zv16}. In case of dark matter and including the RSD effects, 
similar expressions have also been shown in \citep{Bella17}
(up to one-loop).
We have derived the leading contributions 
to the difference of the wiggle and no-wiggle power spectra, 
up to the linear bias $b_1$ and first contribution of the 
first derivative bias $b_\partial$.
Neglected effects (denoted as ``h.o.") would include second 
order bias $b_2$, tidal bias $b_{s^2}$, as well as higher 
order bias effects responsible for shifts of the BAO wiggles.
We note that the effects of e.g.~second order bias $b_2$ in 
case of dark matter halos could be large and comparable to the 
first derivative bias $b_\partial$ even though for simplicity we have dropped it from our analysis.
We also stress that our derivation for the 
BAO feature for dark matter, as well as halos, 
satisfies the Galilean invariance. We note that 
this is one of the novel features introduced in 
our derivation, distinguishing it from all 
previous results for halos in real and redshift space.

In detail, another substantial difference from previous approaches \citep[see e.g.,][]{Crocce08, Matsubara2011, BD2017, Hikage07} is that the previous modeling extends to
the total two-point function while we are considering only the BAO wiggle feature \citep[see also][]{Blas16}. 
In that respect, our critical control parameter is the small $\delta P_{\text{wiggle}} $ spectrum compared to the full power spectrum, 
and the IR resummation affects only this part, while the broadband part is unaffected. 
There are also major differences in the explanation of the physical meaning, interpretation, and estimates of the displacement dispersions 
compared to the previous work. Our displacement dispersion contains only large-scale displacement contributions while 
the previous work dispersions was treated primarily as a free parameter. 
Moreover, it has a well known linear theory redshift dependence via the linear growth rate.
Our approach also offers a systematic and controlled way to compute higher order corrections as well as estimates of theoretical errors due to 
the higher order approximations done in this section.

\subsection{Post-reconstruction power spectrum}\label{subsec:eftpost}

Similarly to above we can also model the power spectrum after density reconstruction.
Let us follow the \textit{Rec-Iso} result from \citet{Seo_etal_16}. 
The reconstructed difference power spectrum is given by 
\eeq{
\df P_{\rm rec}(\vec k, \tau) = \df P_{\rm dd}(\vec k, \tau) - 2 \df P_{\rm sd}(\vec k, \tau) + \df P_{\rm ss}(\vec k, \tau),
}
where, referring to Eq.\eqref{eq:dps}, we have 
\eq{
\df P_{\rm dd}(\vec k, \tau)
&=\int d^3 q ~e^{-i\vec{q} \cdot \vec{k}} ~ e^{-\frac{1}{2} k_ik_jA^s_{{\rm dd},ij}(\vec{q},\tau)}  \Big( 1 - \rm{``bias"} \Big)_{\rm dd}
\df \mathcal A_{\rm dd}^s(\vec k, \vec{q},\tau) \bigg|_{\lambda_1=\lambda_2=0} + \ldots, \non\\
\df P_{\rm sd}(\vec k, \tau)
&=\int d^3 q ~e^{-i\vec{q} \cdot \vec{k}} ~ e^{-\frac{1}{2} k_ik_jA^s_{{\rm sd},ij}(\vec{q},\tau)}  \Big( 1 - \rm{``bias"} \Big)_{\rm sd}
\df \mathcal A_{\rm sd}^s(\vec k, \vec{q},\tau) \bigg|_{\lambda_1=\lambda_2=0} + \ldots, \non\\
\df P_{\rm ss}(\vec k, \tau)
&=\int d^3 q ~e^{-i\vec{q} \cdot \vec{k}} ~ e^{-\frac{1}{2} k_ik_jA^s_{{\rm ss},ij}(\vec{q},\tau)}
\df \mathcal A_{\rm ss}^s(\vec k, \vec{q},\tau) \bigg|_{\lambda_1=\lambda_2=0} + \ldots,
}
and we introduced the labels for the $A^s_{{\rm xx},ij}(\vec{q}) = \llan \Delta^s_{\rm x, i} \Delta^s_{\rm x , j} \rran_c$, 
where 
\eq{
\Delta^s_{{\rm dd},i}(\vec q) &= \psi^s_i(\vec q_2) + \chi^s_i(\vec q_2) - \psi^s_i(\vec q_1) - \chi^s_i(\vec q_1) , \non\\
\Delta^s_{{\rm sd},i}(\vec q) &= \chi_i(\vec q_2) - \psi^s_i(\vec q_1) - \chi^s_i(\vec q_1) , \non\\
\Delta^s_{{\rm ss},i}(\vec q) &= \chi_i(\vec q_2) - \chi_i(\vec q_1),
}
and where, 
\eeq{
 \chi^s_i(\vec k) = \big( \df^K_{ij} + f \hat z_i \hat z_j \big) \lb + i \frac{k_j}{k^2} \mathcal S(k) \frac{\delta_{\rm obs}(\vec k)}{\kappa} \rb,
}
where $\kappa = b (1+\beta \mu^2)$. If we assume the linear theory (Kaiser) result for $\delta_{\rm obs}$ we can write 
$\chi^s_i(\vec k, \tau) = - \mathcal S(k) \big( \delta^K_{ij} + f \hat z_i \hat z_j \big) \psi_j(\vec k) =  \mathcal S(k) \psi^s_i(\vec k)$,
and similarly $\chi_i(\vec k, \tau) = - \mathcal S(k) \psi_i(\vec k)$.
This gives 
\eq{
-\frac{1}{2} k_ik_jA^s_{\rm{dd}, ij}(\vec{q},\tau_1,\tau_2) &= -\frac{1}{2} k_ik_j \llan \Delta^s_{{\rm dd},i} \Delta^s_{{\rm dd},j} \rran_c
= - \frac{1}{3} k^2 \lb 1 + f ( 2 + f) \mu^2  \rb \Big( \xi_{{\rm dd},0}(0) - \xi_{{\rm dd},0}(q) \Big) + \ldots \non\\
-\frac{1}{2} k_ik_jA^s_{\rm{sd}, ij}(\vec{q},\tau_1,\tau_2) &= -\frac{1}{2} k_ik_j
\llan \Delta^s_{{\rm sd},i} \Delta^s_{{\rm sd},j} \rran_c
= - \frac{1}{3} k^2  \lb 1 + f \mu^2 \rb \Big( \xi_{{\rm sd},0}(0) - \xi_{{\rm sd},0}(q) \Big) + \ldots \non\\
-\frac{1}{2} k_ik_jA^s_{\rm{ss}, ij}(\vec{q},\tau_1,\tau_2) &= -\frac{1}{2} k_ik_j \llan \Delta^s_{{\rm ss},i} \Delta^s_{{\rm ss},j} \rran_c
= - \frac{1}{3} k^2 \Big( \xi_{{\rm ss},0}(0) - \xi_{{\rm ss},0}(q) \Big) + \ldots
}
and we have
\eq{
\xi_{{\rm dd},0} (0) &=\int_k ~ \lb 1 - \mathcal S(k) \rb^2 P_L(k), ~~~
~ \xi_{{\rm dd},0} (q) = \int_k ~  \lb 1 - \mathcal S(k) \rb^2 P_L(k)  j_0(qk), ~~~
~ \xi_{{\rm dd},2} (q) =  \int_k ~  \lb 1 - \mathcal S(k) \rb^2 P_L(k)  j_2(qk), \non\\
\xi_{{\rm sd},0} (0) &=  \int_k ~ \frac{1}{2} \lb \mathcal S(k)^2 + \lb 1 - \mathcal S(k) \rb^2 \rb P_L(k), ~~~
~ \xi_{{\rm sd},0} (q) =\int_k ~ \lb 1 - \mathcal S(k) \rb \mathcal S(k) P_L(k)  j_0(qk), ~~~
~ \xi_{{\rm sd},2} (q) = \int_k ~ \lb 1 - \mathcal S(k) \rb \mathcal S(k) P_L(k)  j_2(qk), \non\\
\xi_{{\rm ss},0} (0) &=  \int_k ~ \mathcal S(k)^2 P_L(k),~~~
~ \xi_{{\rm ss},0} (q) =  \int_k ~  \mathcal S(k)^2 P_L(k)  j_0(qk),~~~
~ \xi_{{\rm ss},2} (q) =  \int_k ~  \mathcal S(k)^2 P_L(k)  j_2(qk),
}
where $\int_k = \int \frac{d k}{2\pi^2}$.
It follows that
\eq{
-\frac{1}{2} k_ik_jA^s_{\rm{dd}, ij}(\vec{q},\tau) 
&\simeq - k^2 \lb 1 + f ( 2 + f) \mu^2  \rb \Sigma_{\rm dd}^2(q), \non\\
-\frac{1}{2} k_ik_jA^s_{\rm{sd}, ij}(\vec{q},\tau) 
&\simeq - k^2 \lb 1 + f \mu^2  \rb \Sigma_{\rm sd}^2(q), \non\\
-\frac{1}{2} k_ik_jA^s_{\rm{ss}, ij}(\vec{q},\tau)
&\simeq - k^2 \Sigma_{\rm ss}^2(q),
}
where 
\eq{
\Sigma_{{\rm dd},0}^2(q) &= \frac{1}{3} \int \frac{d k}{2\pi^2} ~ \lb 1 - j_0(qk) \rb  \lb 1 - \mathcal S(k) \rb^2 P_L(k, \tau), \non\\
\Sigma_{{\rm sd},0}^2(q) &= \frac{1}{3} \int \frac{d k}{2\pi^2} ~ \lb  \frac{1}{2} \lb \mathcal S(k)^2 + \lb 1 - \mathcal S(k) \rb^2 \rb  - j_0(qk)  \lb 1 - \mathcal S(k) \rb \mathcal S(k) \rb  P_L(k, \tau), \non\\
\Sigma_{{\rm ss},0}^2(q) &= \frac{1}{3} \int \frac{d k}{2\pi^2} ~ \lb 1 - j_0(qk) \rb  \mathcal S(k)^2 P_L(k, \tau). 
}
Let us now look at the term $A_{\rm dd}^s$
\eq{
 \Big( 1 - \rm{``bias"} \Big) &\df \mathcal A_{\rm dd}^s(\vec k, \vec{q},\tau) \bigg|_{\lambda_1=\lambda_2=0} 
=  \Big( 1 -i \hat c_q (\partial_{\lambda_1} + \partial_{\lambda_2}) - \hat c^2_q \partial_{\lambda_1}\partial_{\lambda_2} \Big) \non\\
&\hspace{5.5cm} \times \lb - \frac{1}{2} k_ik_j \df A^s_{{\rm dd}, ij} - \lambda_1\lambda_2 \df \xi_L 
- (\lambda_1 + \lambda_2 )k_i \df U^{s,10}_{{\rm d}, i} \rb \bigg|_{\lambda_1=\lambda_2=0}  \non\\
& = - \frac{1}{2} k_ik_j \hat s_{in}\hat s_{jm} \df A_{nm} + b_{1}^2 \df \xi_L + 2 i b_{1} k_i \hat s_{in} \df U^{s,10}_{{\rm d}, n} 
+ 2 i b_{\partial}  k_i \hat s_{in} \frac{\partial^2}{k_L^2} \df U^{s,10}_{{\rm d}, n} 
+ 2 b_{1} b_{\partial}  \frac{\partial^2}{k_L^2} \df \xi_L  \ldots
}
We get a similar expression for the $A_{\rm sd}^s$ term.
We then finally have 
\eq{
\df P_{\rm dd}(k, \mu , \tau)
&= e^{- k^2 \lb 1 + f ( 2 + f) \mu^2  \rb \Sigma_{\rm dd}^2(q_{\rm{max}})} 
\Bigg(  \Big( b_1 - \mathcal S(k) +  f \mu^2 \big(1- \mathcal S(k)\big) \Big)^2 \non\\
&\hspace{5.0cm}  + b_{\partial} \lb b_1 - \mathcal S(k) + f \mu^2 \big( 1 - \mathcal S(k) \big) \rb \frac{k^2}{k_L^2} \Bigg)\df P_L(k,\tau) + \rm{h.o.}, \non\\
\df P_{\rm sd}(k, \mu , \tau)
&= - e^{- k^2 \lb 1 + f \mu^2  \rb \Sigma_{\rm sd}^2(q_{\rm{max}})} 
\bigg( b_1 - \mathcal S(k) + f \mu^2 (1-\mathcal S(k)) + \frac{1}{2}b_\partial \frac{k^2}{k_L^2} \bigg)  \mathcal S(k) \df P_L(k,\tau) + \rm{h.o.}
\non\\
\df P_{\rm ss}(k, \mu , \tau)
&= e^{- k^2 \Sigma_{\rm ss}^2(q_{\rm{max}})} \mathcal S(k)^2 \df P_L(k,\tau)
+ \rm{h.o.}.
}
Note that $q_{\rm max}$ should be around $110\rm{Mpc}/h$, i.e. around the BAO peak, but we note also that since the 
$\mathcal S(k)$ filter affects the functional form of terms entering the $\Big( 1 - \rm{``bias"} \Big) \df \mathcal A$ the value 
of $q_{\rm max}$ can change slightly from the pre-reconstruction case. We note this option here but leave a detailed 
investigation for future work. 

If we assume that the $\mathcal S(k)$ filter does not depend on time then all of the $\Sigma^2$ terms above scale simply with 
the linear growth rate $D(\tau)^2$ coming from the linear power spectrum $P_L(k, \tau)$.
Note that in the limit when $\mathcal S(k) \to 0$ we retrieve the expression for the pre-reconstruction case.
The limit $\df P_{\rm sd}$ can be simply compared to the $\llan \df_{\rm dm} \df_{h} \rran$ correlation 
which gives $(b_1+ f \mu^2 + \tfrac{1}{2} b_\partial \tfrac{k^2}{k_L^2})P_L$, which is consistent.

The alternative reconstruction method that we labeled 
\textit{Rec-Cohn} \citep{White2015,Cohn2016} can be obtained in a similar way.
The difference is in the shifted field $\delta_s$ 
because it now contains the RSD effects. We have 
\eq{
-\frac{1}{2} k_ik_jA^s_{\rm{sd}, ij}(\vec{q},\tau_1,\tau_2) 
&= - \frac{1}{3} k^2 \lb 1 + f ( 2 + f) \mu^2  \rb 
\Big( \xi_{{\rm sd},0}(0) - \xi_{{\rm sd},0}(q) \Big) + \ldots
\simeq - k^2 \lb 1 + f ( 2 + f) \mu^2  \rb \Sigma_{\rm sd}^2(q) \non\\
-\frac{1}{2} k_ik_jA^s_{\rm{ss}, ij}(\vec{q},\tau_1,\tau_2) 
& = - \frac{1}{3} k^2 \lb 1 + f ( 2 + f) \mu^2  \rb 
\Big( \xi_{{\rm ss},0}(0) - \xi_{{\rm ss},0}(q) \Big) + \ldots
\simeq - k^2 \lb 1 + f ( 2 + f) \mu^2  \rb \Sigma_{\rm ss}^2(q) ,
}
where the $A^s_{\rm{dd}, ij}$ term stays the same as before,
and we obtain
\eq{
\df P_{\rm sd}(k, \mu , \tau)
&= - e^{- k^2  \lb 1 + f ( 2 + f) \mu^2  \rb \Sigma_{\rm sd}^2(q_{\rm{max}})} 
\bigg( b_1 - \mathcal S(k)  + f \mu^2 (1-\mathcal S(k)) + \frac{1}{2}b_\partial \frac{k^2}{k_L^2} \bigg)  (1+ \mu^2 f) \mathcal S(k)  \df P_L(k,\tau) + \rm{h.o.}
\non\\
\df P_{\rm ss}(k, \mu , \tau)
&= e^{- k^2  \lb 1 + f ( 2 + f) \mu^2  \rb \Sigma_{\rm ss}^2(q_{\rm{max}})} \lb 1+ f \mu^2 \rb^2 \mathcal S(k)^2 \df P_L(k,\tau)
+ \rm{h.o.},
}
where $\df P_{\rm dd}$ is again the same as in the \textit{Rec-Iso}
case.
\end{widetext}
\input{appendixa}

\end{document}

%% file: appendixa.tex
\section{Supplementary figures}
\begin{figure*}
\centering
\includegraphics[width=0.45\linewidth]{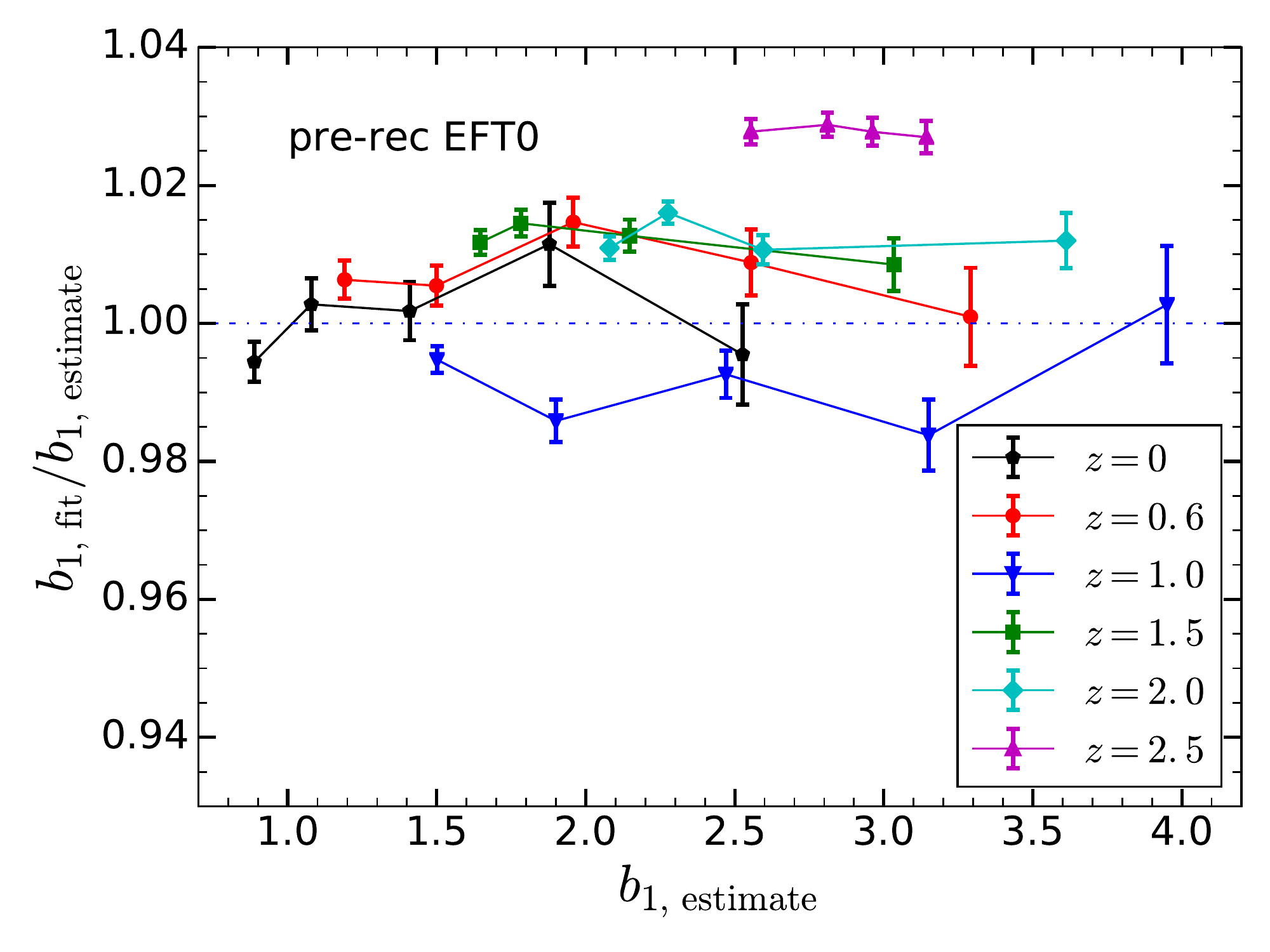}
\includegraphics[width=0.45\linewidth]{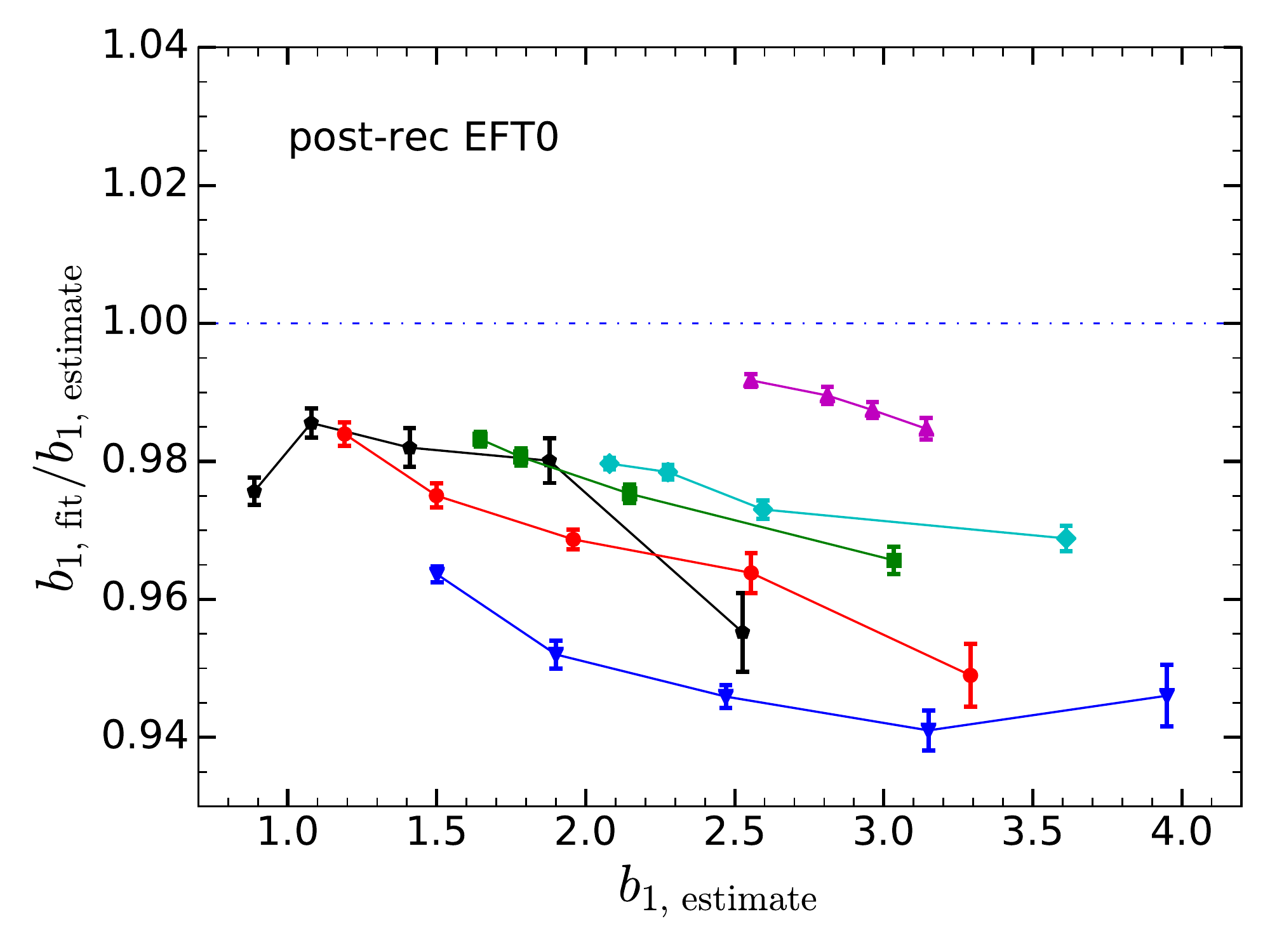}
\caption{Galaxy bias from the best fits using the EFT0 model in comparison to the estimated values from $P_{\rm gg}/P_{\rm gm}$ at large scales ($k \le 0.02\ihMpc$). We derive the fitted and estimated bias from each realization, and compare their mean over all realizations. \textit{Left panel:} pre-reconstruction. \textit{Right panel:} post-reconstruction. It shows that the galaxy bias estimates from two methods have discrepancy less than $3\%$ before reconstruction while the discrepancy tends to increase after reconstruction. After reconstruction, the fitted bias is systematically less than the estimated value from cross power spectrum.}\label{fig:Zmodel_b1}
\end{figure*}

In this section, we show supplementary figures to support the main text. Figure~\ref{fig:Zmodel_b1} shows the difference between the estimates of galaxy/halo bias from the galaxy-matter cross power spectrum and the estimates from the BAO scale fitting. The left panel shows the level of offset is within 3\% before reconstruction and the right panel shows that bias from the BAO fitting tends to be underestimated by as much as 6\%.

\begin{figure*}
\includegraphics[width=0.45\linewidth]{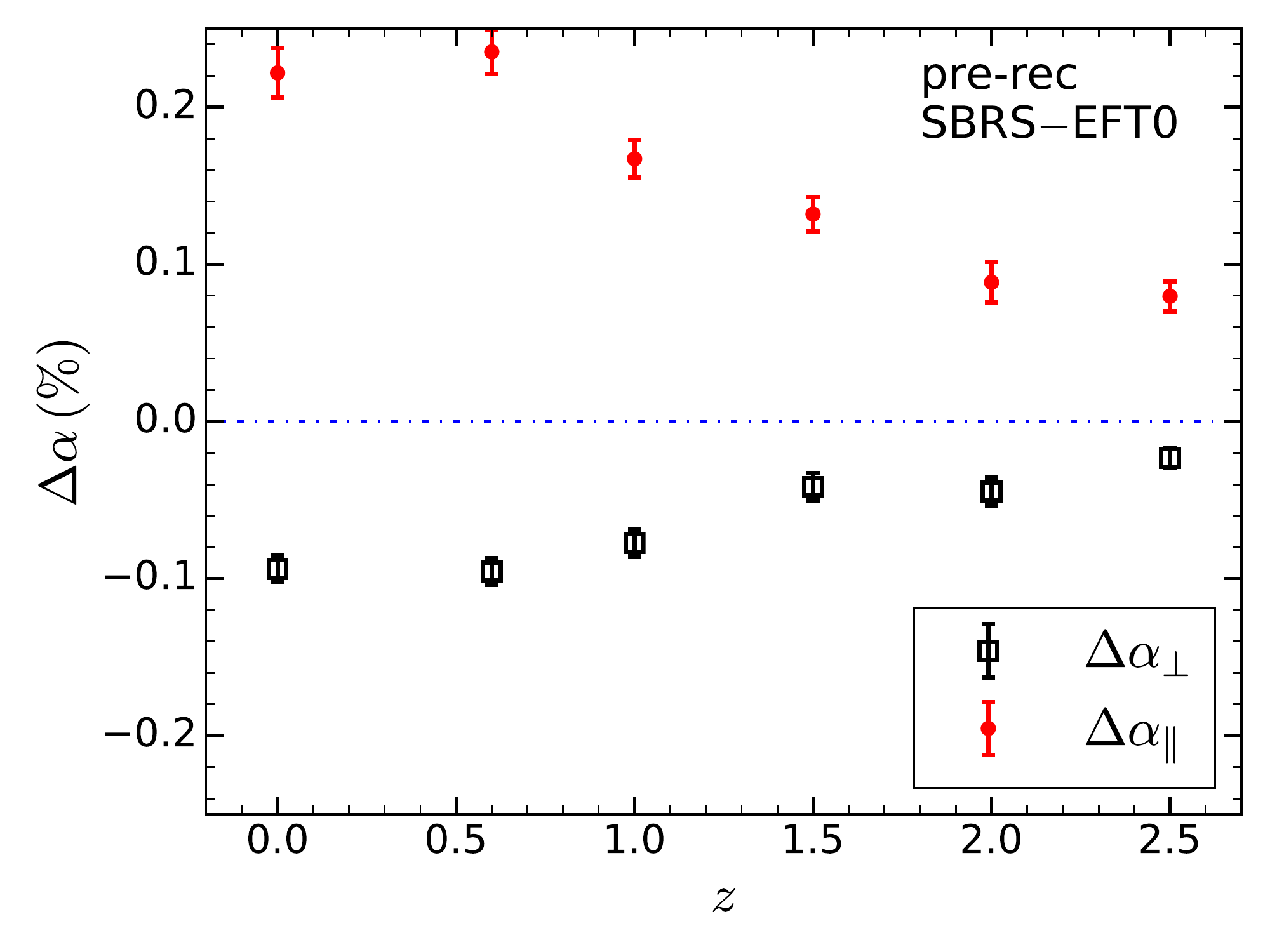}
\includegraphics[width=0.45\linewidth]{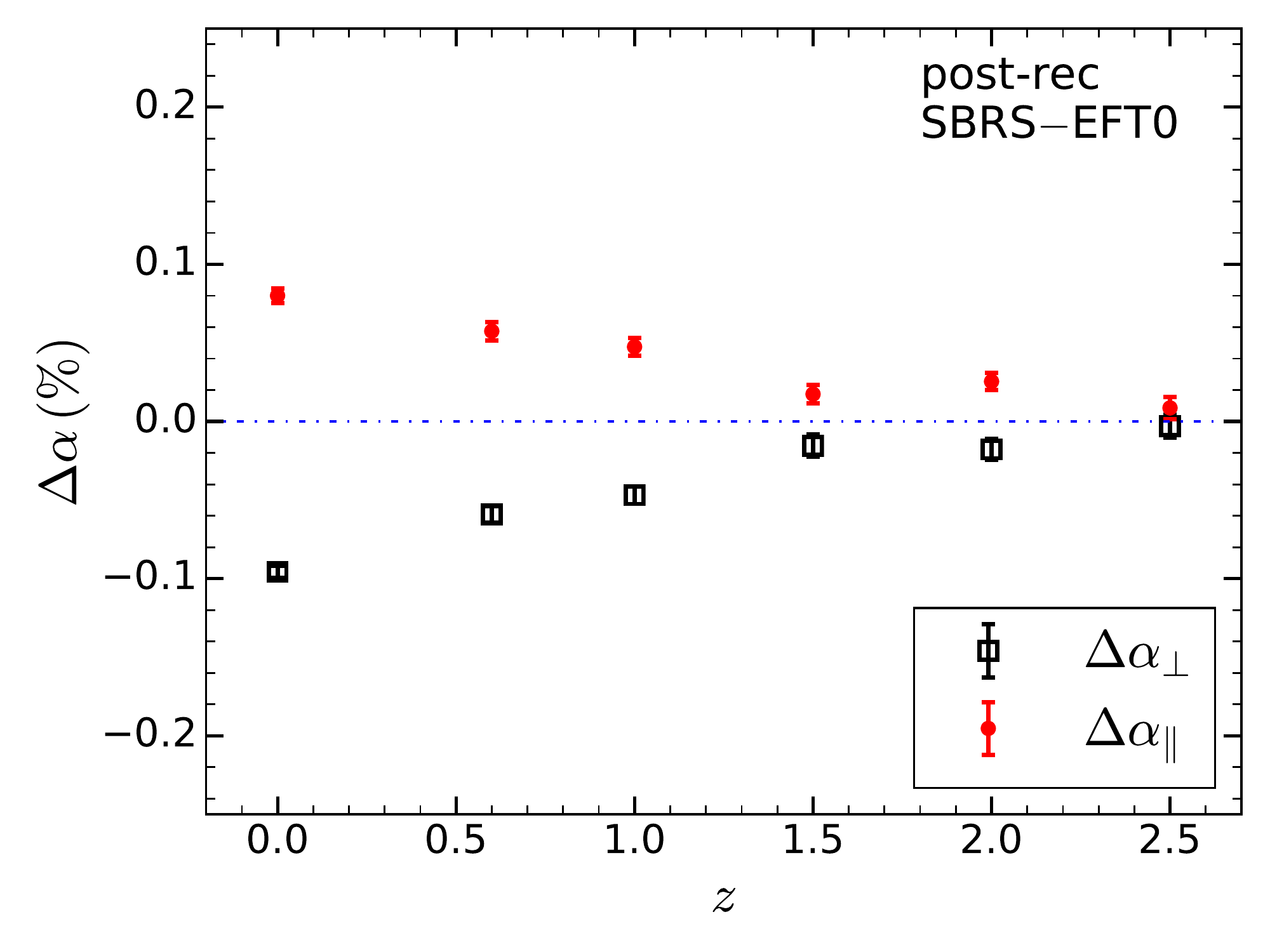}
\includegraphics[width=0.45\linewidth]{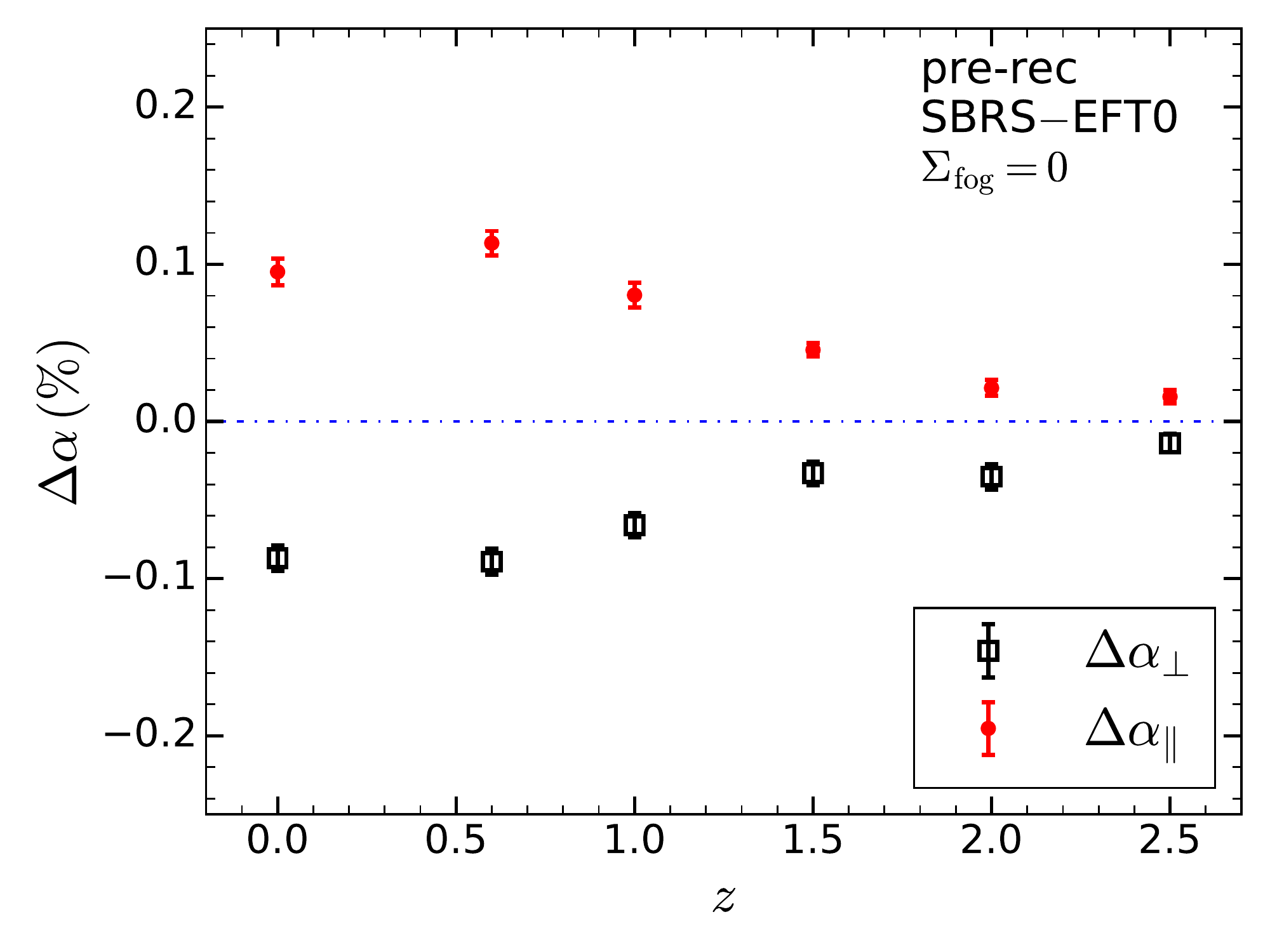}
\includegraphics[width=0.45\linewidth]{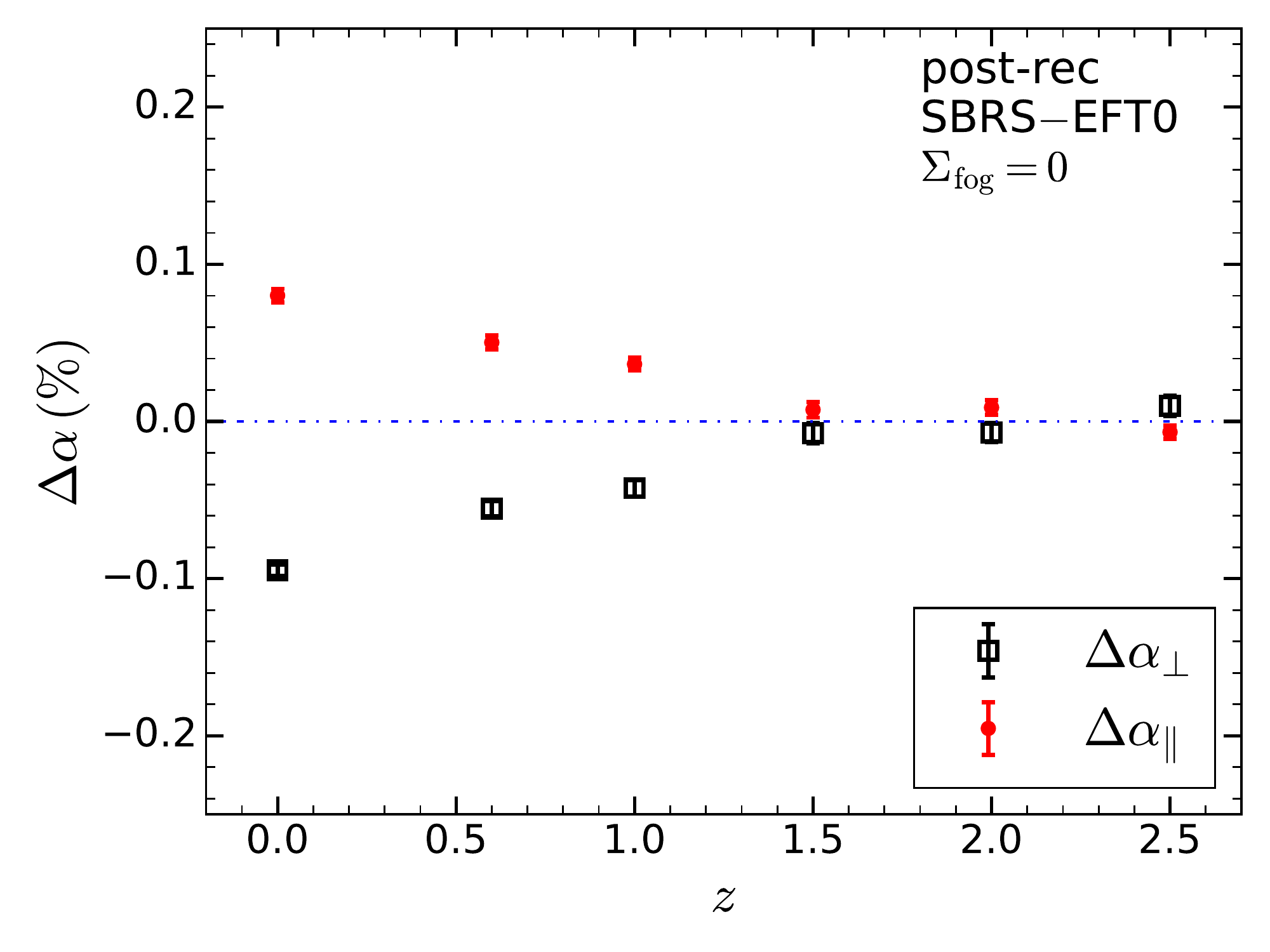}
\caption{The effects of different fitting models on BAO scale systematics for matter focusing on the SBRS model. Supplementary to Figure~\ref{fig:alpha_ZVS}. In the top panels, we show the SBRS model relative to the EFT0 model. In the bottom panels, we show the effect of fixing $\Sfog=0$  in the SBRS model. \textit{Left panel:} pre-reconstruction. \textit{Right panel:} post-reconstruction. The tendency of a pre-reconstruction anisotropic shift decreases when we fix $\Sfog=0$.}\label{fig:SBRSmodel}
\end{figure*}

Figure~\ref{fig:SBRSmodel} is supplementary to Figure~\ref{fig:alpha_ZVS} and shows the difference in the BAO scale shifts of matter when using the SBRS fitting (top panels) instead of the EFT0 model. The degree of offset is close to that of EFT1 model. The bottom panels show the offset we observe using a variant of the SBRS model by fixing $\Sfog=0$. In \S~\ref{subsec:BAOmatter}, we claim that we tend to find anisotropic BAO shifts in the pre-reconstruction cases when using a fitting model with any damping scale free parameter. In this figure, we test if the pre-reconstruction anisotropic BAO shift is reduced when we remove the $\Sfog$ parameter from the SBRS model. We indeed find that the shift for $\apar$ decreases by half at lower redshift.

\begin{figure*}
\centering
\includegraphics[width=0.4\linewidth]{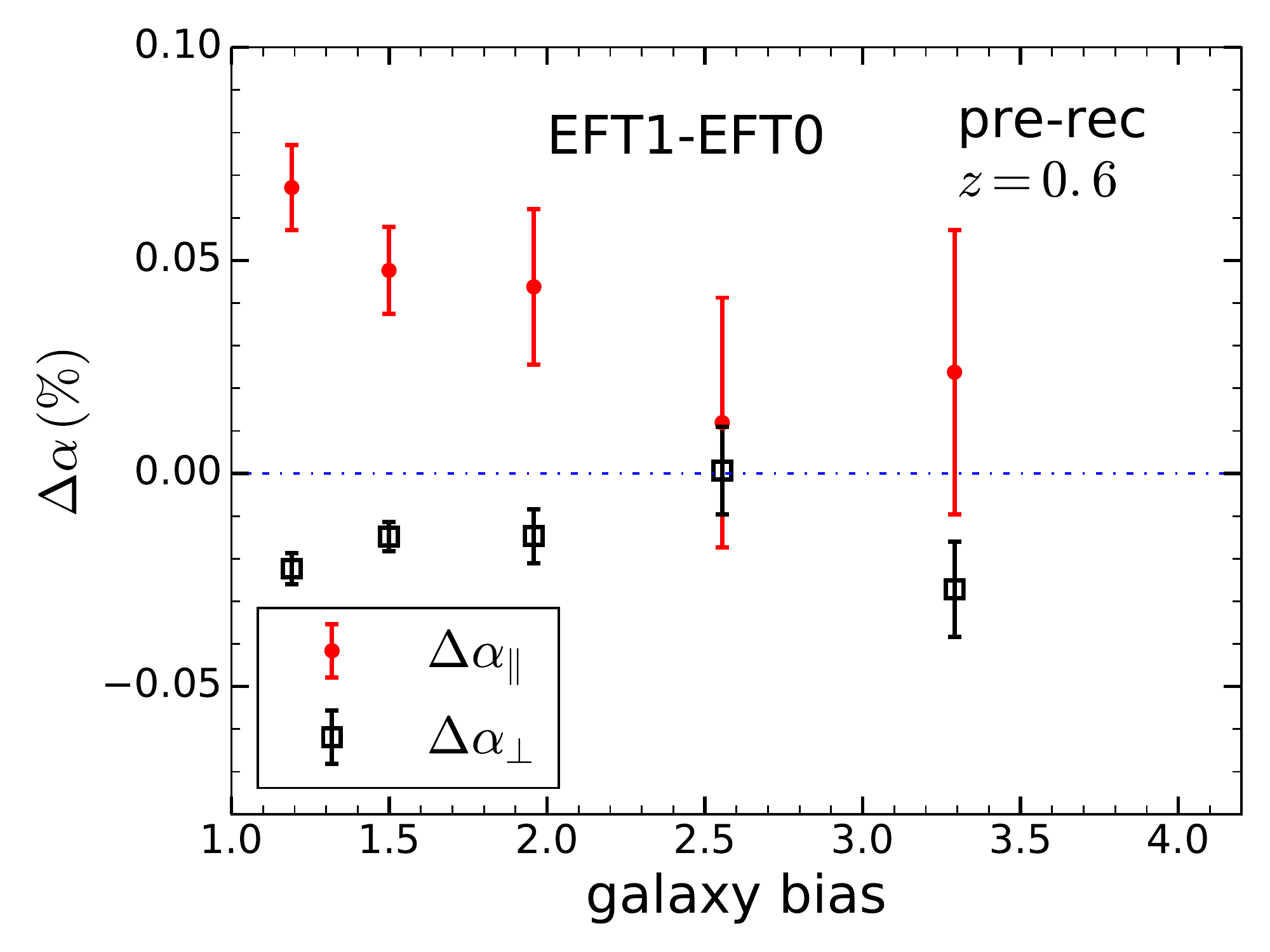}
\includegraphics[width=0.4\linewidth]{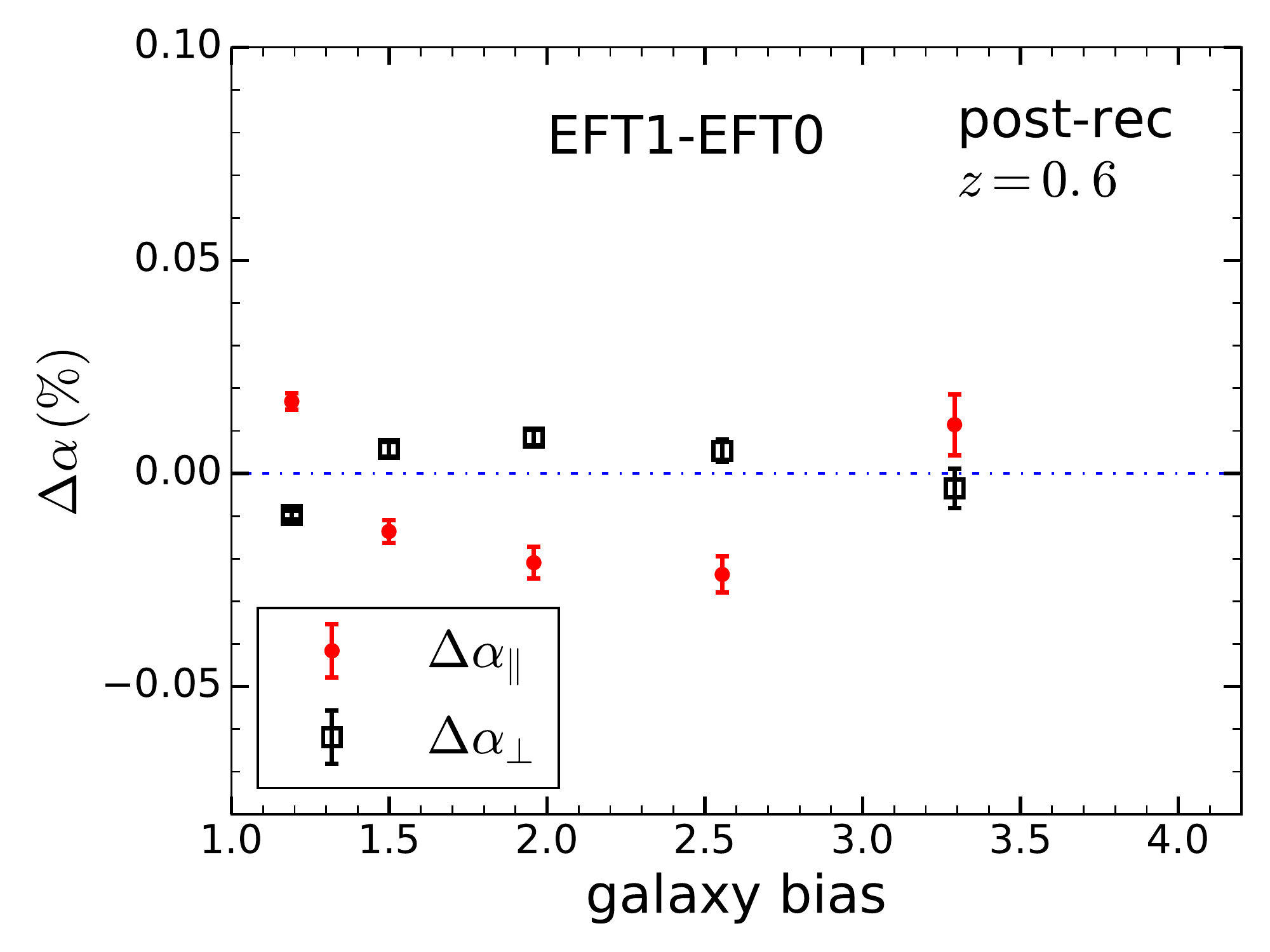}
\includegraphics[width=0.4\linewidth]{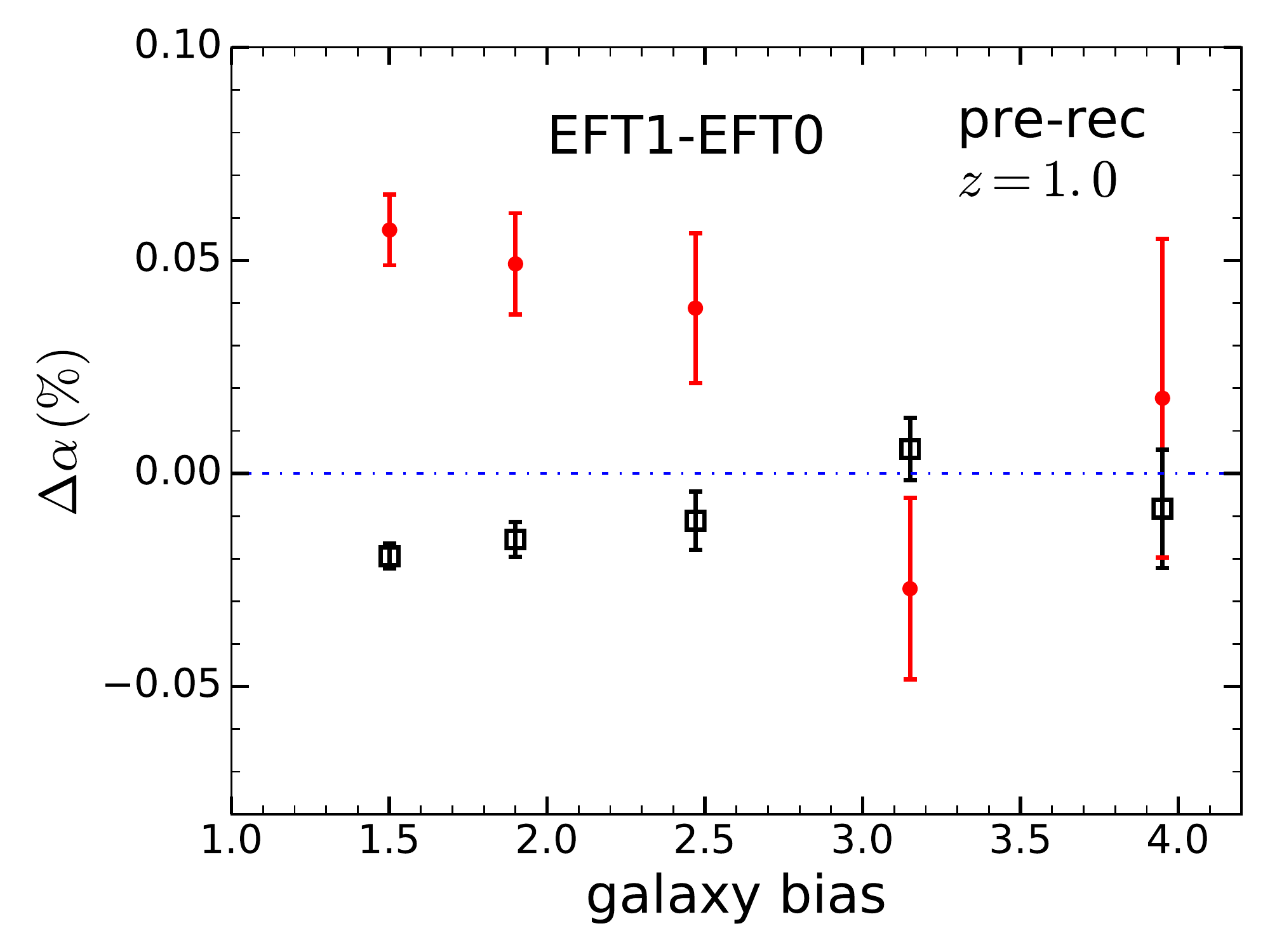}
\includegraphics[width=0.4\linewidth]{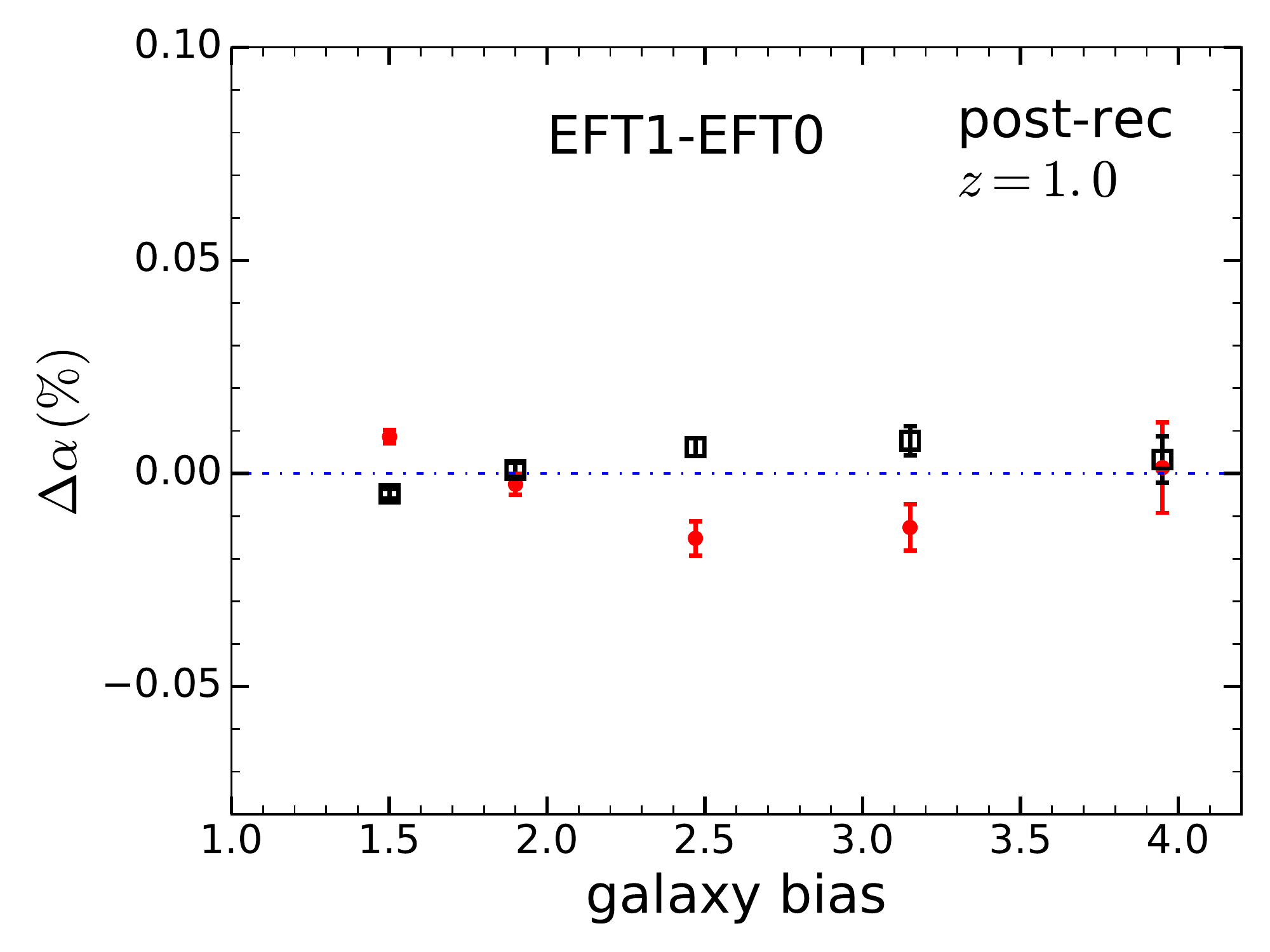}
\includegraphics[width=0.4\linewidth]{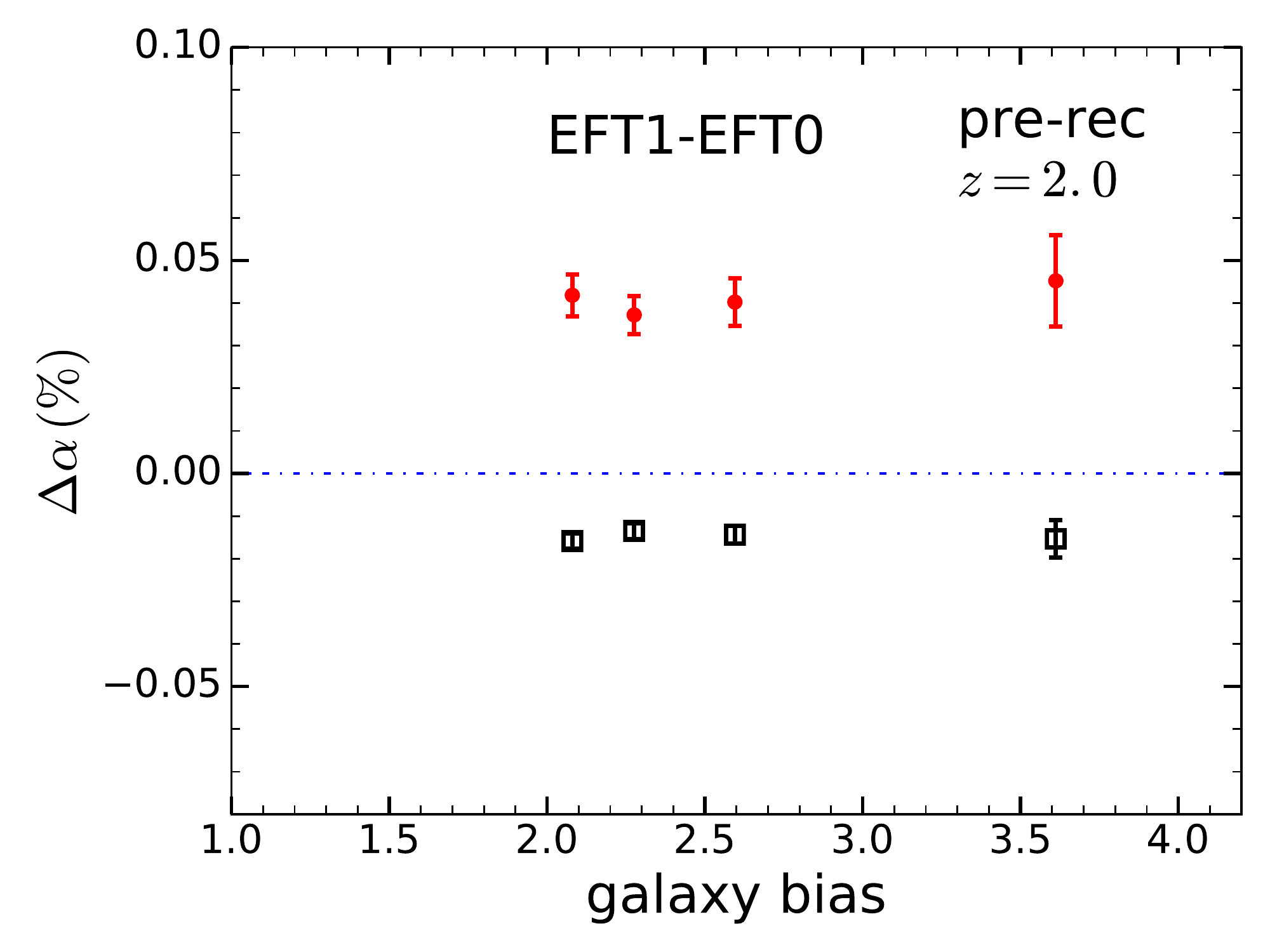}
\includegraphics[width=0.4\linewidth]{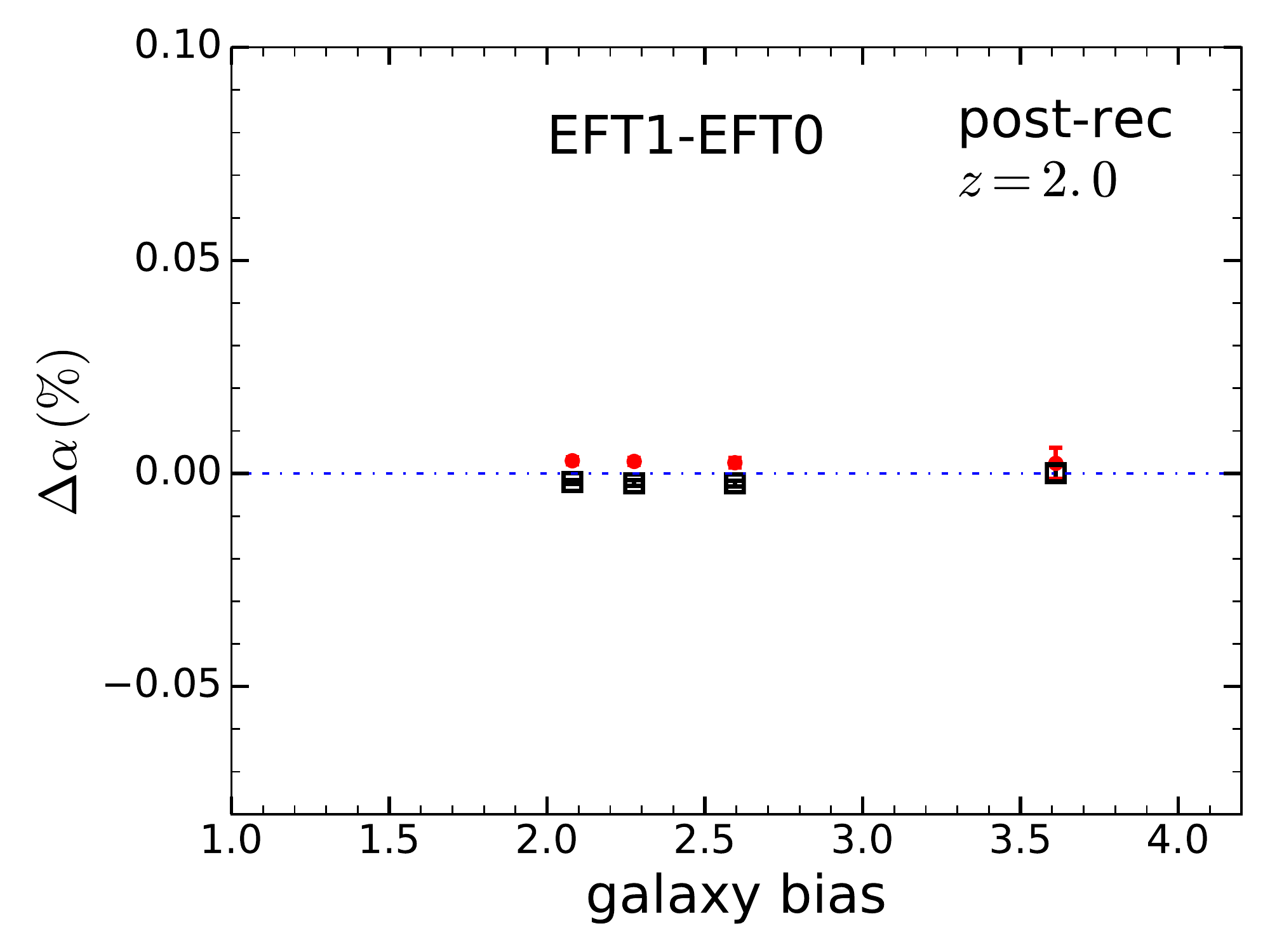}
\caption{The effects of different fitting models for samples with galaxy/halo bias. Supplementary to Figure~\ref{fig:delta_alpha_std}. \textit{Left panels:} pre-reconstruction. \textit{Right panel:} post-reconstruction. Empty square points denote for $\alpe$ and solid circular points for $\alpa$.
We compare the EFT0 model (i.e., with fixed $\Sigxy$) and the EFT1 model (with a free $\Sigxy$ (pre-reconstruction) or a free $\Sigsm$ (post-reconstruction) by taking $\alpha_{\rm EFT1}-\alpha_{\rm EFT0}$ at $z=0.6$ (top panels), $z=1.0$ (middle) and at $z=2.0$ (bottom). We show the mean and dispersion of $\alpha$ {\it differences}; the fitting models are applied to the same sets of simulations, mitigating most of the sample variance effect by taking differences.} \label{fig:delta_alpha_stdS} 
\end{figure*}

Figure~\ref{fig:delta_alpha_stdS} is supplementary to Figure~\ref{fig:delta_alpha_std} showing the difference in the BAO shifts between the EFT0 and the EFT1 model more clearly in the presence of galaxy/halo biases. 
